\documentclass[10pt]{article}
\usepackage[top=2.5cm,bottom=2.5cm,left=1.4cm,right=2.0cm]{geometry}
\usepackage{cite}
\usepackage[T1]{fontenc}
\usepackage[american]{babel}
\usepackage{layout}
\usepackage{amsmath}
\usepackage{amssymb,amstext, amsthm, simplewick}
\usepackage{framed}
\usepackage[svgnames,dvipsnames,x11names]{xcolor}
\usepackage{amsfonts}
\usepackage{color} 
\usepackage{braket}
\usepackage{multicol} 
\usepackage{epsfig}
\usepackage{cancel}
\usepackage{caption}
\usepackage{epsf}
\usepackage{feynmf} 
\usepackage{frontespizio}
\usepackage{rotating}
\usepackage{subfigure}
\usepackage{pstricks}
\usepackage{type1ec}
\usepackage{lettrine}
\usepackage{bbold}
\usepackage{calligra}
\usepackage{tikz}
\usepackage{subfigure}
\usepackage{mathrsfs}
\usepackage{curve2e}
\usepackage{setspace}
\usepackage{indentfirst}
\usepackage{emptypage}
\usepackage{relsize}
\usepackage{mathrsfs}
\usepackage{stackengine}
\usepackage{calc}
\usepackage{hyperref}
\hypersetup{
	colorlinks,
	citecolor=green,
	filecolor=black,
	linkcolor=blue,
	urlcolor=black
}
\numberwithin{equation}{subsection} 
\makeatletter
\@addtoreset{equation}{section}
\makeatother

\def\e{\varepsilon}

\def\de{\delta} 

\def\D{\mathcal{D}}

\def\rg{\sqrt{g}}

\def\lt{\left(}
\def\rt{\right)}

\def\dif{\text{d}}

\def\d4x{\dif^4 x}

\newcommand{\mO}{\mathcal{O}}
\newcommand*{\Scale}[2][4]{\scalebox{#1}{$#2$}}%
\newcommand{\sm}{\mathcal{S}}

\newcommand{\equal}{&\!\!=\!\! &}

\def\bnabla{\bar \nabla}

\newcommand{\hpg}[5]{{}_{#1}\mbox{\rm F}_{\!#2}\!
  \left(\left.\substack{\Scale[1]{ #3} \\[1.5ex] \Scale[1]{#4}}\right| #5 \right) }

\newcommand{\hpgo}[2]{{}_{#1}\mbox{\rm F}_{\!#2}}

\newcommand{\app}[4]{F_{\!#1}\!
  \left(\left.\substack{\Scale[1]{ #2} \\[1.5ex] \Scale[1]{#3}}\right| #4 \right) }
\newcommand{\beq}{\begin{equation}}
\newcommand{\eeq}{\end{equation}}
\let\a=\alpha   \let\b=\beta   \let\g=\gamma   \let\d=\delta
\let\e=\epsilon    \let\h=\eta     
    \let\k=\kappa  \let\l=\lambda  \let\m=\mu
\let\n=\nu      \let\x=\xi     \let\p=\pi      \let\r=\rho
\let\s=\sigma        
\let\c=\chi     \let\y=\psi    
\let\G=\Gamma  \let\D=\Delta   
     \let\P=\Pi

\newcommand{\ro}{\rho}
\newcommand{\nn}{\nonumber}

\newcommand{\be}{\begin{equation}}
\newcommand{\ee}{\end{equation}}
\newcommand{\beqa}{\begin{eqnarray}}
\newcommand{\eeqa}{\end{eqnarray}}
\newcommand{\sdfrac}[2]{\mbox{\small$\displaystyle\frac{#1}{#2}$}}

\newcommand{\pd}{\partial}
\renewcommand{\Re}{\textrm{Re}}
\renewcommand{\Im}{\textrm{Im}}
\newcommand{\figref}[1]{Fig.~\ref{#1}}			% for figures
\newcommand{\secref}[1]{Section~\ref{#1}}		% for sections
\newcommand{\appref}[1]{Appendix~\ref{#1}}		% for appendix references
\def\nbox#1#2{\vcenter{\hrule \hbox{\vrule height#2in
			\kern#1in \vrule} \hrule}}
\def\sq{\,\raise.5pt\hbox{$\nbox{.09}{.09}$}\,}
\def\sqb{\,\raise.5pt\hbox{$\overline{\nbox{.09}{.09}}$}\,}
\def\Box{\sq}
\usepackage{accents}

\begin{document}
\begin{center}
\vspace{1.5cm}
{\Large \bf Conformal field theory in momentum space and anomaly actions in gravity:}
{\Large \bf The analysis of three- and four-point functions\\}
\vspace{0.3cm}
\vspace{1cm}
{\bf $^{(1)}$Claudio Corian\`o and $^{(1,2)}$Matteo Maria Maglio\\}
\vspace{1cm}
{\it  $^{(1)}$ Dipartimento di Matematica e Fisica,  Universit\`{a} del Salento, \\
	and INFN Sezione di Lecce , Via Arnesano 73100 Lecce, Italy\\}
\vspace{0.5cm}
{\it  $^{(2)}$ Galileo Galilei Institute for Theoretical Physics, \\
	Largo Enrico Fermi 2, I-50125 Firenze, Italy\\}
\end{center}
%%%%%%%%%%%%%%%%%%%%%%%%
\begin{abstract}
After a brief outline of general aspects of conformal field theories in coordinate space, in a first part we 
review the solution of the conformal constraints of three- and four-point functions in momentum space in dimensions $d\geq 2$, in the form of conformal Ward identities (CWI's). We center our discussion on the analysis of correlators containing stress-energy tensors $(T)$, conserved currents $(J)$, and scalar operators $(O)$. For scalar four-point functions, we briefly discuss our method for determining the dual conformal solutions of such equations, identified only by the CWI's, and related to the conformal Yangian symmetry, introduced by us in previous work. 
In correlation functions with $T$ tensors, evaluated around a flat spacetime, the conformal anomaly is characterized by the (non-local) exchange of massless poles in specific form factors, a signature that has been investigated both in free field theory and non-perturbatively, by solving the conformal constraints. We discuss the anomaly effective action, and illustrate the derivation of the CWI's directly from its path integral definition and its Weyl symmetry, which is alternative to the standard operatorial approach used in conformal field theories in flat space. For two- and three-point functions, we elaborate on the matching of these types of correlators to free-field theories. Perturbative realizations of CFTs at one-loop provide the simplest expressions of the general solutions identified by the CWI's, for generic operators $T$, $J$, and scalars of specific scaling dimensions, by an appropriate choice of their field content. 
In a technical appendix we offer details on the reconstruction of the $TTO$ and $TTT$ correlators in the approach of  Bzowski, McFadden and Skenderis, and specifically on the secondary Ward identities of the method, in order to establish a complete match with the perturbative description. 
\end{abstract}
\vspace{2cm}
\centerline{\em Physics Reports, in press }

\newpage
\tableofcontents
\newpage
%%%%%%%%%%%%%%%%%%%%%%%%%%%%%%%
\section{Introduction}
%%%%%%%%%%%%%%%%%%%%%%%%%%%%%%%
The study of conformal field theories (CFTs) \cite{Duff:1977ay,Duff:1993wm} has played a central role in theoretical/mathematical physics for over half a century \cite{Kastrup:2008jn,DiFrancesco:1997nk}, with an impact on several different areas, from the theory of critical phenomena to string theory and, more recently, the AdS/CFT (AdS, Anti de Sitter)  correspondence. The latter allows to establish a link between gravitational and gauge forces in a specific supersymmetric setup. In particular, such correspondence \cite{Maldacena:2003nj}, also known as gauge/gravity duality, has shown the importance of conformal symmetry in dimensions $d \geq 2$.
In $d=2$ the symmetry is infinite dimensional, and finite otherwise, being defined by the generators of $SO(d,2)$.  

The relevance of the study of  conformal correlation functions in $d>2$ 
is remarkable in areas as different as cosmology, condensed matter theory and particle phenomenology. For instance, recent experimental advances on Dirac and Weyl semi-metals, have spurred a growing activity on the role played by CFT's and their anomalies (chiral, conformal) in the characterization of the fundamental properties of such materials \cite{Chernodub:2013kya,Ambrus:2019khr,Chernodub:2019tsx,Arjona:2019lxz,Gooth:2017mbd}.  

In our analysis we will consider the case of $d>2$, which is of outmost interest in physics. 
For quantum conformally invariant theories, the general idea is to develop a formalism that allows characterizing the structure of the corresponding correlation functions without resorting to a Lagrangian description. In this more general case, a CFT is determined by a set of operators (primary fields) and their descendants, which close an algebra via an operator product expansion. In principle, this allows to bootstrap correlators of higher points from the lower ones by solving for the corresponding conformal blocks.   

Although in $d>2$ the symmetry is less restrictive, correlation functions of CFT's up to three-point  functions can be determined in their tensorial structures and form factors only modulo few constants. These constraints take the form of conformal Ward identities (CWI's). 

We will show how it is possible to match general CFT results to ordinary Lagrangian formulations, in free field theory, limited to three-point functions. The possibility of such direct match in momentum space, reproduces previous results of coordinate space \cite{Osborn:1993cr} but proceeds in an autonomous way. This approach will allow to investigate this class of theories in a framework which is quite close to the study of the scattering amplitudes in ordinary perturbation theory. At the same time, as we are going to elaborate, it allows a completely new perspective on the role played by the conformal anomaly in the dynamical breaking of such symmetry. 

\subsection{The transition to momentum space and the BMS method}
The solution of the CWI's in position (coordinate, configuration) space has been addressed long ago for three-point  functions \cite{Erdmenger:1996yc,Osborn:1993cr}. Most notably, Osborn and Petkou outlined a method to solve such identities in position space, indicating also the way in which the conformal anomaly could be included in a special and important class of correlators. Their analysis covered correlators containing up to three insertions of stress energy tensors, beside conserved currents and scalar operators all of generic scaling dimensions $(\Delta_i)$. 

The idea of using conformal Ward identities (CWI's) to determine the structure of three-point  functions in momentum space was presented independently in \cite{Coriano:2013jba} and \cite{Bzowski:2013sza}, the second one outlining a method that includes also the tensor case. 
In \cite{Bzowski:2013sza}, Bzowski, McFadden and Skenderis (BMS)  have indicated 
a possible route to identify the solution of the conformal Ward identities (CWI's) of three-point  functions of tensor correlators, the most demanding one being the $TTT$, with three stress energy tensors $(T)$. The method builds the solution of the CWI's starting from the transverse-traceless components of such correlators and their related two-point  functions, such as the $TT$.
We will illustrate the method by working out the $TTO$ and $TTT$cases in a rather detailed way, clarifying all the intermediate technical steps.  We will discuss all the simplifications that a matched perturbative analysis, performed by us in \cite{Coriano:2018bsy,Coriano:2018bbe,Coriano:2018zdo}, provides in the computation of the explicit expression of this and of other similar correlators, that remains valid non-perturbatively. Obviously, this occurs for operators containing insertions of stress energy tensors, conserved currents and scalars of specific scalar dimensions, whose canonical dimensions are simply related to the spacetime dimensions. In this case the exact match is indeed possible, by using general combinations of scalars, fermions and gauge fields running in the loops of the free field theory realization. 

Our interest in this analysis has grown out of previous studies in perturbative field theory, QCD and QED, of similar vertices ($TJJ$) \cite{Giannotti:2008cv,Armillis:2009pq,Armillis:2009im,Armillis:2010qk} where it has been shown that the breaking of conformal and chiral symmetries are associated, for any correlator exhibiting either a chiral or a conformal anomaly, to the specific behavior of one specific form factor extracted from the general parameterization of a tensorial three-point function, that manifests an anomaly pole. It is therefore of interest to investigate if this phenomenon remains valid beyond perturbation theory. In a final section we will show how a similar result holds also for the $TTT$ correlator, although this behavior has been shown to hold, more recently, for the  $4T$ correlator, following a proof that can be directly extended to the $nT$ case \cite{Coriano:2021nvn}.

There are two main reasons why CFT in momentum  space is essential for our understanding of the role of conformal symmetry in field theory. The first  is that it allows to establish a direct connection with ordinary scattering amplitudes, in the analysis of which there has been significant progress up to very high perturbative orders. 
The second is the possibility of investigating the role of the conformal anomalies.  

Indeed, the operatorial expansion of two operators in coordinate space, in any CFT, in general does not address the issue of possible anomalies, since the operators are taken at different spacetime points. In this case the CWI's have to be modified with the addition of ultralocal terms obtained by differentiating the anomaly functional. In other words, the corresponding anomalous CWI's are solved separately in their homogeneous and inhomogeneous forms. At first one solves the equations when the coordinate points of the correlators are kept separate, and then the contribution to their anomaly is added by hand as an ultralocal term, in the coincidence limit.

However, there is no physical understanding of how the breaking of a conformal symmetry by a quantum anomaly occurs, if we stay in coordinate space. In this case the exercise is purely formal, for a phenomenon - the breaking of an important and possibly fundamental symmetry - which is crucial in so many areas of physics.

\subsection{Perturbative breaking in anomaly form factors}
As just mentioned, perturbative analysis in momentum space \cite{Giannotti:2008cv,Armillis:2009pq,Armillis:2009im}, also in supersymmetric theories \cite{Coriano:2014gja}, have shown that the signature of such a breaking is in the appearance of specific massless poles in correlators with a $T$ insertion. The structure that emerges from such virtual exchanges can be directly compared to the free field theory prediction, if we are able to show that the explicit expression of the correlation functions is identical in the two cases. \\
Such a matching has recently been discussed by us \cite{Coriano:2018bsy,Coriano:2018zdo,Coriano:2018bbe}. We have shown how this test can be performed  on a complex correlator such as the $TTT$. For the perturbative matching, in this case one needs two sectors, a scalar and a fermion sector, linearly combined with arbitrary multiplicities of their particle content.  A third sector, with a spin-$1$ running inside the loops, is necessary in order to account for its anomalies. \\  
This mapping of a general, non-perturbative result, to a perturbative one (a simple one-loop matching) allows to proceed with a drastic simplification of the expressions of such correlators, providing the simplest realization of the form factors identified by the BMS decomposition. At this stage, one can identify more clearly the structure of the anomaly contributions 
in the several form factors present in a three-point  function.  \\
In the matched perturbative description, the conformal anomaly emerges from the renormalization of the longitudinal (or semi-local) terms of the correlator, which exhibits a specific pole structure. 
The distinction between such massless interactions, clearly related to renormalization, and other possible massless exchanges, is that they are directly associated to derivatives of the anomaly functional. \\
The pattern, in this case, generalizes what one obtains in the $TJJ$ case. 
We have shown in great detail in \cite{Coriano:2018zdo} that, in this correlator, the generation of  a massless pole is a consequence of the process of regularization of one specific form factor - the anomaly form factor - as the spacetime dimensions $d$ tend to 4 $(d\to 4)$.    \\
It is then natural, from this perspective, to investigate whether a specific conformal anomaly action can account for these (non-local) anomaly contributions which are predicted both in free field theory and non-perturbatively. For this reason we have turned to a description of the non-local Riegert action, which is expected to generate such terms in a direct manner. This aspect will be addressed in the second part of this review. \\
Obviously, studies of 1-particle irreducible effective actions in gravitational backgrounds are common in the literature (see for instance \cite{Bastianelli:2004zp,Bastianelli:2007jv,Bastianelli:2012bz,Coriano:2012wp}), for instance in a $E/m$ expansion, in the presence of massive intermediate states.
More recent investigations of conformal correlators both for chiral and conformal anomalies are those of \cite{Bastianelli:2019zrq} \cite{Bonora:2014qla,Bonora:2017gzz}.  \\
For our goals, we will review the origin of the non-local Riegert action in a pedagogical way, detailing the variational solution of the anomaly constraint by an integration in field space, which takes to such expression.  
\subsection{Massless exchanges as dynamical breaking of the conformal symmetry}
Compared to local actions which introduce extra degrees of freedom - in the form of a dilaton or an axion - unified in a St\"uckelberg supermultiplet in the superconformal case \cite{cc} - the non-local description appears to be dynamical. The pole is the result of a collinear (particle/antiparticle) exchange in the loop, captured by a local spectral density $(\sim\delta(s))$ in the dispersion variable ($s$).
 
The emergence of composite intermediate massless states, characterizing the light-cone dynamics of the field operators in the theory, is also a prediction of the non-local action, which is correctly reproduced by the matched perturbative theory. 
The equivalence between the perturbative and the non-perturbative realizations of specific correlation functions, for operators containing insertions of only $T$'s and $J$'s, and a careful analysis of the renormalization in both approaches, shows that the appearance of massless exchanges in the two cases is not a spurious prediction of perturbation theory, but the signature of the anomaly. 

Its relevance both in the context of condensed matter theory \cite{Chernodub:2019tsx,Arjona:2019lxz,Mottola:2019nui,Chernodub:2017jcp,Rinkel:2019kpo}, in the theory of Dirac and Weyl semimetals and in gravitational waves \cite{Mottola:2016mpl} is what makes it worthy of a close attention.

\subsection{Local duality and dimensional transmutation}
The relevance of both local and non-local anomaly actions has been addressed repeatedly in the past, with results and predictions touching a wide range of phenomena, 
that are gradually being uncovered \cite{Chernodub:2019tsx,Mottola:2019nui,Mottola:2016mpl}. In \cite{Coriano:2019dyc} we have proposed that these actions parametrize the same anomaly phenomenon at two different ends (the UV and the IR) of a renormalization group flow, with the possibility that a non-perturbative dynamics will connect the two. Massless excitations in the UV, captured by the anomaly diagrams, may turn into asymptotic degrees of freedom in the IR in the presence of non-perturbative interactions. Coherent interactions may be the cause of such behavior in condensed matter theory, with the generation of a physical scale. The phenomenon of ``a  massless pole turning into a cut'', with the cut associated to a mass scale, noticed in behavior the spectral densities of anomaly form factors in the past \cite{Giannotti:2008cv,Coriano:2014gja}, is a generic phenomenon of anomalies, likely related to the emergence of a physical scale in a massless theory. \\
 Anomalies come with specific super-convergent sum rules, with remarkable properties of their associated spectral densities. This phenomenon shares a similarity with the property local duality (quark/hadron duality) in QCD, due to the presence, also in this case, of a sum rule and of a dispersion relation connecting the UV and the IR behavior of certain cross sections, once these are integrated over the energy.\\
In the case of a conformal theory, this is essentially linked to dimensional transmutation (DT), 
where a scale invariant theory develops a dynamical scale. \\
In QCD, DT is a non-perturbative effect, related to the breaking of classical scale invariance of the theory and the emergence of $\Lambda_{QCD}$, which is identified, in perturbation theory, from the singularity in momentum space of the running coupling. Obviously, the perturbative prediction has little to say about confinement or chiral symmetry breaking in the theory, which characterize this phenomenon in a far more complex way compared to a simple perturbative analysis.
In the case of quantum scale invariant theories, broken by anomalies, this phenomenon could be different, given the fact that the symmetry holds at quantum level, and could even be simpler.     

\subsection{Organization of this review}
Our review is organized as follows. \\
We will start by introducing some general aspects of CFT in coordinate space, investigating two-point  functions, and moving afterwards to three- and four-point functions. Along the way, we  will also include a brief description of the embedding space formalism. These preliminary sections will provide sufficient information in order to proceed, afterwards, with the construction of the tensorial two- and three-point functions in coordinate space. In particular, the embedding formalism tuns out to be very appropriate in order to simplify this process. This is the content of \secref{chapter1}, which can be skipped by experts.\\
Afterwards, we will begin our discussion of CFT in momentum space, deriving the structure of the conformal and canonical Ward identities first in coordinate and then in momentum space. We will introduce the anomaly effective action, defined by a path integral integration over a conformal sector, illustrating the derivation of the CWI's directly from this formalism.  
The strategy used to solve the CWI's in the method of BMS is exemplified with enough detail presented in an appendix, in order to allow the reader to follow most of the intermediate steps of this method.

%%%%%%%%%%%%%%%%%%%%%%%%%%%%%%%%%%%%%%%%%%%%%%%%%%
%%%%%%%%%%%%%%%%%%%%%%%%%%%%%%%%%
\section{Conformal symmetry in coordinate space: some key aspects}\label{chapter1}
%%%%%%%%%%%%%%%%%%%%%%%%%%%%%%%%%
\subsection{The conformal group}
\label{prima}
We present a brief review of the transformations which identify the conformal group $SO(2,d)$ in $d>2$ dimensions and in a Euclidean space \cite{ DiFrancesco:1997nk,Fradkin:1996is}. \\
Conformal transformations may be defined as those transformations $x_\mu\to x'_\mu(x)$ that preserve the infinitesimal length up to a local factor $(\mu=1,2,\ldots d)$
\begin{equation}
dx_\mu dx^\mu\to dx'_\mu dx'^\mu=\Omega(x)^{-2}dx_\mu dx^\mu.\label{lineint}
\end{equation}
In infinitesimal form they are given by 
\begin{equation}
x'_\mu(x)=x_\mu+a_\mu+\omega_{\mu\nu}x^\nu+\sigma x_\mu+b_\mu x^2-2b\cdot x\,x_\mu, \label{transf}
\end{equation}
where
\begin{equation}
\Omega(x)=1-\l(x),\quad \l(x)=\s-2b\cdot x,\label{Om}
\end{equation}
and $b_\mu$ is a constant $d$-vector. The transformation in \eqref{transf} is composed of the parameters $a_\mu$ for the translations, $\omega_{\mu\nu}=-\omega_{\nu\mu}$ for boosts and rotations, $\sigma$ for the dilatations and $b_\mu$ for the special conformal transformations. The first three define the Poincar\'e subgroup, obtained for $\Omega(x)=1$, which leaves invariant the infinitesimal length. By including the inversion
\begin{equation}
x_\mu\to x'_\mu=\sdfrac{x_\mu}{x^2},\qquad \Omega(x)=x^2,
\end{equation}
we can enlarge the conformal group to $O(2,d)$. Special conformal transformations can be realized by considering a translation preceded and followed by an inversion. \\
 Notice that an infinitesimal transformation
\be
x^\mu(x)\to x'^\mu(x)=x^\mu + v^\mu(x)
\ee
is classified as an isometry if it leaves the metric $g_{\mu \nu}(x)$ invariant in form. If we denote with 
$g'_{\mu\nu}(x')$ the new metric in the coordinate system $x'$, then an isometry is such that
\be
g^\prime_{\mu\nu}(x')=g_{\mu\nu}(x').
\label{met1}
\ee
This condition can be inserted into the ordinary covariant transformation rule for $g_{\mu\nu}(x)$ to give 
\be
g^\prime_{\mu\nu}(x')=\frac{\partial x^\rho  }{\partial x'^\mu }\frac{\partial x^\sigma}{\partial x'^\nu} g_{\rho\sigma}(x)= g_{\mu\nu}(x'),
\ee
from which one derives the Killing equation for the metric
\be
v^\alpha\partial_\alpha g_{\mu\nu} + g_{\mu\sigma} \partial_\nu v^\sigma + 
g_{\sigma \nu} \partial_\mu v^\sigma =0.
 \ee
For a conformal transformation, according to \eqref{lineint}, the metric condition (\ref{met1}) is replaced by the condition 
\be
g'_{\mu\nu}(x')=\Omega^{-2} g_{\mu\nu}(x'),
\ee
generating the conformal Killing equation (with $\Omega(x)= 1-\sigma(x)$)
\be
v^\alpha\partial_\alpha g_{\mu\nu} + g_{\mu\sigma} \partial_\nu v^\sigma + 
g_{\sigma \nu} \partial_\mu v^\sigma=2 \sigma g_{\mu\nu}.
\ee
In the flat spacetime limit this becomes 
\be
\label{sigma}
 \partial_\mu v_\nu + 
\partial_\nu v_\mu=2 \sigma\, \eta_{\mu\nu},\qquad  \sigma=\frac{1}{d} \partial \cdot v.
\ee
From now on we switch to the Euclidean case, neglecting the index positions. Using the fact that every conformal transformation can be written as a local rotation matrix of the form 
\be
\label{rot1}
R^\mu_\alpha=\Omega \frac{\partial x'^\mu}{\partial x^\alpha},
\ee
we can first expand generically $R$ around the identity as 
\be
R=\mathbf{ 1  } + \left[\mathbf{\epsilon}\right] +\ldots,
\ee
with an antisymmetric matrix $\left[\epsilon\right]$, which we can re-expressed in terms of antisymmetric parameters 
($\tau_{\rho\sigma}$) and $1/2 \,d\, (d-1)$ generators $\Sigma_{\rho\sigma}$ of $SO(d)$ as 
\begin{eqnarray}
\left[\epsilon\right]_{\mu\alpha}&=&\frac{1}{2} \tau_{\rho\sigma}\left(\Sigma_{\rho\sigma}\right)_{\mu\alpha},\nonumber \\
\left(\Sigma_{\rho\sigma}\right)_{\mu\alpha}&=&\delta_{\rho\mu}\delta_{\sigma\alpha}-\delta_{\rho\alpha}\delta_{\sigma\mu},
\end{eqnarray}
from which, using also (\ref{rot1}) we derive a constraint between the parameters of the conformal transformation $(v)$ and the parameters $\tau_{\mu\alpha}$ of $R$
\be
R_{\mu\alpha}= \delta_{\mu\alpha} + \tau_{\mu\alpha}=\delta_{\mu\alpha} + \frac{1}{2}\partial_{\,[\alpha }v_{\mu]},
\ee
with $ \partial_{[\alpha }v_{\mu]}\equiv
\partial_{\alpha }v_{\mu}-\partial_{\mu }v_{\alpha}$.\\
As an example, let's consider the action of this symmetry on a vector field.
Denoting with $\Delta_A$ the scaling dimensions of a vector field $A_\mu(x)'$,  its variation under a conformal transformation can be expressed via $R$ in the form 
\begin{eqnarray}
\label{trans1}
A'^\mu(x')=\Omega^{\Delta_A} R_{\mu \alpha} A^\alpha(x)=(1-\sigma+\ldots)^{\Delta_A}\left(\delta_{\mu\alpha}+\frac{1}{2}\partial_{[\alpha }v_{\mu]} +\ldots \right)A^\alpha(x),
\end{eqnarray}
from which one can easily deduce that 
\be
\label{trans2}
\delta A^\mu(x)\equiv A'^\mu(x)-A^\mu(x)=-(v\cdot \partial +\Delta_A \sigma)A^\mu(x) +\frac{1}{2} \partial^{[\alpha }v^{\mu]}A_\alpha(x), 
\ee
which is defined to be the Lie derivative of $A^\mu$ in the $v$ direction
\be
L_v A^\mu(x) \equiv -\delta A^\mu(x).
\ee
As an example, in the case of a generic rank-2 tensor field ($\phi^{I \, K}$) of scaling dimension $\Delta_\phi$, transforming according to a representation $D^I_J(R)$ of the rotation group $SO(d)$, (\ref{trans1}) takes the form 
\be
\phi'^{I\, K}(x')=\Omega^{\Delta_{\phi}} D^I_{I'}(R) D^K_{K'}(R) \phi^{I' \,K'}(x).
\ee
In the case of the stress energy tensor ($D(R)=R$), with scaling (mass) dimension $\Delta_T$  $(\Delta_T=d)$ the analogue of (\ref{trans1}) is 
\begin{eqnarray}
T'^{\mu\nu}(x')&=&\Omega^{\Delta_T} R^\mu_\alpha R^\nu_\beta T^{\alpha\beta}(x)\nonumber \\
 &=&(1- \Delta_T \sigma +\ldots)(\delta_{\mu\alpha}+\frac{1}{2}\partial_{[\alpha }v_{\mu]}+\ldots)
(\delta_{\mu\alpha}+\frac{1}{2}\partial_{[\alpha }v_{\mu]}+\ldots)\,T^{\a\b}(x),
\end{eqnarray}
where $\partial_{[\alpha }v_{\mu]}\equiv\partial_\a v_\mu -\partial_\mu v_\a$. One gets 
\be
\delta T^{\mu\nu}(x)=-\Delta_T\, \sigma\, T^{\mu\nu} -v\cdot \partial \,T^{\mu\nu}(x) +
\frac{1}{2}\partial_{[\alpha }v_{\mu]}\,T^{\alpha\nu} +\frac{1}{2}\partial_{[\nu }v_{\alpha]}T^{\mu\alpha}.
\ee
For a special conformal transformation (SCT) one chooses 
\be
v_{\mu}(x)=b_\mu x^2 -2 x_\mu b\cdot x,
\ee
with a generic parameter $b_\mu$ and  $\sigma=-2 b\cdot x$ (from \ref{sigma}) to obtain
\be
\delta T^{\mu\nu}(x)=-(b^\alpha x^2 -2 x^\alpha b\cdot x )\, \partial_\alpha  T^{\mu\nu}(x)   - \Delta_T \sigma T^{\mu\nu}(x)+
2(b_\mu x_\alpha- b_\alpha x_\mu)T^{\alpha\nu} + 2 (b_\nu x_\alpha -b_\alpha x_\nu)\, T^{\mu\alpha}(x).
\ee
It is sufficient to differentiate this expression with respect to $b_\kappa$ in order to derive the form of the special conformal transformation $K^\kappa$ on $T$ in its finite form 
\begin{eqnarray}
\mathcal{K}^\kappa T^{\mu\nu}(x)&\equiv &\delta_\kappa T^{\mu\nu}(x) =\frac{\partial}{\partial b^\kappa} (\delta T^{\mu\nu})\nonumber \\
&=& -(x^2 \partial_\kappa - 2 x_\kappa x\cdot \partial) T^{\mu\nu}(x) + 2\Delta_T x_\kappa T^{\mu\nu}(x) +
2(\delta_{\mu\kappa}x_\alpha -\delta_{\alpha \kappa}x_\mu) T^{\alpha\nu}(x) \nonumber \\ 
&& + 2 (\delta_{\kappa\nu} x_{\alpha} -\delta_{\alpha \kappa} x_\nu )T^{\mu\alpha}. 
\label{ith}
\end{eqnarray}
%%%%%%%%%%%%%%%%%%%%%%%%%%%%
%%%%%%%%%%%%%%%%%%%%%%%%%%%%
\subsection{From Poincar\'e to the conformal group}
%%%%%%%%%%%%%%%%%%%%%%%%%%%%
%%%%%%%%%%%%%%%%%%%%%%%%%%%%
We can merge the action of the special conformal transformation, of the dilatations and the generators of the Poincar\'e group to construct the Lie algebra of the conformal group.\\
The Poincar\'e subgroup contains the fundamental set of symmetries for any relativistic system, 
	\begin{align}
	i[J^{\mu\nu},J^{\r\s}]&=\d^{\nu\ro}J^{\mu\sigma}-\d^{\mu\ro}J^{\nu\sigma}-\d^{\mu\sigma}J^{\ro\nu}+\d^{\nu\sigma}J^{\ro\mu},\\
	i[P^\mu,J^{\ro\sigma}]&=\d^{\mu\ro}P^\sigma-\d^{\mu\sigma}P^\ro,\\
	[P^\mu,P^\nu]&=0,
	\end{align}
where the $J$'s are the generators of the Lorentz group, $P^\mu$ are the 4 generators of translations. For scale invariant theories, which contain no dimensionful parameter, the Poincar\'e group can be extended by including the dilatation generator $D$, whose commutation relations with the other generators are
\begin{equation}
	\begin{split}
	[P^\mu,D]&=iP^\mu,\\
	[J^{\mu\nu},D]&=0.
	\end{split}
	\label{dil}
\end{equation}
Finally, we include also the conformal transformations, whose generators are denoted as $K^\mu$. The corresponding Lie algebra is
\begin{equation}
	\begin{split}
	[K^\mu,D]&=-iK^\mu,\\
	[P^\mu,K^\nu]&=2i\d^{\mu\nu}D+2iJ^{\mu\nu},\\
	[K^\mu,K^\nu]&=0,\\
	[J^{\ro\sigma},K^\mu]&=i\d^{\mu\ro}K^\sigma-i\d^{\mu\sigma}K^\ro.
	\end{split}
	\label{spec}
\end{equation}
It is clear from \eqref{dil} and \eqref{spec} that scale invariance does not require conformal invariance, but conformal invariance necessarily implies scale invariance. \\
To summarize, having specified the elements of the conformal group, we can define a primary field $\mathcal{O}^i(x)$, where $i$ runs over the representation of the group which the field belongs to, through the transformation property under a conformal transformation $g$ belonging to the conformal group $SO(2,d)$ in the form
\begin{equation}\label{spintransf}
\mathcal{O}^i(x)\xrightarrow{g}\mathcal{O}'^i(x')=\Omega(x)^\Delta\,D^i_j(g)\,\mathcal{O}^j(x),
\end{equation}
where $\Delta$ is the scaling dimension of the field and $D^i_j(g)$ denotes the representation of $O(d)$ acting on $\mathcal{O}'^i$. In the infinitesimal form we have
\begin{equation}
\d_g\mathcal{O}^i(x)=-(L_g\mathcal{O})^i(x),\qquad \text{with}\quad L_g=v\cdot \partial+\Delta\l+\sdfrac{1}{2}\partial_{(\mu}v_{\nu)}\Sigma^{\mu\nu},
\end{equation}
where the vector $v_\mu$ is the infinitesimal coordinate variation $v_\mu=\d_gx_\m=x'_\mu(x)-x_\mu$ and $(\Sigma_{\mu\nu})^i_j$ are the generators of $O(d)$ in the representation of the field $\mathcal{O}^i$. The explicit form of the operators $L_g$ can be obtained from \eqref{transf} and \eqref{Om} and are given by
\begin{align}
\text{translations}\qquad L_g&=a^\mu\partial_\mu,\\
\text{rotations}\qquad L_g&=\sdfrac{\omega^{\mu\nu}}{2}[x_\nu\partial_\mu-x_\mu\partial_\nu]-\Sigma_{\mu\nu},\\
\text{scale transformations}\qquad L_g&=\sigma\,[x\cdot \partial +\Delta],\\
\text{special conformal transformations}\qquad L_g&=b^\mu[x^2\partial_\mu-2x_\mu\,x\cdot \partial-2\Delta\,x_\mu-2x_\nu\Sigma_\mu^{\ \nu}].
\end{align}
Let us now consider the subalgebra of the four-dimensional conformal algebra, corresponding to dilatations and Lorentz transformations. This allows us to label different representations of the conformal algebra with $(\D,j_L,j_R)$, where $\D$ is the scaling dimension and $j_L$, $j_R$ are Lorentz quantum numbers.  For any quantum field theory, unitarity implies that all the states in a representation must have a positive norm, imposing bounds on the unitary representations. If one considers the compact subalgebra $\mathfrak{so}(2)\oplus\mathfrak{so}(4)$ of $\mathfrak{so}(4,2)$, its unitary representations are labeled with $(\D,j_L,j_R)$ and have to satisfy the constraints
\begin{equation}
\begin{split}
\D\ge 1+j_L\quad\text{for}\ \ j_R=0,&\qquad \D\ge 1+j_R\quad\text{for}\ \ j_L=0,\\
\D\ge 2+j_L+j_R,&\quad\text{for both}\ \ j_L,\,j_R\ne0.
\end{split}
\end{equation}
In this context, for scalars we have $\D\ge1$, for vectors $\D\ge3$ and for symmetric traceless tensors $\D\ge4$. These bounds are saturated by a free scalar field $\phi$, a conserved current $J_\m$ and a conserved symmetric traceless tensor $T_{\m\n}$. In $d$ dimensions, the bound for fields of spin-$s$ take the form
	\begin{align}
	&\D\ge \sdfrac{d-2}{2},\hspace{2cm} s=0,\\
	&\D\ge \sdfrac{d-1}{2},\hspace{2cm} s=1/2,\\
	&\D\ge d+s-2,\hspace{1.53cm} s\ge1.
	\end{align}
In a CFT, fields transform under irreducible representations of the conformal algebra. In order to construct its irreducible representations for general dimensions, it is used the method of the induced representations. We briefly comment on this point.\\
First, one analyzes the transformation properties of the field $\phi$ at point $x=0$. Then, with the help of the momentum generator $P^\m$, the argument of the field is shifted to an arbitrary point $x$, in order to obtain the general transformation rule. For Lorentz transformations it is postulated that 
\begin{equation}
[J_{\m\n},\phi(0)]=-\mathcal{J}_{\m\n}\,\phi(0),
\end{equation}
where $\mathcal{J}_{\m\n}$ is a finite-dimensional representation of the Lorentz group, determining the spin of the field $\phi(0)$. In addition, one requires a commutation relations with the dilatation operator $D$, in the conformal algebra, to hold
\begin{equation}
[D,\phi(0)]=-i\D\,\phi(0).
\end{equation}
This relation implies that $\phi$ has the scaling dimension $\D$, i.e. that under dilatations $x\mapsto x'=\l x$ (for a real $\lambda$) it transforms as 
\begin{equation}
\phi(x)\mapsto \phi'(x')=\l^{-\D}\,\phi(x).
\end{equation}
In particular, a field $\phi$, covariantly transforming under an irreducible representation of the conformal algebra, has a fixed scaling dimension and it is therefore an eigenstate of the dilatation operator $D$. In a conformal algebra it is sufficient to consider particular fields, the conformal primary fields, which satisfy the commutation relation
\begin{equation}
[K_\m,\phi(0)]=0.\label{bound}
\end{equation}
By applying the commutation relations of $D$ with $P_\m$ and $K_\m$ to the eigenstates of $D$, one observes that $P_\m$ raises  while $K_\m$ lowers the scaling dimension  since
\begin{align}
[D,[P_\m,\phi(0)]]&=[P_\m,[D,\phi(0)]]+[[D,P_\m],\phi(0)]=-i(\D+1)[P_\m,\phi(0)],\\
[D,[K_\m,\phi(0)]]&=[K_\m,[D,\phi(0)]]+[[D,K_\m],\phi(0)]=-i(\D-1)[K_\m,\phi(0)].
\end{align}
As discussed in the previous section, since in a unitary CFT there is a lower bound on the scaling dimensions of the fields, this implies that any conformal representation must contain operators of lowest dimension which, due to \eqref{bound}, are annihilated by $K_\m$ at $x^\n=0$. \\
In a given irreducible multiplet of the conformal algebra, conformal primary fields are those fields of lowest scaling dimension, determined by the relation \eqref{bound}. All the other fields, \emph{conformal descendants} of $\phi$, are obtained by acting with $P_\m$ on such conformal primary fields. 
We can consider the operator $U(x)=~\exp\left(-i\hat P_\m\,x^\m\right)$ that, acting on $\phi(0)$, gives
\begin{equation}
U(x)\phi(0)U^{-1}(x)=\phi(x).
\end{equation}
Through this operator, we may deduce the commutation relations for a conformal primary field $\phi(x)$, taking into account the relations of the conformal algebra. In order to show the procedure, we consider the case of $[P_\m,\phi(x)]$. Expanding the operator $U(x)$ and using the Haussdorf formula
\begin{equation}
U(x)\,\phi(0)\,U^{-1}(x)=\sum_{n=0}^\infty\,\sdfrac{i^n}{n!}\,x^{\n_1}\dots x^{\n_n}[P_{\n_1},[\dots[P_{\n_n},\phi(0)]\dots]],
\end{equation}
we obtain
\begin{align}
[P_\m,U(x)\,\phi(0)\,U^{-1}(x)]&=[P_\m,U(x)\,\phi(0)\,U^{-1}(x)]\notag\\
&=\left[P_\m\ ,\ \sum_{n=0}^\infty\,\sdfrac{i^n}{n!}\,x^{\n_1}\dots x^{\n_n}[P_{\n_1},[\dots[P_{\n_n},\phi(0)]\dots]]\right]=-i\partial_\m\phi(x),
\end{align}
from which we deduce the commutation relations
\begin{equation}
	\begin{split}
	[P_\m,\phi(x)]&=-i\partial_\m\phi(x),\\
	[D,\phi(x)]&=-i(\D+x^\m\partial_\m)\phi(x),\\
	[J_{\m\n},\phi(x)]&=\left(-\mathcal{J}_{\m\n}+ix_\m\partial_\n-ix_\n\partial_\m\right)\phi(x),\\
	[K_\m,\phi(x)]&=\left[i\left(-x^2\partial^\m+2x_\m, x^\r\partial_\r+2x_\m\D\right)-2x^\n\mathcal{J}_{\m\n}\right]\phi(x).
	\end{split}\label{transform}
\end{equation}
The correlation functions constructed with such primary fields transform covariantly and, in case of scalar conformal primaries we have 
\begin{equation}
\langle\mathcal{O}_1(x'_1)\dots \mathcal{O}_n(x'_n)\rangle=\left|\sdfrac{\partial x'_1}{\partial x_1}\right|^{-\D_1/d}\dots \left|\sdfrac{\partial x'_1}{\partial x_1}\right|^{-\D_n/d}\langle\mathcal{O}_1(x_1)\dots \mathcal{O}_n(x_n)\rangle.\label{invariance}
\end{equation}
In the case of fields of general spin there are extra terms depending on the transformation matrices defined in \eqref{spintransf}. Furthermore, in these theories we assume a unique vacuum state that should be invariant under all global conformal transformations, eigenstate of the dilatation operator with eigenvalue zero. 
The theory can be coupled to a background metric $g_{\m\n}$, which takes the role of an external source for 
an important class of correlation functions. This is intimately related to an important operator of any CFT, which is the stress energy tensor. This operator, due to the constraints 
of conformal invariance, has the properties of being symmetric, traceless and conserved. Once the theory is coupled to a curved background metric $g_{\mu\nu}$, it is natural to consider the action of the Weyl group for which the coordinates, the metric and the scalar primary fields transform as
	\begin{align}
	x'^{\,\mu}&=e^{\sigma(x)}\,x^\mu\label{transformWeylx},\\
	g'_{\mu\nu}(x)&=\,e^{2\sigma(x)}\,g_{\mu\nu}(x),\\
	\mathcal{O}_i'(x)&=e^{-\sigma(x)\,\Delta_i}\,\mathcal{O}_i(x).
	\end{align}
The function $\sigma(x)$ allows to define a subgroup of the group of local Weyl transformation, induced by the conformal transformations, known as the conformal Weyl group. The form of the local function $\sigma(x)$ is written as
\begin{equation}
\sigma(x)=\log\left(\frac{1}{1-2b\cdot x+b^2 x^2}\right),
\end{equation}
with $b$ any constant vector. For $b$ infinitesimal and using the expression of $\sigma(x)$, the transformation \eqref{transformWeylx} corresponds exactly to the conformal transformations given in \eqref{transf}. Every conformal field theory is Weyl invariant, and the scalar correlation function of primary fields given above transforms under the Weyl group as
\begin{equation}
\label{distint}
\langle\mathcal{O}_1(x_1)\dots \mathcal{O}_n(x_n)\rangle_{e^{2\s}g_{\m\n}}=e^{-\s(x_1)\D_1}\dots e^{-\s(x_n)\D_n}\langle\mathcal{O}_1(x_1)\dots \mathcal{O}_n(x_n)\rangle_{g_{\m\n}}.
\end{equation}
Notice also the appearance in the equation above of $x$-dependent scalings $\sigma(x_i)$, functions of the global parameters of a conformal transformation, which is typical of the Weyl symmetry. 
The relation between conformal and Weyl symmetry can be discussed in physical terms by the procedure of Weyl gauging, which leads to Weyl gravity. In this case, in each free-falling frame identified 
in a local region of spacetime, the local flat symmetry in the tangent space is controlled by the generators of the conformal group rather than the Poincar\'e group.\\
It is worth noting that \eqref{distint} is valid at non-coincident point only, i.e. for $x_i\ne x_j$, $i\ne j$. This is due to the choice of the regularization scheme that one has to consider in order to have a defined quantum field theory. The (dimensionful) regulator used in the scheme chosen, may break some of the symmetries of the theory that 
can't be restored once the regulator is removed. For instance,  dimensional regularization, which preserves Lorentz invariance, breaks Weyl invariance and it is manifest in a local manner, i.e. it affects the correlation functions at coincident points only. The effects of the breaking of Weyl symmetry is manifest in the {trace} or {Weyl anomaly}.

In quantum field theory scale and conformal invariance are treated as equivalent, in the sense that all the realistic field theories which are scale invariant are also conformally invariant. Such enhancement of the symmetry (from dilatation invariance to conformal), has been discussed at length in the literature \cite{Nakayama:2013is}, with counter-examples which are quite unrealistic from the point of view of a local, covariant, relativistic field theory. In our case scale and conformal invariance will be taken as equivalent, as is indeed the case in ordinary Lagrangian field theories.  
 %%%%%%%%%%%%%%%%%%%%%%%%%%%%%%%%%%%%%%%%%%%%%%%%%%
\subsection{Correlation functions in CFT\label{secCFT}}
%%%%%%%%%%%%%%%%%%%%%%%%%%%%%%%%%%%%%%%%%%%%%%%%%%
The presence of conformal symmetry in a quantum field theory is very powerful and restrictive. In fact, it imposes very strong constraints on the structure of the correlation functions. Up to three-point functions, the requirement of conformal invariance fixes the structure of the correlator up to constants. 
Starting with the simplest case of the $1$-point functions of a generic scalar operator $\mathcal{O}$, one finds from translational invariance that it has to be a constant. Then, scaling invariance requires that this constant has to vanish for a nonzero scaling dimension of $\mathcal{O}$
\begin{equation}
\label{const}
\braket{\mathcal{O}(x)}=0,\qquad \text{if}\ \Delta\ne0.
\end{equation}
Since the unitarity bound on a CFT requires that $\Delta$ is positive, all the operators, apart from the identity operator, are constrained by \eqref{const}. Therefore, in a conformal theory $\braket{\mathcal{O}(x)}=0$, assuming that $\mathcal{O}$ is not proportional to the identity operator. 
Moving on to the next non-trivial case, we consider two-point functions of two scalar operators $\mO_i$ and $\mO_j$ of conformal dimensions $\D_i$ and $\D_j$ and we construct their general form. From Poincar\'e invariance, these functions depend on their coordinate difference, and imposing scaling invariance one finds 
\begin{equation}
\braket{\mO_i(x_i)\,\mO_j(x_j)}=\frac{C_{ij}}{|x_i-x_j|^{\Delta_i+\Delta_j}}.
\end{equation}
At this stage, we still have to impose the invariance under the special conformal transformations. For such transformations we recall that
\begin{equation}
\left|\sdfrac{\partial x'_i}{\partial x_i}\right|=\sdfrac{1}{(1-2b\cdot x_i+b^2 x_i^2)^d}.
\end{equation} 
Under this transformation the two-point  function transforms as
\begin{align}
\braket{\mO'_1(x')\,\mO'_2(y')}&=\left|\sdfrac{\partial x'}{\partial x}\right|^{\D_1/d}\left|\sdfrac{\partial y'}{\partial y}\right|^{\D_2/d}\braket{\mO_1(x')\,\mO_2(y')}=\sdfrac{(\g_1\g_2)^{(\D_1+\D_2)/2}}{\g_1^{\D_1}\g_2^{\D_2}}\sdfrac{C_{12}}{|x_1-x_2|^{\D_1+\D_2}},
\end{align}
where $\g_i=(1-2b\cdot x_i+b^2 x_i^2)$. The invariance of the correlation function under special conformal transformations induces the constraint
\begin{equation}
\sdfrac{(\g_1\g_2)^{(\D_1+\D_2)/2}}{\g_1^{\D_1}\g_2^{\D_2}}=1,
\end{equation}
which is identically satisfied only if $\D_1=\D_2$. This means that two quasi-primary fields are correlated only if they have identical scaling dimensions
\begin{equation}
\braket{\mO_1(x_1)\,\mO_2(x_2)}=\frac{C_{12}}{|x_1-x_2|^{2\D_1}}\d_{\D_1\D_2}.\label{2poin}
\end{equation}
This result can also be obtained analysing only the behaviour of the correlation function under the operation of coordinate inversion. It follows from an important property of the special conformal transformations, i.e. that they can be written as a composition of two inversions $I^\m(x)$ and a translation by a vector $b^\mu$, as follows
\begin{equation}
I^\m(x)=\sdfrac{x^\m}{x^2},\qquad \sdfrac{x'^\m}{x^2}=I^\m(x)+b^\m.\label{invers}
\end{equation}
Therefore, in order to analyse the implications of the action of the special conformal transformation, in many cases it is enough to consider the action of $I^\m(x)$. \\
A similar method can be applied to the three-point  functions. Poincar\'e invariance requires that
\begin{equation}
\braket{\mO_1(x_1)\,\mO_2(x_2)\,\mO_3(x_3)}=f\big(\,|x_1-x_2|\, ,\, |x_1-x_3|\,,\,|x_2-x_3|\,\big),
\end{equation}
and the invariance under scale transformations 
\begin{equation}
f\big(\,|x_1-x_2|\, ,\, |x_1-x_3|\,,\,|x_2-x_3|\,\big)=\l^{\D_1+\D_2+\D_3}f\big(\,\l|x_1-x_2|\, ,\, \l|x_1-x_3|'\,,\,\l|x_2-x_3|'\,\big)
\end{equation}
imposes another constraint that forces a generic three-point  function to take the following form
\begin{equation}
\braket{\mO_1(x_1)\,\mO_2(x_2)\,\mO_3(x_3)}=\frac{C_{123}}{x_{12}^a\ x_{23}^b\ x_{13}^c}\label{3point},
\end{equation}
where $x_{ij}=|x_i-x_j|$. $C_{123}$ is a constant and the coefficients $a$, $b$, $c$ must satisfy the relation
\begin{equation}
a+b+c=\D_1+\D_2+\D_3.
\end{equation}
The invariance of \eqref{3point} under special conformal transformations implies that
\begin{equation}
\sdfrac{C_{123}}{\g_1^{\D_1}\g_2^{\D_2}\g_3^{\D_3}}\sdfrac{(\g_1\g_2)^{a/2}(\g_2\g_3)^{b/2}(\g_1\g_3)^{c/2}}{x_{12}^a\ x_{23}^b\ x_{13}^c}=\frac{C_{123}}{x_{12}^a\ x_{23}^b\ x_{13}^c}.
\end{equation}
To solve this constraint, all the factors involving the transformation parameter $b^\m$ must disappear, which leads to the following set of equations
\begin{equation}
\g_1^{a/2+c/2-\D_1}=1,\qquad\g_2^{a/2+b/2-\D_2}=1,\qquad\g_3^{b/2+c/2-\D_3}=1,
\end{equation}
that can be solved in terms of the conformal dimensions of the operators $\mO_j$
\begin{equation}
a=\D_1+\D_2-\D_3,\qquad b=\D_2+\D_3-\D-1,\qquad c=\D_3+\D_1-\D_2.
\end{equation}
Therefore, the correlator of three quasi-primary fields is given by
\begin{equation}
\braket{\mO_1(x_1)\,\mO_2(x_2)\,\mO_3(x_3)}=\frac{C_{123}}{x_{12}^{\D_1+\D_2-\D_3}\ x_{23}^{\D_2+\D_3-\D-1}\ x_{13}^{\D_3+\D_1-\D_2}}.
\end{equation}
$C_{123}$ and $\Delta_i$'s are usually referred to as ``the conformal data''. The strong constraints on the two- and three-point functions previously discussed are not enough to fix four-point functions and correlators of higher orders. In this case, indeed, one can construct combinations of the coordinates $\{x_j\}$ that are invariant under the symmetry.  For instance, the Poincar\'e invariance implies that the variables appearing in the correlation function must be of the form $x_{ij}=|x_i-x_j|$, while to generate scale invariant combinations  one has to construct ratios of such variables. We recall that, under the special conformal transformation, the coordinate distance $x_{i j}$ is mapped to $x'_{i j}$ in the form
	\begin{equation}
	|x'_i-x_j'|=\frac{|x_i-x_j|}{(1-2b\cdot x_i+b^2x_i^2)^{1/2}(1-2b\cdot x_j+b^2x_j^2)^{1/2}}.
	\end{equation}
Therefore, the simplest objects that are conformally invariant are
\begin{equation}
u_{ijkl}\equiv\frac{x_{ij}\,x_{kl}}{x_{ik\,}x_{jl}},\qquad v_{ijkl}=\frac{x_{ij}\,x_{kl}}{x_{il}\,x_{jk}},
\end{equation}
known as {anharmonic ratios} or {cross-ratios}. These ratios are well-defined as far as the points are distinct. Every function of these cross-ratios will be conformally invariant, and for this reason $4$- and higher-point functions are determined up to an arbitrary function of the cross-ratios, and are not uniquely fixed by conformal invariance. 
For instance, the four-point function can be written as
\begin{equation}
\braket{\mO_1(x_1)\,\mO_2(x_2)\,\mO_3(x_3)\,\mO_4(x_4)}=f\left(\frac{x_{12}x_{34}}{x_{13}x_{24}},\frac{x_{12}x_{34}}{x_{23}x_{14}}\right)\ \prod_{1\le i<j\le 4}^{4}\,x_{ij}^{\D_t/3-\D_i-\D_j},
\end{equation}
where $\D_t=\sum_{i=1}^4\D_i$ and $f$ is an undetermined function. In the case of four-point functions there are two independent conformal ratios. In the case of $n$-point functions there are $n(n-3)/2$ independent conformal ratios. 
%%%%%%%%%%%%%%%%%%%%%%
\subsection{The embedding space formalism}
%%%%%%%%%%%%%%%%%%%%%%
The non-linear action of the conformal group becomes simpler and more transparent in the formalism of the {embedding space} \cite{Simmons_Duffin_2014, Simmons-Duffin:2016gjk}. In fact, taking a Euclidean CFT in $d$-dimensions with the conformal group acting 
non-linearly on $\mathbb{R}^d$, it is well known that it is locally isomorphic to $SO(d+1,1)$ and therefore it acts linearly on the space $\mathbb{R}^{d+1,1}$, called the {embedding space}. The procedure consists in embedding the target space $\mathbb{R}^d$ into $\mathbb{R}^{d+1,1}$.\\  
We introduce the light-cone coordinates in $\mathbb{R}^{d+1,1}$ as
\begin{equation}
X^A=(X^+,X^-,X^a)\in\mathbb{R}^{d+1,1},
\end{equation}
where $a=1,\dots,d$ with the inner product
\begin{equation}
X\cdot X=\eta_{AB}\,X^A\,X^B=-X^+\,X^-+X_a\,X^a.
\end{equation}
Using such coordinates, the condition $X^2=0$ defines a $SO(d+1,1)$ invariant subspace of dimension $d+1$, the null cone. Imposing the gauge condition $X^+=1$, that defines the {Poincar\'e section} of the embedding, we identify the projective null-cone with $\mathbb{R}^d$, and the null vectors take the form
\begin{equation}
X=(1,x^2,x^\m)\label{poinc}.
\end{equation}
It is worth to note that the transformation of $X$ by $SO(d+1,1)$ is just a matrix multiplication, and the transformation
\begin{equation}
X\to h\,X/(h\,X)^+, \qquad h\in SO(d+1,1)
\end{equation}
defines the non-linear action of the conformal group on $\mathbb{R}^d$. Finally, taking two points $X_i$ and $X_j$, setting $X_{ij}=X_i-X_j$, on the Poincar\'e section where $X^2=0$, one obtains
\begin{equation}
X_{ij}^2=-2X_i\cdot X_j=(x_i-x_j)^2=x_{ij}^2.\label{xxx}
\end{equation}
Now, we want to lift the fields defined on $\mathbb{R}^d$ to the embedding space. The first step is to consider a primary scalar field $\mO_d(x)$ in $\mathbb{R}^d$ with dimension $\D$, for which one can define a scalar on the entire null-cone, with the requirement
\begin{equation}
\mO_{d+1,1}(\l X)=\l^{-\D}\mO_{d+1,1}(X),
\end{equation}
with the dimension of $\mO_d$ reflected in the degree of $\mO_{d+1}$.
For any such fields in $\mathbb{R}^{d+1,1}$ one can always perform a projection on $\mathbb{R}^d$ as
\begin{equation}
\mO_{d}(x)=\left(X^+\right)^{\Delta}\mO_{d+1,1}(X(x)).
\end{equation}
The field $\mO_{d+1,1}(X)$ then transforms in a simple way under the conformal transformation $\mO_{d+1,1}(X)\to \mO_{d+1}(hX)$. 

Conformal invariance requires that the correlators containing $\mO_{d+1,1}(X)$ are invariant under $SO(d+1,1)$. 
In order to derive the structure of a correlator in the embedding space, one starts by 
writing down its most general form, covariant under  $SO(d+1,1)$, satisfying the appropriate scaling conditions, according to to \eqref{scale}. Then we substitute \eqref{poinc} and use \eqref{xxx} in order to derive its expression in ordinary coordinates. For example, the two-point  function of operators of dimension $\D$ is fixed by conformal invariance, homogeneity, and the null condition $X_i^2=0$, to take the form
\begin{equation}
\braket{\mO(X_1)\,\mO(X_2)}=\frac{C_{12}}{(-2X_1\cdot X_2)^\D}.
\end{equation}
We notice that $-2X_1\cdot X_2$ is the only Lorentz invariant that we can construct out of two points, since on the projective null-cone the condition $X_j^2=0$ is satisfied. Now, using \eqref{xxx} we obtain
\begin{equation}
\braket{\mO(x_1)\,\mO(x_2)}=\frac{C_{12}}{(X^2_{12})^\D}=\frac{C_{12}}{|x_1-x_2|^{2\D}},
\end{equation}
which is \eqref{2poin}. 
The next step is to extend the embedding formalism to tensor operators \cite{Ferrara:1973yt,Weinberg:2010fx,Costa:2011mg,SimmonsDuffin:2012uy}. We are going to illustrate the method for totally symmetric and traceless tensors, since such tensors transform with the simplest irreducible representation of $SO(d+1,1)$. 

Considering a tensor field of $SO(d+1,1)$, denoted as $\mathcal{O}_{A_1\dots A_n}(X)$, with the properties
\begin{itemize}
\item defined on the null-cone $X^2=0$,
\item traceless and symmetric,
\item homogeneous of degree $-\Delta$ in $X$, i.e., $\mathcal{O}_{A_1\dots A_n}(\lambda\,X)=\lambda^{-\Delta} \mathcal{O}_{A_1\dots A_n}(X)$,
\item transverse $X^{A_i}\mathcal{O}_{A_1\dots  A_{i}\dots A_n}(X)=0$, with $i=1,\dots,n$,
\end{itemize}
it is clear that those are conditions rendering $\mathcal{O}_{A_1\dots A_n}(X)$ manifestly invariant under $SO(d+1,1)$. In order to find the corresponding tensor in $\mathbb{R}^d$, one has to restrict $\mathcal{O}_{A_1\dots A_n}(X)$ to the Poincar\'e section and project the indices as
\begin{equation}
\mathcal{O}_{\mu_1\dots\mu_n}(x)=\frac{\partial\,X^{A_1}}{\partial x^\mu_1}\dots\frac{\partial\,X^{A_n}}{\partial x^\mu_n}\mathcal{O}_{A_1\dots A_n}(X).\label{transfemb1}
\end{equation}
For example, the most general form of the two-point  function of two operators with spin-$1$ and dimension $\D$ can be derived as
\begin{equation}
\braket{\mO^A(X_1)\mO^B(X_2)}=\frac{C_{12}}{(X_1\cdot X_2)^\D}\left[\h^{AB}+\a\frac{X_2^A\,X_1^B}{X_1\cdot X_2}\right]\label{2pointemb},
\end{equation}
according to the rules given above. The transverse condition 
\begin{equation}
X_1^A\braket{\mO^A(X_1)\mO^B(X_2)}=\frac{C_{12}}{(X_1\cdot X_2)^\D}\left[X_1^B+\a X_1^B\right]=0,
\end{equation}
implies that $\a=-1$. The projection on $\mathbb{R}^d$ using \eqref{transfemb1} leads to define the two-point  function in $\mathbb{R}^d$ as
\begin{equation}
\braket{\mO^\m(x_1)\mO^\n(x_2)}=\frac{\partial X_1^A}{\partial x_1^\m}\frac{\partial X_2^B}{\partial x_2^\n}\braket{\mO^A(X_1)\mO^B(X_2)}.\label{transfemb}
\end{equation}
Finally, one needs to compute the terms
\begin{equation}
\h_{AB}\frac{\partial X_1^A}{\partial x_1^\m}\frac{\partial X_2^B}{\partial x_2^\n},\qquad X_2^A\,X_1^B\frac{\partial X_1^A}{\partial x_1^\m}\frac{\partial X_2^B}{\partial x_2^\n},
\end{equation}
that are explicitly given as
	\begin{align}
	\h_{AB}\sdfrac{\partial X_1^A}{\partial x_1^\m}\sdfrac{\partial X_2^B}{\partial x_2^\n}&=-\sdfrac{1}{2}\left(\sdfrac{\partial X_1^+}{\partial x_1^\m}\sdfrac{\partial X_2^-}{\partial x_2^\n}+\sdfrac{\partial X_1^+}{\partial x_1^\m}\sdfrac{\partial X_2^-}{\partial x_2^\n}\right)+\d_{ab}\sdfrac{\partial X_1^a}{\partial x_1^\m}\sdfrac{\partial X_2^b}{\partial x_2^\n}=\d_{\m\n},\\
	X_2^A\sdfrac{\partial X_1^A}{\partial x_1^\m}&=-\sdfrac{1}{2}\left(X_2^+\sdfrac{\partial X_1^-}{\partial x_1^\m}+X_2^-\sdfrac{\partial X_1^+}{\partial x_1^\m}\right)+\d_{ab}X_2^a\sdfrac{\partial X_1^b}{\partial x_1^\m}=x_2^\m-x_1^\m,\\
	X_1^B\sdfrac{\partial X_2^B}{\partial x_2^\n}&=-\sdfrac{1}{2}\left(X_1^-\sdfrac{\partial X_2^+}{\partial x_2^\n}+X_1^+\sdfrac{\partial X_2^-}{\partial x_2^\n}\right)+\d_{ab}X_1^a\sdfrac{\partial X_2^b}{\partial x_2^\n}=x_1^\n-x_2^\n,
	\end{align}
which allow to define the following projective rules
\begin{equation}
X_1^B\mapsto x_1^\n-x_2^\n,\qquad X_2^A\mapsto x_2^\m-x_1^\m,\qquad \h^{AB}\mapsto\d^{\m\n},\qquad X_1\cdot X_2\mapsto-1/2\,x_{12}^2.
\end{equation}
Therefore, from \eqref{transfemb} we infer the equation
\begin{equation}
\braket{\mO^\m(x_1)\mO^\n(x_2)}=\frac{C_{12}\,I^{\m\n}(x_{12})}{x_{12}^{2\D}},\label{finres}
\end{equation}
where 
\begin{equation}
I^{\m\n}(x)=\d^{\m\n}-\frac{2x^\m\,x^\n}{x^2}\label{Itensor},
\end{equation}
and $C_{12}$ is an undetermined constant. The $I^{\m\n}$ tensor appearing in \eqref{finres} plays an important role in the definition of the conformal structure. In fact, it is important in realizing the operation of inversion in a CFT, and appears in all the correlation functions of tensor operators \cite{Osborn:1993cr, Erdmenger:1996yc}.\\
Using the same strategy, one can easily construct correlation functions of higher spin, and for a traceless spin-$2$ operator of dimension $\D$, we find
\begin{equation}
\braket{\mO^{\m\n}(x_1)\mO^{\r\s}(x_2)}=\sdfrac{C_{12}}{x_{12}^{2\D}}\left[I^{\m\r}(x_{12})I^{\n\s}(x_{12})+I^{\m\s}(x_{12})I^{\n\r}(x_{12})-\sdfrac{2}{d}\d^{\m\n}\d^{\r\s}\right].\label{2poinembed}
\end{equation}
It is worth mentioning that expressions \eqref{finres} and \eqref{2poinembed} satisfy the requirement of being traceless and symmetric, with the possibility of further specialization of this result. In fact, from \eqref{finres}, in the case of conserved currents $J^\mu$, the conservation Ward identities \eqref{Ward1}
\begin{equation}
\frac{\partial}{\partial x_1^{\mu_1}}\,\braket{J^{\mu_1}(x_1)J^{\mu_2}(x_2)}=0,
\end{equation}
impose a strong condition on the conformal dimension. We can use in this expression the full form of the two-point  function given in \eqref{finres}, to derive the relation
\begin{align}
0&=\sdfrac{\partial}{\partial x_1^\m}\left(\sdfrac{C_{12}\,I^{\m\n}(x_{12})}{x_{12}^{2\D}}\right)=C_{12}\left(-2\D\sdfrac{I^{\m\n}(x_{12})}{x_{12}^{2(\D+1)}}(x_{12})_\m+\sdfrac{1}{x_{12}^{2\D}}\sdfrac{\partial\,I^{\m\n}(x_{12})}{\partial x_1^\m}\right)\notag\\
&=\sdfrac{C_{12}}{x_{12}^{2\D}}\left[-2\D\sdfrac{(x_{12})_\m}{x_{12}^2}\left(\d^{\m\n}-2\sdfrac{x_{12}^\m\,x_{12}^\n}{x_{12}^2}\right)-2(d-1)\sdfrac{x_{12}^\n}{x_{12}^2}\right]=-2(d-\D-1)\,C_{12}\,\sdfrac{x_{12}^\n}{x_{12}^{2(\D+1)}}.
\end{align}
This shows that a conformal primary operator of spin-$1$ is conserved if an only if its dimension is $\D=d-1$. A similar result holds for conserved energy momentum tensors $T^{\mu_i\nu_i}(x_i)$, by using the condition \eqref{transWardF}
\begin{equation}
\frac{\partial}{\partial x_1^{\mu_1}}\,\braket{T^{\mu_1\nu_1}(x_1)T^{\mu_2\nu_2}(x_2)}=0.
\end{equation}
In this case one has to use the relation 
\begin{align}
0&=\sdfrac{\partial}{\partial x_1^\m}\left\{\sdfrac{C_{12}}{x_{12}^{2\D}}\left[I^{\m\r}(x_{12})I^{\n\s}(x_{12})+I^{\m\s}(x_{12})I^{\n\r}(x_{12})-\sdfrac{2}{d}\d^{\m\n}\d^{\r\s}\right]\right\}\notag\\
&=-2\D\sdfrac{C_{12}}{x_{12}^{2\D+2}}(x_{12})_\m\left[I^{\m\r}(x_{12})I^{\n\s}(x_{12})+I^{\m\s}(x_{12})I^{\n\r}(x_{12})-\sdfrac{2}{d}\d^{\m\n}\d^{\r\s}\right]\notag\\
&\hspace{2cm}+\sdfrac{C_{12}}{x_{12}^{2\D}}\sdfrac{\partial}{\partial x_1^\m}\left[I^{\m\r}(x_{12})I^{\n\s}(x_{12})+I^{\m\s}(x_{12})I^{\n\r}(x_{12})-\sdfrac{2}{d}\d^{\m\n}\d^{\r\s}\right].\label{3A}
\end{align}
Using the results
\begin{align}
\sdfrac{\partial}{\partial x_1^\m}\,I^{\m\a}(x_{12})&=-2(d-1)\sdfrac{x_{12}^\a}{x_{12}^2},\\
(x_{12})_{\m}\,I^{\m\a}(x_{12})&=-x_{12}^\a,\\
\sdfrac{\partial}{\partial x_1^\m}\,I^{\n\a}(x_{12})&=-\sdfrac{2}{x_{12}^2}\left(\d^\n_\m\,x_{12}^\a+\d^\a_\m\,x_{12}^\n-2\sdfrac{(x_{12})_\m\,x_{12}^\n\,x_{12}^\a}{x_{12}^2}\right),
\end{align}
we can simplify \eqref{3A} in the form
\begin{align}
0&=-2\D\sdfrac{C_{12}}{x_{12}^{2\D+2}}\left[-x_{12}^\r\,I^{\n\s}(x_{12})-x_{12}^\s\,I^{\n\r}(x_{12})-\sdfrac{2}{d}\d^{\r\s}x_{12}^\n\right]-2\sdfrac{C_{12}}{x_{12}^{2(\D+1)}}(d-1)\left[x_{12}^\r\,I^{\n\s}(x_{12})+x_{12}^\s\,I^{\n\r}(x_{12})
\right]\notag\\
&\hspace{0.1cm} -2\sdfrac{C_{12}}{x_{12}^{2(\D+1)}}\left[I^{\m\r}(x_{12})\left(\d^\n_\m\,x_{12}^\s+\d^\s_\m\,x_{12}^\n-2\sdfrac{(x_{12})_\m\,x_{12}^\n\,x_{12}^\s}{x_{12}^2}\right)+I^{\m\s}(x_{12})\left(\d^\n_\m\,x_{12}^\r+\d^\r_\m\,x_{12}^\n-2\sdfrac{(x_{12})_\m\,x_{12}^\n\,x_{12}^\r}{x_{12}^2}\right)\right]\notag\\
&=-2\sdfrac{C_{12}}{x_{12}^{2(\D+1)}}\bigg\{(d-\D)\left[x_{12}^\r\,I^{\n\s}(x_{12})+x_{12}^\s\,I^{\n\r}(x_{12})\right]-\sdfrac{2\D}{d}\d^{\r\s}x_{12}^\n+2\,I^{\r\s}(x_{12})x_{12}^\n+4\sdfrac{x_{12}^\n\,x_{12}^\s\,x_{12}^\r}{x_{12}^2}\bigg\}\notag\\
&=-2\sdfrac{C_{12}}{x_{12}^{2(\D+1)}}\bigg\{(d-\D)\left[x_{12}^\r\,I^{\n\s}(x_{12})+x_{12}^\s\,I^{\n\r}(x_{12})\right]-2\left(\sdfrac{\D}{d}-1\right)\d^{\r\s}x_{12}^\n\bigg\}.\label{A5}
\end{align}
In the square bracket of \eqref{A5} there are independent tensor terms that do not add to zero. For this reason the only way to satisfy \eqref{A5} is when the coefficients of all the independent tensor structures are zero, implying that $\D=d$. We conclude that a primary operator of spin-$2$ is conserved if and only if its dimension is $\D=d$. 
These results ensure that the dimensions of $J^\m$ and $T^{\m\n}$ are protected in any CFT from any quantum correction. We will show explicitly the details of these properties in \secref{ConfWIMomentumSpace}. 
For the three-point  function one can proceed in a similar way, and observe that the problem can be greatly simplified by identifying all the possible tensor structures that can appear in a given correlation function \cite{Osborn:1993cr,Erdmenger:1996yc,Costa:2011mg}.

%%%%%%%%%%%%%%%%%%%%%%%%%%%%%%%%%%%%%%
\subsection{Conformal symmetries and Ward identities  in coordinate space}
%%%%%%%%%%%%%%%%%%%%%%%%%%%%%%%%%%%%%%
In this section we turn to a brief illustration of the derivation of the CWI's to be imposed on such correlators.
Considering a generating functional $Z[\phi_0^{(j)}]$, defined by
\begin{equation}
Z\left[\phi_0^{(j)}\right]=\int\,D\Phi\,\exp\left(-S-\sum_{j}\int d^dx\,\phi_0^{(j)}(x)\,\mathcal{O}_j(x)\right),
\end{equation}	
one can define the correlation function by functional differentiation with respect to the sources as
\begin{align}
\braket{\mathcal{O}_1(x_1)\dots \mathcal{O}_n(x_n)}=(-1)^n\left.\frac{\delta^n\,Z[\phi_0^{(j)}]}{\delta \phi_0^{(1)}(x_1)\dots\delta\phi_0^{(n)}(x_n)}\right|_{\phi_0^{(j)}=0}=\int\,D\Phi\,\mathcal{O}_1(x_1)\dots \mathcal{O}_n(x_n)\,e^{-S}.
\end{align}
Assuming that this correlation function is invariant under a symmetry $g$ - and the generating functional as well since the measure is invariant under such a transformation - one obtains the Ward identities for correlation functions of the form
\begin{equation}
\sum_{i=1}^n\braket{\mathcal{O}_1(x_1)\dots\delta_g\mathcal{O}_i(x_i) \dots\mathcal{O}_n(x_n)}=0,\label{globalW}
\end{equation} 
where for a general infinitesimal conformal transformation defined by $g(x)$ the variation $\delta_g\mathcal{O}_i(x_i)$ is defined as
\begin{equation}
\delta_g\mathcal{O}_i(x_i)=\left(-g\cdot\partial-\frac{\Delta}{d}\,\partial\cdot g-\frac{1}{2}\,\partial_{\,[\mu}g_{\,\nu]}\mathcal{J}_{\mu\nu}\right)\mathcal{O}_i(x_i),\label{transfSpecial}
\end{equation}
according to \eqref{transform}. This allows us to reobtain in a slightly more general framework the results of \secref{prima}. 
 For instance, in the case of translations, the Ward identity takes the form
\begin{equation}
0=\sum_{j=1}^n\sdfrac{\partial}{\partial x_j^\m}\braket{\mO_1(x_1)\dots\mO_n(x_n)},
\end{equation}
implying that the correlation function depends on the differences $x_i-x_j$ only. Then the dilatation WI's can be easily constructed by using the dilatation generator, for which we find
\begin{equation}
\left[\sum_{j=1}^n\,\D_j+\sum_{j=1}^n\,x_j^\a\sdfrac{\partial}{\partial x_j^\a}\right]\braket{\mO_1(x_1)\dots\mO_n(x_n)}=0,
\end{equation}
where here we indicate with $\D_1,\dots,\D_n$ the dimensions of the conformal primary operators $\mO_1,\dots,\mO_n$. Analogously to the previous cases, the special  CWI's, corresponding to special conformal transformations, can be derived for the scalar case, and in particular we have
\begin{equation}
\sum_{j=1}^n\left(2\D_j\,x_j^\k+2x_j^\k\,x_j^\a\sdfrac{\partial}{\partial x_{j\a}}-x_j^2\sdfrac{\partial}{\partial x_{j\k}}\right)\braket{\mO_1(x_1)\dots\mO_n(x_n)}=0,\label{special}
\end{equation}
where $\k$ is a free Lorentz index. In the case of tensor operators one needs to add an additional term to the previous equation, the contribution $\mathcal{J}_{\mu\nu}$ in \eqref{transfSpecial}, the finite-dimensional representation of the Lorentz group determining the spin of the field considered.\\
For example, if we assume that the tensor $\mO_j$ has $r_j$ Lorentz indices, i.e. $\mO_j=\mO_j^{\m_{j_1}\dots\m_{j_{r_j}}}$, for $j=1,2,\dots,n$, in this case we need to add the extra term
\begin{align}
2\sum_{j=1}^n\sum_{h=1}^{r_j}\left[(x_j)_{\a_{j_h}}\d^{\k\m_{j_h}}-x_j^{\m_{j_h}}\d^\k_{\a_{j_h}}\right]\braket{\mO_1^{\m_{1_{\scalebox{0.5}{1}}}\m_{1_{\scalebox{0.5}{2}}}\dots\m_{1_{r_{\scalebox{0.4}{ 1}}}}}(x_1)\dots\  \mO_j^{\m_{j_{\scalebox{0.5}{1}}}\dots\a_{j_{\scalebox{0.5}{h}}}\dots\m_{j_{r_{\scalebox{0.4}{j}}}}}(x_j)\ \dots\ \mO_n^{\m_{n_{\scalebox{0.5}{1}}}\m_{n_{\scalebox{0.5}{2}}}\dots\m_{n_{r_{\scalebox{0.4}{n}}}}}(x_n)}\label{spinpart}
\end{align}
to the left-hand side of \eqref{special}. Finally, we can consider the WI associated with rotations in coordinate space, by taking for $\delta_g$ a Lorentz transformation in \eqref{transfSpecial}. As in the previous case, we first consider the $n$-point function of scalar operators, obtaining
\begin{equation}
\sum_{j=1}^n\left(x_j^\n\sdfrac{\partial}{\partial x_{j\m}}-x_j^\m\sdfrac{\partial}{\partial x_{j\n}}\right)\braket{\mO_1(x_1)\dots\mO_n(x_n)}=0.\label{rot1x}
\end{equation}
For tensor operators one needs to add to the left-hand side of the previous equation the contribution
\begin{align}
\sum_{j=1}^n\sum_{h=1}^{r_j}\left[\d_{\n\a_{j_h}}\d^{\m_{j_h}}_\m-\d_{\m\a_{j_h}}\d_\n^{\m_{j_h}}\right]\braket{\mO_1^{\m_{1_{\scalebox{0.5}{$1$}}}\m_{1_{\scalebox{0.5}{$2$}}}\dots\m_{1_{r_{\scalebox{0.4}{$1$}}}}}(x_1)\dots\  \mO_j^{\m_{j_{\scalebox{0.5}{$1$}}}\dots\a_{j_{\scalebox{0.5}{$h$}}}\dots\m_{j_{r_{\scalebox{0.4}{$j$}}}}}(x_j)\ \dots\ \mO_n^{\m_{n_{\scalebox{0.5}{$1$}}}\m_{n_{\scalebox{0.5}{$2$}}}\dots\m_{n_{r_{\scalebox{0.4}{$n$}}}}}(x_n)},\label{rot2x}
\end{align}
that plays an important role in constraining the expression of the operator in momentum space.
%%%%%%%%%%%%%%%%%%%%%%%%%%%%%%%%%%%%%%%%%%%%%%%%%%%%%%%%%%
\section{The conformal Ward identities in momentum space }\label{ConfWIMomentumSpace}
%%%%%%%%%%%%%%%%%%%%%%%%%%%%%%%%%%%%%%%%%%%%%%%%%%%%%%%%%% 
As already mentioned in the previous sections, most of the current and past analysis in 
CFT has been centered around coordinate space. This is the natural domain where primary operators are introduced in order to discuss the fluctuations of physical systems around a certain critical point, described by a certain correlation function, as a function of distance. The operator algebra of a CFT  is endowed with the operator product expansion, which allows to express the product of two operators at separate points in terms of an infinite series of of local operators \cite{ Ferrara:1973yt,Dolan:2000ut, Poland:2018epd, Poland:2016chs}. A problem arises at coincident points, i.e. when two or more primary operators in the product approach the same spacetime point. For stress energy tensors, this limit introduces an anomaly in the operator algebra. The stress energy tensor is not traceless any more, as expected in a CFT, and the conformal anomaly is the manifestation of the breaking of the quantum conformal symmetry.\\
As we move to momentum space, due to the Fourier transform acting on the spacetime points of a correlation function, the integration over its coordinates includes also domains in which such coordinates coincide. Therefore, the contributions from the anomaly, generated when the coordinates of the correlators coincide, are naturally included in the expression of a correlator in momentum space \cite{Capper:1975ig, Deser:1976yx, Riegert:1984kt, Coriano:2017mux}. For instance, in the case of correlators of stress energy tensors, 
multiple traces of the original $n$-point correlator are affected by the anomalies of those of lower orders 
($n-1$, $n-2$ and so on). This means that the anomaly of the original $n$-point function is generated not only  when all the external points go into coincidence, but also when a subset of them does it (see \cite{Coriano:2021nvn} for four-point functions of stress energy tensors). 
  
%%%%%%%%%%%%%%%%%%%%%
%%%%%%%%%%%%%%%%%%%%%
\subsection{The dilatation equation}
%%%%%%%%%%%%%%%%%%%%%
%%%%%%%%%%%%%%%%%%%%%
We are now going, as a first step, to reformulate the conformal constraints (i.e. the conformal Ward identities) in  momentum space, proceeding from the scalar case and then moving to the tensor case.
We will be using some condensed notations in order to shorten the expressions 
of the transforms in momentum space. We will use the following conventions
\begin{align}
\label{conv}
& \Phi(\underline{x})\equiv \langle \phi_1(x_1)\phi_2(x_2)\ldots \phi_n(x_n)\rangle, &&\hspace{-2cm} e^{i \underline{p x}}\equiv e^{i(p_1 x_1 + p_2 x_2 + \ldots p_n x_n)},  \notag\\
& \underline{d p}\equiv dp_1 dp_2 \ldots d p_n, &&\hspace{-2cm}
\Phi(\underline{p})\equiv \Phi(p_1,p_2,\ldots, p_n), 
\end{align}
with $\Phi$ denoting an $n$-point correlation function of primary operators $\phi_i$.
It will also be useful to introduce the total momentum $P=\sum_{j=1}^{n} p_j$.\\
The momentum constraint is enforced via a delta function $\delta(P)$ under integration. For instance, translational invariance of $\Phi(\underline{x})$ gives 
\be
\label{ft1}
\Phi(\underline{x})=\int \underline{dp}\,\delta(P) \,e^{i\underline{p x}} \,\Phi(p_1,\dots,p_n).
\ee
In general, for an $n$-point function $\Phi(x_1,x_2,\ldots, x_n)=\langle \phi_1(x_1)\phi_2(x_2)...\phi_n(x_n)\rangle $, the condition of translational invariance  
\be
\langle 
 \phi_1(x_1)\phi_2(x_2),\ldots, \phi_n(x_n)\rangle = \langle\phi_1(x_1+a )\phi_2(x_2+a)\ldots \phi_n(x_n+a)\rangle
 \ee 
 generates the expression in momentum space of the form (\ref{ft1}), from which we can remove one of the momenta, conventionally the last one $p_n$, which is replaced by 
 $\overline{p}_n=-(p_1+p_2 +\ldots p_{n-1})$, giving
 \be
 \Phi(x_1,x_2,\ldots,x_n)=\int dp_1 dp_2... dp_{n-1}e^{i(p_1 x_1 + p_2 x_2 +...p_{n-1} x_{n-1} + 
 \overline{p}_n x_n)}\Phi(p_1,p_2,\ldots,\overline{p}_n).
 \ee
We start by considering the dilatation WI. The condition of scale covariance for the fields $\phi_i$ of scale dimensions $\Delta_i$ (in mass units)
 \be
 \label{scale1}
 \Phi(\lambda x_1,\lambda x_2,\ldots,\lambda x_n)=\lambda^{-\Delta} \Phi(x_1,x_2,\ldots, x_n), \qquad 
 \Delta=\Delta_1 +\Delta_2 +\ldots \Delta_n,
 \ee
 after setting $\lambda=1 +\epsilon$ and Taylor expanding up to $O(\epsilon)$ gives the 
 scaling relation 
 \be
 \label{ft2}
(D_n + \Delta)\Phi\equiv \sum_{j=1}^n \left(x_j^\alpha \frac{\partial}{\partial x_j^\alpha} +\Delta_j\right) \Phi(x_1,x_2,\ldots,x_n) =0,
 \ee 
 with 
 \be
 D_n=\sum_{j=1}^n x_j^\alpha\frac{\partial}{\partial x_j^\alpha}. 
 \ee
The expression of the dilatation equation in momentum space can be obtained either by a Fourier transform 
of (\ref{ft2}), or more simply, exploiting directly (\ref{scale1}). In the latter case, using the translational invariance of the correlator under the integral, by removing the $\delta$-function constraint, one obtains
 \begin{align}
 \Phi(\lambda x_1,\lambda x_2,\ldots,\lambda x_n)&= \int d^d p_1 d^d p_2\ldots d^d p_{n-1} e^{i\lambda (p_1 x_1 + p_2 x_2 +...p_{n-1} x_{n-1} + 
 \overline{p}_n x_n)}\Phi(p_1,p_2,\ldots,\overline{p}_n)\nonumber \\
&=\lambda^{-\Delta}  \int d^d p_1 d^d p_2\ldots d^d p_{n-1} e^{i(p_1 x_1 + p_2 x_2 +...p_{n-1} x_{n-1} + 
 \overline{p}_n x_n)}\Phi(p_1,p_2,\ldots,\overline{p}_n).
 \end{align}
We perform the change of variables $ p_i=p'_i/\lambda$ on the right-hand-side (rhs) of the equation above (first line) with $d p_1...d p_{n-1}=(1/\lambda)^{d(n-1)} d^d p'_1 \ldots d^d p'_{n-1}$ to derive the relation 
 \begin{equation}
 \frac{1}{{\lambda}^{d(n-1)}}\Phi\left(\frac{p_1}{\lambda},\frac{p_2}{\lambda},\ldots,\frac{\overline{p}_n}{\lambda}\right)=\lambda^{-\Delta} \Phi(p_1,p_2,\ldots,\overline{p}_n).
\end{equation}
Setting $\lambda=1/s$ this generates the condition
\be
 s^{(n-1) d-\Delta}\Phi(s p_1,s p_2,\ldots, s \overline{p}_n)=\Phi(p_1,p_2,\ldots,\overline{p}_n),
\ee
and with $s\sim 1 +\epsilon$, expanding at $O(\epsilon)$ we generate the equation 
\be
\label{sc1}
\left[\sum_{j=1}^n \Delta_j  -(n-1) d -\sum_{j=1}^{n-1}p_j^\alpha \frac{\partial}{\partial p_j^\alpha}\right]
\Phi(p_1,p_2,\ldots,\overline{p}_n)=0.
\ee
There are some important comments to be made. The action of any differential operator which is separable on each of the coordinates $x_i$, once transformed to momentum space, violates the Leibniz rule if we want to differentiate only the independent momenta. This is because of momentum conservation, which is a consequence of the translational invariance of the correlator. This point has been illustrated at length in \cite{Coriano:2018bbe}, to which we refer for further details. Notice that in \eqref{sc1} the sum runs over the first $n-1$ momenta. The equations, though, must be reduced to a scalar form, and at that stage, their hypergeometric structure will appear clear with the procedure first presented in \cite{Coriano:2013jba, Bzowski:2013sza}.
%%%%%%%%%%%%%%%%%%%%%%%%%%%%%%%%
%%%%%%%%%%%%%%%%%%%%%%%%%%%%%%%%
\subsection{Special conformal WI's for scalar correlators} 
%%%%%%%%%%%%%%%%%%%%%%%%%%%%%%%%
%%%%%%%%%%%%%%%%%%%%%%%%%%%%%%%%
We now turn to the analysis of the special conformal transformations in momentum space. 
Also in this case we discuss both the symmetric and the asymmetric forms of the equations, focusing our attention first on the scalar case. 
The Ward identity in the scalar case is given by
\be
\sum_{j=1}^{n} \left(- x_j^2\frac{\partial}{\partial x_{j\kappa}}+ 2 x_j^\kappa x_j^\alpha \frac{\partial}
{\partial x_j^\alpha} +2 \Delta_j x_j^\kappa\right)\Phi(x_1,x_2,\ldots,x_n) =0,
\ee
which in momentum space, using 
\be
x_j^\alpha\to -i \frac{\partial}{\partial p_j^\alpha}, \qquad \frac{\partial}{\partial x_{j\kappa}}\to i p_j^\kappa,
\ee
becomes 
\be
\sum_{j=1}^n \int \underline {d^d p}\left(p_j^\kappa \frac{\partial^2}{\partial p_j^\alpha \partial p_{j\alpha}} -
2 p_j^\alpha \frac{\partial^2}{\partial p_j^\alpha \partial p_{j\kappa}}  -2 \Delta_j\frac{\partial}{\partial p_{j\kappa}}\right) 
e^{i\underline{p\cdot x}} \,\delta^d(P)\phi(\underline{p})=0,
\ee
where the action of the operator is only on the exponential. At this stage we integrate by parts, bringing the derivatives from the exponential to the correlator and on the Dirac $\delta$ function obtaining 
\be
\int \underline{d^d p}\,e^{i \underline{p x}} \,K_s^\kappa\,\Phi(\underline{p})\delta^d(P), +\delta'_\textrm{term}  =0,
\ee
in the notations of \eqref{conv}, where we have introduced the differential operator acting on a scalar correlator in a symmetric form
\be
 \,K_s^\kappa=\sum_{j=1}^n\left(p_j^\kappa \frac{\partial^2}{\partial p_j^\alpha\partial p_j^\alpha} + 2(\Delta_j- d)\frac{\partial}{\partial p_{j\kappa}}-2 p_j^\alpha\frac{\partial^2}{\partial p_{j\kappa }\partial p_j^\alpha}\right).
\ee
Using some distributional identities derived in \cite{Coriano:2018bbe}
 \begin{align}
 \label{uno}
\,K_s^\k \delta^d(P)&=\left(P^\k\frac{\partial^2}{\partial P^\alpha \partial P_\alpha}  -2 P^\alpha \frac{\partial^2}{\partial P^\alpha\partial P_\k} 
+2 (\Delta- n\, d)\frac{\partial }{\partial P_\k}\right)\delta^d(P)\nn\\
&=2 d( d \,n -  d -\Delta)P^\k\frac{\delta^d(P)}{P^2}\nn\\
&=-2 ( d \,n -  d -\Delta)\frac{\partial}{\partial P_\k} {\delta^d(P)},
\qquad \qquad \Delta=\sum_{j=1}^n\Delta_j,
\end{align}
we obtain
\begin{align}
\delta'_\textrm{term}&= \int {\underline{d^d p}}\,e^{i\underline{p\cdot x}}
\left[ \frac{\partial}{\partial P^\alpha} \delta^d(P)\sum_{j=1}^n\left(p_j^\alpha\frac{\partial}{\partial p_{j\kappa}}- p_j^\kappa\frac{\partial}{\partial p_{j\alpha}}\right)\Phi(\underline{p})  + 2 \frac{\partial}{\partial P_\kappa} \delta^d(P)\left(\sum_{j=1}^n \left(\Delta_j -  p_j^\alpha 
\frac{\partial}{\partial p_j^\alpha} \right)- (n-1) d \right)\Phi(\underline{p})\right].
\end{align}
Notice that such terms vanish \cite{Coriano:2018bbe} by using the rotational invariance of the scalar correlator 
\be
\sum_{j=1}^3\left(p_j^\alpha\frac{\partial}{\partial p_{j\kappa}}- p_j^\kappa\frac{\partial}{\partial p_{j\alpha}}\right)\Phi(\underline{p})=0,
\ee
as a consequence of the SO(4) symmetry 
 \be
\sum_{j=1}^3 L^{\mu\nu}(x_j) \langle \phi(x_1)\phi(x_2)\phi(x_3)\rangle =0, 
\ee
with 
\be
L_{\mu\nu}(x)=i\left(x^\mu\partial^\nu - x^\nu\partial^\mu \right),
\ee
and the symmetric scaling relation,
\be
\left(\sum_{j=1}^n \Delta_j - \sum_j^{n-1} p_j^\alpha 
\frac{\partial}{\partial p_j^\alpha} - (n-1) d \right)\Phi(\underline{p})=0.
\ee
Using (\ref{uno}) and the vanishing of the $\delta'_\textrm{term}$ terms, the structure of the CWI on the correlator $\Phi(p)$ then takes the symmetric form 
 \be
 \label{sm1}
  \sum_{j=1}^n \int \underline {d^d p}e^{i\underline{p\cdot x}}\left(p_j^\kappa \frac{\partial^2}{\partial p_{j\alpha} \partial p_j^\alpha} -
2 p_j^\alpha \frac{\partial^2}{\partial p_j^\alpha \partial p_{j\kappa}}  + 2( \Delta_j-d)\frac{\partial}{\partial p_{j\kappa}}\right) 
 \phi(\underline{p})\delta^d(P)=0.
\ee
This {\em symmetric} expression is the starting point in order to proceed with the elimination of one of the momenta, say $p_n$. 
Also in this case, one can proceed by following the same procedure used in the derivation of the dilatation identity, dropping the contribution coming from the dependent momentum $p_n$, thereby obtaining the final form of the equation

\begin{equation}
\sum_{j=1}^{n-1}\left(p_j^\kappa \frac{\partial^2}{\partial p_j^\alpha\partial p_{j\alpha}} + 2(\Delta_j- d)\frac{\partial}{\partial p_{j\kappa}}-2 p_j^\alpha\frac{\partial^2}{\partial p_{j\kappa }\partial p_j^\alpha}\right)\Phi(p_1,\ldots p_{n-1},\bar{p}_n)=0.\label{SpCWIs}
\end{equation}
Also in this case the differentiation with respect to $p_n$ requires the chain rule. For a certain sequence of scalar single particle operators 
\be
 \Phi(p_1,\ldots p_{n-1},\bar{p}_n)=\langle \phi(p_1)\ldots \phi (\bar{p}_n)\rangle,
\ee
the Leibniz rule is therefore violated. As we have already mentioned, the system of scalar equations obtained starting from the tensor one are, however, symmetric. For further details 
concerning this point and the arbitrariness in the choice of the independent momentum of the correlator we refer to  \cite{Coriano:2018bbe}. 

\subsection{Conformal constraints on two-point  functions\label{TwoPointSection}}
For two-point functions, the differential equations simplify considerably, being expressed in terms 
of just one independent momentum $p$. We consider two operators $O_1$ and $O_2$ and the corresponding correlation function
 $G^{ij}(p) \equiv \langle \mathcal O_1^i(p) \mathcal O_2^j(-p) \rangle$, 
where the indices $i$ and $j$ run over a specific representation of the Lorentz group. In this case we have 
\begin{equation}
\label{ConformalEqMomTwoPoint}
 \left( - p_{\mu} \, \frac{\partial}{\partial p_{\mu}}  + \Delta_1 + \Delta_2 - d \right) G^{ij}(p) = 0 \,, 
\end{equation}
for the dilatation and 
\begin{equation}
 \left(  p_{\mu} \, \frac{\partial^2}{\partial p^{\nu} \partial p_{\nu}}  - 2 \, p_{\nu} \, \frac{\partial^2}{ \partial p^{\mu} 
\partial p_{\nu} }    + 2 (\Delta_1 - d) \frac{\partial}{\partial p^{\mu}}\right) G^{ij}(p)  + 2 (\Sigma_{\mu\nu})^{i}_{k} \frac{\partial}{\partial 
p_{\nu}}  G^{kj}(p)   = 0 \,,
\label{duedue}
\end{equation}
for the special conformal Ward identities. The equation \eqref{ConformalEqMomTwoPoint} dictates the scaling behavior of the correlation function, while special conformal 
invariance allows a non-zero result only for equal scale dimensions of the two operators $\Delta_1 = \Delta_2$, as we know from the 
corresponding analysis in coordinate space. \\
For the correlation function $G_S$ of two scalar quasi primary fields, the invariance under the Poincar\'{e} group obviously 
implies that $G_S \equiv G_S(p^2)$, so that the derivatives with respect to the momentum $p_\mu$ can be easily recast in terms of 
the variable $p^2$. \\
The invariance under scale transformations implies that $G_S(p^2)$ is a homogeneous function of degree 
$\alpha = \frac{1}{2}(\Delta_1 + \Delta_2 - d)$. 
At the same time, it is easy to show that \eqref{duedue} can be satisfied only if $\Delta_1 = \Delta_2$. 
Therefore conformal symmetry fixes the structure of the scalar two-point function up to an arbitrary overall constant $C$ as
\begin{equation}
\label{TwoPointScalar}
G_S(p^2) = \langle \mathcal O_1(p) \mathcal O_2(-p) \rangle = \delta_{\Delta_1 \Delta_2}  \, C\, (p^2)^{\Delta_1 - d/2} \, .
\end{equation}
If we redefine
\be
C=c_{S 12} \,  \frac{\pi^{d/2}}{4^{\Delta_1 - d/2}} \frac{\Gamma(d/2 - \Delta_1)}{\Gamma(\Delta_1)},
\ee
in terms of the new integration constant $c_{S 12}$, the two-point function takes the form
\be
\label{TwoPointScalar2}
G_S(p^2) =  \delta_{\Delta_1 \Delta_2}  \, c_{S 12} \,  \frac{\pi^{d/2}}{4^{\Delta_1 - d/2}} \frac{\Gamma(d/2 - \Delta_1)}{\Gamma(\Delta_1)} 
(p^2)^{\Delta_1 - d/2} \,,
\ee
and after a Fourier transform in coordinate space it takes the familiar form
\be
\langle \mathcal O_1(x_1) \mathcal O_2(x_2) \rangle \equiv \mathcal{F.T.}\left[ G_S(p^2) \right] =  \delta_{\Delta_1 \Delta_2} \,  c_{S 12} 
\frac{1}{(x_{12}^2)^{\Delta_1}} \,,
\ee
where $x_{12} = x_1 - x_2$. 
The ratio of the two Gamma functions relating the two integration constants $C$ and $c_{S 12}$ correctly reproduces the ultraviolet singular behavior of the correlation function and plays a role in the discussion of the origin of the scale anomaly.

Now we turn to the vector case where we define $G_V^{\alpha \beta}(p) \equiv \langle V_1^\alpha(p) V_2^\beta(-p) \rangle$. If the 
vector current is conserved, $\partial^\mu V_\mu = 0$, then the tensor structure of the two-point correlation function is entirely fixed as 
\begin{equation}
\label{TwoPointVector0}
G_V^{\alpha \beta}(p) =  \pi^{\alpha\beta}(p) \, f_V(p^2)\,, \quad   \text{with} \quad 
\pi^{\alpha\beta}(p) = \eta^{\alpha \beta} -\frac{p^\alpha p^\beta}{p^2}, 
\end{equation}
where $f_V$ is a function of the invariant square $p^2$ whose form, as in the scalar case, is determined by the conformal constraints. 
Following the same reasoning discussed previously we find that
\begin{equation}
\label{TwoPointVector}
G_V^{\alpha \beta}(p) = \delta_{\Delta_1 \Delta_2}  \, c_{V 12}\, 
\frac{\pi^{d/2}}{4^{\Delta_1 - d/2}} \frac{\Gamma(d/2 - \Delta_1)}{\Gamma(\Delta_1)}\,
\left( \eta^{\alpha \beta} -\frac{p^\alpha p^\beta}{p^2} \right)\
(p^2)^{\Delta_1-d/2} \,,
\end{equation}
with $c_{V12}$ being an arbitrary constant. 
We recall that \eqref{duedue} gives consistent results for the two-point function in \eqref{TwoPointVector} only when the scale dimension $\Delta_1 = d - 1$. 
To complete this short excursus, we present the solution of the conformal constraints for the two-point function built out of two energy momentum tensor operators which are symmetric, conserved and traceless
\be
\label{EMTconditions}
T_{\mu\nu} = T_{\nu\mu} \,,  \qquad \partial^{\mu} T_{\mu\nu} = 0 \,,  \qquad {T_{\mu}}^{\mu} = 0 \,.
\ee
Exploiting the conditions defined in \eqref{EMTconditions} we can unambiguously define the tensor structure of the correlation 
function $G^{\alpha\beta\mu\nu}_T(p) = \Pi_{d}^{\alpha\beta\mu\nu}(p) \, f_T(p^2)$ with
\be 
\label{TT}
\Pi^{\alpha\beta\mu\nu}_{d}(p) = \frac{1}{2} \bigg[ \pi^{\alpha\mu}(p) \pi^{\beta\nu}(p) + \pi^{\alpha\nu}(p) \pi^{\beta\mu}(p) 
\bigg] 
- \frac{1}{d-1} \pi^{\alpha\beta}(p) \pi^{\mu\nu}(p) \,,
\ee
and the scalar function $f_T(p^2)$ determined as usual, up to a multiplicative constant, by requiring the invariance under 
dilatations and special conformal transformations. We obtain
\begin{equation}
\label{TwoPointEmt}
G^{\alpha\beta\mu\nu}_T(p) = \delta_{\Delta_1 \Delta_2}  \, 
c_{T 12}\,\frac{\pi^{d/2}}{4^{\Delta_1 - d/2}} \frac{\Gamma(d/2 - \Delta_1)}{\Gamma(\Delta_1)}\, 
\Pi^{\alpha\beta\mu\nu}_{d}(p) \, (p^2)^{\Delta_1 - d/2} \,.
\end{equation}
As for the conserved vector currents, also for the energy momentum tensor the scaling dimension is fixed by \eqref{duedue} and it is given by $\Delta_1 = d$. This particular value ensures that $\partial^\mu T_{\mu\nu}$ is also a quasi primary (vector) field. 

\subsection{More about two-point  functions}
\label{AppTwoPoint}
In this section we provide some details on the solutions of the conformal constraints for the two-point functions with conserved vector and tensor operators. \\
In the first case the tensor structure of the two-point function is uniquely fixed by the condition $\partial^\mu V_\mu=0$ as
\be
G_V^{\alpha \beta}(p) = f(p^2) t^{\alpha \beta}(p)\,, \qquad \mbox{with} \quad t^{\alpha\beta}(p) = p^2 \eta^{\alpha\beta} - p^{\alpha} p^{\beta} \,.
\ee 
For simplicity, we have employed in the previous equation a slightly different notation with respect to \eqref{TwoPointVector0}, that can be recovered, anyway, with the identification $f(p^2) = f_V(p^2)/p^2$. \\
In order to exploit the invariance under scale and special conformal transformations it is useful to compute first and second order derivatives of the $t^{\alpha \beta}$ tensor structure. In particular we have
\begin{eqnarray}
\label{TDerivatives}
t_1^{\alpha \beta, \mu}(p) &\equiv& \frac{\partial }{\partial p_\mu} t^{\alpha \beta}(p) = 2 \, p^{\mu} \eta^{\alpha \beta} - p^{\alpha} \eta^{\mu \beta} - p^{\beta} \eta^{\mu \alpha} \,, \nn \\
t_2^{\alpha \beta, \mu \nu}(p) &\equiv& \frac{\partial^2 }{\partial p_\mu \, \partial p_\nu} t^{\alpha \beta}(p) = 2 \, \eta^{\mu \nu} \eta^{\alpha \beta} - \eta^{\nu \alpha} \eta^{\mu \beta} - \eta^{\nu \beta} \eta^{\mu \alpha} \,, 
\end{eqnarray}
with the properties
\begin{eqnarray}
&& p_\mu t_1^{\alpha \beta, \mu}(p) = 2 \, t^{\alpha \beta}(p) \,, \qquad     t_1^{\alpha \beta, \alpha}(p) = - (d - 1) p^\beta \,, \nn \\
&& p_\mu t_2^{\alpha \beta, \mu \nu}(p) = t_1^{\alpha \beta, \nu}(p) \,, \qquad    t_2^{\alpha \beta, \mu \mu}(p) = 2(d-1) \eta^{\alpha \beta} \,. 
\end{eqnarray}
As we have already mentioned, the invariance under scale transformations implies that
\be
\label{FandLambda}
f(p^2) = (p^2)^\lambda , \qquad \mbox{with} \quad \lambda = \frac{\Delta_1 + \Delta_2 - d}{2} -1 \,,
\ee
which can be easily derived from the first order differential equation in \eqref{ConformalEqMomTwoPoint} by using \eqref{TDerivatives}. Having determined the structure of the scalar function $f(p^2)$, one can compute the derivatives appearing in \eqref{duedue}, namely the constraint following from invariance under the special conformal transformations
\begin{eqnarray}
\label{GDerivatives}
\frac{\partial}{\partial p_\mu} G_V^{\alpha \beta}(p) &=& (p^2)^{\lambda - 1} \bigg[ 2 \lambda \, p^\mu t^{\alpha \beta}(p) + p^2 \, t_1^{\alpha\beta, \mu}(p) \bigg] \,, \nn \\
\frac{\partial^2}{\partial p_\mu \, \partial p_\nu} G_V^{\alpha \beta}(p) &=&(p^2)^{\lambda - 2} \bigg[ 4 \lambda (\lambda -1) p^{\mu} p^{\nu} t^{\alpha \beta}(p)  + 2 \lambda p^2 \eta^{\mu\nu}   t^{\alpha \beta}(p) + 2 \lambda  p^2 p^{\mu} t_1^{\alpha\beta,\nu}(p)  \nn \\
&& \qquad +  \, 2 \lambda  p^2 p^{\nu} t_1^{\alpha\beta,\mu}(p) + (p^2)^2 t_2^{\alpha\beta, \mu\nu}(p) 
\bigg] \,,
\end{eqnarray}
where we have used the definitions in \eqref{TDerivatives}.
Concerning the spin dependent part in \eqref{duedue}, we use the spin matrix for the vector field, that in our conventions is given by
\be
( \Sigma_{\mu\nu}^{(V)})^{\alpha}_{\beta} = \delta_{\mu}^{\alpha} \, \eta_{\nu \beta} - \delta_{\nu}^{\alpha} \, \eta_{\mu \beta} \,,
\ee
and obtain
\be
\label{SigmaPart}
2( \Sigma_{\mu\nu}^{(V)})^{\alpha}_{\rho} \frac{\partial}{\partial p_\nu} G_V^{\rho\beta}(p) = - (p^2)^{\lambda- 1} \bigg[ 2 \lambda \, p^\alpha {t_{\mu}}^{\beta}(p) + (d-1)p^2 p^\beta \delta_\mu^\alpha  + p^2 {t_{1 \, \mu}}^{\beta, \alpha}(p) \bigg] \,.
\ee
Using the results derived in \eqref{GDerivatives} and \eqref{SigmaPart}, we have fully determined the special conformal constraint on the two-point vector function.
Then we can project \eqref{duedue} onto the three independent tensor structures, and set $\lambda$ to the value given in \eqref{FandLambda}, obtaining three equations for the scale dimensions $\Delta_i$ of the vector operators
\be
\begin{cases}
(\Delta_1 - \Delta_2) (\Delta_1 + \Delta_2 - d) = 0 \,, \nn \\
\Delta_1 - d +1 = 0 \,, \nn \\
\Delta_2 - d +1 = 0 \,. \nn
\end{cases}
\\
\ee
The previous system of equations can be consistently solved only for $\Delta_1 = \Delta_2 = d -1$, as expected. This completes our derivation of the vector two-point function which, up to an arbitrary multiplicative constant, can be written as in \eqref{TwoPointVector}.

The characterization of the two-point function with a symmetric, traceless and conserved rank-2 tensor follows the same lines of reasoning of the vector case. These conditions (see \eqref{EMTconditions}) fix completely the tensor structure of the two-point function as
\be
G_T^{\alpha\beta\mu\nu}(p) = g(p^2) \, T^{\alpha\beta\mu\nu}(p),
\ee
with
\be
T^{\alpha\beta\mu\nu}(p) =  \frac{1}{2} \bigg[ t^{\alpha\mu}(p) t^{\beta\nu}(p) + t^{\alpha\nu}(p) t^{\beta\mu}(p) 
\bigg] 
- \frac{1}{d-1} t^{\alpha\beta}(p) t^{\mu\nu}(p) \,.
\ee
For consistency with the convention used in \secref{TwoPointSection} we have set $g(p^2) \equiv f_T(p^2)/(p^2)^2$. \\
As in the previous case, we give the first and second order derivatives of the $T^{\alpha\beta\mu\nu}(p)$ tensor structure
\begin{align}
\label{TTDerivatives}
T_1^{\alpha\beta\mu\nu, \rho}(p) &\equiv\frac{\partial}{\partial p_\rho} T^{\alpha\beta\mu\nu}(p) = \frac{1}{2} \bigg[ t_1^{\alpha\mu, \rho}(p) t^{\beta\nu}(p) 
+ t^{\alpha\mu}(p) t_1^{\beta\nu, \rho}(p) + \left( \mu \leftrightarrow \nu \right) \bigg] \nn \\
& - \frac{1}{d-1} \bigg[ t_1^{\alpha\beta, \rho}(p) t^{\mu\nu}(p) + t^{\alpha\beta}(p) t_1^{\mu\nu, \rho}(p) \bigg] \,, \nn \\
T_2^{\alpha\beta\mu\nu, \rho\sigma}(p) &\equiv \frac{\partial}{\partial p_\rho \, \partial p_\sigma} T^{\alpha\beta\mu\nu}(p) = \frac{1}{2} \bigg[ 
t_2^{\alpha\mu, \rho \sigma}(p) t^{\beta\nu}(p) + t_1^{\alpha\mu, \rho}(p) t_1^{\beta\nu, \sigma}(p) + t_1^{\alpha\mu, \sigma}(p) t_1^{\beta\nu, \rho}(p) \nn \\
& +  \, t^{\alpha\mu}(p) t_2^{\beta\nu, \rho \sigma}(p)+ \left( \mu \leftrightarrow \nu \right) \bigg] - \frac{1}{d-1} \bigg[  t_2^{\alpha\beta, \rho\sigma}(p) t^{\mu\nu}(p) +   t_1^{\alpha\beta, \rho}(p) t_1^{\mu\nu, \sigma}(p) + (\mu\nu) \leftrightarrow (\alpha\beta)  \bigg] \,,
\end{align}
together with some of their properties
\begin{eqnarray}
p_\rho T_1^{\alpha\beta\mu\nu, \rho}(p) = 4 \, T^{\alpha\beta\mu\nu}(p) \,, \qquad
p_\rho T_2^{\alpha\beta\mu\nu, \rho \sigma}(p) = 3 \, T_1^{\alpha\beta\mu\nu, \sigma}(p)  \,.
\end{eqnarray}
As we have already stressed, \eqref{ConformalEqMomTwoPoint} defines the scaling behavior of the two-point function, providing, therefore, that the functional form of $g(p^2)$ is given by
\be
g(p^2) = (p^2)^\lambda, \qquad \mbox{with} \quad \lambda = \frac{\Delta_1 +\Delta_2 -d}{2} -2 \,.
\ee
On the other hand, \eqref{duedue}, which represents the constraint from the special conformal transformations, fixes the scaling dimensions of the tensor operators. In this case the spin connection is given by
\be
(\Sigma_{\mu\nu}^{(T)})^{\alpha \beta}_{\rho \sigma } = \left( \delta_{\mu}^{\alpha} \, \eta_{\nu \rho} - \delta_{\nu}^{\alpha} \, \eta_{\mu \rho} \right) \delta_{\sigma}^{\beta}
+ \left( \delta_{\mu}^{\beta} \, \eta_{\nu \sigma} - \delta_{\nu}^{\beta} \, \eta_{\mu \sigma} \right) \delta_{\rho}^{\alpha} \,.
\ee
The algebra is straightforward but rather cumbersome due to the proliferation of indices. Here we give only the final result, which can be obtained by projecting \eqref{ConformalEqMomTwoPoint} and by using \eqref{TTDerivatives} in all the different independent tensor structures, giving
\begin{equation}
\begin{cases}
(\Delta_1 - \Delta_2) (\Delta_1 + \Delta_2 - d) = 0 \,, \nn \\
\Delta_1 - d = 0 \,, \nn \\
\Delta_2 - d = 0 \,, \nn
\end{cases}
\\
\end{equation}
which implies that $\Delta_1 = \Delta_2 = d$, as described in \eqref{TwoPointEmt}.
%%%%%%%%%%%%%%%%%%%%%%%%%%%%%%%%%%%%%%%%%
\subsection{Conformal Ward identities from the vector to the scalar form}
%%%%%%%%%%%%%%%%%%%%%%%%%%%%%%%%%%%%%%%%% 
We now come to illustrate the procedure for obtaining the conformal constraints in the form of partial differential equation with respect to scalar invariants.\\
We consider first the case of three-point function of scalar primary operators. Due to the translational invariance, the correlator can be expressed as a function of three invariants, i.e. the magnitudes of the momenta, defined as ${p}_i=\sqrt{p_i^2}$
\begin{equation}
\braket{\mathcal{O}_1(p_1)\mathcal{O}_2(p_2)\mathcal{O}_3(\bar{p}_3)}=\Phi(p_1,p_2,p_3).
\end{equation} 
All the conformal WI's can be re-expressed in scalar form using the chain rules
\begin{equation}
\label{chainr}
\frac{\partial \Phi}{\partial p_i^\mu}=\frac{p_i^\mu}{  p_i}\frac{\partial\Phi}{\partial  p_i} 
-\frac{\bar{p}_3^\mu}{  p_3}\frac{\partial\Phi}{\partial   p_3}, \quad i=1,2,
\end{equation}
where $\bar{p}_3^\mu=-p_1^\mu-p_2^\mu$ and $p_3=\sqrt{(p_1+p_2)^2}$. By using this equation, we re-write the covariant differential operator 
\begin{equation}
\sum_{i=1}^2\,p_i^\mu\,\frac{\partial }{{p_i}^\mu}\Phi(p_1,p_2,p_3)=\left(
{p}_1\frac{ \partial }{\partial   p_1} +   p_2\frac{ \partial }{\partial   p_2} +   p_3\frac{ \partial }{\partial   p_3}\right)\Phi(p_1,p_2,p_3).
\end{equation}
Therefore, the scale equation becomes 
\begin{equation}
\label{scale}
\left(\sum_{i=1}^3\Delta_i -2 d - \sum_{i=1}^3    p_i \frac{ \partial}{\partial   p_i}\right)\Phi(p_1,p_2,\bar{p}_3)=0.
\end{equation}
It takes a straightforward but lengthy computation to show that the special conformal Ward identities in $d$ dimensions take the form
\begin{align}
0&={K}_{scalar}^{\kappa}\Phi(p_1,p_2,p_3)=\bigg(p_1^\kappa\,K_1+p_2^\kappa\,K_2+\bar{p}_3^\kappa\,K_3\bigg)\Phi(p_1,p_2,p_3)\notag\\
&=p_1^\kappa\Big(K_1-K_3\Big)\Phi(p_1,p_2,p_3)+p_2^\kappa\Big(K_2-K_3\Big)\Phi(p_1,p_2,p_3),\label{SCWIs}
\end{align}
where we have used the conservation of the total momentum, with the $K_i$ operators defined as
\begin{equation}
{ K}_i\equiv \frac{\partial^2}{\partial    p_i \partial    p_i} 
+\frac{d + 1 - 2 \Delta_i}{   p_i}\frac{\partial}{\partial   p_i}, \quad i=1,2,3. 
\end{equation}
The equation \eqref{SCWIs} is satisfied if every coefficient of the independent four-momenta $p_1^\mu,\,p_2^\mu$ is equal to zero. This condition leads to the scalar form of the special conformal constraints 
\begin{equation}
\frac{\partial^2\Phi}{\partial   p_i\partial   p_i}+
\frac{1}{  p_i}\frac{\partial\Phi}{\partial  p_i}(d+1-2 \Delta_1)-
\frac{\partial^2\Phi}{\partial   p_3\partial   p_3} -
\frac{1}{  p_3}\frac{\partial\Phi}{\partial  p_3}(d +1 -2 \Delta_3)=0\qquad i=1,2,
\label{3k1}
\end{equation}
and defining 
\begin{equation}
\label{kij}
K_{ij}\equiv {K}_i-{K}_j,
\end{equation}
\eqref{3k1} can be written in the concise form 
\begin{equation}
\label{3k2}
K_{13}\,\Phi(p_1,p_2,p_3)=0, \qquad K_{23}\,\Phi(p_1,p_2,p_3)=0.
\end{equation}
Notice that in the derivation of \eqref{SCWIs} one needs at an intermediate step the derivative of the scaling WI 
\begin{equation}
  p_1\frac{\partial^2\Phi}{\partial   p_3\partial  p_1} 
+   p_2\frac{\partial^2\Phi}{\partial   p_3\partial  p_2}=(\Delta -2 d -1)
\frac{\partial\Phi}{\partial  p_3} -  p_3\frac{\partial^2\Phi}{\partial   p_3\partial   p_3}.
\end{equation}

%%%%%%%%%%%%%%%%%%%%%%%%%%%%%%%%%%%%%%%%%%%%%%%%%%%%%%%%%%%%%%%%%%%%
\section{Hypergeometric systems: the scalar case}\label{Sol3Point}
%%%%%%%%%%%%%%%%%%%%%%%%%%%%%%%%%%%%%%%%%%%%%%%%%%%%%%%%%%%%%%%%%%%%

In this section, we illustrate the hypergeometric character of the CWI's, proven in the approach presented in \cite{Coriano:2013jba, Coriano:2018bsy, Coriano:2018bbe, Maglio:2019grh}. An independent analysis performed in \cite{Bzowski:2013sza} has connected the solutions of such equations to 3K (or triple-K) integrals. The analysis of \cite{Coriano:2013jba} proved that the fundamental basis for the most general solutions of such equations is given by Appell functions $F_4$.
We present the analysis of scalar three-point functions, then discuss the analogous four-point functions, elaborating on our extension, which is contained in \cite{Maglio:2019grh}. Details of our approach for four-point functions can be found in our work, recently reviewed by us in \cite{Coriano:2020ccb}.

\subsection{The case of scalar three-point  functions}
\label{fuchs}
The hypergeometric character of the CWI's emerges in various ways. One may proceed from \eqref{SpCWIs} and introduce the change of variables - as originally done in
\cite{Coriano:2013jba} -
\begin{eqnarray}
\frac{\partial}{\partial p_{1}^{\mu}}  &=&   2 (p_{1\, \mu} + p_{2 \, \mu}) \frac{\partial}{\partial p_3^2} + \frac{2}{p_3^2}\left( 
(1- x) p_{1 \, \mu}  - x  \,  p_{2 \, \mu} \right) \frac{\partial}{\partial x} - 2  (p_{1\, \mu} + p_{2 \, \mu}) \frac{y}{p_3^2} 
\frac{\partial}{\partial y} \,, \nn \\
\frac{\partial}{\partial p_{2}^{\mu}}  &=& 2 (p_{1\, \mu} + p_{2 \, \mu}) \frac{\partial}{\partial p_3^2}   -   2  (p_{1\, \mu} + 
p_{2 \, \mu}) \frac{x}{p_3^2} \frac{\partial}{\partial x}   + \frac{2}{p_3^2}\left( (1- y) p_{2 \, \mu}  - y  \,  p_{1 \, \mu} 
\right) \frac{\partial}{\partial y}. \, 
\end{eqnarray}
with $x=\frac{p_1^2}{p_3^2}$ and $y=\frac{p_2^2}{p_3^2}$, having chosen $p_3$ as the ``pivot'' in the expansion. In the future we are going to replace $p_3$ with $p_1$, being the two choices equivalent. The results of the general expressions obtained in the two cases are related by some known inversion formulae of the hypergeometric function $F_4$.  

Consider the case of the scalar correlator 
$\Phi(p_1,p_2, p_3)$, which is simpler, defined by the two homogeneous conformal equations
\begin{equation}
K_{31}\Phi=0,  \qquad K_{21}\Phi=0,
\end{equation}
obtained by subtracting the relations in \eqref{3k2}, and combined with the scaling equation 
\begin{equation}
\sum_{i=1}^3 p_i\frac{\partial}{\partial p_i} \Phi=(\Delta_t-2 d) \Phi, 
\end{equation}
with $\Delta_t=\Delta_1+\Delta_2+\Delta_3$. As shown in \cite{Coriano:2013jba}, the ansatz for the solution can be taken of the form 
\begin{equation}
\label{ans}
\Phi(p_1,p_2,p_3)=p_1^{\Delta_t - 2 d} x^{a}y^{b} F(x,y).
\end{equation}
We require that $\Phi$ is homogeneous of degree $\Delta_t-2 d$ under a scale transformation, according to (\ref{scale}). In (\ref{ans}) this condition is taken into account by the factor $p_1^{\Delta - 2 d}$. This procedure will be used extensively in the search of hypergeometric solutions also of other correlators, even in four-point functions, as shown by us for the dual conformal/conformal (dcc) solutions that we will discuss below.

The use of the scale invariant variables $x$ and $y$, now defined as
\begin{equation}
x=\frac{p_2^2}{p_1^2},\qquad y=\frac{p_3^2}{p_1^2},
\end{equation} 
reduces the equations to a generalized hypergeometric form 
\begin{align}
&K_{21}\Phi(p_1,p_2,p_3) =\notag\\
&= 4 p_1^{\Delta -2d -2} x^a y^b
\left(  x(1-x)\frac{\partial }{\partial x \partial x}  + (A x + \gamma)\frac{\partial }{\partial x} -
2 x y \frac{\partial^2 }{\partial x \partial y}- y^2\frac{\partial^2 }{\partial y \partial y} + 
D y\frac{\partial }{\partial y} + \left(E +\frac{G}{x}\right)\right)F(x,y)=0,
\end{align}
with the parameters of the equations given by
\begin{align}
&A=D=\Delta_2 +\Delta_3 - 1 -2 a -2 b -\frac{3 d}{2} \qquad \gamma(a)=2 a +\frac{d}{2} -\Delta_2 + 1,
\notag\\
& G=\frac{a}{2}(d +2 a - 2 \Delta_2),
\notag\\
&E=-\frac{1}{4}(2 a + 2 b +2 d -\Delta_1 -\Delta_2 -\Delta_3)(2 a +2 b + d -\Delta_3 -\Delta_2 +\Delta_1).
\end{align}
Similar constraints are obtained from the equation $K_{31}\Phi=0$, with the obvious exchanges $(a,b,x,y)\to (b,a,y,x)$
\begin{align}
&K_{31}\Phi(p_1,p_2,p_3) =\notag\\
&= 4 p_1^{\Delta -2 d -2} x^a y^b
\left(  y(1-y)\frac{\partial }{\partial y \partial y}  + (A' y + \gamma')\frac{\partial }{\partial y} -
2 x y \frac{\partial^2 }{\partial x \partial y}- x^2\frac{\partial^2 }{\partial x \partial x} + 
D' x\frac{\partial }{\partial x} + \left(E' +\frac{G'}{y}\right)\right)F(x,y)=0,
\end{align}
with
\begin{equation}
\begin{aligned}
&A'=D'= A,   &&\hspace{1cm}\gamma'(b)=2 b +\frac{d}{2} -\Delta_3 + 1,\\
& G'=\frac{b}{2}(d +2 b - 2 \Delta_3), && \hspace{1cm}E'= E.
\end{aligned}
\end{equation}
In the mathematical analysis of hypergeometric systems of Appell type, one encounters 4 possible values for the ``indices'' $a$ and $b$ of the ansatz that we have introduced above. As shown by us in \cite{Coriano:2018bbe}, such values are exactly those that set the $1/x , 1/y$ terms of the equations, in the new $x,y$ variables to zero.  This gives 
\begin{equation}
\label{cond1}
a=0\equiv a_0 \qquad \textrm{or} \qquad a=\Delta_2 -\frac{d}{2}\equiv a_1.
\end{equation}
From the equation $K_{31}\Phi=0$ we obtain a similar condition for $b$ by setting $G'/y=0$, thereby fixing the two remaining indices
\begin{equation}
\label{cond2}
b=0\equiv b_0 \qquad \textrm{or} \qquad b=\Delta_3 -\frac{d}{2}\equiv b_1.
\end{equation}
The four independent solutions of the CWI's will all be characterized by the same 4 pairs of indices $(a_i,b_j)$ $(i,j=1,2)$.
Setting 
\begin{equation}
\alpha(a,b)= a + b + \frac{d}{2} -\frac{1}{2}(\Delta_2 +\Delta_3 -\Delta_1), \qquad \beta (a,b)=a +  b + d -\frac{1}{2}(\Delta_1 +\Delta_2 +\Delta_3),
\label{alphas}
\end{equation}
then
\begin{equation}
E=E'=-\alpha(a,b)\beta(a,b), \qquad A=D=A'=D'=-\left(\alpha(a,b) +\beta(a,b) +1\right).
\end{equation}
the solutions take the form 
\begin{align}
\label{F4def}
F_4(\alpha(a,b), \beta(a,b); \gamma(a), \gamma'(b); x, y) = \sum_{i = 0}^{\infty}\sum_{j = 0}^{\infty} \frac{(\alpha(a,b), {i+j}) \, 
	(\beta(a,b),{i+j})}{(\gamma(a),i) \, (\gamma'(b),j)} \frac{x^i}{i!} \frac{y^j}{j!},
\end{align}
where $(\alpha,i)=\Gamma(\alpha + i)/ \Gamma(\alpha)$ is the Pochammer symbol. We will refer to $\alpha\ldots \gamma'$ as to the first,$\ldots$, fourth parameters of $F_4$.\\ 
This hypergeometric function is an Appell function \cite{APPELL,Bateman:100233,Bateman:1935,Slater:1966}. The 4 independent solutions are then all of the form $x^a y^b F_4$, where the 
hypergeometric functions will take some specific values for their parameters, with
$a$ and $b$ fixed by (\ref{cond1}) and (\ref{cond2}). Specifically we have
\begin{equation}
\Phi(p_1,p_2,p_3)=p_1^{\Delta-2 d} \sum_{a,b} c(a,b,\vec{\Delta})\,x^a y^b \,F_4(\alpha(a,b), \beta(a,b); \gamma(a), \gamma'(b); x, y),
\label{compact}
\end{equation}
where the sum runs over the four values $a_i, b_i$ $i=0,1$ with arbitrary constants $c(a,b,\vec{\Delta})$, with $\vec{\Delta}=(\Delta_1,\Delta_2,\Delta_3)$. Notice that \eqref{compact} is a very compact way to write down the solution. However, once these types of solutions of a homogeneous hypergeometric system are inserted into an inhomogeneous system of equations, the sum over $a$ and $b$ 
needs to be made explicit. For this reason it is convenient to define 
\begin{align} 
&\alpha_0\equiv \alpha(a_0,b_0)=\frac{d}{2}-\frac{\Delta_2 + \Delta_3 -\Delta_1}{2},\, && \beta_0\equiv \beta(b_0)=d-\frac{\Delta_1 + \Delta_2 +\Delta_3}{2},  \nn \\
&\gamma_0 \equiv \gamma(a_0) =\frac{d}{2} +1 -\Delta_2,\, &&\gamma'_0\equiv \gamma(b_0) =\frac{d}{2} +1 -\Delta_3.
\end{align}
to be the 4 basic (fixed) hypergeometric parameters, and define all the remaining ones by shifts with respect to these. The 4 independent solutions can be re-expressed in terms of the parameters above as 
\begin{align}
S_1(\alpha_0, \beta_0; \gamma_0, \gamma'_0; x, y)&\equiv F_4(\alpha_0, \beta_0; \gamma_0, \gamma'_0; x, y) = \sum_{i = 0}^{\infty}\sum_{j = 0}^{\infty} \frac{(\alpha_0,i+j) \, 
(\beta_0,i+j)}{(\gamma_0,i )\, (\gamma'_0,j)} \frac{x^i}{i!} \frac{y^j}{j!}, \\
S_2(\alpha_0, \beta_0; \gamma_0, \gamma'_0; x, y) &= x^{1-\gamma_0} \, F_4(\alpha_0-\gamma_0+1, \beta_0-\gamma_0+1; 2-\gamma_0, \gamma'_0; x,y) \,,  \\
S_3(\alpha_0, \beta_0; \gamma_0, \gamma'_0; x, y) &= y^{1-\gamma'_0} \, F_4(\alpha_0-\gamma'_0+1,\beta_0-\gamma'_0+1;\gamma_0,2-\gamma'_0 ; x,y) \,,  \\
S_4(\alpha_0, \beta_0; \gamma_0, \gamma'_0; x, y) &= x^{1-\gamma_0} \, y^{1-\gamma'_0} \, 
F_4(\alpha_0-\gamma_0-\gamma'_0+2,\beta_0-\gamma_0-\gamma'_0+2;2-\gamma_0,2-\gamma'_0 ; x,y) \, .
\end{align}
Notice that in the scalar case, one is allowed to impose the complete symmetry of the correlator under the exchange of the 3 external momenta and scaling dimensions, as discussed in \cite{Coriano:2013jba}. This reduces the four
constants to just one. 

\subsection{The case of four-point functions and the dcc (dual conformal/conformal)/ conformal Yangian symmetry}
We are going to extend the analysis presented above to scalar four-point functions. 
As already discussed in \secref{secCFT}, the structure of the four-point functions in coordinate space are not completely constrained by the conformal symmetry. In that case one identifies the two cross ratios 
\begin{equation}
\label{uv}
u(x_i)=\frac{x_{12}^2 x_{34}^2}{x_{13}^2 x_{24}^2}, \qquad v(x_i)=\frac{x_{23}^2 x_{41}^2}{x_{13}^2 x_{24}^2},
\end{equation}
and the general solution can be written in the form 
\begin{equation}
\label{general}
\langle \mO_1(x_1)\mO_2(x_2)\mO_3(x_3)\mO_4(x_4)\rangle=  \frac{h(u(x_i),v(x_i))}{\left(x_{12}^2\right)^\frac{\Delta_1 + \Delta_2}{2}\left(x_{3 4}^2\right)^\frac{\Delta_3 + \Delta_4}{2}},
\end{equation}
where $h(u(x_i),v(x_i))$ remains unspecified. 
In momentum space, the correlator depends on six invariants that we will normalize as
$p_i=|\sqrt{p_{i}\,{\hspace{-0.09cm}}^2}|$, $i=1,\dots,4$, representing the magnitudes of the momenta, and $s=|\sqrt{(p_1+p_2)^2}|$, $t=|\sqrt{(p_2+p_3)^2}|$ the two Mandelstam invariants redefined by a square root, for which 
\begin{equation}
\braket{O(p_1)\,O(p_2)\,O(p_3)\,O(\bar{p}_4)}=\Phi(p_1,p_2,p_3,p_4,s,t).\label{invariant}
\end{equation}
This correlation function, to be conformally invariant, has to verify the dilatation Ward identity
\begin{equation}
\left[\sum_{i=1}^4\D_i-3d-\sum_{i = 1}^3p_i^{\mu}\frac{\partial}{\partial p_i^\mu}\right]\Phi(p_1,p_2,p_3,p_4,s,t)=0,
\end{equation}
and the special conformal Ward identities
\begin{equation}
\sum_{i=1}^3\left[2(\D_i-d)\frac{\partial}{\partial p_{i\,\k}}-2p_i^\alpha\frac{\partial^2}{\partial p_i^\alpha\partial p_i^\kappa}+p_i^\kappa\frac{\partial^2}{\partial p_i^\alpha\partial p_{i\,\alpha}}\right]\Phi(p_1,p_2,p_3,p_4,s,t)=0.
\end{equation}
One can split these equations in terms of the invariants of the four-point function written in \eqref{invariant}, by using the chain rules
\begin{align}
\frac{\partial}{\partial p_{1\,\mu}}&=\frac{p_1^\mu}{p_1}\frac{\partial}{\partial p_1}-\frac{\bar{p}_4^\mu}{p_4}\frac{\partial}{\partial p_4}+\frac{p_1^\mu+p_2^\mu}{s}\frac{\partial}{\partial s},\\
\frac{\partial}{\partial p_{2\,\mu}}&=\frac{p_2^\mu}{p_2}\frac{\partial}{\partial p_2}-\frac{\bar{p}_4^\mu}{p_4}\frac{\partial}{\partial p_4}+\frac{p_1^\mu+p_2^\mu}{s}\frac{\partial}{\partial s}+\frac{p_2^\mu+p_3^\mu}{t}\frac{\partial}{\partial t},\\
\frac{\partial}{\partial p_{3\,\mu}}&=\frac{p_3^\mu}{p_3}\frac{\partial}{\partial p_3}-\frac{\bar{p}_4^\mu}{p_4}\frac{\partial}{\partial p_4}+\frac{p_2^\mu+p_3^\mu}{t}\frac{\partial}{\partial t},
\end{align}
where $\bar{p}_4^\mu=-p_1^\mu-p_2^\mu-p_3^\mu$. From this prescription the dilatation WI becomes
\begin{align}
\bigg[(\D_t-3d)-\sum_{i=1}^4p_i\frac{\partial}{\partial p_i}-s\frac{\partial}{\partial s}-t\frac{\partial}{\partial t}\bigg]\Phi(p_1,p_2,p_3,p_4,s,t)=0,\label{Dilatation4}
\end{align}
with $\Delta_t=\sum_{i=1} ^4\Delta_i$ is the total scaling,
and the special CWI's can be written as
\begin{equation}
\sum_{i=1}^3\ p_i^\kappa\, C_i=0\label{primary},
\end{equation}
where the coefficients $C_i$ are differential equations of the second order with respect to the six invariants previously defined. Being $p_1^\k,\ p_2^\k,\ p_3^\k$, in \eqref{primary} independent variables, 
we derive three scalar second order equations for each of the three $C_i$, which must vanish independently.\\
At this stage the procedure in order to simplify the corresponding equations is similar to the one described in \cite{Coriano:2018bsy,Coriano:2018bbe}. A lengthy computation allows to rewrite the equations in the form
\begin{align}
C_1&=\bigg\{\frac{\partial^2}{\partial p_1^2}+\frac{(d-2\D_1+1)}{p_1}\frac{\partial}{\partial p_1}-\frac{\partial^2}{\partial p_4^2}-\frac{(d-2\D_4+1)}{p_4}\frac{\partial}{\partial p_4}\notag\\[1.5ex]
&\qquad+\frac{1}{s}\frac{\partial}{\partial s}\left(p_1\frac{\partial}{\partial p_1}+p_2\frac{\partial}{\partial p_2}-p_3\frac{\partial}{\partial p_3}-p_4\frac{\partial}{\partial p_4}\right)+\frac{(\D_3+\D_4-\D_1-\D_2)}{s}\frac{\partial}{\partial s}\notag\\[1.5ex]
&\qquad+\frac{(p_2^2-p_3^2)}{st}\frac{\partial^2}{\partial s\partial t}\bigg\}\,\Phi(p_1,p_2,p_3,p_4,s,t)=0\label{C1},
\end{align}
for $C_1$ and similarly for the other coefficients. \\
It is quite clear that these equations involve a differential operator which is of hypergeometric type, such as the $K_{ij}$ operators discussed before, combined with other terms. Notice that such operators are not separable in the $s, t$ and $p_i^2$ variables.
%%%%%%%%%%%%%%%%%%%%%%%%%%%%%%
\subsection{Enhancing the symmetry: the dcc solutions} 
%%%%%%%%%%%%%%%%%%%%%%%%%%%%%%
Now we turn to consider possible solutions of the conformal constraints \eqref{C1} which are built around specific dual conformal ans\"atze \cite{Maglio:2019grh}. It is worth mentioning that the property of an object to be conformally and dual conformally invariant is also known as conformal Yangian (CY) invariance.

The number of independent equations in \eqref{primary}, by using the ansatz that we are going to present below, will then reduce from three down to two. We choose  the ansatz
		\begin{equation}
		\Phi(p_i,s,t)=\big(s^2t^2\big)^{n_s}\,F(x,y)\label{ansatz},
		\end{equation}
		where $n_s$ is a coefficient (scaling factor of the ansatz) that we will fix below by the dilatation WI, and the variables $x$ and $y$ are defined by the quartic ratios
		\begin{equation}
		x=\frac{p_1^2\,p_3^2}{s^2\,t^2},\qquad y=\frac{p_2^2\,p_4^2}{s^2\,t^2}.
		\end{equation}
		By inserting the ansatz \eqref{ansatz} into the dilatation Ward identities, and turning to the new variables $x$ and $y$, after some manipulations we obtain from \eqref{Dilatation4} the condition
		\begin{equation}
		\label{dil1}
		\bigg[(\D_t-3d)-\sum_{i=1}^4p_i\frac{\partial}{\partial p_i}-s\frac{\partial}{\partial s}-t\frac{\partial}{\partial t}\bigg]\big(s^2t^2\big)^{c}\,F(x,y)=\big(s^2t^2\big)^{n_s}\big[(\D_t-3d)-4 n_s\big] \,F(x,y)=0,
		\end{equation}
		which determines $n_s=(\D_t-3d)/4$, giving
		\begin{equation}
		\Phi(p_i,s,t)=\big(s^2t^2\big)^{(\D_t-3d)/4}\,F(x,y).\label{ansatz2}
		\end{equation}
		The functional form of $F(x,y)$ will then be furtherly constrained.

\subsubsection{Equal scaling solutions}

	We start investigating the solutions of these equations by assuming, as a first example, that the scaling dimensions of all the scalar operators are  equal $\D_i=\D$, $i=1,\dots,4$.
	The special conformal Ward identities, re-expressed in terms of $x$ and $y$, are given in the form
	\begin{equation}
		\left\{\begin{aligned}
		&\bigg[y(1-y)\partial_{yy} -2x\,y\,\partial_{xy}-x^2\partial_{xx}-(1-2n_s)x\,\partial_x+\left(1-\D+\frac{d}{2}-y(1-2n_s)\right)\,\partial_y-n_s^2\bigg] F(x,y)=0\\[2ex]
		&\bigg[x(1-x)\partial_{xx} -2x\,y\,\partial_{xy}-y^2\partial_{yy}-(1-2n_s)y\,\partial_y+\left(1-\D+\frac{d}{2}-x(1-2n_s)\right)\,\partial_x-n_s^2\bigg] F(x,y)=0
		\end{aligned}\right.\label{EqualScal},
	\end{equation}
	where we recall that $n_s$ is the scaling of the correlation function under dilatations, now given by
	\begin{equation}
	\label{scaled}
	n_s=\D-\frac{3d}{4},
	\end{equation}
	since $\D_t=4\, \D$. One easily verifies that Eqs. \eqref{EqualScal} correspond to  a hypergeometric system and its solutions can be expressed as linear combinations of four Appell functions $F_4$ of two variables $x$ and $y$, as in the case of three-point  functions. The general solution of such system is expressed as
\begin{align}
\Phi(p_i,s,t) &=\big(s^2t^2\big)^{(\D_t-3d)/4}\, \sum_{a,b} c(a,b,\Delta) x^a y^b F_4\left(\a(a,b),\b(a,b),\g(a),\g'(b);x, y\right).\label{solutionEq}
\end{align}
Notice that the solution is similar to that of the three-point  functions given by \eqref{compact}, discussed above.\\
The general solution \eqref{solutionEq} has been written as a linear superposition of these, with independent constants $c(a,b)$ labeled by the exponents $a,b$ 
\begin{align}
&a=0,\,\D-\frac{d}{2},&&b=0,\,\D-\frac{d}{2},\label{FuchsianPoint}
\end{align}
which fix the dependence of the $F_4$  
\begin{align}
\label{s2}
&\a(a,b)=\frac{3}{4}d-\D+a+b,&&\b(a,b)=\frac{3}{4}d-\D+a+b,\notag\\
&\Gamma(a)=\frac{d}{2}-\D+1+2a,&&\g'(b)=\frac{d}{2}-\D+1+2b.
\end{align}

\subsubsection{More general conditions}
The solution that we have identified in the equal scaling case can be extended by relaxing the conditions on the scaling dimensions, in the form
\begin{equation}
\D_1=\D_3=\D_x,\qquad \D_2=\D_4=\D_y.
\end{equation}
 In this case the CWI's give the system of equations
\begin{equation}
\left\{\begin{aligned}
&\bigg[y(1-y)\partial_{yy} -2x\,y\,\partial_{xy}-x^2\partial_{xx}-(1-2n_s)x\,\partial_x+\left(1-\D_y+\frac{d}{2}-y(1-2n_s)\right)\,\partial_y-n_s^2\bigg] F(x,y)=0\\[2ex]
&\bigg[x(1-x)\partial_{xx} -2x\,y\,\partial_{xy}-y^2\partial_{yy}-(1-2n_s)y\,\partial_y+\left(1-\D_x+\frac{d}{2}-x(1-2n_s)\right)\,\partial_x-n_s^2\bigg] F(x,y)=0
\end{aligned}\right.,
\end{equation}
where now $n_s$ is defined as 
\begin{equation}
n_s=\frac{\D_x}{2}+\frac{\D_y}{2}-\frac{3}{4}d,
\end{equation}
whose solutions are expressed as
\begin{equation}
\label{ress}
\Phi(p_i,s,t)=\big(s^2t^2\big)^{(\D_t-3d)/4}\,\sum_{a,b} c(a,b,\vec\Delta_t) x^a y^bF_4\left(\a(a,b),\b(a,b),\g(a),\g'(b);x,y\right),
\end{equation}
with $\vec\Delta_t=(\Delta_x,\Delta_y,\Delta_x,\Delta_y)$, $\Delta_t=2 \Delta_x + 2 \Delta_y$ and the Fuchsian points are fixed by the conditions
\begin{align}
&a=0,\,\D_x-\frac{d}{2},&&b=0,\,\D_y-\frac{d}{2},\notag\\
&\a(a,b)=\frac{3}{4}d-\frac{\D_x}{2}-\frac{\D_y}{2}+a+b,&&\b(a,b)=\frac{3}{4}d-\frac{\D_x}{2}-\frac{\D_y}{2}+a+b,\notag\\
&\g(a)=\frac{d}{2}-\D_x+1+2a,&&\g'(b)=\frac{d}{2}-\D_y+1+2b.
\end{align}
We pause for a moment to discuss the domain of convergence of such solutions. Such domain, for $F_4$ is bounded by the relation 
\begin{equation}
\sqrt{x}+\sqrt{y}< 1, 
\end{equation}
which is satisfied in a significant kinematic region, and in particular at large energy and momentum transfers. Notice that the analytic continuation of \eqref{ress} in the physical region can be simply obtained by sending $t^2\to -t^2$ (with $t^2<0$) and leaving all the other invariants untouched. In this case we get
\begin{equation}
 \sqrt{p_1^2 p_3^2}  +\sqrt{p_2^2 p_4^2} < \sqrt{- s^2 t^2}.
\end{equation}
At large energy and momentum transfers the correlator exhibits a power-like behavior of the form 
\begin{equation}
\Phi(p_i,s,t)\sim \frac{1}{(- s^2 t^2)^{(3 d - \Delta_t)/4}}. 
\end{equation}
Given the connection between the function $F_4$ and the triple-K integrals, we will reformulate this solution in terms of such integrals. They play a key role in the solution of the CWI's for tensor correlators, as discussed in \cite{Bzowski:2013sza}, for three-point  functions.

The link between $3$- and four-point functions outlined in the previous section allows to re-express the solutions in terms of a class of parametric integrals of 3 Bessel functions, as done in the case of the  scalar and tensor correlators  \cite{Bzowski:2013sza}, with the due modifications.
We consider the case of the solutions characterized by $\D_1=\D_2=\D_3=\D_4=\D$ or $\D_1=\D_3=\D_x$ and\  $\D_2=\D_4=\D_y$. We will show that the solution can be written in terms of triple-K integrals which are connected to the Appell function $F_4$ by the relation 
\begin{align}
& \int_0^\infty d x \: x^{\alpha - 1} K_\lambda(a x) K_\mu(b x) K_\nu(c x) =\frac{2^{\alpha - 4}}{c^\alpha} \bigg[ B(\lambda, \mu) + B(\lambda, -\mu) + B(-\lambda, \mu) + B(-\lambda, -\mu) \bigg], \label{3K}
\end{align}
where
\begin{align}
B(\lambda, \mu) & = \left( \frac{a}{c} \right)^\lambda \left( \frac{b}{c} \right)^\mu \Gamma \left( \frac{\alpha + \lambda + \mu - \nu}{2} \right) \Gamma \left( \frac{\alpha + \lambda + \mu + \nu}{2} \right) \Gamma(-\lambda) \Gamma(-\mu) \times \notag\\
& \qquad \times F_4 \left( \frac{\alpha + \lambda + \mu - \nu}{2}, \frac{\alpha + \lambda + \mu + \nu}{2}; \lambda + 1, \mu + 1; \frac{a^2}{c^2}, \frac{b^2}{c^2} \right), \label{3Kplus}
\end{align}
valid for
\begin{equation}
\Re\, \alpha > | \Re\, \lambda | + | \Re \,\mu | + | \Re\,\nu |, \qquad \Re\,(a + b + c) > 0, \nn
\end{equation}
with the Bessel functions $K_\nu$ satisfying the equations 
\begin{equation}
\begin{split}
\frac{\partial}{\partial p}\big[p^\b\,K_\b(p\,x)\big]&=-x\,p^\b\,K_{\b-1}(p x),\\
K_{\b+1}(x)&=K_{\b-1}(x)+\frac{2\b}{x}K_{\b}(x). 
\end{split}\label{der}
\end{equation}
In particular  the solution can be written as
\begin{equation}
I_{\a\{\b_1,\b_2,\b_3\}}(p_1\,p_3; p_2\,p_4;s\,t)=\int_0^\infty\,dx\,x^\a\,(p_1\,p_3)^{\b_1}\,(p_2\,p_4)^{\b_2}\,(s\,t)^{\b_3}\,K_{\b_1}(p_1\,p_3\,x)\,K_{\b_2}(p_2\,p_4\,x)\,K_{\b_3}(s\,t\,x).\label{trekappa}
\end{equation}
 Using  \eqref{der} one can derive several relations, 
such as 
\begin{align}
\frac{\partial^2}{\partial p_1^2}I_{\a\{\b_1,\b_2,\b_3\}}&=-\,p_3^2\,I_{\a+1\{\b_1-1,\b_2,\b_3\}}+p_1^2\,p_3^4\,\,I_{\a+2\{\b_1-2,\b_2,\b_3\}},
\end{align}
which generate identities such as 
\begin{align}
p_1^2\,p_3^2\,I_{\a+2\{\b_1-2,\b_2,\b_3\}}&=I_{\a+2\{\b_1,\b_2,\b_3\}}-2(\b_1-1)\,I_{\a+1\{\b_1-1,\b_2,\b_3\}}.
\end{align}
We refer to \appref{AppendixJ} for more details. Using these relations, the dilatation WI's \eqref{Dilatation4} become
\begin{equation}
(\D_t-3d) I_{\a\{\b_1,\b_2,\b_3\}}+2p_1^2p_3^2\ I_{\a+1\{\b_1-1,\b_2,\b_3\}}+2p_2^2p_4^2\ I_{\a+1\{\b_1,\b_2-1,\b_3\}}+2s^2t^2\ I_{\a+1\{\b_1,\b_2,\b_3-1\}}=0,
\end{equation}
where the arguments of the $I_{\a\{\b_1\b_2\b_3\}}$ function, written explicitly in \eqref{trekappa}, have been omitted for simplicity. The $I$ integrals satisfy the differential equations
\begin{align}
\frac{1}{s}\frac{\partial}{\partial s}\left(p_1\frac{\partial}{\partial p_1}+p_2\frac{\partial}{\partial p_2}-p_3\frac{\partial}{\partial p_3}-p_4\frac{\partial}{\partial p_4}\right)I_{\a\{\b_1,\b_2,\b_3\}}&=0,\\
\frac{1}{t}\frac{\partial}{\partial t}\left(p_1\frac{\partial}{\partial p_1}+p_4\frac{\partial}{\partial p_4}-p_2\frac{\partial}{\partial p_2}-p_3\frac{\partial}{\partial p_3}\right)I_{\a\{\b_1,\b_2,\b_3\}}&=0,
\end{align} 
which can be checked using the relations given in the same appendix, and we finally obtain
\begin{equation}
(\D_t-3d+2\a+2-2\b_t) I_{\a\{\b_1,\b_2,\b_3\}}=0,
\end{equation}
where $\b_t=\b_1+\b_2+\b_3$. In order to satisfy this equation the $\a$ parameter has to be equal to a particular value given by 
\begin{equation}
\tilde{\a}\equiv \frac{3}{2}d+\b_t-1-\frac{\D_t}{2}.
\end{equation}
 In the particular case of $\D_i=\D$, the special conformal Ward identities are given by
\begin{equation}
\left\{\begin{aligned}&\\[-1.2ex]
&\bigg [\frac{\partial^2}{\partial p_1^2}+\frac{(d-2\D+1)}{p_1}\frac{\partial}{\partial p_1}-\frac{\partial^2}{\partial p_3^2}-\frac{(d-2\D+1)}{p_3}\frac{\partial}{\partial p_3}+\frac{(p_1^2-p_3^2)}{st}\frac{\partial^2}{\partial s\partial t}\bigg ]\,I_{\tilde\a\{\b_1,\b_2,\b_3\}}=0\\[1ex]
&\bigg[\frac{\partial^2}{\partial p_2^2}+\frac{(d-2\D+1)}{p_2}\frac{\partial}{\partial p_2}-\frac{\partial^2}{\partial p_4^2}-\frac{(d-2\D+1)}{p_4}\frac{\partial}{\partial p_4}+\frac{(p_2^2-p_4^2)}{st}\frac{\partial^2}{\partial s\partial t}\bigg ]\,I_{\tilde\a\{\b_1,\b_2,\b_3\}}=0\\[1ex]
&\bigg[\frac{\partial^2}{\partial p_3^2}+\frac{(d-2\D+1)}{p_3}\frac{\partial}{\partial p_3}-\frac{\partial^2}{\partial p_4^2}-\frac{(d-2\D+1)}{p_4}\frac{\partial}{\partial p_4}+\frac{(p_2^2-p_1^2)}{st}\frac{\partial^2}{\partial s\partial t}\bigg ]\,I_{\tilde\a\{\b_1,\b_2,\b_3\}}=0\\[-0.7ex]
\label{neweq}
\end{aligned}\right.
\end{equation}
and using the properties of Bessel functions they can be rewritten in a simpler form. The first equation, for instance, can be written as
\begin{equation}
\label{oone}
(p_1^2-p_3^2)\bigg( (d-2\D+2\b_1)\,I_{\tilde\a+1\{\b_1-1,\b_2,\b_3\}}-2\b_3\,I_{\tilde\a+1\{\b_1,\b_2,\b_3-1\}} \bigg)=0,
\end{equation}
which is identically satisfied if the conditions 
\begin{equation}
\b_1=\D-\frac{d}{2},\qquad\b_3=0
\end{equation}
hold. In the same way we find that the second equation takes the form
\begin{equation}
\label{otwo}
(p_2^2-p_4^2)\bigg((d-2\D+2\b_2)\,I_{\tilde\a+1\{\b_1,\b_2-1,\b_3\}}-2\b_3\,I_{\tilde\a+1\{\b_1,\b_2,\b_3-1\}}\bigg)=0,
\end{equation}
and it is satisfied if 
\begin{equation}
\b_2=\D-\frac{d}{2},\qquad \b_3=0.
\end{equation}
One can check that the third equation 
\begin{equation}
p_2^2(d-2\D+2\b_2)\,I_{\tilde\a+1\{\b_1,\b_2-1,\b_3\}}-p_1^2(d-2\D+2\b_1)\,I_{\tilde\a+1\{\b_1-1,\b_2,\b_3\}}-2(p_2^2-p_1^2)\b_3\,I_{\tilde\a+1\{\b_1,\b_2,\b_3-1\}}=0,
\end{equation}
generates the same conditions given by \eqref{oone} and \eqref{otwo}.
After some computations, finally the solution for the four-point function, in this particular case, can be written as
\begin{equation}
\braket{O(p_1)\,O(p_2)\,O(p_3)\,O(\bar{p}_4)}=\,\bar{\a} \,I_{\frac{d}{2}-1\left\{\D-\frac{d}{2},\D-\frac{d}{2},0\right\}}(p_1\, p_3;p_2\,p_4; s\,t),
\end{equation}
where $\bar{\a}$ is an undetermined constant. 

In the case $\D_1=\D_3=\D_x$ and $\D_2=\D_4=\D_y$, the special CWI's can be written as
\begin{equation}
\left\{\begin{aligned}&\\[-1.9ex]
&\bigg [\frac{\partial^2}{\partial p_1^2}+\frac{(d-2\D_x+1)}{p_1}\frac{\partial}{\partial p_1}-\frac{\partial^2}{\partial p_3^2}-\frac{(d-2\D_x+1)}{p_3}\frac{\partial}{\partial p_3}+\frac{(p_1^2-p_3^2)}{st}\frac{\partial^2}{\partial s\partial t}\bigg ]\,I_{\tilde\a\{\b_1,\b_2,\b_3\}}=0\\[1ex]
&\bigg[\frac{\partial^2}{\partial p_2^2}+\frac{(d-2\D_y+1)}{p_2}\frac{\partial}{\partial p_2}-\frac{\partial^2}{\partial p_4^2}-\frac{(d-2\D_y+1)}{p_4}\frac{\partial}{\partial p_4}+\frac{(p_2^2-p_4^2)}{st}\frac{\partial^2}{\partial s\partial t}\bigg ]\,I_{\tilde\a\{\b_1,\b_2,\b_3\}}=0\\[1ex]
&\bigg[\frac{\partial^2}{\partial p_3^2}+\frac{(d-2\D_x+1)}{p_3}\frac{\partial}{\partial p_3}-\frac{\partial^2}{\partial p_4^2}-\frac{(d-2\D_y+1)}{p_4}\frac{\partial}{\partial p_4}+\frac{(p_2^2-p_1^2)}{st}\frac{\partial^2}{\partial s\partial t}\bigg ]\,I_{\tilde\a\{\b_1,\b_2,\b_3\}}=0\\[1.3ex]
\end{aligned}\right.
\end{equation}
whose solution is
\begin{align}
\braket{O(p_1)\,O(p_2)\,O(p_3)\,O(\bar{p}_4)}&=\,\bar{\bar{\a}} \,I_{\frac{d}{2}-1\left\{\D_x-\frac{d}{2},\D_y-\frac{d}{2},0\right\}}(p_1\, p_3;p_2\,p_4; s\,t),
\end{align}
which takes a form which is typical of the three-point function.
\subsection{Symmetric solutions as \texorpdfstring{$F_4$}{F4} hypergeometrics or triple-K integrals. The equal scalings case}
The derivation of symmetric expressions of such correlators requires some effort, and can be obtained either by 
using the few known relations available for the Appell function $F_4$ or, alternatively (and more effectively), by resorting to the formalism of the triple-K integrals \cite{Bzowski:2015yxv}. \\ 
A solution which is symmetric with respect to all the permutation of the momenta $p_i$, expressed in terms of 3 of the four constants $c(a,b)$, after some manipulations, can be expressed in the form
\begin{align}
\braket{O(p_1)O(p_2)O(p_3)O(p_4)}&=\notag\\
&\hspace{-2cm}=\sum_{a,b}c(a,b)\Bigg[\,(s^2\,t^2)^{\D-\frac{3}{4}d}\left(\frac{p_1^2p_3^2}{s^2t^2}\right)^a\left(\frac{p_2^2p_4^2}{s^2t^2}\right)^bF_4\left(\a(a,b),\b(a,b),\g(a),\g'(b),\frac{p_1^2p_3^2}{s^2t^2},\frac{p_2^2p_4^2}{s^2t^2}\right)\notag\\
&\hspace{-1.5cm}+\,(s^2\,u^2)^{\D-\frac{3}{4}d}\,\left(\frac{p_2^2p_3^2}{s^2u^2}\right)^{a}\left(\frac{p_1^2p_4^2}{s^2u^2}\right)^{b}F_4\left(\a(a,b),\b(a,b),\g(a),\g'(b),\frac{p_2^2p_3^2}{s^2u^2},\frac{p_1^2p_4^2}{s^2u^2}\right)\notag\\
&\hspace{-1.5cm}+\,(t^2\,u^2)^{\D-\frac{3}{4}d}\,\left(\frac{p_1^2p_2^2}{t^2u^2}\right)^{a}\,\left(\frac{p_3^2p_4^2}{t^2u^2}\right)^{b}\,F_4\left(\a(a,b),\b(a,b),\g(a),\g'(b),\frac{p_1^2p_2^2}{t^2u^2},\frac{p_3^2p_4^2}{t^2u^2}\right)
\Bigg],
\label{fform}
\end{align}
where the four coefficients $c(a,b)$'s given in \eqref{fform} are reduced to three by the constraint 
\begin{align}
c\left(0,\D-\frac{d}{2}\right)=c\left(\D-\frac{d}{2},0\right).
\end{align}

Additional manipulations, in order to reduce even further the integration constants, are hampered by the absence of known 
relations among the Appell functions $F_4$. As already mentioned above, it is possible, though, to bypass the problem by turning to the triple-K formalism. Equation \eqref{fform} can be further simplified using this formalism. 
In order to show this, \eqref{fform} can be written in terms of a linear combination of triple-K integrals as
\begin{align}
\braket{O(p_1)O(p_2)O(p_3)O(p_4)}&= C_1\,I_{\frac{d}{2}-1\{\D-\frac{d}{2},\D-\frac{d}{2},0\}}(p_1\,p_3,p_2\,p_4,s\,t)\notag\\[1.2ex]
&\hspace{-1.5cm}+C_2\, \,I_{\frac{d}{2}-1\{\D-\frac{d}{2},\D-\frac{d}{2},0\}}(p_2\,p_3,p_1\,p_4,s\,u)+C_3\, \,I_{\frac{d}{2}-1\{\D-\frac{d}{2},\D-\frac{d}{2},0\}}(p_1\,p_2,p_3\,p_4,t\,u),
\end{align}
by an explicit symmetrization of the momenta in the parametric integrals.  It is now much simpler to show that the symmetry under permutations forces the $C_i$ to take the same value, and the final symmetric result is given by
\begin{align}
\braket{O(p_1)O(p_2)O(p_3)O(p_4)}&= C\bigg[\,I_{\frac{d}{2}-1\{\D-\frac{d}{2},\D-\frac{d}{2},0\}}(p_1\,p_3,p_2\,p_4,s\,t)\notag\\
&\hspace{-1.5cm}+\, \,I_{\frac{d}{2}-1\{\D-\frac{d}{2},\D-\frac{d}{2},0\}}(p_2\,p_3,p_1\,p_4,s\,u)+\, \,I_{\frac{d}{2}-1\{\D-\frac{d}{2},\D-\frac{d}{2},0\}}(p_1\,p_2,p_3\,p_4,t\,u)\bigg],
\end{align}
written in terms of only one arbitrary constant overall, $C$. We can use the relation between the triple-K integrals and the $F_4$ written in \eqref{3K} and \eqref{3Kplus}, to re-express the final symmetric solution, originally given in \eqref{fform}, in terms of a single constant in the form
\begin{align}
\braket{O(p_1)O(p_2)O(p_3)O(p_4)}&=2^{\frac{d}{2}-4}\ \ C\,\sum_{\l,\m=0,\D-\frac{d}{2}}\x(\l,\m)\bigg[\big(s^2\,t^2\big)^{\D-\frac{3}{4}d}\left(\frac{p_1^2 p_3^2}{s^2 t^2}\right)^\l\left(\frac{p_2^2p_4^2}{s^2t^2}\right)^\m\nonumber\\
&\hspace{-3cm}\times\,F_4\left(\frac{3}{4}d-\D+\l+\m,\frac{3}{4}d-\D+\l+\m,1-\D+\frac{d}{2}+\l,1-\D+\frac{d}{2}+\m,\frac{p_1^2 p_3^2}{s^2 t^2},\frac{p_2^2 p_4^2}{s^2 t^2}\right)\notag\\
&+\big(s^2\,u^2\big)^{\D-\frac{3}{4}d}\left(\frac{p_2^2 p_3^2}{s^2 u^2}\right)^\l\left(\frac{p_1^2p_4^2}{s^2u^2}\right)^\m\notag\\
&\hspace{-3cm}\times\,F_4\left(\frac{3}{4}d-\D+\l+\m,\frac{3}{4}d-\D+\l+\m,1-\D+\frac{d}{2}+\l,1-\D+\frac{d}{2}+\m,\frac{p_2^2 p_3^2}{s^2 u^2},\frac{p_1^2 p_4^2}{s^2 u^2}\right)\notag\\
&+\big(t^2\,u^2\big)^{\D-\frac{3}{4}d}\left(\frac{p_1^2 p_2^2}{t^2 u^2}\right)^\l\left(\frac{p_3^2p_4^2}{t^2u^2}\right)^\m\notag\\
&\hspace{-3cm}\times\,F_4\left(\frac{3}{4}d-\D+\l+\m,\frac{3}{4}d-\D+\l+\m,1-\D+\frac{d}{2}+\l,1-\D+\frac{d}{2}+\m,\frac{p_1^2 p_2^2}{t^2 u^2},\frac{p_3^2 p_4^2}{t^2 u^2}\right)\bigg].\label{finalSol}
\end{align}
where the coefficients $\x(\l,\m)$ are explicitly given by
\begin{equation}
\begin{split}
\x\left(0,0\right)&=\left[\Gamma\left(\frac{3}{4}d-\D\right)\right]^2\left[\Gamma\left(\D-\frac{d}{2}\right)\right]^2,\\
\x\left(0,\D-\frac{d}{2}\right)&=\x\left(\D-\frac{d}{2},0\right)=\left[\Gamma\left(\frac{d}{4}\right)\right]^2\Gamma\left(\D-\frac{d}{2}\right)\Gamma\left(\frac{d}{2}-\D\right),\\
\x\left(\D-\frac{d}{2},\D-\frac{d}{2}\right)&=\left[\Gamma\left(\D-\frac{d}{4}\right)\right]^2\left[\Gamma\left(\frac{d}{2}-\D\right)\right]^2.
\end{split}\label{xicoef}
\end{equation}
The solution found in \eqref{finalSol} is explicitly symmetric under all the possible permutations of the momenta and it is fixed up to one undetermined constant $C$. Equation \eqref{finalSol} gives the final expression of the solution obtained from the dual conformal ansatz \eqref{ansatz2}.
%%%%%%%%%%%%%%%%%%%%%%%%%%%%%%%%%
\section{Path Integral formulation and the effective action}
%%%%%%%%%%%%%%%%%%%%%%%%%%%%%%%%%
In a classical conformally invariant theory, such as massless Quantum Electrodynamics (QED), the dilatation operator generates a symmetry of the Lagrangian. Due to the renormalization procedure, this symmetry is broken. As a result, the stress energy tensor acquires a non-vanishing trace, a trace anomaly contribution which can be generated by 
an effective action which is purely gravitational. In this approach, the matter contribution is integrated out, and the action provides a semi-classical description of the interaction of the gravitational field with ordinary matter. 

An anomaly action is not unique. It is an action which accounts for the trace/conformal anomaly but it is defined modulo traceless contributions that remain, obviously, arbitrary, from the point of view of the solution of the variational problem. 

These actions may differ  by the number of asymptotic degrees of freedom introduced in the action itself. With the term ``asymptotic'' we refer to fields which are part of the effective action, but not of the original theory.  This is the case, for instance, for local anomaly actions constructed by the inclusion of a dilaton using the Weyl gauging procedure, that we are going to review below.\\ 
It is natural, indeed, to introduce a dilaton as a Goldstone mode that couples to the divergence of the conformal current $(J_c)$, which is not conserved at quantum level. In the case of a conformal anomaly, this coupling is at most quartic in $d=4$ and can be worked out by the Noether method \cite{Coriano:2013nja,Coriano:2013xua}. 

All the local forms of such actions introduce one extra degree of freedom, in the form of a dilaton field. 
In the case of supersymmetric theories the dilaton field turns into a mutiplet with a dilaton, an axion and an axino and can be described by a St\"uckelberg-like Lagrangian \cite{Coriano:2010ws}. 

The steps that take to the conformal anomaly may be quite diverse, 
for instance one can rely on direct perturbative computations, which allow to identify on a diagrammatic basis some components of the anomaly functional. 
In this approach one can consider correlators with several insertions of stress energy tensors and/or currents, which are sensitive to different components of the anomaly functional. For instance, the trace of the $TJJ$ accounts for the $F^2$ part of the anomaly, where $F$ is the field strength of the gauge field coupled to the $J$ current, but it is insensitive to other components, which come from gravity. For this reason a complete picture of the trace anomaly requires other, more complex correlators containing multiple insertions of stress energy tensors.\\
One advantage of this approach, which can be performed in momentum space, is to bring us close to the core of an interacting theory, with the emergence of dynamical degrees of freedom. Ultimately, quantum field theory is a theory of massive and massless particles propagating in spacetime either as real or virtual 
(interpolating) states. Such effective interaction can be worked out directly in a traditional Feynman expansion. 

A complementary/alternative approach is to start from the vacuum persistence amplitude, which takes to the DeWitt-Schwinger or heat-kernel method. One important feature of the method is to provide an expression for the anomaly action which allows to identify 
the general structure of the anomaly functional. This is clearly advantageous compared to the standard perturbative approach in momentum space, and for this reason we are going 
to briefly summarize it in this section. 

%%%%%%%%%%%%%%%%%%%
\subsection{The anomaly action}
%%%%%%%%%%%%%%%%%%%
The conformal anomaly effective action plays an important role in the identification of the impact of conformal symmetry at a certain physical scale, having integrated out a certain number of degrees of freedom in the functional integral. For a given Lagrangian CFT, $\mathcal{S}(g)$ (for former discussions see \cite{Cappelli:1988vw,Cappelli:2001pz,Erdmenger:1998xv,Asorey:2003uf,Asorey:2006rm,Buchbinder:1992rb, Hamada}) is a functional of the external metric  $g$, and its expansion in the fluctuations around a given background $\bar{g}$ ($g= \bar{g} + \delta g)$ allows to define the $n$-point correlation functions of stress-energy tensors for any $n$.  In general it is generated, for instance, by integrating out some conformal matter in the path integral, leaving the external metric arbitrary. The simplest example is provided, for instance, by scalar, fermionic and spin-$1$ free field theories coupled to gravity  \cite{Coriano:2018bbe,Coriano:2018bsy,Serino:2020pyu}, at least at $d=4$.  

Integrating out conformal matter in the path integral, allows to identify its back-reaction on 
gravity, following an approach that has some similarity with Sakharov's theory of induced gravity, where integration over ordinary matter, for a generic metric $g$, is expected to generate terms of the form  \cite{Visser:2002ew}
\begin{equation} 
\label{st}
\sm(g)\sim \int d^4 x \sqrt{g}\left( \Lambda + c_1(g) R + c_2\,``\,R^2\,"\right).
\end{equation}
They correspond to a cosmological constant, the Einstein-Hilbert action and to generic higher derivative invariant terms. The integration over conformal matter does not introduce any scale, if the result of the integration turns out to be finite, and the structure of \eqref{st} simplifies. $\mathcal{S}(g)$ is, in this case, non-local and Weyl invariant. 

The process of renormalization, as $d\to 4$ induces a non-vanishing variation of $\mathcal{S}(g)$ under a Weyl rescaling of the metric. This can be attributed, in a rather simple way, to the need of including a dimensional constant $\mu$ in order to balance the mass-dimensions of the counterterms that are necessary in order to render the integration on the quantum degrees of freedom, finite.
This appears as a balancing factor $\mu^{-\epsilon}$ - with $\epsilon=d-4$ - in the structure of the counterterms, which are expressed in terms of Weyl-invariant operators in $d=4$. 
The renormalized action then acquires a $\log k^2/\mu^2$ dependence, where $k$ is a generic momentum, determining the breaking of dilatation invariance. 

We recall that in an ordinary field theory the relation between the partition function and the functional of all the connected correlators is obviously given by the functional relation
\begin{equation}
\label{defg}
e^{-\mathcal{S}(g)}=Z(g) \leftrightarrow \mathcal{S}(g)=-\log Z(g).
\end{equation}
 As mentioned, $Z(g)$ can be thought of as related to a functional integral in which we integrate the action of a generic CFT  over a field $(\phi)$, or in general, a collection of fields,  in a background metric $g_{\mu\nu}$, 
\begin{equation} 
\label{induced}
Z(g)=N \int D\phi e^{-S_0(g,\phi)},   \qquad Z(\eta)=1.
\end{equation}
Its logarithm, $\mathcal{S}(g)$, is our definition of the effective action, while $S_0(g,\phi)$ is the classical action. As usual $Z(g)$, in the Feynman diagrammatic expansion, will contain both connected and disconnected graphs, while $\mathcal{S}(g)$ collects only connected graphs. It is easy to verify that this 
collection corresponds also to 1PI (1 particle irreducible) graphs only in the case of free field theories embedded in external (classical) gravity.  

The emergence of bilinear mixing on the external graviton lines, as we are going to realize at the end of our analysis, should then be interpreted as a dynamical response 
of the theory, induced by the process of renormalization, with the generation of an intermediate dynamical degree of freedom propagating with the $1/\square$ operator. For this reason, the presence of such terms does not invalidate the 1PI nature of this functional.

\subsection{The scalar case}
As a reference for our discussion,  we may assume that $S(\phi,g)$ describes, for instance, a free scalar field $\phi$ in a generic background. The action, in  this case, is given by 
\beq
\label{phi}
S_0=\frac{1}{2}\int\, d^dx\,\sqrt{-g}\left[g^{\mu\nu}\nabla_\mu\phi\nabla_\nu\phi-\chi\, R\,\phi^2\right],
\eeq
where we have included a conformal coupling $\chi(d)=\frac{1}{4}\frac{(d-2)}{(d-1)}$, and $R$ is the scalar curvature. This choice of $\chi(d)$ guarantees the conformal invariance of this action in $d$ dimensions and generates a term of improvement for the stress-energy tensor in the flat limit, which becomes symmetric and traceless in this limit. We refer to 
 \cite{Coriano:2011zk,Coriano:2011ti} for a general perturbative analysis of the role of this term as we couple the Standard Model to gravity, which requires a conformally coupled Higgs sector. One can verify explicitly that the ordinary counterterms which renormalize the Lagrangian of the Standard Model, renormalize also insertions of the stress energy tensor of the theory, only if the Higgs sector is conformally coupled.\\
The stress energy tensor, in this case takes the form
  \beqa
 \label{defT}
T^{\mu\nu}_{scalar}
&\equiv&\frac{2}{\sqrt{g}}\frac{\delta S_0}{\delta g_{\mu\nu}}\nonumber \\
&=&\nabla^\mu \phi \, \nabla^\nu\phi - \frac{1}{2} \, g^{\mu\nu}\,g^{\alpha\beta}\,\nabla_\alpha \phi \, \nabla_\beta \phi
+ \chi \bigg[g^{\mu\nu} \Box - \nabla^\mu\,\nabla^\nu + \frac{1}{2}\,g^{\mu\nu}\,R - R^{\mu\nu} \bigg]\, \phi^2 .
\eeqa
A conformal free field theory realization stays conformal at quantum level, except for the appearance  of an anomaly in even dimensions. For an interacting theory, on the other hand, anomalous dimensions appear as soon as we switch-on an interaction in \eqref{phi}.\\
Coming to the definition of our correlators, in our conventions, $n$-point correlation functions will be defined as
\begin{equation}
\label{exps1}
\langle T^{\mu_1\nu_1}(x_1)\ldots T^{\mu_n\nu_n}(x_n)\rangle \equiv\frac{2}{\sqrt{g_1}}\ldots \frac{2}{\sqrt{g_n}}\frac{\delta^n \sm[g]}{\delta g_{\mu_1\nu_1}(x_1)\delta g_{\mu_2\nu_2}(x_2)\ldots \delta g_{\mu_n\nu_n}(x_n)} ,
\end{equation}
with $\sqrt{g_1}\equiv \sqrt{|\textrm{det} \, g_{{\mu_1 \nu_1}}(x_1)} $ and so on. \\
$\sm$ collects all the connected contributions of the correlation functions in the expansion with respect to the metric fluctuations, and may as well be expressed in a covariant expansion as 
\begin{equation}
\label{exps2}
\sm(g)=\sm(\bar{g})+\sum_{n=1}^\infty \frac{1}{2^n n!} \int d^d x_1\ldots d^d x_n \sqrt{g_1}\ldots \sqrt{g_n}\,\langle T^{\mu_1\nu_1}\ldots \,T^{\mu_n\nu_n}\rangle_{\bar{g}}\delta g_{\mu_1\nu_1}(x_1)\ldots \delta g_{\mu_n\nu_n}(x_n).
\end{equation}
Diagrammatically, for a scalar theory in a flat background, it takes the form
\begin{align}
\label{figg}
\sm(g)=& \sum_n \quad\raisebox{-6.9ex}{{\includegraphics[width=0.2\linewidth]{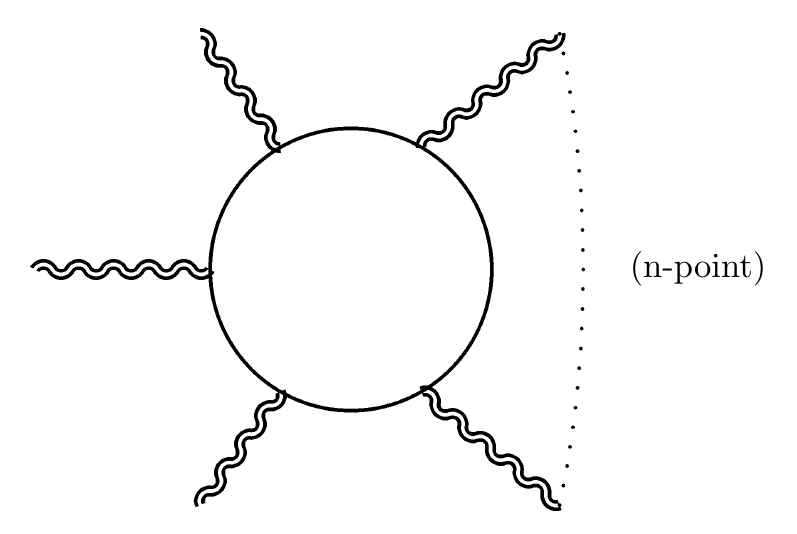}}}\,,
\end{align}
and defines the bare action in $d$ spacetime dimensions. As $d\to 4$ it will develop singularities in the form of single poles in $1/\epsilon$ that will be removed by the action of the counterterms.  
Here, the external weavy lines represent gravitational fluctuations, in terms of contributions that are classified as tadpoles, two-point , three-point  and $n$-point correlation functions of stress energy tensors. Tadpoles are removed in dimensional regularization (DR) in flat space, and the sum in \eqref{figg} starts from two-point  functions. The anomaly contributions start from three-point  functions.\\ 
Indeed, in the case of the scalar free field theory presented above, the topological contributions coming from the 4-$T$ are just summarized by the vertices
\begin{align}
\label{lighter}
\mathcal{S}_4(g)=\quad\raisebox{-4.9ex}{{\includegraphics[width=0.20\linewidth]{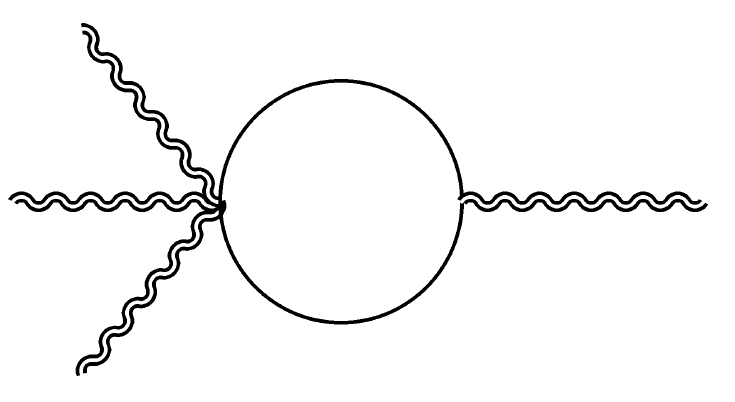}}}+\quad\raisebox{-4.5ex}{{\includegraphics[width=0.15\linewidth]{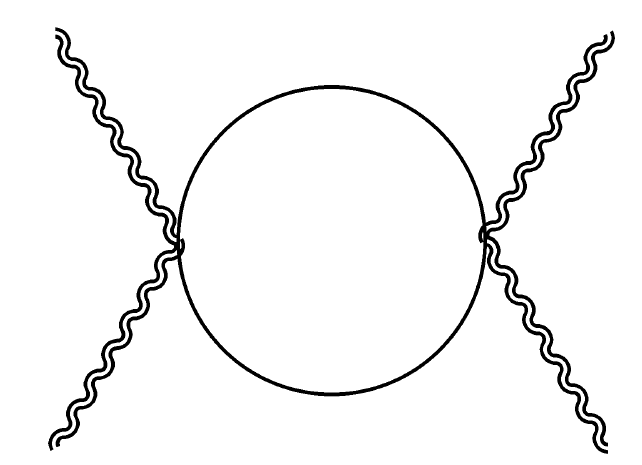}}}+\quad\raisebox{-3.5ex}{{\includegraphics[width=0.15\linewidth]{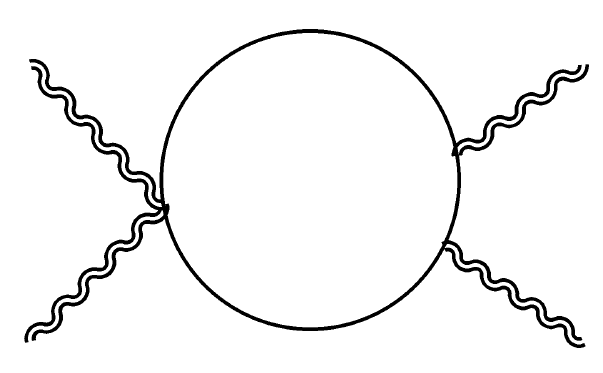}}}+\quad\raisebox{-6ex}{{\includegraphics[width=0.2\linewidth]{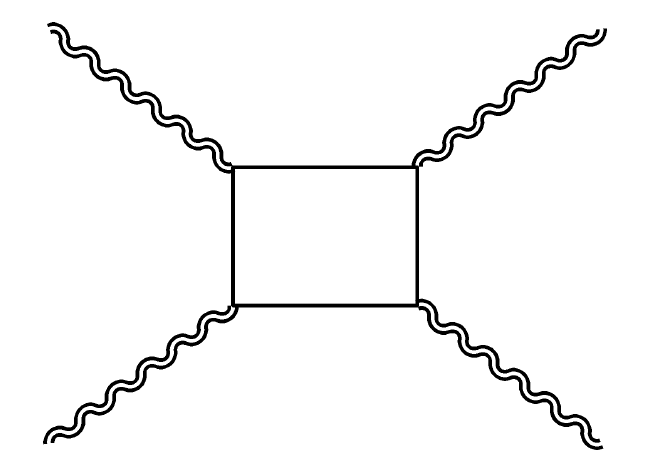}}}\nonumber \\
\end{align}
which can be directly computed in perturbation theory \cite{Serino:2020pyu}. 
Notice that the process of renormalization will require the inclusion of a counterterm action. An example is provided by ${S}_0$ as given in \eqref{phi}. In this case, the counterterm action will be determined by two gravitational counterterms, which will be defined below \eqref{counter}, and the renormalized partition action will be defined as 
\begin{equation}
Z[g]=\mathcal{N}\int D\phi e^{i(S_0(g,\phi) + S_{ct}(g))}=\mathcal{N}e^{i S_{ct}(g)}\int D\phi e^{i(S_0(g,\phi)}, 
\end{equation} 
where 
$\mathcal{N}$ is a normalization constant. By taking the logarithm of Z[g] one finds that 
 
 \begin{equation}
 \mathcal{S}_{ren}(g)=\mathcal{S}(g) + S_{ct}(g), 
 \end{equation}
 where $\mathcal{S}(g)$ is given in \eqref{exps2}.
 
In the example of the scalar theory presented above, this implies that the loop contributions are renormalized just using the vertices obtained by the functional differentiation of $S_{ct}$ with respect to the metric. the choice of different backgrounds allows to make contact with cosmology, if, for instance, $\bar{g}$ is chosen to be Weyl-flat, giving the opportunity for studying the conformal back reaction in De Sitter cosmology.

\subsection{From flat to more general backgrounds}
The derivation of the CWIs that we have discussed in flat space can all be re-obtained using the formalism of the effective action, moving from a general background to a flat one. The flat spacetime limit corresponds to the simplest non-trivial case and the transition to momentum space is, in this case, rather straightforward. 
Results obtained in this case, using the gravitational formulation of $\sm[g]$, 
are equivalent to those obtained for ordinary conformal field theory, if we specialize our results to flat space. This is rather obvious, since a CFT can be naturally defined without any reference to gravity. \\
In general, the variation $\sm[g]$ generates the semi-classical relation 
\begin{equation}
\label{anomx}
\frac{\delta \sm}{\delta \sigma(x)}=\sqrt{g} \,g_{\mu\nu}\,\langle T^{\mu\nu}\rangle, 
\qquad\textrm{where}\qquad  \langle T^{\mu\nu}\rangle = \frac{2}{\sqrt{g}}\frac{\delta \sm}{\delta g_{\mu\nu}},
\end{equation} 
is the quantum average of the stress energy tensor. If we introduce the infinitesimal Weyl variation of the metric 
  \begin{equation}
\delta_\sigma g_{\mu\nu}= 2 \sigma g_{\mu\nu}, 
\end{equation}
and its variation under diffeomorphisms ($x^\mu\to x^\mu+ \epsilon(x)$)  
\begin{equation} 
\delta_\epsilon g_{\mu\nu}=-\nabla_{\mu}\epsilon_{\nu}- \nabla_{\nu}\epsilon_{\mu}.
\end{equation}
the conditions of Weyl and diffeomorphism invariance of $\sm$ 
\begin{equation} 
\label{eww}
\delta_\sigma \sm=0, \qquad \delta_\epsilon \sm=0,
\end{equation}
are then summarised by the relations 
 \begin{equation}
\label{comby}
\langle T^\mu_\mu\rangle=0, \qquad \textrm{and}\qquad  \nabla_\mu\langle T^{\mu\nu}\rangle=0.
\end{equation}
Trace and conformal WI's can be derived from the equations above by functional differentiations of $\sm[g]$ with respect to the metric background, and are modified in the presence of an anomaly, which is introduced by the renormalization procedure. 
 The anomalous trace WIs can be derived by allowing for an anomaly contribution on the rhs of the $\sigma$ variation in \eqref{eww}
\begin{equation}
\label{plus}
\delta_\sigma \sm=\int d^4 x\sqrt{g} \bar{\mathcal{A}}(x), \qquad \langle T^\mu_\mu\rangle =\bar{\mathcal{A}}(x),
\end{equation}
which violates Weyl invariance. Here, $\sqrt{g} \bar{\mathcal{A}}(x)$ is the anomaly functional, whose structure we will be reviewing in a following section, using the heat-kernel (De Witt-Schwinger) method.\\
One can verify that there is no trace of the regulator $\mu$ in the expression of the anomaly functional, while this will characterize the breaking of the dilatation WI for a certain correlator. The clearest way to identify such behavior is to use free field theory realizations in various dimensions, and we will offer some explicit examples of this behavior in the next sections, by analysing the $TT$ correlator in $d=2$ and $d=4$, before moving to the $TTT$. 
  
The functional differentiations of the anomalous relations derived above generates the hierarchy of trace WIs which are an important part of the reconstruction program \cite{Coriano:2021nvn}.
%%%%%%%%%%%%%%%%%%%%%%%%%%%%%%%%%%%%%%%
\subsection{Spacetime symmetries and conserved currents}
%%%%%%%%%%%%%%%%%%%%%%%%%%%%%%%%%%%%%%%
It is interesting to identify the implications of the presence of spacetime symmetries which leave the effective action invariant. Two example are provided by metrics which allow Killing (KV) or conformal Killing vectors (CKV). The requirement that the action $\sm$ is invariant under such symmetries allows to define some conserved currents. These are constructed by contracting the stress energy tensor derived from $\sm$ with such 4-vector fields. 
The relations that one derives are valid at quantum level, since the stress energy tensor derived by varying $\sm$ is a quantum average. In particular, the use of conformal Killing vectors allows to define a conformal current $J_c$ which plays a key role in the derivation of the CWI's and of the corresponding CWI's in the flat spacetime limit. \\
We recall that in Minkowski space, scale invariance of a certain theory implies that the corresponding stress energy tensor has zero trace, and one can then define the conserved current
\begin{equation} 
J^\mu=x_\nu T^{\mu\nu}, \qquad \partial \cdot J=T^\mu_\mu=0.
\end{equation} 
At quantum level, in a curved background, we can generalize this approach by defining similar currents 
\begin{equation} 
\label{confcur}
\langle J^\mu\rangle =\epsilon_\nu^{(K)} \langle T^{\mu\nu}\rangle, 
\end{equation}
where $\epsilon_\mu^{(K)}(x)$ is a Killing or a conformal Killing vector field of the metric $g$. The proof of their conservation at quantum level follows closely the derivation at classical level. For instance, if we assume that $g$ allows vector isometries $\epsilon_\mu^{(K)}(x)$, which for changes of coordinates $x\to  x + \epsilon^{(K)}_\mu$ leave the metric invariant 
\begin{equation}
\delta_\sigma g_{\mu\nu}=-(\nabla_\mu \epsilon^{(K)}_\nu + \nabla_\nu \epsilon^{(K)}_\mu)=0,
\end{equation}
then the requirement of diffeomorphism invariance of the effective action $\sm[g]$ in that metric implies that the quantum average of $T^{\mu\nu}$  is conserved
\begin{equation} 
\nabla_\mu \langle T^{\mu\nu}\rangle=0.
\end{equation}
Combining this condition with the requirement that $\epsilon^{(K)}_\mu$ are Killing vectors,
then $J$ in \eqref{confcur} is conserved at quantum level
\begin{equation}
\label{iso}
\nabla\cdot \langle J\rangle =0. 
\end{equation}
A conformal current $J_c$ can be defined analogously to \eqref{confcur} by assuming that
the background metric $g$ allows conformal Killing vectors. 
In this more general case, we recall that the CKVs are solution of the equation
\begin{equation}
(ds')^2 = e^{2 \sigma(x)}(ds)^2 \qquad \leftrightarrow\qquad  \nabla_\mu \epsilon_\nu^{(K)} + \nabla_\nu \epsilon_\mu^{(K)} = 2 \sigma \delta_{\mu\nu}\qquad \sigma=\frac{1}{4}\nabla\cdot\epsilon^{(K)}.
\end{equation}
We remark that if we introduce a conformal current, now  using the CKVs of the background metric as in \eqref{confcur} in order to define $J_c$, if conditions \eqref{comby} are respected by $\sm$, then $J_c$ is conserved as in the isometric case \eqref{iso}. \\
On the other hand, anomalous CWI's are generated if we allow a Weyl-variant term in $\sm$, which takes place in $d=4$, after renormalization, as in the case of an anomaly action. \\
In this case the anomaly induces a non-zero trace, and modifies the semi-classical condition \eqref{iso} into the new form
\begin{eqnarray}
\nabla\cdot \langle J_c\rangle =\frac{1}{4}\nabla \cdot \epsilon^{(K)} \langle T^\mu_\mu\rangle  +\epsilon^{(K)}_\nu \nabla_{\mu} \langle T^{\mu\nu}\rangle=\frac{1}{4}\nabla \cdot \epsilon^{(K)} \langle T^\mu_\mu\rangle .
\end{eqnarray}
This relation can still be used for the derivation of the special CWI's of $n$-point functions, as shown in \cite{Coriano:2017mux,Coriano:2021nvn}.
Notice that $\sigma(x)$ is, at the beginning, a generic scalar function, which in a Taylor expansion around a given point $x^\mu$ is characterized by an infinite and arbitrary number of constants. Their number gets drastically reduced if we require that the spacetime manifold with metric $g$ allows a tangent space at each of its points, endowed with a flat conformal symmetry. 

In an equivalent, more general definition, one introduces an infinite dimensional abelian Weyl group $\mathcal{G}$ of transformations, with generators $J_x$ acting on the space of fields 
\begin{equation}
\label{jx}
J_x=-2 g^{\mu\nu}\frac{\delta}{\delta g^{\mu\nu}(x)} - \Delta_f  \Phi_f(x),
\end{equation}
summed over the fields $\Phi_f$, where $\Delta_f$ denote their scaling dimensions.
 The finite action of an element of $\mathcal{G}$ can be expressed as
 \begin{equation}
 e^{\sigma(x)\cdot J_x},\qquad \textrm{where} \qquad {\sigma(x)\cdot J_x}\equiv \int d^4 x \sigma(x) J_x,
 \end{equation}
with Weyl invariance of a generic action action expressed in the form 
\begin{equation} J_x\mathcal{S}_0[g,\Phi_f]=0. 
\end{equation}
In the case of the anomaly action $\Gamma[g]$, the expression \eqref{jx} includes only the variation of the metric, since all the remaining fields $\Phi_f$ have been integrated out, 
\begin{equation}
J_x \sm[g]=0.
\end{equation}
In flat space, the conformal Killing equation identify CKVs $\epsilon^\mu$ which are at most quadratic in $x$, are expressed in terms of the fifteen parameters $(a^\mu,\omega^{\mu\nu}, \lambda_s, b^\mu)$ of the conformal group, indicated as $K^\mu(x)$ 
\begin{equation}
\label{Kil}
\epsilon^{\mu}(x)\vline_{flat}\to K^\mu(x)= a^\alpha + \omega^{\mu\nu} x^\nu +\lambda_s x^\mu + b^\mu x^2 -2 x^\mu b\cdot x.
\end{equation}
Using such CKVs, the derivation of the special CWIs, following the approach of \cite{Coriano:2017mux}, can be performed directly in $d=4$, and takes to anomalous special CWIs. 
%%%%%%%%%%%%%%%%%%%%%%%%%%
\subsection{Polynomiality of the counterterms}
%%%%%%%%%%%%%%%%%%%%%%%%%%
Around a flat metric background, if we use a  mass-independent regularization scheme, the structure of the counterterms is polynomial in momentum space. \\
Indeed, in general, the breaking of Weyl invariance takes to the anomalous variation
\begin{equation}
 \delta_\sigma \Gamma=\frac{1}{(4 \pi)^2}\int d^4 x \sqrt{g}\left(c_1 R_{\mu\nu\rho\sigma}R^{\mu\nu\rho\sigma} + c_2 R_{\mu\nu}R^{\mu\nu} +c_3 R^2\right),
\end{equation}
which is constrained by the Wess-Zumino consistency condition 
\begin{equation}
\label{WZ}
\left[\delta_{\sigma_1},\delta_{\sigma_2}\right]\Gamma=0,
\end{equation}
to assume the form 
\beq
\label{anom1}
\delta_\sigma \Gamma=\frac{1}{(4\pi)^2}\int d^4 x\sqrt{g}\sigma\left( b_1 C^{(4)}_{\mu\nu\rho\sigma}C^{(4)\mu\nu\rho\sigma} + b_2 E_{4} +b_3 \square R\right),
\eeq
given in terms of the dimension-4 curvature invariants
\begin{align}
\label{fourd}
E_4&\equiv R_{\mu\nu\alpha\beta}R^{\mu\nu\alpha\beta}-4R_{\mu\nu}R^{\mu\nu}+R^2  ,\\
( C^{(4)})^2&\equiv R_{\mu\nu\alpha\beta}R^{\mu\nu\alpha\beta}-2R_{\mu\nu}R^{\mu\nu}+\frac{1}{3}R^2, \label{fourd2}
\end{align}
which are the Euler-Gauss-Bonnet (GB) invariant and the square of the Weyl conformal tensor, respectively, in $d=4$. 
The definition of $C^2$, in dimensional regularization, requires an expansion of $d$ around 4. For this purpose, the parametric dependence of $C^2$ on $d$ allows to introduce   
both a $(C^{(d)})^2$,  and a $(C^{(4)})^2$ operator, both defined in $d$ dimensions. This means that the contraction of the indices in both cases are all performed in $d$ spacetime dimensions while the scalar coefficients in their definition  {\em parametrically} dependent on $d$ 
\beq
C^{(d) \alpha\beta\gamma\delta}C^{(d)}_{\alpha\beta\gamma\delta}
=
R^{\alpha\beta\gamma\delta}R_{\alpha\beta\gamma\delta} -\frac{4}{d-2}R^{\alpha\beta}R_{\alpha\beta}+\frac{2}{(d-2)(d-1)}R^2,
\eeq
where
\beq
C^{(d)}_{\alpha\beta\gamma\delta} = R_{\alpha\beta\gamma\delta} -
\frac{2}{d-2}( g_{\alpha\gamma} \, R_{\delta\beta} + g_{\alpha\delta} \, R_{\gamma\beta}
- g_{\beta\gamma} \, R_{\delta\alpha} - g_{\beta\delta} \, R_{\gamma\alpha} ) +
\frac{2}{(d-1)(d-2)} \, ( g_{\alpha\gamma} \, g_{\delta\beta} + g_{\alpha\delta} \, g_{\gamma\beta}) R\, .
\eeq

We prefer to separate the parametric dependence from the range of variability of the tensor indices of this operator, since different choices of their range (and of the parametric dependence) induce finite renormalization of the effective action and change the anomaly by local terms. Here the term ``local'' refers to 
terms which are obtained from the Weyl variation of a local action. The remaining contributions, usually termed ``non-local'', refer to operators appearing in the anomaly functional which can be derived by varying a non-local action. An example of such action is the Riegert action, that we are going to discuss next.
In general one could define 
\begin{align}
& C^{(d)}_{\lambda\mu\nu\rho}\qquad     \lambda,\mu,\nu,\rho=0,1\ldots d-1,\notag \\
& C^{(4)}_{\lambda\mu\nu\rho}\qquad     \lambda,\mu,\nu,\rho=0,1\ldots d-1,\notag \\
& C^{(4) r}_{\lambda\mu\nu\rho},\qquad  C^{(d) r}_{\lambda\mu\nu\rho}  \qquad \lambda,\mu,\nu,\rho=0,1,2,3 \qquad  \textrm{restricted forms},
\end{align}
as possible operators appearing in the counterterm action that renormalizes the effective action. After differentiation of these expression with respect to the metric, only a contraction of the result with the metric itself will set a difference between these different definitions. If the indices of the differentiation are left open, there will be no extra dependence on $d$ which is generated in the various cases. \\
The operators above  have a {\em parametric} dependence on the dimension {\em and} an index variability, and both  can be 4 or $d$. In all the three cases, their Weyl-scaling 
\beq \label{GAUGED} g_{\mu \nu} = e^{2 \phi} \bar g_{\mu \nu}. \eeq

can be shown to be the same 
\begin{equation}  
C{}_{\lambda\mu\nu\rho}=e^{2 \sigma} \bar{C}_{\lambda\mu\nu\rho},\qquad C^{\lambda\mu\nu\rho}=e^{-6  \sigma} \bar{C}^{ \lambda\mu\nu\rho},
\end{equation}
and corresponds to Weyl covariance. Notice that this property of Weyl covariance holds both for $C^{(d)}$ and for $C^{(4)}$.
By contraction, one gets in all the cases 
\begin{align}
C^2\equiv g^{\alpha\mu}g^{\beta\nu}C^{\lambda}_{\gamma\alpha\beta}C^{\gamma}_{\lambda\mu\nu}=& e^{-4\sigma(x)}\bar{C}^{\lambda}_{\gamma\alpha\beta}\bar{C}^{\gamma}_{\lambda\mu\nu}=e^{-4\sigma(x)}\bar{C}^2.  
\end{align}
Notice that any version of the Weyl tensor caries the same symmetries of the Riemann tensor and satisfies a Bianchi identity
\beq
C_{\mu\nu\rho\sigma}=C_{\rho\sigma\mu\nu},
\eeq
\beq
C_{\mu\nu\rho\sigma}=-C_{\nu\mu\rho\sigma}=-C_{\mu\nu\sigma\rho},
\eeq
\beq
C_{\mu\nu\rho\sigma}+C_{\mu\sigma\nu\rho}+C_{\mu\rho\sigma\nu}=0.
\eeq
The choice of the Weyl counterterm is affected by the prescription dependence. For instance, one could choose a counterterm of the form 
\begin{equation} 
\label{choice1}
\frac{1}{\epsilon}V_{C^2}=\frac{\mu^{-\epsilon}}{\epsilon}\int d^d x\sqrt{g} (C^{(d)})^2,
\end{equation}
 or choose a variant of the form 
\begin{equation} 
\frac{1}{\epsilon}\tilde{V}_{C^2}=\frac{\mu^{-\epsilon}}{\epsilon}\int d^d x\sqrt{g} (C^{(4)})^2.
\end{equation}
We can, briefly, take a look at the differences between the two choices. Setting $d=4-\epsilon$ in the integrand of \eqref{choice1}, we obtain the relation
\beq
C^{(d)\alpha\beta\gamma\de}C_{(d)\alpha\beta\gamma\de}=R^{\alpha\beta\gamma\de}R_{\alpha\beta\gamma\de}-2R^{\alpha\beta}R_{\alpha\beta}+\frac{1}{3}R^2+\epsilon\lt -R^{\alpha\beta}R_{\alpha\beta}+\frac{5}{18}R^2 \rt,
\eeq
that is
\begin{equation}
\label{interm}
(C^{(4-\epsilon)})^2=(C^{(4)})^2 +\epsilon\left( - (R_{\mu\nu})^2 + \frac{5}{18}R^2\right).
\end{equation}
Using the property that the integration measure scales in $d$ dimensions as 
\begin{equation} 
g=\textrm{det} g_{\alpha\beta} =\epsilon^{\mu_1\ldots \mu_d}g_{0 \mu_1}\ldots g_{0\mu_d}\to e^{2 d \sigma} g ,
\end{equation}
 one obtains 
\begin{equation} 
\frac{\delta}{\delta \sigma(x)}\int d^d x \sqrt{-g}(C^{4)})^2(x)=\epsilon \sqrt{g}(C^{(4)})^2,
\end{equation}
as well as 
\begin{equation} 
\frac{\delta}{\delta \sigma(x)}\int d^d x \sqrt{-g}(C^{d)})^2(x)=\epsilon \sqrt{g}(C^{(d)})^2.
\end{equation}
As already mentioned, the use of \eqref{interm} before the Weyl variation in $\sigma$, generates a different result, since the operation of expanding in $d$ and the variation do not commute. This will always be true whenever the variations in the metric is accompanied by a contraction with the metric itself, as in a $\sigma$ variation. If we separate the two variations in the form \eqref{interm}, we need to compute separately the variation of the 
$O(\epsilon)$ contribution which is given by  
\begin{equation} 
\frac{\delta}{\delta \sigma(x)}\int d^d x \sqrt{-g}\left( - R_{\mu\nu}^2 + \frac{5}{18}R^2\right)=
-\frac{2}{3}\epsilon \sqrt{g} \Box R.
\end{equation}
In this way we obtain
\beq\label{C2 piu boxR}
\frac{1}{d-4}\frac{\de}{\de\sigma(y)}\int d^dx \rg (C^{(4-\epsilon)})^2=\rg\lt (C^{(4)})^2 -\frac{2}{3}\Box R\rt.
\eeq
In the first two cases, all the traces are performed in $d$ spacetime dimensions, as in ordinary dimensional regularization (DR), and the renormalization of the entire functional is then obtained by the addition of the counterterm action
\begin{align}
\label{counter}
\sm_{ct}&=-\frac{\mu^{-\varepsilon}}{\varepsilon}\,\int\,d^dx\,\sqrt{-g}\left(b\,(C^{(d)})^2+b'\,E\right),
\end{align}
corresponding to the Weyl tensor squared and the Euler density in $d=4$ as in \eqref{fourd} and \eqref{fourd2}. Here, $\mu$ is a renormalization scale. Notice that the Gauss-Bonnet term does not carry any parametric dependence in $d$ and it is quadratic in the curvatures at $d=4$, but its tensorial structure is expanded in $d$ dimensions, according to the regularization. The scaling of $E$ is more involved compared to $C^2$ and clearly depends on the dimension rather non-trivially
\beq\label{GaussB mode ex}
\rg E=\sqrt{\bar g} e^{(d-4)\phi}\biggl \{ \bar E+(d-3)\bar\nabla_\mu \bar J_1^\mu+(d-3)(d-4) \bar J_2  \biggl \} ,
\eeq
where we have defined
\beq \label{GBexJ}
\bar J_1^\mu=8\bar R^{\mu\nu}\bar\nabla_\nu\phi-4\bar R\bar \nabla^\mu\phi+4(d-2)(\bar\nabla^\mu\phi\bar \Box \phi-\bar \nabla^\mu\bar\nabla^\nu\phi\bar \nabla_\nu\phi+\bar\nabla^\mu\phi\bar\nabla_\lambda\phi\bar\nabla^\lambda\phi),
\eeq
\beq \label{GBexK}
\bar J_2=4\bar R^{\mu\nu}\bar\nabla_\mu\phi\bar\nabla_\nu\phi-2\bar R\bar\nabla_\lambda\phi\bar\nabla^\lambda\phi+4(d-2)\bar\Box\phi\bar\nabla_\lambda\phi\bar\nabla^\lambda\phi+(d-1)(d-2)(\bnabla_\lambda \phi\bnabla^\lambda \phi)^2.
\eeq
and all the barred operators are computed respect to the fiducial metric $\bar{g}_{\mu\nu}$.
 We are going to see through simple examples 
that the contribution to the regularization of the effective action given by $V_E$, with 
\begin{equation}
V_E=\int d^d x\sqrt{g} E,
\end{equation}
generates terms which are evanescent in $d=4$ and $d=2$. In $d=2$ $V_E$ is linear in the scalar curvature and given by the Einstein Hilbert action. \\
Notice that, being covariant under Weyl transformations, the Weyl tensor can be used to build a conformally invariant action at $d=4$ in the form
\beq
S_G=\alpha\int d^4 x \rg (C^{(4)})^2,
\label{azione gravitazionale conforme di weyl}
\eeq
which is the action of Weyl gravity. In the context of anomaly actions, the derivative of $V_{C^2}$ (i.e. $V'_{C^2}$) with respect to the dimension $d$ will be part of the renormalized action and the approach can be modified by the inclusion of a finite renormalization. 
 From now on we will be choosing $(C^{(d)})^2$ as the operator appearing in the counterterm action, and we will be using the variation
\begin{equation}
\label{dif1}
\frac{\delta}{(d-4)\delta \sigma (x) }\int d^d x \sqrt{-g} (C^{(d)})^2 =\sqrt{-g} (C^{(d)})^2,
\end{equation}
which differs, as shown above, from the analogous one
\begin{equation}
\label{nn1}
\frac{\delta}{(d-4)\delta \sigma (x) }\int d^d x \sqrt{-g} (C^{(4)})^2 =\sqrt{-g}\Bigg( (C^{(4)})^2
-\frac{2}{3}\square R\Bigg) ,
\end{equation}
introduced in some of the literature on the conformal anomaly. Therefore, unless explicitly stated, in all the equations below we will be using $C^2$ to refer to $(C^{(d)})^2$.
We remark that the number of Weyl invariants operators depends on the dimension. For instance, the case of $d=6$ is discussed in several works \cite{Bastianelli:2000hi,Ferreira:2015lna,Ferreira:2017wqz}.\\
One can derive an anomaly action of the Wess-Zumino form starting from those invariants, using the Weyl gauging approach \cite{Iorio:1996ad,Codello:2012sn}. An example, in the case of $d=6$, can be found in \cite{Coriano:2013nja}.  \\
The anomaly action that one introduces via the regularization process, is derived using counterterms which are defined  for any even dimension ($2 k$), and analytically continued around the integer value $k$.
 
 In generic (even) spacetime dimensions, the structure of the counterterm Lagrangian is modified accordingly, with the Euler/ GB  density given by
\beqa
\label{Ed}
E_d = \frac{ 1}{2^{d/2}}
\delta_{\mu_1 \cdots \mu_d}^{\nu_1 \cdots \nu_d}
{R^{\mu_1 \mu_2}}_{\nu_1 \nu_2} \cdots
{R^{\mu_{d-1} \mu_d}}_{\nu_{d-1} \nu_d} \ ,
\eeqa
which, for $d=4$, is quadratic in the curvatures, and is indeed given by \eqref{fourd}. The topological nature 
of $V_E$, which is metric independent, is evident only if $d=2k$, but the regularization procedure will modify the effective action with the inclusion of a finite contribution ($V'_E$) which is metric-dependent, since the integration of the Euler density is performed in $d=2 k +\epsilon$ dimensions.
%%%%%%%%%%%%%%%%%%%%%%%%%%%%%%%%%%%%%%%%%%
 \subsection{The finite renormalization induced by the Gauss-Bonnet term}
 %%%%%%%%%%%%%%%%%%%%%%%%%%%%%%%%%%%%%%%%%%
It is quite obvious that the inclusion of $V_E$ induces a finite renormalization of the effective action in $d=4$. This can be simply shown by noticing that both $V_E$ and $V_{C^2}$ manifest an explicit dependence on $\epsilon$, i.e. 
\beq
V_{E/C^2}\equiv V_{E/C^2}(d),
\end{equation} 
and the counterterm contributions can be expanded around $d=4$. 
We adopt the following notation. Given $f(d)$, it will be convenient to denote its Taylor expansion around $d=4$ in the form 
\begin{equation} 
f(d)=f(4) + \varepsilon f'(4) ,
\end{equation}
with coefficients which are square bracketed once they are computed at $d=4$. The expansion of $V_{E/C^2}(d)$ then takes the form 
\begin{equation} 
\label{expand}
\frac{1}{\varepsilon}V_{E/C^2}(d)=\frac{\mu^{-\varepsilon}}{\varepsilon}\left( V_{E/C^2}(4) + \varepsilon 
V_{E/C^2}'(4) +O(\varepsilon^2) \right),
\end{equation}
where the first correction to the residue at the pole in $\varepsilon$ comes from the derivative with respect to the dimension $d$. 
Notice that $V_E(4)$ 
\begin{equation}
V_E(4)=\int d^4 x \sqrt{g}E(x)=4\pi \chi(\mathcal{M}),
\end{equation}
is a topological term, typical of the spacetime manifold $(\mathcal{M})$,  and it is therefore metric-independent. Its value is related to the global topology of the spacetime and it is therefore a pure number. 
It is clear that the $1/\varepsilon V_E(4)$ term in \eqref{expand} will not contribute to the renormalization of the bare $nT$ vertex, since each possible counterterm coming from the differentiation of  $V_E(4)$ respect to the metric will vanish.
The only contribution of the $V_E$ as $\varepsilon\to 0$ to the renormalized anomaly action will be related to $V'_E(4)$, and it is indeed finite, as a 0/0 contribution in $\varepsilon$. \\
Therefore, the inclusion of $V_E$ in the expression of the anomaly action, will induce only a finite renormalization of the bare $nT$  vertex, and henceforth of the entire effective action, since this result remains valid to all orders.\\
Finally, we can relate $V'_E(4)$ to the anomaly by the equation 
\begin{equation}
\sqrt{g} E(x)=2 \lim_{d\to 4} g_{\alpha\beta}{V'_E}^{\alpha\beta},
\end{equation} 
where the indices of ${V'_E}^{\alpha\beta}$, after the limit, will be running only over the physical 4 dimensions. 

For a generic $nT$ correlator, the only counterterm needed for its renormalization, is obtained by the inclusion of a classical gravitational vertex generated by the differentiation of \eqref{counter} $n$ times. The coefficients $b$ and $b'$ will be fixed by the field content of the quantum corrections, i.e.  by the number of scalars, fermions and spin-$1$ field running inside the loops.\\
The renormalized effective action $\mathcal{S}_R$ is then defined by the sum of the two terms
\begin{equation}
\sm_R[g]=\sm[g] + \sm_{ct}[g],
\end{equation}
\begin{align}
\label{cct1}
\sm_{ct}=\raisebox{-0.8ex}{\includegraphics[width=0.15\linewidth]{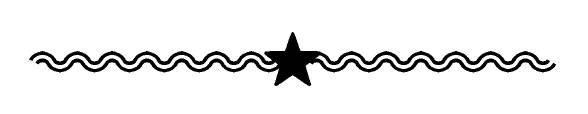}}+\raisebox{-5ex}{\includegraphics[width=0.15\linewidth]{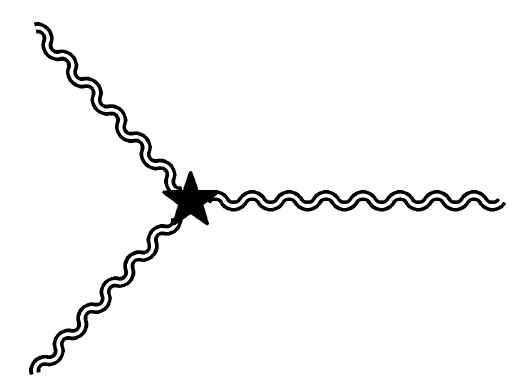}}+\raisebox{-6ex}{\includegraphics[width=0.15\linewidth]{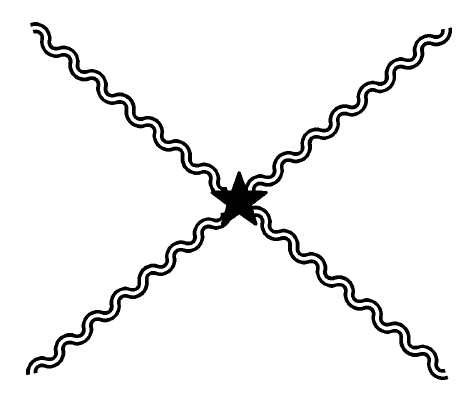}}+...
\end{align}
with $\sm_{ct}$ shown in \eqref{cct1}. Both terms of $\sm_R[g]$ are expanded in the metric fluctuations as in \eqref{cct1}. If we resort to a path integral definition of a certain CFT,  it is clear that any renormalized correlation function appearing in the expansion of $\sm_R[g]$ would be expressed in terms of a bare contribution accompanied by a counterterm vertex. Its structure is summarized by the equation 
\begin{equation}
\label{simp}
\sm_R(g)=\sm_f (4)  + V'_E(4) + V'_{C^2}(4), 
\end{equation}
 where $\sm_f (4)$ is the finite contribution obtained by subtracting to the bare effective action $\mathcal{S}$, evaluated in $d$ dimension the (single pole) singularities generated as $d\to 4$.   
The correlation functions extracted by the renormalized action can be expressed as the sum of a finite $(f)$  correlator and of an anomaly term ($anomaly$) in the form
\begin{align}
\label{cct}
 \braket{T^{\mu_1\nu_1}T^{\mu_2\nu_2}\ldots T^{\mu_n\nu_n}}_{Ren}=
&\bigg[\braket{T^{\mu_1\nu_1}T^{\mu_2\nu_2}\ldots T^{\mu_n\nu_n}}_{bare}+\braket{T^{\mu_1\nu_1}T^{\mu_2\nu_2}\ldots T^{\mu_n\nu_n}}_{count}\bigg]_{d\to4}=\notag\\
&=\braket{T^{\mu_1\nu_1}T^{\mu_2\nu_2}\ldots T^{\mu_n\nu_n}}^{(d=4)}_{f}+\braket{T^{\mu_1\nu_1}T^{\mu_2\nu_2}\ldots T^{\mu_n\nu_n}}^{(d=4)}_{anomaly}.
\end{align} 
The renormalized correlator shown above satisfies anomalous CWIs. 

\subsection{The inclusion of the anomaly}
To characterize the anomaly contribution to each correlation function, we start from the 1-point function. In a generic background $g$, the 1-point function is decomposed as 
\begin{equation}
\langle T^{\mu\nu}\rangle_{R}=\frac{2}{\sqrt{g}}\frac{\delta \sm_{R}}{\delta g_{\mu\nu}} =\langle T^{\mu\nu} \rangle_A  + \langle \overline{T}^{\mu\nu}\rangle_f,
\end{equation}
with
\begin{equation}
g^{\mu\nu}\frac{\delta \Gamma}{\delta g^{\mu\nu}} = g^{\mu\nu}\frac{\delta \sm_A}{\delta g^{\mu\nu}}\equiv \frac{\sqrt{g}}{2} g_{\mu\nu} \langle T^{\mu\nu} \rangle_A ,
\end{equation}
being the trace anomaly equation, and $\langle \overline{T}^{\mu\nu}\rangle_f$ is the Weyl-invariant (traceless) term. We will omit below the subscript $R$ from the quantum average $\langle T\rangle_R$.

Following the discussion in \eqref{anom1}, these scaling violations may be written for the 1-point function in the form
\begin{equation}
\label{anomeq}
\braket{T^{\mu}_{\ \ \mu}(x)}=\mathcal{A}(x),
\end{equation} 
- having dropped the suffix {\em Ren} from the renormalized stress energy tensor - 
where the finite terms on the right hand side of this equation denote the anomaly contribution
with
\begin{equation}
\label{AF}
\mathcal{A}(x)=\sqrt{-g(x)}\bigg[b\,C^2(x)+b'E(x)\bigg],
\end{equation}
being the anomaly functional. We will be needing several differentiation of this functional, 
evaluated in the flat limit. This procedure generates expressions which are  polynomial in the momenta, that can be found in the Appendix.
 In general, one also finds additional dimension-4 local invariants $\mathcal{L}_i$, if there are couplings to other background fields, as for instance in the QED and QCD cases, with coefficients related to the $\beta$ functions of the corresponding gauge couplings. \\
For $n$-point functions the trace anomaly, as well as all the other CWIs, are far more involved, and take a hierarchical structure. \\
For all the other WIs, in DR the structure of the equations can be analyzed in two different frameworks. \\
In one of them, we are allowed to investigate the correlators directly in $d$ spacetime dimensions, deriving ordinary (anomaly-free) CWI's, which are then modified by the inclusion of the 4-dimensional counterterm as $d\to 4$. In this limit, the conformal constraints become anomalous and the hierarchical equations are modified by the presence of extra terms which are anomaly-related. \\
Alternatively, it is possible to circumvent this limiting procedure by working out the equations directly in d=4, with the inclusion of the contributions coming from the anomaly functional, as we are going to show below. 

%%%%%%%%%%%%%%%%%%%%%%%%%%%%%%%%%%%%%%%%%%%%%%%%%%%
%%%%%%%%%%%%%%%%%%%%%%%%%%%%%%%%%%%%%%%%%%%%%%%%%%%
\section{The non-local action}
The non-local form of the action was originally introduced by Riegert \cite{1984PhLB..134...56R}. Discussions of such action in a cosmological context can be found in \cite{Hamada,Antoniadis:2011ib,Antoniadis:2006wq,Ferreira:2017wqz}. Local and non-local formulations of such actions, with the possible inclusions of extra degrees of freedom in the form of a dilaton, have been recently reviewed in \cite{Coriano:2019dyc}.\\
 In the following sections, we will study the origin of the trace anomaly as obtained in \cite{Birrell:1982ix}, with its explicit form first proposed in \cite{Deser:1976yx}. Then, using the approach of \cite{Riegert:1987kt}, we will construct a covariant, non-local action, which can be expanded around flat space in such a way that the trace of its  stress energy tensor takes the general form of the anomaly functional. Finally, using this anomaly effective action, we will show that the massless poles in the three-point correlation function $\braket{TTT}$ for a CFT in $d=4$, are precisely those obtained by the solution of the CWI's directly in flat space, by the methods of \cite{Bzowski:2013sza}, that we review in the appendix. This equivalence is the content 
 of \cite{Coriano:2017mux}. \\
A discussion of the effective action using the Schwinger-DeWitt expansion and the heat-kernel regularization. can be found in \cite{Birrell:1982ix}. 
In this section we are going to derive and discuss the non-local anomalous effective action in $d=4$ proposed in \cite{Riegert:1984kt}. \\
We have shown in the previous section that, in a four-dimensional system of massless fields in interaction with external gravitational  and gauge fields, the quantum corrections induce on the trace $T_A$ of the energy-momentum tensor $T_{\mu\nu}$ an anomaly of the general form
\begin{equation}
T_A=\left(bC^2+b'E+b''\,\square\,R+dR^2+cF^2\right),\label{G}
\end{equation}
where the coefficients $b$, $b'$, $b''$, $c$, $d$ depend on the specific particle content (fermion, scalar, vector) and interactions of the theory, as described by the corresponding Lagrangian. The main idea in \cite{Riegert:1984kt} is to extract information about the exact effective action $\mathcal{S}_{\scalebox{0.5}{$eff$}}$, which is related to the trace of $T_{\mu\nu}$ as
\begin{equation}
\frac{2}{\sqrt{-g}}g_{\m\n}\frac{\d\mathcal{S}_{\scalebox{0.5}{$eff$}}}{\d g_{\m\n}}=T\label{trace1},
\end{equation}
using only the form of the trace anomaly equation \eqref{G}. This is a general procedure and it does not depend on the explicit form of the coefficients in \eqref{G}. We define the part of the effective action which generates the anomalous trace $T_A$ , $\mathcal{S}_{\scalebox{0.6}{anom}}$, the {anomaly effective action}. For example, the contribution $\square R$ in \eqref{G},  can be generated as a variation of 
\begin{equation}
\sdfrac{2}{\sqrt{-g}}\,g_{\m\n}\sdfrac{\d}{\d g_{\m\n}}\int d^4x\,\sqrt{-g}\,R^2=-12\,\square \,R,\label{varRsq}
\end{equation}.
With a suitable renormalization factor this is given by $c/(192\p^2)\int d^4x\sqrt{-g}R^2$. However, the other terms in \eqref{G} can't be obtained from the trace of the functional variation of an integral of simple scalar geometric quantities. For this reason, we proceed first by using a non-covariant approach in order to find such terms, and then we will try to identify the covariant non-local form of the anomaly effective action. We use the property of the Weyl group, and we consider a local conformal parametrization of the metric of the form $g_{\m\n}=e^{2\s}\bar g_{\m\n}$ for arbitrary $\sigma(x)$, where $\bar g_{\m\n}$ has a fixed determinant.  
Inserting this expression into the trace anomaly equation we obtain
\begin{equation}
\frac{\d\mathcal{S}_{\scalebox{0.6}{anom}}}{\d\s}=\sqrt{-\bar{g}}\,e^{4\s}\,T_A.\label{an}
\end{equation}
The trace anomaly, by using the identity $E-C^2=-2(R^{\a\b}R_{\a\b}-\frac{1}{3}R^2)$, can be rewritten in the form
\begin{equation}
T_A=\left[(b+b')C^2-2b'\left(R^{\a\b}R_{\a\b}-\frac{1}{3}R^2\right)+b''\square\,R+dR^2+cF^2\right],\label{TA}
\end{equation}
and this simplifies the analysis. In fact, by conformal invariance, $C^\l_{\s\m\n}(g)=C^\l_{\s\m\n}(\bar g)\equiv\bar C^\l_{\s\m\n}$ and the gauge field  $F^i_{\m\n}$ is metric independent. Therefore we find that the $C^2$ and $F^2$ terms become
\begin{equation}
\begin{split}
C^2&\equiv g^{\a\m}g^{\b\n}C^\l_{\g\a\b}C^\g_{\l\m\n}=e^{-4\s}\bar g^{\a\m}\bar g^{\b\n}\bar C^\l_{\g\a\b}\bar C^\g_{\l\m\n}\equiv e^{-4\s}\bar C^2,\notag\\
F^2&\equiv g^{\a\m}g^{\b\n}F^i_{\a\b}F_{i\m\n}=e^{-4\s}\bar g^{\a\m}\bar g^{\b\n}F^i_{\a\b}F^i_{\m\n}\equiv e^{-4\s}\bar 
F^2,\notag
\end{split}
\end{equation}
Notice that for the fil
and the equation to be solved to account for these terms is given by \eqref{an}
\begin{equation}
\frac{\d\mathcal{S}_{\scalebox{0.6}{anom}}^{C^2+F^2}}{\d\s}=\sqrt{-\bar{g}}\bigg[(b+b')\bar{C}^2+c\bar{F}^2\bigg].
\end{equation}
It admits the solution
\begin{equation}
\mathcal{S}_{\scalebox{0.6}{anom}}^{C^2+F^2}=\int d^4x\sqrt{-\bar{g}}\ \left[(b+b')\bar C^2+e\bar F^2\right]\s,\label{act}
\end{equation}
modulo an arbitrary functional, independent of $\s$, which in any 
case can be added to the non-anomalous part of the effective action.
The remaining piece of the anomaly to study is $R^{\a\b}R_{\a\b}-\frac{1}{3}R^2$ that, under a local conformal parametrization of the metric, behaves like
\begin{align}
\frac{1}{2}\sqrt{-g}( C^2 - E)&= \sqrt{-g}\left(R^{\a\b}R_{\a\b}-\frac{1}{3}R^2 \right)\notag\\
&=\sqrt{-\bar g}\left(\bar R^{\a\b}\bar R_{\a\b}-\frac{1}{3}\bar R^2 -4\bar R^{\a\b}\left(\bar\nabla_{\a}\bar\nabla_\b\s-\bar\nabla_{\a}\s\bar\nabla_\b\s\right)+2\bar R\bar\square\,\s \right.\notag\\
&\qquad -4(\bar\square\,\s)^2-4\bar\square\,\s\bar\nabla_\a\s\bar\nabla^\a\s+4\bar\nabla_\a\bar\nabla_\b\s\bar\nabla^\a\bar\nabla^\b\s-8\bar\nabla^\a\nabla^\b\s\bar\nabla_\a\s\bar\nabla_\b\s\Big)\label{var1},
\end{align}
where the barred notation indicates that the dependence is on the metric $\bar{g}_{\mu\nu}$. In the next section we will illustrate a constructive method by which one can build an action, given its equation of motion, and we will apply it to \eqref{var1} in order to find the last missing piece of the anomaly effective action in its local form, before obtaining, at the last step, its non-local expression.  

%%%%%%%%%%%%%%%%%%%%%%%%%%%%%%%%%%%%%%%%%%%%%%%%%%%%%%%%%%%%%%%%%%%
\subsection{Reconstruction of the local anomaly effective action}
%%%%%%%%%%%%%%%%%%%%%%%%%%%%%%%%%%%%%%%%%%%%%%%%%%%%%%%%%%%%%%%%%%%

If we consider an action $A[\phi]$ which depends on a set of fields $\phi_\k(x)$, then its equations of motion are obtained through the stationary principle, with the variation of the action written as
\begin{equation}
\delta A=\int\,dx\,\sdfrac{\delta A}{\delta \phi_\k(x)}\delta\phi_\k(x).
\end{equation}

Now suppose that $\delta A/\delta\phi_\k(x)$ is a given function of the fields and their derivatives. We can use the information about $\delta A/\delta\phi_k(x)$ to reconstruct the action $A[\phi]$. First of all we have to choose a reference configuration $\bar{\phi}_\k(x)$ of the fields. Then, in order to evaluate $A[\psi]$ for an arbitrary choice of fields $\phi_\k(x)=\psi_\k(x)$, we choose a ``path'' $\phi_\k(x,\l)$, $0\le\l\le1$, that begins at the reference point $\phi_\k(x,0)=\bar{\phi}_\k(x)$ and leads to the desired final point, $\phi_\k(x,1)=\psi_\k(x)$. For instance, a straight line path 
\[\phi_\k(x,\l)=\l\,\psi_\k(x)+(1-\l)\bar{\phi}_\k(x),\]
may be a convenient choice. Now consider the action $A[\phi(\l)]$ evaluated along the path; its derivative with respect to $\l$ is 
\begin{equation}
\sdfrac{d}{d\l}A[\phi(\l)]=\int dx\left.\sdfrac{\delta A}{\delta\phi_\k(x)}\right|_{\phi=\phi(\l)}\sdfrac{\partial \phi_\k(x,\l)}{\partial\l},
\end{equation}
and the integration of this equation from $\l=0$ to $\l=1$ gives
\begin{equation}
A[\psi]=\int dx\,\int_0^1\,d\l\,\left.\sdfrac{\delta A}{\delta\phi_\k(x)}\right|_{\phi=\phi(\l)}\sdfrac{\partial \phi_\k(x,\l)}{\partial\l}\,+\,A[\bar{\phi}].
\end{equation}
This equation determines $A[\psi]$, up to an arbitrary additive constant $A[\bar\phi]$. 
It is worth mentioning that it may be impossible to find an action that identically reproduces a given set of equations of motion as its Euler-Lagrange equations. Nevertheless, one can sometimes change the form of the equations without changing their content in such a way that the modified equations do follow from an action principle. 
{ We apply this method to find the corresponding term of \eqref{var1} in the anomaly effective action. The equation of motion that we consider is written in the form
\begin{align}
\frac{\d\mathcal{S}_{\scalebox{0.6}{anom}}^{C^2-E}}{\d\s}&=\sqrt{-\bar g}\left(\bar R^{\a\b}\bar R_{\a\b}-\frac{1}{3}\bar R^2 -4\bar R^{\a\b}\left(\bar\nabla_{\a}\bar\nabla_\b\s-\bar\nabla_{\a}\s\bar\nabla_\b\s\right)+2\bar R\bar\square\,\s \right.\notag\\
&\qquad -4(\bar\square\,\s)^2-4\bar\square\,\s\bar\nabla_\a\s\bar\nabla^\a\s+4\bar\nabla_\a\bar\nabla_\b\s\bar\nabla^\a\bar\nabla^\b\s-8\bar\nabla^\a\nabla^\b\s\bar\nabla_\a\s\bar\nabla_\b\s\Big)\label{an1}.
\end{align}
Following the method above, in order to solve the equation \eqref{an1} and find $\mathcal{S}_{\scalebox{0.6}{anom}}^{C^2-E}[\s]$, we choose a path $\s(x,\l)$, $0\le\l\le1$ that begins from a reference point $\s(x,0)=0$ and goes to the final point $\s(x,1)=\tilde\s$. The simplest choice we are going to consider is a straight line, that in a parametric form is $\s(x,\l)=\l\tilde\s$. The variation of the action $\mathcal{S}_{\scalebox{0.5}{\text{anom}}}^{C^2-E}[\s]$ along this path is given by
\begin{equation}
\sdfrac{d}{d\l}\mathcal{S}_{\scalebox{0.6}{anom}}^{C^2-E}[\s(\l)]=\int d^4x\left.\frac{\d\mathcal{S}_{\scalebox{0.6}{anom}}}{\d \s}\right|_{\s=\s(\l)}\sdfrac{\partial \s(x,\l)}{\partial \l}.
\end{equation}
Integrating in the range of variations of the parameter $0\le\l\le1$ the action becomes
\begin{align}
\mathcal{S}_{\scalebox{0.6}{anom}}^{C^2-E}[\tilde{\s}]&=\int d^4x\int^1_0d\l\, \sqrt{-\bar g}\left(\bar R^{\a\b}\bar R_{\a\b}-\frac{1}{3}\bar R^2 +4\bar R^{\a\b}\left(\l\bar\nabla_{\a}\bar\nabla_\b\tilde\s-\l^2\bar\nabla_{\a}\tilde\s\bar\nabla_\b\tilde\s\right)-2\l\bar R\bar\square\,\tilde\s \right.\notag\\
&\qquad -4\l^2(\bar\square\,\tilde\s)^2-4\l^3\bar\square\,\tilde\s\bar\nabla_\a\tilde\s\bar\nabla^\a\tilde\s+4\l^2\bar\nabla_\a\bar\nabla_\b\tilde\s\bar\nabla^\a\bar\nabla^\b\tilde\s-8\l^3\bar\nabla^\a\nabla^\b\tilde\s\bar\nabla_\a\tilde\s\bar\nabla_\b\tilde\s\Big)\tilde\s\label{CE},
\end{align}
and the integration over $\lambda$ is simple to achieve and generates extra terms which are independent of $\sigma$. We integrate by parts, and by using the properties
\begin{equation}
\bar\nabla_\m\bar\square\,\tilde\s\equiv\bar\square\,\bar\nabla_\m\tilde\s+\bar R_{\m\n}\bar\nabla^\n\tilde\s,\qquad \bar\nabla_\m\bar R^{\m\n}\equiv\frac{1}{2}\bar\nabla^\n \bar R,
\end{equation} 
we find the final form of the action for the term $1/2 (C^2-E)$ in the trace anomaly as
\begin{align}
\mathcal{S}_{\scalebox{0.6}{anom}}^{C^2-E}[\s]&=\int d^4x\sqrt{-\bar g}\left[\left(\bar R^{\a\b}\bar R_{\a\b}-\frac{1}{3}\bar R^2\right)\s -2\bigg(\bar R^{\a\b}-\frac{1}{2}g^{\a\b}\bar R\bigg)\bar\nabla_{\a}\s\bar\nabla_\b\s\right.+2\bar\nabla_\a\s\bar\nabla^\a\s\bar\square\,\s+(\bar\nabla_\a\s\bar\nabla^\a\s)^2\bigg]\label{CEF},
\end{align}
having redefined $\tilde{\sigma}\to \sigma$ after the parametric integration.
It is straightforward to verify that the variation with respect to $\s$ of $\mathcal{S}_{\scalebox{0.6}{anom}}^{C^2-E}[\s]$, obtained in \eqref{CEF}, produces exactly the trace relation \eqref{an1}.

The remaining piece of the trace anomaly $T_A$ to consider is the $R^2$ term in \eqref{TA}. Following the same procedure discussed above, one can obtain the form of the action as
\begin{equation}
\mathcal{S}_{\scalebox{0.6}{anom}}^{R^2}=\int d^4 x\sqrt{-\bar g}\left\{\bar R^2\s+12\bar R\left(\frac{\s}{2}\bar\square\,\s+\frac{\s}{3}\bar\nabla^\a\s\bar\nabla_\a\s\right)+36\left[\frac{\s}{3}(\bar\square\,\s)^2+\frac{\s}{2}\bar\square\,\s\bar\nabla^\a\s\bar\nabla_\a\s+\frac{\s}{5}(\bar\nabla^\a\s\bar\nabla_\a\s)^2\right]\right\}.\label{SR2}
\end{equation}
However, this action, once it is functionally differentiated, does not reproduce the $R^2$ contribution in \eqref{TA}, and this issue is related to the fact that the functional derivative of \eqref{SR2} is not a symmetric kernel. Hence, we conclude that there exists no action, \emph{local} or \emph{non-local}, which has $R^2$ as its trace.\\
Finally, we have found an action that reproduces all the parts of the trace anomaly, except for the $R^2$ piece, and it is written in a local non-covariant form 
\begin{equation}
\mathcal{S}_{anom, loc}=\mathcal{S}^{C^2-E}_{anom} +\mathcal{S}^{C^2 + F^2}_{anom}.
\end{equation}

Now we will describe the procedure to obtain the non-local and covariant form of the anomaly action. 
First of all, we consider the Weyl transformation $g_{\mu\nu}=e^{2\s}\bar g_{\mu\nu}$ of $\sqrt{-g} E$ and $\sqrt{-g} \square\, R$, commonly used in the Weyl gauging 
\cite{Coriano:2013nja,Coriano:2013xua,Codello:2012sn}, written explicitly as
\begin{align}
\sqrt{-g}\,E&=\sqrt{-\bar g}\left[\bar E+8\bar R^{\a\b}\left(\bar\nabla_\a\bar\nabla_\b\s-\bar\nabla_\a\s\bar\nabla_\b\right)-4\bar R\bar\square\,\s-8(\bar\square\,\s)^2\right.\notag\\
&\qquad\qquad\qquad\qquad\qquad-\left.8\bar\square\,\s\bar\nabla^\a\s\bar\nabla_\a\s+8\bar\nabla^\a\bar\nabla^\b\s\bar\nabla_\a\bar\nabla_\b\s-16\bar\nabla^\a\bar\nabla^\b\s\bar\nabla_\a\s\bar\nabla_\b\s\right]\label{E1},\\
\sqrt{-g}\,\square\, R&=\sqrt{-\bar g}\left[\bar\square\,\bar R-12R^{\a\b}\bar\nabla_\a\s\bar\nabla_\b\s-2
\bar R\bar\square\,\s-2\bar\nabla_\a\bar R\bar\nabla^\a\s-6\bar\square^2\s\right.\notag\\
&\left.\qquad\qquad +12(\bar\square\,\s)^2+12\bar\square\,\s\bar\nabla^\a\s\bar\nabla_\a\s+24\bar\nabla^\a\bar\nabla^\b\s\bar\nabla_\a\s\bar\nabla_\b\s-12\bar\nabla^\a\bar\nabla^\b\s\bar\nabla_\a\bar\nabla_\b\s\right]\label{R1},
\end{align}
and we observe that the combination 
\begin{equation}
\frac{1}{4}\sqrt{-g}\left(E-\frac{2}{3}\square\, R\right)=\sqrt{-\bar g}\left[\frac{1}{4}\left(\bar E-\frac{2}{3}\bar\square\, \bar R\right)+\bar\D_4\s\right]\label{point}
\end{equation}
is no longer of order $\sigma^3$, but it is linear in $\sigma$ and $\D_4$ is a fourth order differential operator that behaves under Weyl transformations as 
\begin{align}
\D_4&\equiv \square^2+2 R^{\m\n}\nabla_\m\nabla_\n-\frac{2}{3}R\,\square\,+\frac{1}{3}(\nabla^\m\,R)\nabla_\m,\\
\sqrt{-g}\,\D_4&=\sqrt{-\bar g}\,\bar{\D}_4,\label{point2}
\end{align}
 \cite{Riegert:1987kt,Antoniadis:1992xu,Antoniadis:1991fa,Mazur:2001aa}. Consequently, the operator $\Delta_4$, is the (unique) fourth-order conformal covariant differential operator acting on a field of zero scale dimension. Furthermore, $\Delta_4$ has the property of being self-adjoint 
\begin{equation}
\int d^4x\sqrt{-g}\,\y(\D_4\x)=\int d^4x\sqrt{-g}\,(\D_4\y)\x\label{point3},
\end{equation}
where $\x$ and $\psi$ are scalar fields of zero scaling dimensions. The second property implies the existence of an action  
\begin{equation}
S[\xi]=\int\,d^4x\sqrt{-g}\left[\sdfrac{1}{2}(\square\,\xi)^2+R^{\mu\nu}\,\nabla_\mu\xi\,\nabla_\nu\xi-\sdfrac{1}{3}R\,\nabla^\mu\xi\,\nabla_\mu\xi\,\right],
\end{equation}
whose variation gives $\D_4\xi$. The cornerstone in the construction of the non-local and covariant form of the anomaly effective action is in this operator. We define the Green's function $D_4(x,y)$ inverse of $\D_4$ by
\begin{equation}
(\sqrt{-g}\,\D_4)_xD_4(x,y)=\d^4(x-y),\label{point4}
\end{equation}
that is conformally invariant, by virtue of the independence of $\d^4(x-y)$ with respect to the metric. We invert \eqref{point} using the properties of the operator $\Delta_4$ to find the explicit form of the function $\sigma(x)$, and we obtain
\begin{equation}
\s(x)=\int d^4y\,D_4(x,y)\left[\frac{1}{4}\sqrt{-g}\left(E-\frac{2}{3}\square\, R\right)\right]_y+\text{$\s$ independent terms}.
\end{equation}
Using the explicit expression of $\sigma$, we finally find the non-local but manifestly covariant anomaly effective action as
\begin{equation}
\mathcal{S}_{\rm anom}^{^{NL}}[g] =\frac {1}{4}\!\int \!d^4x\sqrt{-g_x}\, \left(E - \frac{2}{3}\sq R\right)_{\!x} 
\int\! d^4x'\sqrt{-g_{x'}}\,D_4(x,x')\left[\frac{b'}{2}\, \left(E - \frac{2}{3}\sq R\right) +  b\,C^2\right]_{x'}
\label{Snonl},
\end{equation}
with the condition $b''=-2/3b'$ in \eqref{TA}, and we have used the superscript $NL$ to indicate that this is the non-local form of the anomaly effective action. Finally, we can impose the condition on the coefficient $c=0$ in \eqref{TA} since a non-zero $R^2$ in this basis cannot be obtained from any effective action (local or not) \cite{Antoniadis:1992xu, Bonora:83, Mazur:2001aa}. 
}
%%%%%%%%%%%%%%%%%%%%%%%%%%%%%%%%%%%%%%%%%%%%%%%%%%%%%%%%%%%%
\subsection{Anomalous trace Ward identities}
\label{Sec:TraceWI}
%%%%%%%%%%%%%%%%%%%%%%%%%%%%%%%%%%%%%%%%%%%%%%%%%%%%%%%%%%%%%
{
In this and the next sections we will focus just on the gravitational part of the anomaly effective action. As we have previously seen, the renormalization procedure gives rise to a non-vanishing trace of the energy momentum tensor. If one considers any correlation function involving at least two stress energy tensors in $d=4$, the trace Ward identity \eqref{tracettt} has to be modified in order to consider the anomalous trace contributions. As in the case of the conservation Ward identities, the trace identities for the $n$-point functions may be derived by successive
variation of the fundamental trace identity of the one-point function. In fact, considering only the $b$ and $b'$ terms, and rewriting \eqref{trace1} in the form
\begin{equation}
2\, g_{\m\n}(x)\,\frac{\d \mathcal{S}[g]}{\d g_{\m\n}(x)} =\mathcal{A} \equiv \sqrt{-g}\, \Big\{ b\, C^2 + b' \big(E - \tfrac{2}{3}\sq R\big) \Big\},
\label{Adef}
\end{equation}
by varying again with respect to the metric, and finally evaluating with a flat space Minkowski metric $\eta_{\m\n}$ gives
\begin{equation}
\eta_{\m_1\n_1}\braket{T^{\m_1\n_1}(x_1)T^{\m_2\n_2}(x_2)} = 2\, \frac{\d \mathcal{A}(x_1)}{\d g_{\m_2\n_2}(x_2)}\Bigg\vert_{flat},
\label{twoptrace}
\end{equation}
for the two-point function, and 
\begin{align}
\eta_{\m_1\n_1}\braket{T^{\m_1\n_1}(x_1)T^{\m_2\n_2}(x_2)T^{\m_3\n_3}(x_3)}&=
- 2\,\Big\{\d^4 (x_1-x_2) + \d^4(x_1-x_3)\Big\}\, \braket{T^{\m_2\n_2}(x_2)T^{\m_3\n_3}(x_3)} \notag\\
&\hspace{2cm}+ \, 4 \, \frac{\d^2\mathcal{A}(x_1)}{\d g_{\m_2\n_2}(x_2)\d g_{\m_3\n_3}(x_3)}\Big|_{flat}
\label{threeptrace},
\end{align}
for the three-point function. The corresponding results in momentum space are obtained by Fourier transforming the previous relations giving
\begin{equation}
\eta_{\a_1\b_1} \, \braket{T^{\a_1\b_1}(p)T^{\m_2\n_2}( -p)} =  2\, \tilde{\mathcal{A}}_1^{\m_2\n_2} (p)
\label{twoptr},
\end{equation}
for the two-point function and 
\begin{align}
&\d_{\a_1\b_1}\braket{T^{\a_1\b_1}(p_1)T^{\m_2\n_2}(p_2)T^{\m_3\n_3}(p_3)}\notag\\
&\hspace{2cm}= - 2\,\braket{T^{\m_2\n_2}(p1+p2)T^{\m_3\n_3}(p_3)} -2\, \braket{T^{\m_2\n_2}(p_2)T^{\m_3\n_3}(p_1 + p_3) }
+ \, 4 \, \tilde {\mathcal{A}}_2^{\m_2\n_2\m_3\n_3} (p_2,p_3),
\label{threeptr}
\end{align}
for the three-point function, where 
\begin{align}
&(2\pi)^4 \,\d^4 (p_1+ \dots + p_{n+1})\, \tilde{ \mathcal{A}}_{n}^{\m_2\n_2\dots\m_{n+1}\n_{n+1}}(p_2,\dots ,p_{n+1}) \notag\\
& \left.\qquad\qquad \equiv \int d^4x_1\dots d^4x_{n+1} \ e^{i p_1\cdot x_1 + \dots + i p_{n+1}\cdot x_{n+1}} \, 
\frac{\d^{n}\mathcal{A}(x_1)}{\d g_{\m_2\n_2}(x_2)\dots \d g_{\m_{n+1}\n_{n+1}}(x_{n+1})}\right|_{flat}
\label{Avardef},
\end{align}
is the Fourier transform of the $n^{th}$ variation of the anomaly in the flat space limit.} The locality of $\mathcal{A}(x_1)$ implies 
that $\tilde{\mathcal{A}}_{n}^{\m_2\n_2\dots\m_{n+1}\n_{n+1}}$ is a polynomial with only positive powers of the $p_j$, containing no 
$1/p_j^2$ pole terms or logarithms. We note also that if $b'' \neq 0$, the $b'' \sq R$ term may easily be included in $\mathcal{A}$, 
giving an additional local contribution to $\tilde {\mathcal{A}}_{n}$.

The trace identity (\ref{threeptr}) for the three-point function contains two terms involving the two-point  function, which would
usually be considered ``non-anomalous'', since they are present even if $\tilde {\mathcal{A}}_2 =0$, notwithstanding the fact that the two-point  correlation function
itself carries an implicit dependence upon the first variation $\tilde{ \mathcal{A}}_1$ through (\ref{twoptr}). In addition, (\ref{threeptr}) contains the  
last term involving the second variation $\tilde {\mathcal{A}}_2$, which is anomalous. Clearly one may take additional variations of the fundamental trace identity 
(\ref{Adef}) with respect to the metric in order to obtain trace identities for higher $n+1$-point functions, and this pattern will continue 
with the hierarchy of trace WI's, each implicitly dependent upon the $(n-1)-${th} and on the lower variations, as well as on their non-anomalous parts. At each order we encounter an explicit new anomalous 
term involving the $n-${th} variation of the trace anomaly $\tilde {\mathcal{A}}_n$.

%%%%%%%%%%%%%%%%%%%%%%%%%%%
\subsection{The total effective action}
\label{Sec:AnomAct}
%%%%%%%%%%%%%%%%%%%%%%%%%%%
It is worth mentioning that the anomaly effective action found in \eqref{Snonl} satisfies the Wess-Zumino consistency condition, for which 
\begin{equation}
\mathcal{S}_{\rm anom}[e^{2\s}\bar g] = \mathcal{S}_{\rm anom}[\bar g] + \G_{WZ}[\bar g;\s]
\label{WZGamma},
\end{equation} 
for an arbitrary Weyl transformation of the metric $g_{\m\n}(x) = e^{2\s (x)} \,\bar g_{\m\n}(x)$, and whose variation is the anomaly, i.e. \eqref{Adef}. The latter equation can be written in the  equivalent form
\begin{equation}
\frac{\d \G_{WZ}[\bar g;\s]}{\d \s (x)}=\sqrt{-g}\, \Big\{ b\, C^2 + b' \big(E - \tfrac{2}{3}\sq R\big) \Big\}\Big\vert_{g = e^{2\s}\bar g}.
\label{traceiden}
\end{equation}
Moreover, the general form of the anomaly is the consequence of the locality of the underlying QFT,  and of the Wess-Zumino consistency condition exposed above. As shown in the previous section, one constructs the Wess-Zumino functional in (\ref{WZGamma}), quartic in $\sigma$, as 
\begin{align}
\G_{_{\!W\!Z}}[\bar g;\s] &= 2 b'\! \int\,d^4x\,\sqrt{-\bar g}\ \s\,\bar\D_4\,\s + \int\,dx\ \overline{\!\mathcal{A}} \, \s\notag\\
& = b' \int\,d^4x\,\sqrt{-\bar g}\,\Big[2\,\s\,\bar\Delta_4\,\s + \left(\bar E - \tfrac{2}{3} \sqb \bar R\right)\s \Big]
+ b \int\,d^4x\,\sqrt{-\bar g}\, \bar C^2\,\s.
\label{WZfour}
\end{align}
We have seen that inverting the equation \eqref{point} in order to express $\s$ in terms of the Green's function of the fourth differential operator $\Delta_4$, we can write the Wess-Zumino functional in the form
\begin{equation}
\G_{_{\!W\!Z}}[\bar g;\s] = \mathcal{S}_{\rm anom}^{^{NL}}[g=e^{2\s}\bar g] - \mathcal{S}_{\rm anom}^{^{NL}}[\bar g],
\label{Weylshift}
\end{equation}
with $\mathcal{S}_{\rm anom}^{^{NL}}[g]$ the non-local form of the exact quantum 1PI anomaly action defined in \eqref{Snonl}. This part of the anomaly is the part which is not Weyl-invariant, and cannot be removed by any addition of local terms in the effective action, such as the $\int R^2$ associated with $b''$ term in \eqref{G} by (\ref{varRsq}). Moreover, there is the possibility of adding an arbitrary Weyl-invariant term (local or not) in the effective action, producing no changes in the structure of \eqref{Adef}, but dropping out of the difference in (\ref{Weylshift}). In addition, these kind of terms cannot remove the
non-locality in the Weyl non-invariant part of the anomaly action (\ref{Snonl}).
For instance, adding the non-local Weyl invariant term 
\begin{equation}
\sdfrac {b^2}{8b'}\!\int \!d^4x\sqrt{-g_x}\, \big(C^2\big)_{x} \int\! d^4x'\sqrt{-g_{x'}}\,D_4(x,x') \big(C^2\big)_{x'}\,,
\label{SaddW}
\end{equation}
to \eqref{Snonl}, we will obtain an anomaly effective action written as
\begin{equation}
\mathcal{S}_{\rm anom}^{^{NL}}[g]+\sdfrac {b^2}{8b'}\!\int \!d^4x\sqrt{-g_x}\, \big(C^2\big)_{x} \int\! d^4x'\sqrt{-g_{x'}}\,D_4(x,x') \big(C^2\big)_{x'}= \sdfrac {1}{8b'}\int d^4x \int d^4x'\, \mathcal{A}(x)\, D_4(x,x')\, \mathcal{A}(x'),
\label{Snonlsq}
\end{equation}
that still satisfies the trace relation and is symmetrical in the invariants $E$ and $C^2$. 
As shown in \cite{Mottola:2006ew,Shapiro:1994ww}, one can derive a local form of the effective action by introducing a single new scalar field $\varphi$, to get
\begin{align}
&&\hspace{-1.5cm} \mathcal{S}_{\rm anom}[g;\varphi] \equiv -\sdfrac{b'}{2} \int d^4x\,\sqrt{-g}\, \Big[ (\sq \varphi)^2 - 2 \big(R^{\m\n} - \tfrac{1}{3} R g^{\m\n}\big)
(\nabla_\m\varphi)(\nabla_\n \varphi)\Big]\notag\\
&& \hspace{1.5cm} +\, \sdfrac{1}{2}\,\int d^4x\,\sqrt{-g}\  \Big[b'\big(E - \tfrac{2}{3}\sq R\big) + b\,C^2 \Big]\,\varphi,
\label{Sanom}
\end{align}
where $\varphi$ has canonical mass dimension zero. One can observe that, by using the form of $\mathcal{S}_{\rm anom}[g; \varphi]$, linear shifts in the spacetime scalar $\varphi$ are related to conformal transformations of the spacetime metric, and indeed
the Wess-Zumino consistency condition \eqref{Weylshift} implies the non-trivial relation
\begin{equation}
\mathcal{S}_{\rm anom}[g; \varphi + 2 \s] = \mathcal{S}_{\rm anom}[e^{-2 \s} g; \varphi] + \mathcal{S}_{\rm anom}[g; 2 \s] .
\label{SanomWZ}
\end{equation}
 Note that although 
$\varphi$ is closely related to and couples to the conformal part of the metric tensor, $\varphi$ is an independent spacetime
scalar field and the local action (\ref{Sanom}) is fully coordinate invariant, unlike $\G_{_{\!W\!Z}}$ in (\ref{WZfour}), which 
depends separately upon $\bar g_{\m\n}$ and $\s$, and is therefore dependent on the conformal frame.
So far we have studied different part of the effective action and finally we can assert that the exact 1PI quantum effective action for a CFT is written as
\begin{equation}
\mathcal{S} = \mathcal{S}_{\rm local}[g] + \mathcal{S}_{\rm inv}[g] + \mathcal{S}_{\rm anom}[g;\varphi]
\label{genS},
\end{equation}
where $\mathcal{S}_{\rm local}[g]$ contains the local $\int R^2$ term, whose conformal variation (\ref{varRsq}) 
is associated with the $b'' \sq R$ term in (\ref{trace}),
while $\mathcal{S}_{\rm inv}[g]$ is an arbitrary Weyl-invariant term
\begin{equation}
\mathcal{S}_{\rm inv}[e^{2\s}g] = \mathcal{S}_{\rm inv}[g],
\end{equation}
analogous to \eqref{SaddW}, previously used to obtain a symmetric form of the anomaly effective action. In general, this term is non-local and its expansion around flat space is responsible for the CWI's, instead of  $\mathcal{S}_{\rm anom}[g;\varphi]$ given by \eqref{Sanom} which is responsible for the anomalous trace \eqref{traceiden}. The form (\ref{genS}) of the decomposition 
of the quantum effective action was obtained in \cite{Mazur:2001aa} by considerations on the abelian group of local Weyl shifts, 
and its cohomology. The local and Weyl-invariant terms are elements of the trivial cohomology of the local Weyl group, while (\ref{Snonl}) or $\mathcal{S}_{\rm anom}[g;\varphi]$ is an element of the non-trivial cocycles of this cohomology, which is uniquely specified by the $b$ and $b'$ anomaly coefficients \cite{Bonora:83,Antoniadis:1992xu,Karakhanian:1994yd,Arakelian:1995ye,Mazur:2001aa}. 
A non-trivial test of the effective action \eqref{genS} and the correctness of the anomaly 
action \eqref{Sanom} is obtained by the reconstruction algorithm of \cite{Bzowski:2013sza} for the $\braket{TTT}$ in CFTs in flat spacetime. One can show that the anomalous trace Ward identities obeyed by $\braket{ TTT}$, must come entirely from the variation of $\mathcal{S}_{\rm anom}[g;\varphi]$, as pointed out in \cite{Mottola:2010, Coriano:2018bsy}.

\subsection{Expansion of the effective action to the third order}
\label{Sec:VarAct}
In this section we discuss the anomaly part of the correlator $\braket{TTT}$ in CFTs in flat spacetime as computed directly from the effective action \eqref{genS}. In the next section we will show that this is exactly what is obtained from the explicit calculation in perturbation theory \cite{Coriano:2018bsy} and from the method proposed in \cite{Bzowski:2013sza}.
In order to obtain the contributions of the anomaly effective action to the three-point function we require the expansion of $\mathcal{S}_{\rm anom}[g;\varphi]$ to third order in deviations from flat space. The expansion can be performed by using the local form of the action \eqref{Sanom}, or equivalently the non-local version \eqref{Snonl}, as proposed in \cite{Coriano:2017mux}. The final covariant result does not depend on this choice, as expected. 
In this case we use the local form \eqref{Sanom}, and the consistent expansion  of $\mathcal{S}_{\rm anom}[g;\varphi]$ around flat space is 
expressed in terms of both the metric $g^{\m\n}$ and of $\varphi$
\begin{align}
g^{\m\n} &= g^{\m\n (0)} + g^{\m\n{(1)}} + g^{\m\n{(2)}} + \dots \equiv \eta^{\m\n} - h^{\m\n} + h^{\m\n (2)} + \dots\,,\\
\varphi &= \varphi^{(0)} +  \varphi^{(1)} +  \varphi^{(2)}  + \dots\,,
\end{align}
in \eqref{Sanom}. The expansion of the scalar field $\varphi$ can be performed order by order through its equation of motion 
\begin{equation}
\sqrt{-g}\,\Delta_4\,\varphi =\sqrt{-g}\left[\frac{1}{2}\left(E-\frac{2}{3}\,\square R\right)+\frac{b}{2b'}\,C^2\right],
\end{equation}
for which we have the equations
	\begin{align}
&\hspace{4cm}\sqb^2 \varphi^{(0)} = 0 \label{eom0},\\
&\hspace{-1.5cm}(\sqrt{-g} \D_4)^{(1)} \varphi^{(0)} + \sqb^2 \varphi^{(1)} = \left[\sqrt{-g}
\left( \sdfrac{E}{2}- \sdfrac{\!\sq R\!}{3} + \sdfrac{b}{2b'}\, C^2 \right)\right]^{(1)}
= - \sdfrac{\!1\!}{3}\, \sqb R^{(1)} \label{eom1},\\
&\hspace{-2cm}(\sqrt{-g} \D_4)^{(2)} \varphi^{(0)} + (\sqrt{-g} \D_4)^{(1)} \varphi^{(1)} + \sqb^2 \varphi^{(2)} =
\left[\sqrt{-g}\left(\sdfrac{E}{2}- \sdfrac{\!\sq R\!}{3} + \sdfrac{b}{2b'}\, C^2 \right)\right]^{(2)} \notag\\
&\hspace{2cm}= \sdfrac{1}{2}E^{(2)} - \sdfrac{1}{3}\, [\sqrt{-g}\sq R]^{(2)} + \sdfrac{b}{2b'}\, [C^2]^{(2)} \label{eom2},
\end{align}
where $\sqb$ is the d'Alembertian wave operator in flat Minkowski spacetime, and we have used the fact that $E$ and $C^2$ are of second order in the curvature invariants while the Ricci scalar $R$ has an expansion that starts at first order. The equation of motion at zero order \eqref{eom0} has the trivial solution $\varphi^{(0)}=0$ that corresponds to a choice of the boundary conditions appropriate for the standard Minkowski space vacuum. This condition is equivalent to taking $\braket{T^{\m\n}}_\eta = 0$ in flat Minkowski spacetime with no boundary effects. In this manner one solves the equations \eqref{eom1} and \eqref{eom2} recursively to get $\varphi^{(1)}$, $\varphi^{(2)}$ and $\varphi^{(n)}$ at all orders in the expansion. 
We have now gathered all the building blocks in order to express the expansion of the anomaly action at the third order, in order to proceed with a comparison against the anomaly part of the $\braket{TTT}$. It is given in an implicit form by the relation
\begin{align}
&\mathcal{S}_{\rm anom}^{(3)} =  - \sdfrac{b'}{2} \int d^4x \, \left\{2\,\varphi^{(1)} \sqb^2 \varphi^{(2)} +\varphi^{(1)} \big(\sqrt{-g} \D_4\big)^{(1)} \,\varphi^{\!(1)} \right\}\notag\\
&\hspace{1.5cm} + \sdfrac{b'}{2} \int d^4x \left\{\left( - \sdfrac{2}{3} \sqb R^{(1)}\right) \varphi^{(2)} + \left(E^{(2)} - \sdfrac{2}{3}\, \sqrt{-g}\sq R\right)^{\!(2)} \varphi^{(1)} \right\}
+  \sdfrac{b}{2} \int d^4x \, (C^2)^{(2)}\,\varphi^{(1)},
\label{Sanom3a}
\end{align}
where 
\begin{equation}
\big(\sqrt{-g} \D_4\big)^{\!(1)} =  \big(\sqrt{-g} \sq^2\big)^{\!(1)} + 2\, \partial_{\m} \left(R^{\m\n} - \sdfrac{1}{3} \eta^{\m\n} R\right)^{\!(1)}\partial_{\n}.
\end{equation}
After a lengthy but straightforward calculation we end up with the final expression
\begin{align}
&\hspace{-5mm} \mathcal{S}_{\rm anom}^{(3)} =
\sdfrac{b'}{9} \int\! d^4x \int\!d^4x'\!\int\!d^4x''\!\left\{\big(\partial_{\m} R^{(1)})_x\left(\sdfrac{1}{\sqb}\right)_{\!xx'}  
\!\left(R^{(1)\m\n}\! - \!\sdfrac{1}{3} \eta^{\m\n} R^{(1)}\right)_{x'}\!
\left(\sdfrac{1}{\sqb}\right)_{\!x'x''}\!\big(\partial_{\n} R^{(1)})_{x''}\right\}\notag\\
&\hspace{-5mm}- \sdfrac{1}{6} \int\! d^4x\! \int\!d^4x'\! \left(b'\, E^{\!(2)} + b\,  [C^2]^{(2)}\right)_{\!x}\! \left(\sdfrac{1}{\sqb}\right)_{\!xx'} \!R^{(1)}_{x'}
+ \sdfrac{b'}{18}  \int\! d^4x\, R^{(1)}\left(2\, R^{\!(2)} + (\sqrt{-g})^{(1)} R^{(1)}\right),
\label{S3anom3}
\end{align}
where the last term is purely local. We observe that in spite of the presence of a double coincident pole $(\bar\square^2)^{-1}$ in $\varphi^{(2)}$ once \eqref{eom2} is inverted, the final expression of the anomaly contribution of the $\braket{TTT}$ has no double propagator
terms. The last term in \eqref{S3anom3} may be recognized as the expansion up to third order of the covariant local action
\begin{equation}
\sdfrac{b'}{18}  \int\! d^4x\, \sqrt{-g} \,R^2,
\label{bprimelocal}
\end{equation}
which, if subtracted from $S_{\rm anom}$ in (\ref{Sanom}), would cancel the $- \frac{2b'}{3} \sq R$ contribution to the conformal anomaly resulting from $S_{\rm anom}$, upon using (\ref{varRsq}), and leaving just $b'\, E + b \,C^2$ for the trace. 
As previously mentioned, the result \eqref{S3anom3} may as well be derived from the non-local form of the anomaly action \eqref{Snonl}. We refer to \cite{Coriano:2017mux} for more details.

\subsection{The prediction of the anomaly action for the \texorpdfstring{$TTT$}{ttt}}
\label{Sec:AnomTTT}

We have obtained the expansion at third order of the anomaly action in \eqref{S3anom3}, and we are going to calculate now the anomaly contribution to the $\braket{TTT}$ directly from \eqref{S3anom3}. Since by \eqref{fourd} and \eqref{fourd2}, both $E$ and $C^2$ are second order in curvature tensors, it suffices in (\ref{S3anom3}) to compute the Riemann tensor to first order in the metric variation $h_{\m\n}$
\begin{equation}
R_{\m\a\n\beta}^{(1)} = \sdfrac{1}{2}\, \Big\{\!- \partial_\a\partial_\beta h_{\m\n}- \partial_\m\partial_\n h_{\a\beta} + \partial_\a\partial_\n h_{\b\m}
+ \partial_\b\partial_\m h_{\a\n}\Big\}.
\end{equation}
All the contractions may be carried out with the use of the lowest order
flat space metric $\eta^{\m\n}$. In momentum space the expression above takes the form
\begin{equation}
\int d^4x \, e^{ip\cdot x} \, R_{\m\a\n\b}^{(1)}(x) \equiv \big[R_{\m\a\n\b}^{(1)}\big]^{\m_1\n_1}(p)\, \tilde h_{\m_1\n_1}(p),
\label{RieFour}
\end{equation}
which serves to define the tensor polynomial
\begin{equation}
\big[R_{\m\a\n\b}^{(1)}\big]^{\m_1\n_1}(p) = \sdfrac{1}{2}\, \Big\{\d^{(\m_1}_\a\, \d^{\n_1\!)\hspace{-4pt}}\,_\b\,p_\m\, p_\n
+ \d^{(\m_1}_\m\, \d^{\n_1)}_\n\,p_\a\, p_\b  - \d_\b\,^{\hspace{-2pt}(\m_1}\, \d^{\n_1)}_\m\,p_\a \,p_\n - \d^{(\m_1}_\a\, \d^{\n_1)}_\n\,p_\b \,p_\m  \Big\},
\label{Riemom}
\end{equation}
which has the contractions
\begin{equation}
\big[R^{(1)}_{\m\n}\big]^{\m_1\n_1}(p) = \d^{\a\b}\, \big[R_{\m\a\n\b}^{(1)}\big]^{\m_1\n_1}(p)
= \sdfrac{1}{2}\,\Big\{\d^{\m_1\n_1}\,p_\m\, p_\n  + \d^{(\m_1}_\m\, \d^{\n_1)}_\n\,p^2 
- p^{(\m_1}\, \d^{\n_1)}_\m\,p_\n- p^{(\m_1} \d^{\n_1)}_\n\,p_\m \Big\} ,
\label{Riccip}
\end{equation}
and
\begin{equation}
\big[R^{(1)}\big]^{\m_1\n_1}(p) = \d^{\m\n} \big[R^{(1)}_{\m\n}\big]^{\m_1\n_1}(p) = p^2 \d^{\m_1\n_1} - p^{\m_1}p^{\n_1}
= p^2\,  \pi^{\m_1\n_1}(p),
\label{Ricscalp}
\end{equation}
defined in an analogous fashion to (\ref{RieFour}). We also require the squared contractions 
\begin{align}
&\big[R_{\m\a\n\b}^{(1)}R^{(1)\m\a \n\b}\big]^{\m_1\n_1\m_2\n_2} (p_1, p_2) \equiv
\big[R_{\m\a\n\b}^{(1)}\big]^{\m_1\n_1} (p_1) \big[R^{(1)\m\a \n\b}\big]^{\m_2\n_2}(p_2) 
\notag\\
&\hspace{1.5cm}= (p_1 \cdot p_2)^2\, \eta^{\m_1(\m_2}\eta^{\n_2)\n_1}
- 2\, (p_1\cdot p_2)\, p_1\,^{\hspace{-4pt}(\m_2}\eta^{\n_2)(\n_1}p_2\,^{{\hspace{-2pt}}\m_1)}
+ p_1^{\m_2}\,p_1^{\n_2}\,p_2^{\m_1}\,p_2^{\n_1}
\label{Riemsq},
\end{align}
and
\begin{align}
&\big[R_{\m\n}^{(1)}R^{(1)\m\n}\big]^{\m_1\n_1\m_2\n_2} (p_1, p_2) \equiv
\big[R_{\m\n}^{(1)}\big]^{\m_1\n_1} (p_1) \big[R^{(1)\m\n}\big]^{\m_2\n_2}(p_2) 
\notag\\
&\hspace{1.5cm}=\sdfrac{1}{4}\, p_1^2 \, \Big(p_2^{\m_1}\,  p_2^{\n_1}\, \eta^{\m_2\n_2}  
-  2\, p_2\,^{\hspace{-4pt}(\m_1}\eta^{\n_1)(\n_2}  p_2\,^{\hspace{-2.5pt}\m_2)}\Big)
+ \sdfrac{1}{4}\, p_2^2\, \Big(p_1^{\m_2}\,  p_1^{\n_2}\, \eta^{\m_1\n_1} 
-  2\,p_1\,^{\hspace{-4pt}(\m_1}\eta^{\n_1)(\n_2}  p_1\,^{\hspace{-2.5pt}\m_2)}\Big)\notag\\
&\hspace{1.6cm}+ \sdfrac{1}{4}\, p_1^2\ p_2^2\, \eta^{\m_1(\m_2}\eta^{\n_2)\n_1}
+ \sdfrac{1}{4}\, (p_1\cdot p_2)^2\, \eta^{\m_1\n_1}\eta^{\m_2\n_2} 
+ \sdfrac{1}{2} \, p_1^{(\m_1}\,p_2^{\n_1)}\,p_1^{(\m_2}\,p_2^{\n_2)}\notag\\
&\hspace{1.6cm}+  \sdfrac{1}{2} \, (p_1\cdot p_2)\,
\Big( p_1\,^{\hspace{-4pt}(\m_1}\, \eta^{\n_1)(\n_2}  p_2\,^{\hspace{-2.5pt}\m_2)}
-\eta^{\m_1\n_1}  \, p_1^{(\m_2}\,  p_2^{\n_2)} -\eta^{\m_2\n_2}  \, p_1^{(\m_1}\,  p_2^{\n_1)}\Big)\,.
\end{align}
With these expressions at hand, together with the relaton
\begin{equation}
\big[(R^{(1)})^2\big]^{\m_1\n_1\m_2\n_2} (p_1, p_2) \equiv
\big[R^{(1)}\big]^{\m_1\n_1} (p_1) \big[R^{(1)}\big]^{\m_2\n_2}(p_2)
= p_1^2\, p_2^2 \, \pi^{\m_1\n_1}(p_1)\, \pi^{\m_2\n_2}(p_2),
\label{Riccscalsq}
\end{equation}
we may express the third order anomaly action and its contribution to the 
three-point correlator in momentum space in the form
\begin{align}
&\braket{T^{\m_1\nu_1}(p_1)T^{\m_2\nu_2}(p_2)T^{\m_3\nu_3}(p_3)}_{anom}=\sdfrac{8}{3} \Big\{\pi^{\m_1\nu_1}(p_1)
\,\left[b'E^{(2)}+b(C^2)^{(2)}\right]^{\m_2\nu_2\m_3\nu_3}(p_2,p_3)+(\text{cyclic})\Big\}\notag\\
&\hspace{5cm} -\sdfrac{16b'}{9}\Big\{ \pi^{\m_1\nu_1}(p_1)\,Q^{\m_2\nu_2}(p_1,p_2,p_3)\, \pi^{\m_3\nu_3}(p_3)+(\text{cyclic})\Big\}\notag\\
&\hspace{4cm} +\sdfrac{16b'}{27\,}\, \pi^{\m_1\nu_1}(p_1)\,\pi^{\m_2\nu_2}(p_2)\,\pi^{\m_3\nu_3}(p_3)\,\Big\{p_3^2\, p_1\cdot p_2+(\text{cyclic})\Big\}
+ ({\rm local}) \label{AS3},
\end{align}
having taken into account a ($2^3 = 8$) normalization factor in the definition of the correlator for $n=3$, and having summed over the $3$ cyclic permutations of the 
indices $(1,2,3)$. In (\ref{AS3})
\begin{align}
&Q^{\m_2\nu_2}(p_1,p_2,p_3) \equiv p_{1\m}\, [R^{\m\nu}]^{\m_2\nu_2}(p_2)\,p_{3\n} \notag\\
&= \sdfrac{1}{2} \,\Big\{(p_1\cdot p_2)(p_2\cdot p_3)\, \eta^{\m_2\n_2} 
+ p_2^2 \ p_1^{(\m_2}\, p_3^{\n_2)} - (p_2\cdot p_3) \, p_1^{(\m_2}\, p_2^{\n_2)} - (p_1\cdot p_2) \, p_2^{(\m_2}\, p_3^{\n_2)}\Big\},
\end{align}
by (\ref{Riccip}). We have used
\begin{subequations}
	\begin{align}
	&\hspace{-1cm}\big[E^{(2)}\big]^{\m_i\nu_i\m_j\nu_j} =\big[R_{\m\a\n\b}^{(1)}R^{(1)\m\a \n\b}\big]^{\m_i\nu_i\m_j\nu_j}
	-4\,\big[R_{\m\n}^{(1)}R^{(1)\m\n}\big]^{\m_i\nu_i\m_j\nu_j}
	+\big[ \big(R^{(1)}\big)^2\big]^{\m_i\nu_i\m_j\nu_j},\\
	&\hspace{-1cm} \big[(C^2)^{(2)}\big]^{\m_i\nu_i\m_j\nu_j}= \big[R_{\m\a\n\b}^{(1)}R^{(1)\m\a \n\b}\big]^{\m_i\nu_i\m_j\nu_j}
	-2\,\big[R_{\m\n}^{(1)}R^{(1)\m\n}\big]^{\m_i\nu_i\m_j\nu_j}
	+ \sdfrac{1}{3}\,\big[(R^{(1)})^2\big]^{\m_i\nu_i\m_j\nu_j},
	\end{align}
\end{subequations}
together with (\ref{Riemsq})-(\ref{Riccscalsq}). The term labeled as "(local)" in \eqref{AS3}
refers to the third variation of (\ref{bprimelocal}), the purely local last term in (\ref{S3anom3}). 
{At this point, we show how to express \eqref{AS3} in a form that will be useful for its comparison with the perturbative one. It is a straightforward exercise 
in tensor algebra using (\ref{Riccip})-(\ref{Riccscalsq}) to verify that
\begin{align}
&\eta_{\a_1\b_1}\,\braket{T^{\a_1\b_1}(p_1)T^{\m_2\n_2}(p_2)T^{\m_3\n_3}(p_3)}_{anom}\Big\vert_{p_3 = -(p_1 + p_2)} = \tilde{\mathcal{A}}^{\m_2\n_2\m_3\n_3}(p_2,\bar{p}_3)\notag\\
&\hspace{2cm}=8 b\, \big[(C^2)^{(2)}\big]^{\m_2\n_2\m_3\n_3} (p_2, \bar{p}_3) + \,8b'\, \big[E^{(2)}\big]^{\m_2\n_2\m_3\n_3} (p_2, \bar{p}_3) ,
\label{traceS3}
\end{align}
where $\bar{p}_3=-p_1-p_2$,  giving on the right hand side the second variation of the trace anomaly. This is consistent with the explicitly anomalous contribution to the trace
identity presented in (\ref{threeptr}), provided again that the $\sq R$ contribution from the local term (\ref{bprimelocal}) 
is neglected. 
Taking an additional trace of (\ref{traceS3}), we find 
\begin{align}
&\eta_{\a_1\b_1}\eta_{\a_3\b_3}\,\braket{T^{\a_1\b_1}(p_1)T^{\m_2\m_2}(p_2)T^{\a_3\b_3}(p_3)}\big\vert_{p_3 = -(p_1 + p_2)} 
=  \, \eta_{\a_3\b_3}\tilde{\mathcal{A}}^{\m_2\n_2\a_3\b_3}(p_2,\bar{p}_3) \notag\\
& \qquad = 16b'\, Q^{\m_2\n_2}(p_1,p_2,\bar{p}_3) \, + \ 8b'\,p_2^2\, \left( p_1^2  + p_1\cdot p_2\right)\pi^{\m_2\n_2}(p_2) .
\label{dbltrace3}
\end{align}
Finally, we calculate the triple trace of \eqref{traceS3} to obtain
\begin{align}
\eta_{\a_1\b_1}\eta_{\a_2\b_2}\eta_{\a_3\b_3}\,\braket{T^{\a_1\b_1}(p_1)T^{\a_2\b_2}(p_2)T^{\a_3\b_3}(p_3)}_{\scalebox{0.5}{$anom$}}\big\vert_{p_3 = -(p_1 + p_2)} &= \, \eta_{\a_2\b_2}\eta_{\a_3\b_3}\tilde{\mathcal{A}}^{\a_2\b_2\a_3\b_3}(p_2,\bar{p}_3)\notag\\
&=
16b' \left[ p_1^2\,p_2^2 - (p_1\cdot p_2)^2\right].
\label{triptrace3}
\end{align}
Using these relations, we can write the anomaly contribution \eqref{AS3} to the $\braket{TTT}$ as 
\begin{align}
&\braket{T^{\m_1\n_1}(p_1)T^{\m_2\n_2}(p_2)T^{\m_3\n_3}(p_3)}_{\scalebox{0.5}{$anom$}}=\left[\sdfrac{1}{3}\, \pi^{\m_1\n_1}(p_1)\,\eta_{\a_1\b_1}\,\braket{T^{\a_1\b_1}(p_1)T^{\m_2\n_2}(p_2)T^{\m_3\n_3}(p_3)}_{\scalebox{0.5}{$anom$}}+(\text{cyclic})\right]\notag\\
&\quad-\left[\sdfrac{1}{9}\,\pi^{\m_1\n_1}(p_1)\,\pi^{\m_2\n_2}(p_2)\,\eta_{\a_1\b_1}\eta_{\a_2\b_2}\,\braket{T^{\a_1\b_1}(p_1)T^{\a_2\b_2}(p_2)T^{\m_3\n_3}(p_3)}_{\scalebox{0.5}{$anom$}} +(\text{cyclic})\right]\notag\\
&\quad+\sdfrac{1}{27}\,\pi^{\m_1\n_1}(p_1)\pi^{\m_2\n_2} (p_2)\pi^{\m_3\n_3}(p_3)\,\eta_{\a_1\b_1}\eta_{\a_2\b_2}\eta_{\a_3\b_3}\,\braket{T^{\a_1\b_1}(p_1)T^{\a_2\b_2}(p_2)T^{\a_3\b_3}(p_3)}_{\scalebox{0.5}{$anom$}}\,
\label{fin1},
\end{align}
and we will show that this is exactly what is expected from the reconstruction method in \cite{Bzowski:2013sza} and from the explicit perturbative calculation in \cite{Coriano:2018bsy}. 

It is worth noting that the result in \eqref{fin1} generalizes the one obtained in several perturbative analysis in free-field theories for specific correlators such as the $TJJ$.} The latter defines the most significant gravitational correction to a two-point function (in this case the photon propagator), at phenomenological level.   
The structure of the anomaly action, in this case, corresponds to the $F^2$ part -or gauge part - of the non-local anomaly functional and it is given by the expression
\begin{equation}
 \label{pole}
\mathcal{S}_{pole}= - \frac{e^2}{ 36 \pi^2}\int d^4 x d^4 y \left(\square h(x) - \partial_\mu\partial_\nu h^{\mu\nu}(x)\right)  \square^{-1}_{x\, y} F_{\alpha\beta}(y)F^{\alpha\beta}(y).
\end{equation}
More details concerning this result can be found in the original works \cite{Giannotti:2008cv,Armillis:2009pq,Armillis:2009im,Armillis:2010qk} in the QED and QCD cases and in \cite{Coriano:2014gja} for supersymmetry, where the same pattern emerges from the analysis of the superconformal anomaly multiplet \cite{Coriano:2019dyc}.
%%%%%%%%%%%%%%%%%%%%%%%%%%%%%%%%%%%%%%%%%%%%%%%%%%
\section{Renormalization and anomaly in \texorpdfstring{$d=2$}{d2} and \texorpdfstring{$d=4$}{d4}: the \texorpdfstring{$TT$}{tt} case } \label{renren}
%%%%%%%%%%%%%%%%%%%%%%%%%%%%%%%%%%%%%%%%%%%%%%%%%%
We have already pointed out that $V_E$ induces only 
a finite renormalization on the $n$-point function. Its presence in the structure of the anomaly action, is requested by the Wess-Zumino consistency condition, but from the point of view of regulating the effective action $\mathcal{S}(g)$ as we take the $d\to 4$ limit, it amounts to an evanescent contribution.  \\
We are going to illustrate this subtle point in the case of the $TT$ first in $d=2$ and then in $d=4$.\\
In $d=2$ the $TT$ does not exhibit any singularity, since a $1/\varepsilon$ behavior in a kinematic prefactor is accompanied by a tensor structure which vanishes as $d\to 2$. We describe the steps in this case, and then move to the case of $d=4$ after that. The inclusion of a topological $V_E$ counterterm, allows to derive a consistent description of this correlator.  
A one-loop calculation yields in this case 
\begin{align}
	\braket{T^{\mu_1\nu_1}(p)T^{\mu_2\nu_2}(-p)}= \frac{ c(d)}{(d-2)}\left(p^2\right)^{d/2}\Pi^{\mu_1\nu_1\mu_2\nu_2}_{(d)}(p)\label{TT2dim},
\end{align}
where $\Pi_{(d)}$ is the transverse traceless projector defined in general $d$ dimensions and the constant $c(d)$ is defined as
\begin{align}
	c(d)=4\,c_T \left(\frac{\pi}{4}\right)^{d/2}\frac{(d-1)\Gamma\left(2-\frac{d}{2}\right)}{\Gamma\left(d+2\right)},
\end{align}
with $c_T$ depending on the matter field realization of the conformally invariant action. Notice that the constant $c(d)$ is finite for any $d>1$. 

In dimensional regularization \eqref{TT2dim} the correlator in $d=2(1+\varepsilon)$ takes the following form
\begin{align}
	\braket{T^{\mu_1\nu_1}(p)T^{\mu_2\nu_2}(-p)}= \frac{ c(2+2\varepsilon)}{2\varepsilon}\left(p^2\right)^{1+\varepsilon}\Pi^{\mu_1\nu_1\mu_2\nu_2}_{(d=2+2\varepsilon)}(p)\label{TTreg},
\end{align}
with the appearance of the UV divergence as $1/\varepsilon$ pole after the expansion around $d=2$.
However the limit $\varepsilon\to 0$ is ambiguous due to the fact that the transverse traceless projector in two dimensions goes to zero. This properties is the consequence of the tensor degeneracy when $n=d$ for an $n$-point function. The way out of this indeterminacy is to 
introduce a $n-p$ decomposition of the Kronecker $\delta_{\mu\nu}$  and perform the limit 
 to $d=2$ of the entire tensor structure in order to prove the finiteness of the result. We provide a discussion of this basis in the $d=4$ case in appendix \ref{nnp}.
 
Indeed, for the case of the two-point function in $d=2$, one could define the independent momentum $n^\mu$ using the Levi-Civita tensor $\epsilon^{\mu\nu}$ as
\begin{align}
	n^{\mu}=\epsilon^{\mu\nu}p_\nu,\label{indMom}
\end{align}
 orthogonal to the other momentum $p_\mu$. Having such two independent momenta $p_\mu,\,n_\mu$, then the metric is not an independent tensor and we can rewrite it as
\begin{equation}
	\delta^{\mu\nu}=\sum_{j,k=1}^2q_j^\mu\,q_k^\nu\big(Z^{-1}\big)_{kj},
\end{equation}
with $q_1^\mu=p^\mu$ and $q_2^\mu=n^\mu$, and $Z$ is the gram matrix, i. e. $Z_{ij}=(q_i\cdot q_j)^2_{i,j=1}$. The Gram matrix is trivially $Z=p^2 \mathbb 1$, because $n^2=p^2$. Then, the metric tensor can be written as
\begin{align}
	\delta^{\mu\nu}=\frac{1}{p^2}\left(p^\mu p^\nu+n^\mu n^\nu\right),
\end{align}
for which the transverse projector takes the form
\begin{equation}
	\pi^{\mu\nu}_{(d=2)}(p)\equiv\delta^{\mu\nu}-\frac{p^{\mu}p^{\nu}}{p^2}=\frac{n^\mu n^\nu}{p^2},
\end{equation}
and the transverse traceless projector in $d=2$ vanishes as
\begin{equation}
	\Pi^{\mu_1\nu_1\mu_2\nu_2}_{(d=2)}(p)=\pi^{\mu_1(\mu_2}\pi^{\nu_2)\nu_1}-\pi^{\mu_1\nu_1}\pi^{\mu_2\nu_2}=0.
\end{equation}
At this s,tage from the transverse traceless projector defined in $d$ dimensions and taking into account this degeneracy in $d=2$, the projector around $d=2$ takes the form 
\begin{equation}
	\Pi^{\mu_1\nu_1\mu_2\nu_2}_{d=2+2\varepsilon}(p)=\pi^{\mu_1(\mu_2}\pi^{\nu_2)\nu_1}-\frac{1}{1+2\varepsilon}\pi^{\mu_1\nu_1}\pi^{\mu_2\nu_2}=\frac{2\varepsilon}{1+2\epsilon}\frac{n^{\mu_1}n^{\nu_1}n^{\mu_2}n^{\nu_2}}{p^4},
\label{ths}
\end{equation}
making the limit $\varepsilon\to0$ well defined. We are expanding the parametric 
dependence of the projector in $\varepsilon$ and then the tensor structure in terms of the non-degenerate $n-p$ basis. \\
 Inserting \eqref{ths} in \eqref{TT2dim} one derives a finite result as $\varepsilon\to 0$ 
 \begin{align}
	\braket{T^{\mu_1\nu_1}(p)T^{\mu_2\nu_2}(-p)}_{(d=2)}= \frac{ c(2+2\varepsilon)}{1+2\varepsilon}\left(p^2\right)^{\varepsilon-1}\,n^{\mu_1}n^{\nu_1}n^{\mu_2}n^{\nu_2}\,\,\overset{\varepsilon\to0}{=}\,\,c(2)\frac{n^{\mu_1}n^{\nu_1}n^{\mu_2}n^{\nu_2}}{p^2}.
\end{align} 
The expression above has a non-zero trace, manifesting a trace anomaly relation. Here, we are not relying on the introduction of any counterterm, rather, we are exploiting the degeneracy of the tensor structures in 
$d=2$, extracting a renormalized expression of the correlation function. Notice also the absence of any scale dependence in the result. We are going to reproduce the same features of this result in a different approach, by introducing a topological counterterm from the beginning.

\subsection{Dimensional regularization}
The result above can be obtained by a different approach following the standard procedure of dimensional regularization scheme and renormalization. We will be using the same method also in $d=4$.\\
 We start from the regulated expression of the $TT$ obtained by an expansion of \eqref{TTreg} in power of $\varepsilon$, 
\begin{align}
	\braket{T^{\mu_1\nu_1}(p)T^{\mu_2\nu_2}(-p)}_{Reg}=& \frac{ c(2)}{2\varepsilon}\left(p^2\right)\,\Pi^{\mu_1\nu_1\mu_2\nu_2}_{(2+2\varepsilon)}(p)+\frac{c(2)}{2}\Pi^{\mu_1\nu_1\mu_2\nu_2}_{(2)}\,p^2\log p^2+p^2c^\prime(2)\,\Pi^{\mu_1\nu_1\mu_2\nu_2}_{(2)}+O(\varepsilon),\label{RegTT}
\end{align}
In this case we are not making use of the tensor degeneracies in $d=2$ yet. The only possible counterterm action that we can define in $d=2$ is given by
\begin{align}
	S_{ct}=-\frac{1}{\varepsilon}\,\beta_c\,\int d^dx\, \sqrt{-g}\,\mu^{d-2}\,R,
\end{align}
which will be included in the process of renormalization. We differentiate twice this action with respect to the metric, and take the flat space limit before going to momentum space
\begin{align}
	\braket{T^{\mu_1\nu_1}(p)T^{\mu_2\nu_2}(-p)}_{Count}=-\frac{\beta_c\,p^2\,\mu^{d-2}}{2\varepsilon}\,\left(\Pi^{\mu_1\nu_1\mu_2\nu_2}_{(d)}(p)-\frac{(d-2)}{(d-1)}\pi^{\mu_1\nu_1}(p)\pi^{\mu_2\nu_2}(p)\right), 
\end{align}
whose expansion in power of $\varepsilon$ is 
\begin{align}
	\braket{T^{\mu_1\nu_1}(p)T^{\mu_2\nu_2}(-p)}_{Count}=-\frac{\beta_c\,p^2}{2\varepsilon}\,\left(\Pi^{\mu_1\nu_1\mu_2\nu_2}_{(2+2\varepsilon)}(p)-\frac{2\varepsilon}{(1+2\varepsilon)}\pi^{\mu_1\nu_1}(p)\pi^{\mu_2\nu_2}(p)\right)\left(1+\varepsilon \log\mu^2\right).\label{CountTT}
\end{align}
At this point we add the counterterm contribution \eqref{CountTT} to the regularized correlator \eqref{RegTT}, and by choosing $\beta_c=c(2)$ we remove the divergence. We obtain a finite renormalized result in the limit $\varepsilon\to0$ given by
\begin{align}
	\braket{T^{\mu_1\nu_1}(p)T^{\mu_2\nu_2}(-p)}_{(d=2)}^{Ren}=&\frac{c(2)}{2}\Pi^{\mu_1\nu_1\mu_2\nu_2}_{(2)}\,p^2\log\left( \frac{p^2}{\mu^2}\right)+p^2c^\prime(2)\,\Pi^{\mu_1\nu_1\mu_2\nu_2}_{(2)}+c(2)\,p^2\pi^{\mu_1\nu_1}(p)\pi^{\mu_2\nu_2}(p).\label{TTRen}
\end{align}
This result is not traceless, but is characterized by non-zero anomalous trace 
\begin{equation}
	\braket{T^{\mu_1}_{\ \ \ \mu_1}(p)T^{\mu_2\nu_2}(-p)}_{(d=2)}^{Ren}=c(2)\,p^2\,\pi^{\mu_2\nu_2}(p),
\end{equation}
which coincides with the result of taking functional derivative of trace anomaly of the two-point function in two dimensions
\begin{align}
	\mathcal{A}=c(2)\,\sqrt{-g}\,R.
\end{align}
We have reproduced the correct structure of the anomaly, as in the previous section, 
but we need to recover  the scale independence of \eqref{TTRen}, which is not apparent from that equation. Notice that $c(2)$ identifies a topological contribution and therefore it is independent of any scale. \\
The anomalous scale-dependence of the two-point function is
\begin{align}
\label{rr}
	\mu\frac{\partial}{\partial\mu}	\braket{T^{\mu_1\nu_1}(p)T^{\mu_2\nu_2}(-p)}=-c(2)\,p^2\,\Pi^{\mu_1\nu_1\mu_2\nu_2}_{(2)},
\end{align}
and manifests an apparent contradiction. It is sufficient, at this stage, to use the degeneracy of the tensor structure $\Pi^{\mu_1\nu_1\mu_2\nu_2}_{(2)}$ to immediately realize that the 
the rhs of \eqref{rr} vanishes. 
The inclusion of the topological counterterm in $d=2$, even if it is zero at the integer value of the dimensions on which we are going to project, of course, it allows to obtain the correct expression of the renormalized correlator with open indices, while at the same time induces the correct expression of its anomaly.\\
 The advantage of following this procedure is evident especially for multi-point functions. 
\subsection{Longitudinal projectors in \texorpdfstring{$d=4$}{d4}}
Longitudinal projectors in multi-point correlation functions in $d=4$ are naturally induced by the renormalization. To illustrate this point, we start from the case of the $TT$, then move to the $3T$ and conclude our discussion with the $4T$. 
The simplest context in which to discuss the renormalization of the $TT$ is in free field theory, and include three independent sectors with $n_S$ scalars, $n_F$ fermions and $n_G$ gauge fields. A direct computation in perturbation theory gives 
\begin{align}
\braket{T^{\mu_1\nu_1}(p)T^{\mu_2\nu_2}(-p)}=&-\frac{\p^2\,p^4}{4(d-1)(d+1)}\,B_0(p^2)\,\Pi^{\mu_1\nu_1\mu_2\nu_2}(p)\Big[2(d-1)n_F+(2d^2-3d-8)n_G+n_S\Big]\notag\\
&+\frac{\p^2\,p^4\,n_G}{8(d-1)^2}(d-4)^2(d-2)\p^{\mu_1\nu_1}(p)\p^{\mu_2\nu_2}(p)\,B_0(p^2)\label{TTddim},
\end{align}
where $B_0(p^2)$ is the scalar two-point  function defined as
\begin{equation}
B_0(p^2)=\frac{1}{\pi^\frac{d}{2}}\,\int\,d^d\ell\,\frac{1}{\ell^2\,(\ell-p)^2}.
\end{equation}
\eqref{TTddim} shows the separation of the result into a transverse-traceless $(\Pi)$ and longitudinal part $(\pi^{\mu_1\nu_1})$. Around $d=4$, the projectors are expanded using the relation 
\begin{equation}
\label{pexp1}
\Pi^{(d)\,\mu_1\nu_1\mu_2\nu_2}(p)=\Pi^{(4)\,\mu_1\nu_1\mu_2\nu_2}(p)-\frac{2}{9}\varepsilon\,\pi^{\mu_1\nu_1}(p)\,\pi^{\mu_2\nu_2}(p)+O(\varepsilon^2), 
\end{equation}
performed on the parametric dependence of the projector $\Pi(d)$. 
As usual in DR, the tensor indices are continued to $d$ dimensions and contracted with a d-dimensional Euclidean metric $(\delta^\mu_\mu=d)$.

Using \eqref{pexp1} in \eqref{TTddim}, the latter takes the form
\begin{align}
\label{result}
\braket{T^{\mu_1\nu_1}(p)T^{\mu_2\nu_2}(-p)}&=-\frac{\p^2\,p^4}{4}\,\bigg(\frac{1}{\varepsilon}+\bar{B}_0(p^2)\bigg)\,\bigg(\Pi^{(4)\,\mu_1\nu_1\mu_2\nu_2}(p)-\frac{2}{9}\varepsilon\,\pi^{\mu_1\nu_1}(p)\,\pi^{\mu_2\nu_2}(p)+O(\varepsilon^2)\bigg)\notag\\
&\hspace{-2cm}\times\Bigg[\bigg(\frac{2}{5}+\frac{4}{25}\varepsilon+O(\varepsilon^2)\bigg)n_F+\bigg(\frac{4}{5}-\frac{22}{25}\varepsilon+O(\varepsilon^2)\bigg)n_G+\bigg(\frac{1}{15}+\frac{16}{225}\varepsilon+O(\varepsilon^2)\bigg)n_S\Bigg]\notag\\
&\hspace{-1cm}+\frac{\p^2\,p^4\,n_G}{8}\p^{\mu_1\nu_1}(p)\p^{\mu_2\nu_2}(p)\,\bigg(\frac{1}{\varepsilon}+\bar{B}_0(p^2)\bigg)\bigg[\frac{8}{9}\varepsilon^2+\frac{8}{27}\varepsilon^3+O(\varepsilon^4)\bigg],
\end{align}
where $\Pi^{ (4)\,\,\mu_1\nu_1\mu_2\nu_2}(p)$ is the transverse and traceless projector in $d=4$ and $\bar{B}_0(p^2)= 2 - \log(p^2)$ is the finite part at $d=4$ of the scalar integral in the $\overline{MS}$ scheme. The last term of \eqref{result}, generated by the addition of a non-conformal sector ($\sim n_G$), vanishes separately as $\varepsilon\to 0$. Finally, combining all the terms we obtain the regulated ($reg$) expression of the $TT$ around $d=4$ in the form
\begin{align}
\braket{T^{\mu_1\nu_1}(p)T^{\mu_2\nu_2}(-p)}_{reg}&=-\frac{\p^2\,p^4}{60\,\varepsilon}\Pi^{(4)\,\mu_1\nu_1\mu_2\nu_2}(p)\left(6 n_F + 12 n_G + n_S\right)\notag\\
&\hspace{-3cm}+\frac{\p^2\,p^4}{270}\p^{\mu_1\nu_1}(p)\p^{\mu_2\nu_2}(p)\left(6 n_F + 12 n_G + n_S\right)-\frac{\p^2\,p^4}{300}\bar{B}_0(p^2)\Pi^{\mu_1\nu_1\mu_2\nu_2}(p)\left(30n_F+60n_G+5n_S\right)\notag\\
&\hspace{-3cm}-\frac{\p^2\,p^4}{900}\Pi^{\mu_1\nu_1\mu_2\nu_2}(p)\left(36n_F-198 n_G+16n_S\right)+O(\varepsilon).
\end{align}
The divergence in the previous expression can be removed through the one loop counterterm Lagrangian $\mathcal{S}_{ct}$. In fact, the second functional derivative of $\mathcal{S}_{ct}$ \eqref{count} with respect to the background metric gives 
\begin{align}
\braket{T^{\mu_1\nu_1}(p)T^{\mu_2\nu_2}(-p)}_{count}&\equiv -\sdfrac{\mu^{-\varepsilon}}{\varepsilon}\bigg(4b\,\big[\sqrt{-g}\,C^2\big]^{\m_1\n_1\m_2\n_2}(p,-p)\bigg)=-\frac{8(d-3)\,\mu^{-\varepsilon}\,b}{(d-2)\,\varepsilon}p^4\Pi^{(d)\,\mu_1\nu_1\mu_2\nu_2}(p),
\end{align}
having used the relation $V_{E}^{\m_1\n_1\m_2\n_2}(p,-p)=0$. In particular, expanding around $d=4$ and using again \eqref{pexp1} we obtain
\begin{align}
&\braket{T^{\mu_1\nu_1}(p)T^{\mu_2\nu_2}(-p)}_{count}=\notag\\
&\hspace{1cm}-\frac{8\,b\,p^4}{\varepsilon}\bigg(\Pi^{(4)\,\mu_1\nu_1\mu_2\nu_2}(p)-\frac{2}{9}\varepsilon\,\pi^{\mu_1\nu_1}(p)\,\pi^{\mu_2\nu_2}(p)+O(\varepsilon^2)\bigg)\,\bigg(\frac{1}{2}-\frac{\varepsilon}{2}\left(\frac{1}{2}+\log\mu\right)+O(\varepsilon^2)\bigg)\notag\\
&\hspace{1cm}=-\sdfrac{4\,b}{\varepsilon}p^4\,\Pi^{(4)\,\mu_1\nu_1\mu_2\nu_2}(p)+4\,b\, p^4\bigg[\Pi^{(4)\,\mu_1\nu_1\mu_2\nu_2}(p)+\frac{2}{9}\p^{\mu_1\nu_1}(p)\p^{\mu_2\nu_2}(p)\bigg]+O(\varepsilon),
\end{align}
which cancels the divergence arising in the two-point function, if one chooses the parameter $b$ as 
\begin{equation}
\label{choiceparm1}
b=-\frac{3\pi^2}{720}n_S-\frac{9\pi^2}{360}n_F-\frac{18\pi^2}{360}n_G,
\end{equation}
The renormalized two-point  function using  \eqref{choiceparm1} then takes the form  \begin{align}
\braket{T^{\mu_1\nu_1}(p)T^{\mu_2\nu_2}(-p)}_{Ren}&=\braket{T^{\mu_1\nu_1}(p)T^{\mu_2\nu_2}(-p)}+\braket{T^{\mu_1\nu_1}(p)T^{\mu_2\nu_2}(-p)}_{count}\notag\\
&=-\frac{\p^2\,p^4}{60}\bar{B}_0\left(\frac{p^2}{\mu^2}\right)\Pi^{\mu_1\nu_1\mu_2\nu_2}(p)\left(6n_F+12n_G+n_S\right)\notag\\
&\quad-\frac{\p^2\,p^4}{900}\Pi^{\mu_1\nu_1\mu_2\nu_2}(p)\big(126n_F-18n_G+31n_S\big).
\label{tren1}
\end{align}
Notice that the final renormalized expression is transverse and traceless. Obviously, this result holds in the case in which we choose a counterterm in such a way that \eqref{dif1} is satisfied.  If we had chosen the $C^2$ counterterm action to satisfy \eqref{nn1}, we would have found the relation 
\begin{equation}
\delta_{\mu_1\nu_1}\langle T^{\mu_1\nu_1}(p_1)T^{\mu_1\nu_2}(-p_1)\rangle=
\mathcal{A}^{\mu_2\nu_2}(p_1),
\end{equation}
where on the right hand side we have the contribution of the $\square R$ term that can be removed by adding a local term $R^2$ in the effective action, obtaining a finite renormalization procedure. 
We will see that the same choice of parameters $b$ given in \eqref{choiceparm1} and for $b'$ as
\begin{equation}\label{choiceparm2}
	b'=\frac{\pi^2}{720}n_S+\frac{11\pi^2}{720}n_F+\frac{31\pi^2}{360}n_G.
\end{equation}
removes the divergences in the three-point function, as we are going to discuss below. In fact, as we have seen, the expression of $b$ is related to the renormalization of the two-point function, instead $b'$ is intrinsically related to the renormalization of the three-point  function.
Notice that the renormalized result of the two-point function \eqref{tren1} does not contain any trace anomaly contributions but, due to the explicit $\mu$-dependence, it acquires an anomalous dilatation WI of the form
\begin{align}
\mu\frac{\partial}{\partial\mu}\braket{T^{\mu_1\nu_1}(p)T^{\mu_2\nu_2}(-p)}_{Ren}=-2\left[\frac{\pi^2\,p^4}{60}\,\Pi^{\mu_1\nu_1\mu_2\nu_2}(p)\left(6n_F+12n_G+n_S\right)\right].
\end{align}
%%%%%%%%%%%%%%%%%%%%%%%%%%%%%%%%%%%%%%%%%%%%%%%%%%%%%%%%%%%%
\section{Perturbative results in CFT}\label{Perturbative}
%%%%%%%%%%%%%%%%%%%%%%%%%%%%%%%%%%%%%%%%%%%%%%%%%%%%%%%%%%%%
In this section we show the correspondence between the perturbative realizations of CFT correlators and the general solution obtained by solving the CWI's \cite{Coriano:2018bsy, Coriano:2018zdo, Coriano:2018bbe}. 
Moreover, we consider the correlation functions $\braket{TTT}$ and $\braket{TJJ}$, and we also study the behavior of these correlators at $d = 4$, where a breaking of conformal invariance appears, made manifest by the trace anomaly.
\subsection{The \texorpdfstring{$TTT$}{ttt} case in free field theory}
Our analysis of the matching is performed in dimensional regularization and we will adopt the $\overline{MS}$ renormalization scheme. We search for a free field theory with a field content that can be matched to the general solutions of the CWI's. While this could be expected on general grounds, the details of the matching are very important, since the general solution is rather complex and requires a renormalization procedure on the triple-K integrals which is far from being straighforward. This approach allows to simplify all the form factors that appear in the final result, showing that the $TTT$ of a general CFT is, ultimately, reproduced by an expression containing standard scalar one-loop integrals $B_0$, $C_0$, corresponding to self-energies and scalar triangle diagrams. These appear in the result combined with numerical factors counting the different massless degrees of freedom needed in order to perform the match with the general solution. 
$C_0$ is the only dilogarithmic expression appearing in $d=4$, which is given by

\begin{equation}
C_0 ( p_1^2,p_2^2,p_3^2) = \frac{ 1}{p_3^2} \Phi (x,y),
\end{equation}
where the function $\Phi (x, y)$  is given by
\cite{Usyukina:1993ch}
\begin{equation}
\Phi( x, y) = \frac{1}{\lambda} \biggl\{ 2 [Li_2(-\rho  x) + Li_2(- \rho y)]  +
\ln \frac{y}{ x}\ln \frac{1+ \rho y }{1 + \rho x}+ \ln (\rho x) \ln (\rho  y) + \frac{\pi^2}{3} \biggr\},
\label{Phi}
\end{equation}
with
\begin{align}
&\lambda(x,y) = \sqrt {\Delta},
&&\hspace{-2cm}\Delta=(1-  x- y)^2 - 4  x  y,
\label{lambda} \\
&\rho( x,y) = 2 (1-  x-  y+\lambda)^{-1},
&&\hspace{-2cm}x=\frac{p_1^2}{p_3^3} \, ,\qquad \qquad y= \frac {p_2^2}{p_3^2}\, .
\end{align}
We consider scalar and fermion sectors in the actions
\begin{align}
S_{scalar}&=\sdfrac{1}{2}\int\, d^dx\,\sqrt{-g}\left[g^{\m\n}\nabla_\m\phi\nabla_\n\phi-\c\, R\,\phi^2\right],\\
S_{fermion}&=\sdfrac{i}{2}\int\, d^dx\,e\,e^{\m}_a\left[\bar{\psi}\g^a(D_\m\psi)-(D_\m\bar{\psi})\g^a\psi\right],
\end{align}
where $\c =(d-2)/(4d-4)$ for a conformally coupled scalar in $d$ dimensions, and $R$ is the Ricci scalar. $e_\m^a$ is the vielbein and $e$ its determinant, with the covariant derivative $D_\m$ given by 
\begin{equation}
D_\m=\partial_\m+\Gamma_\m=\partial_\m+\sdfrac{1}{2}\Sigma^{ab}\,e^\s_a\nabla_\m\,e_{b\,\s}.
\end{equation}
The $\Sigma^{ab}$ are the generators of the Lorentz group in the spin-$1/2$ representation. Latin indices are related to the flat space-time and the Greek indices to the curved space-time by the vierbein $e^\s_a\nabla_\m$. In $d=4$, a third sector is also possible, and corresponds to the spin-$1$. In the case of the $TTT$, however, two integration constants allow to characterize the non-perturbative solution and the corresponding anomalies. The role of the spin-$1$ sector is relevant in other diagrams, such as the $TJJ$. We have
\begin{equation}
S_{abelian}=S_{M}+S_{gf}+S_{gh},
\end{equation}
where the three contributions are the Maxwell action, the gauge fixing contribution and the ghost action
\begin{align}
S_M&=-\sdfrac{1}{4}\int d^4x\,\sqrt{-g}\,F^{\m\n}F_{\m\n},\\
S_{gf}&=-\sdfrac{1}{\xi}\int d^4x\,\sqrt{-g}\,(\nabla_\m A^\m)^2,\\
S_{gh}&=\int d^4x\,\sqrt{-g}\,\,\partial^\m\bar c\,\partial_\m \,c.
\end{align}  
We have included the explicit expressions of the vertices for convenience in \figref{vertices} and in Appendix \ref{Appendix1}.
\begin{figure}[t]
	\centering
	\vspace{-2cm}
	\subfigure{\includegraphics[scale=0.14]{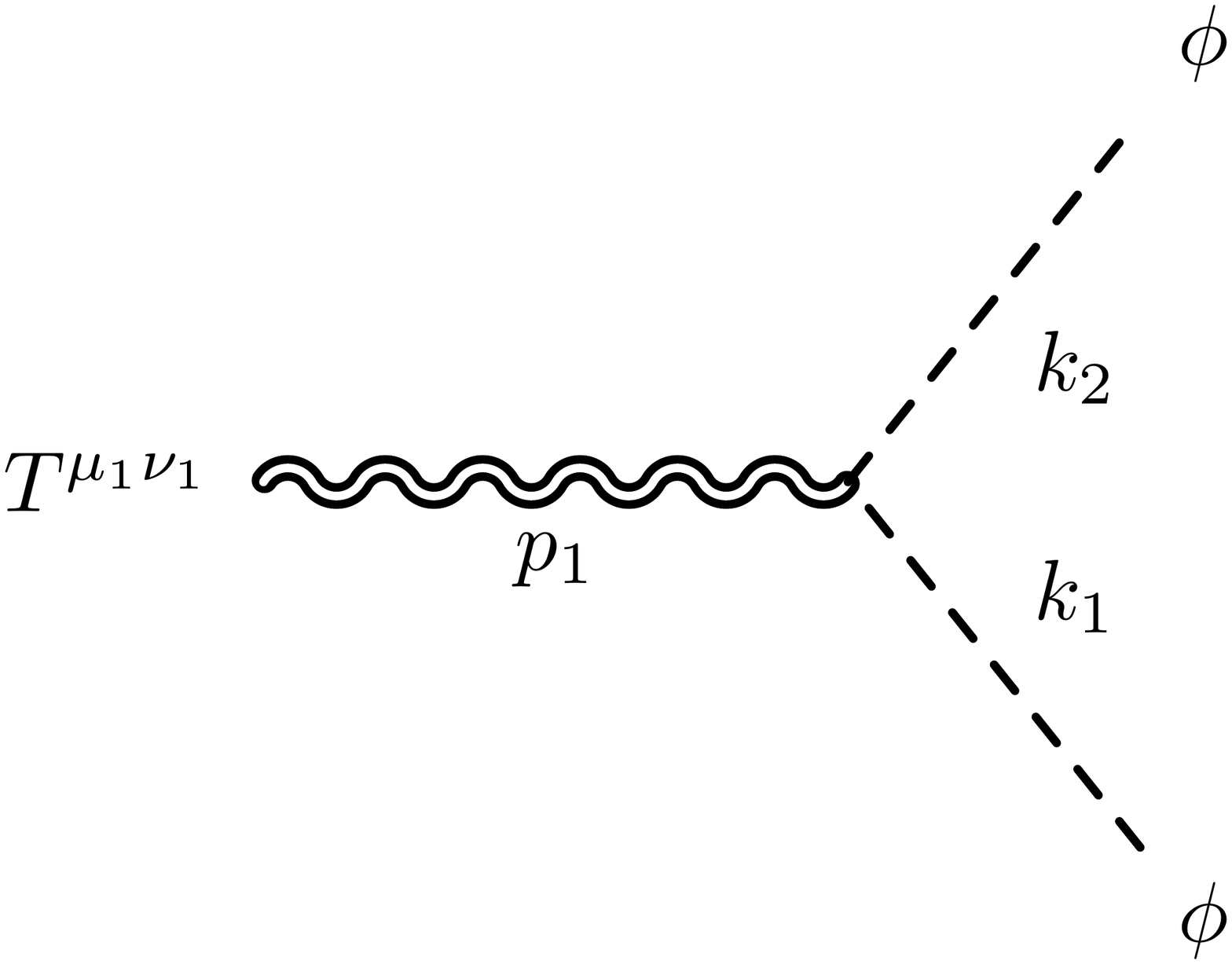}} \hspace{.3cm}
	\subfigure{\includegraphics[scale=0.14]{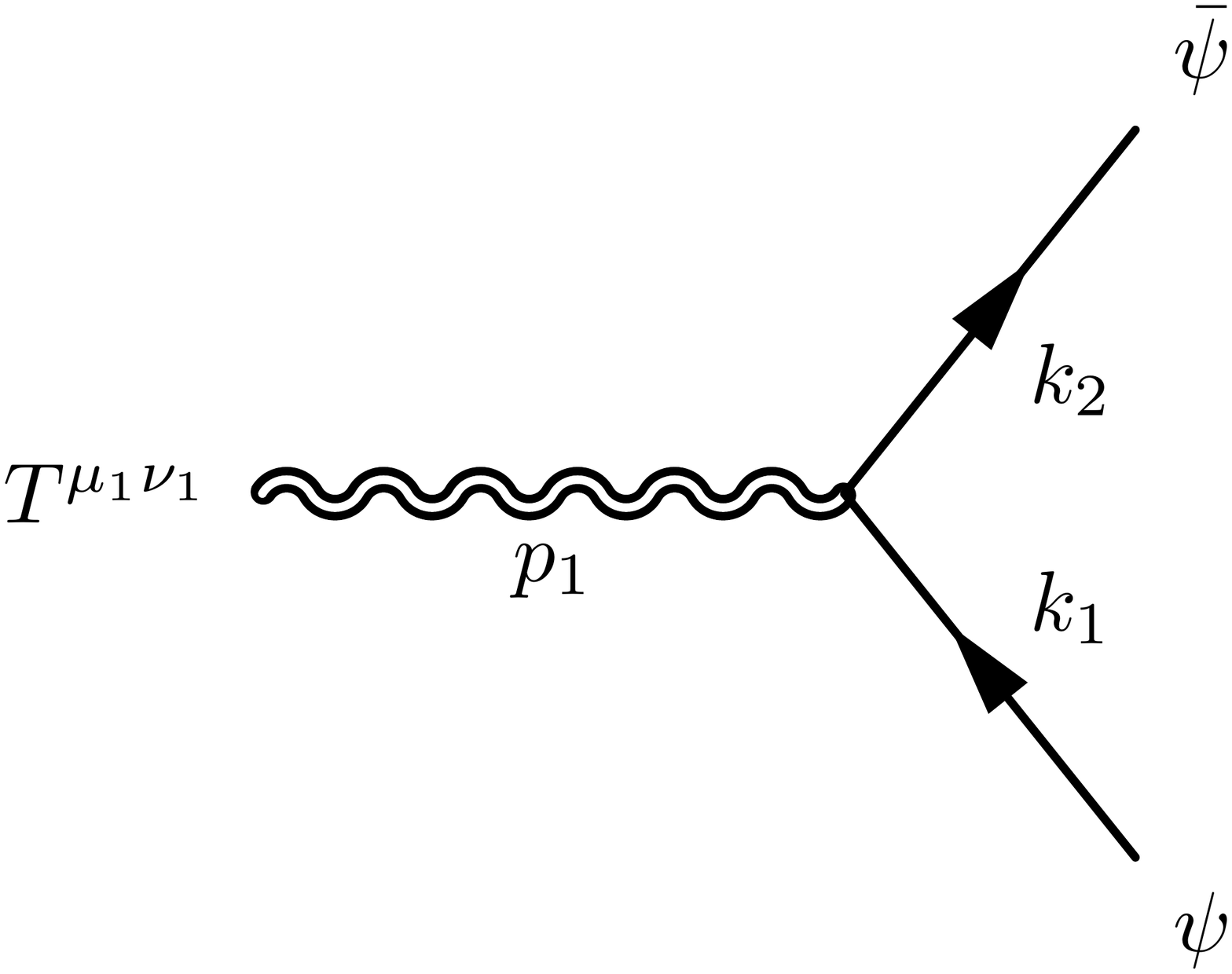}} \hspace{.3cm}
	\subfigure{\includegraphics[scale=0.14]{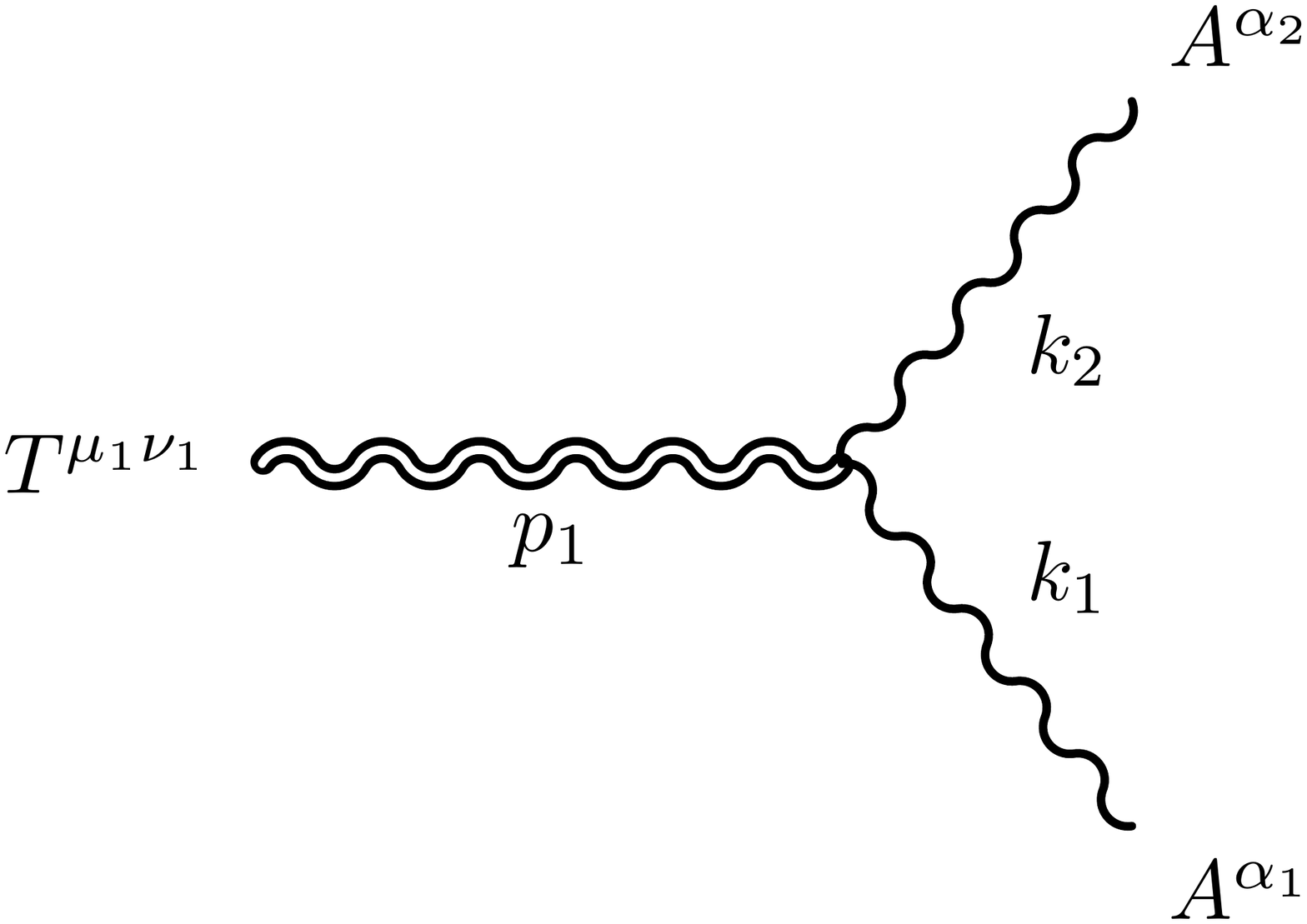}} \hspace{.3cm}
	\subfigure{\includegraphics[scale=0.14]{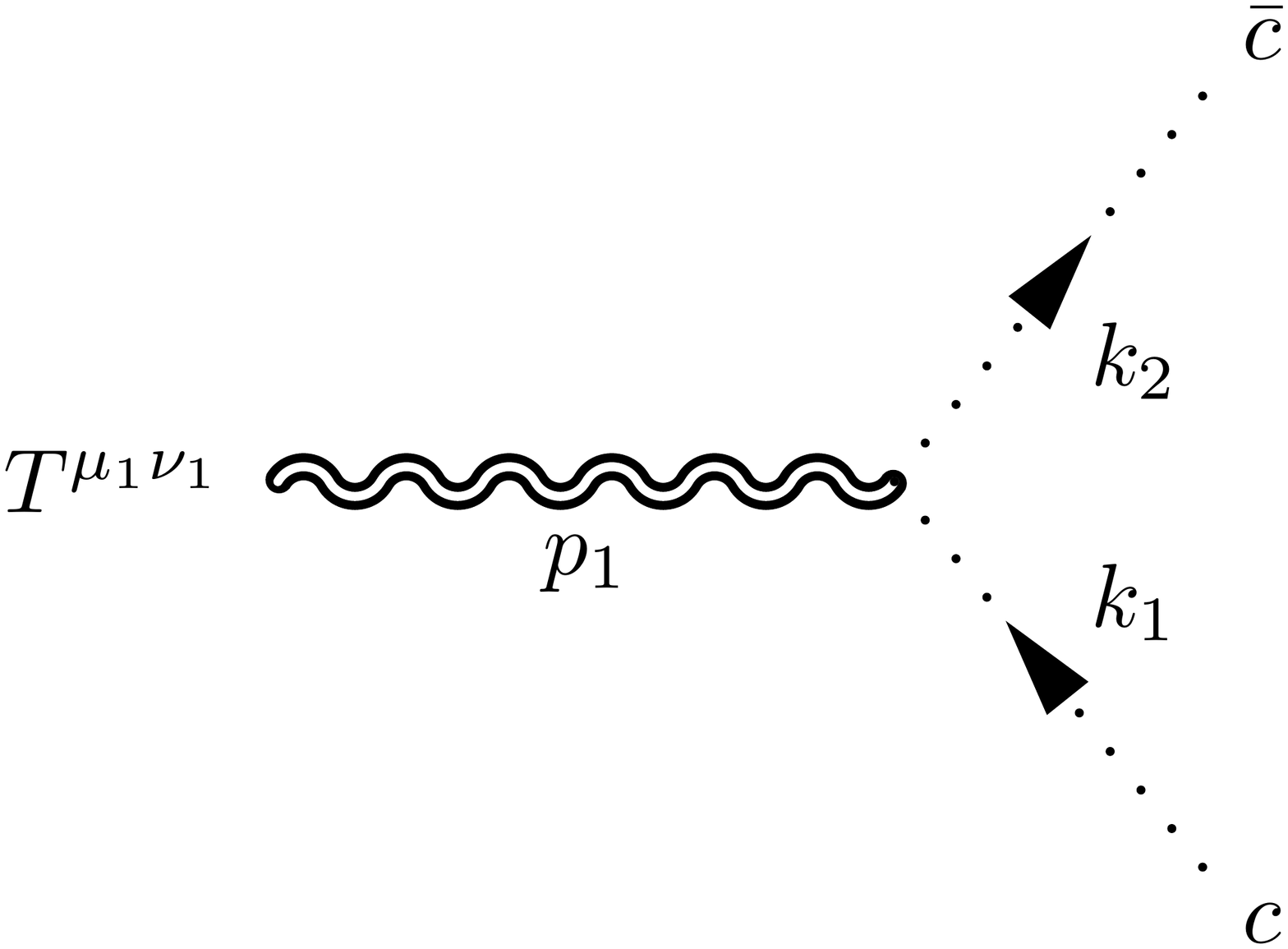}}
	\\
	\vspace{-1.5cm}
	\subfigure{\includegraphics[scale=0.14]{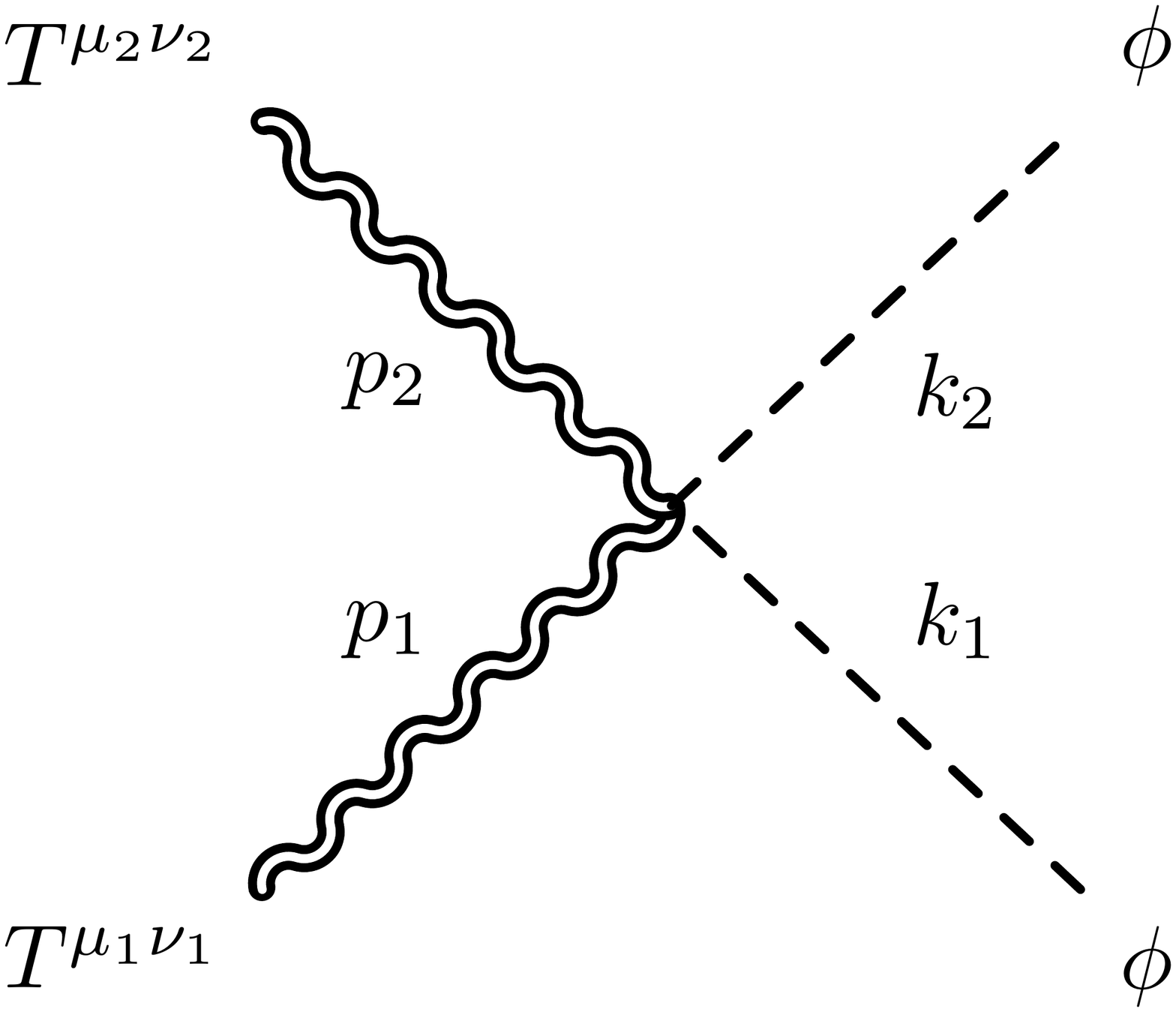}} \hspace{.3cm}
	\subfigure{\includegraphics[scale=0.14]{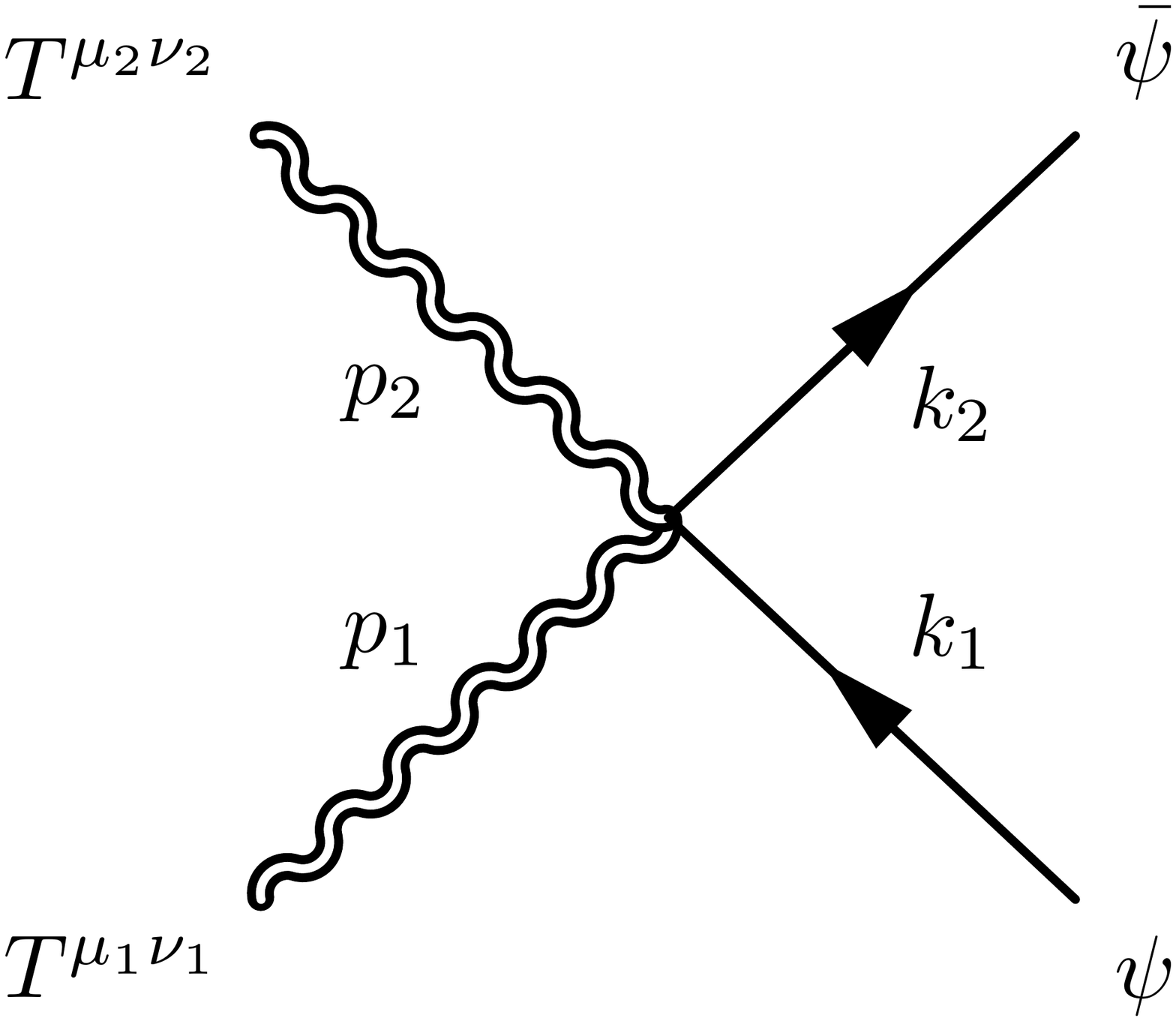}} \hspace{.3cm}
	\subfigure{\includegraphics[scale=0.14]{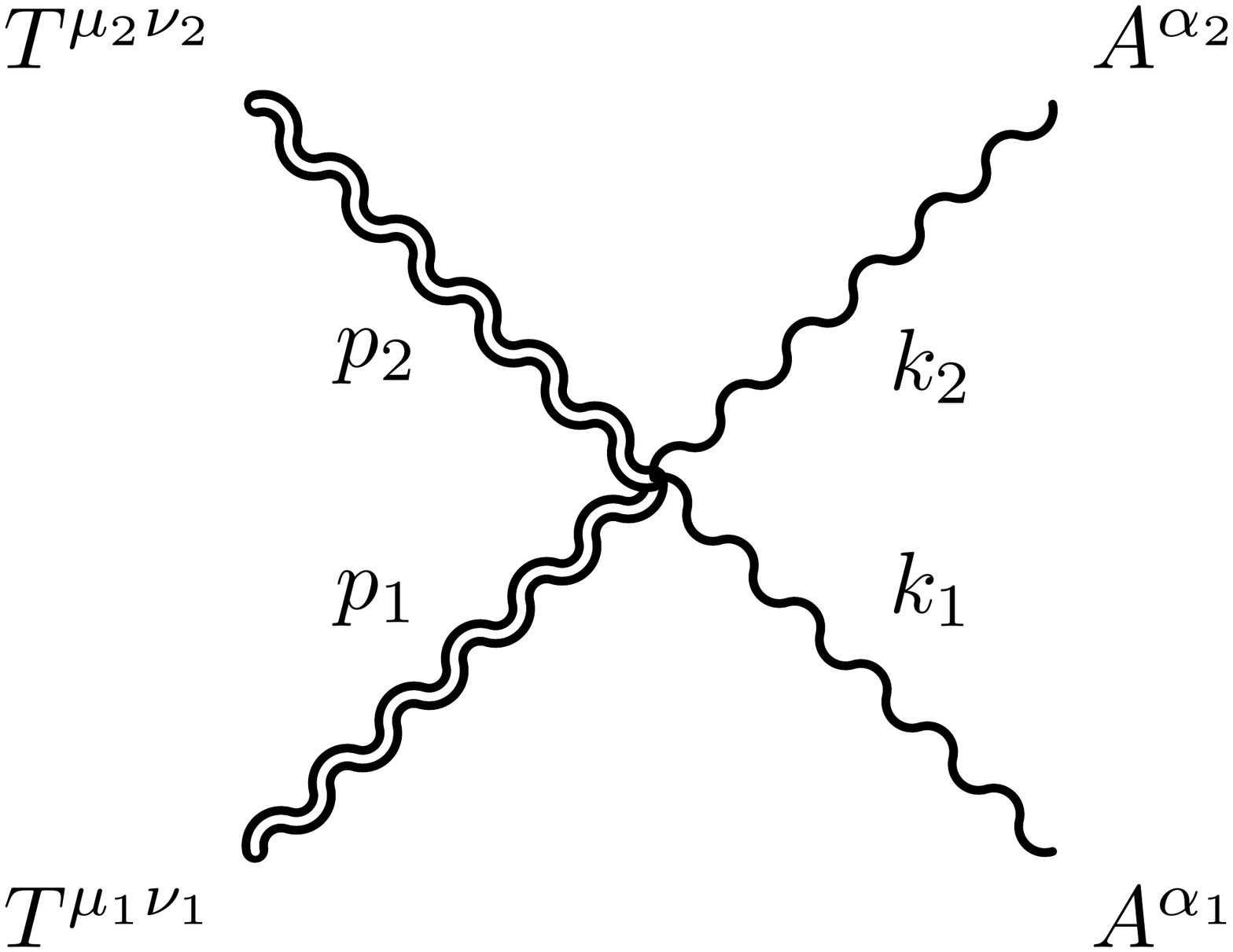}} \hspace{.3cm}
	\subfigure{\includegraphics[scale=0.14]{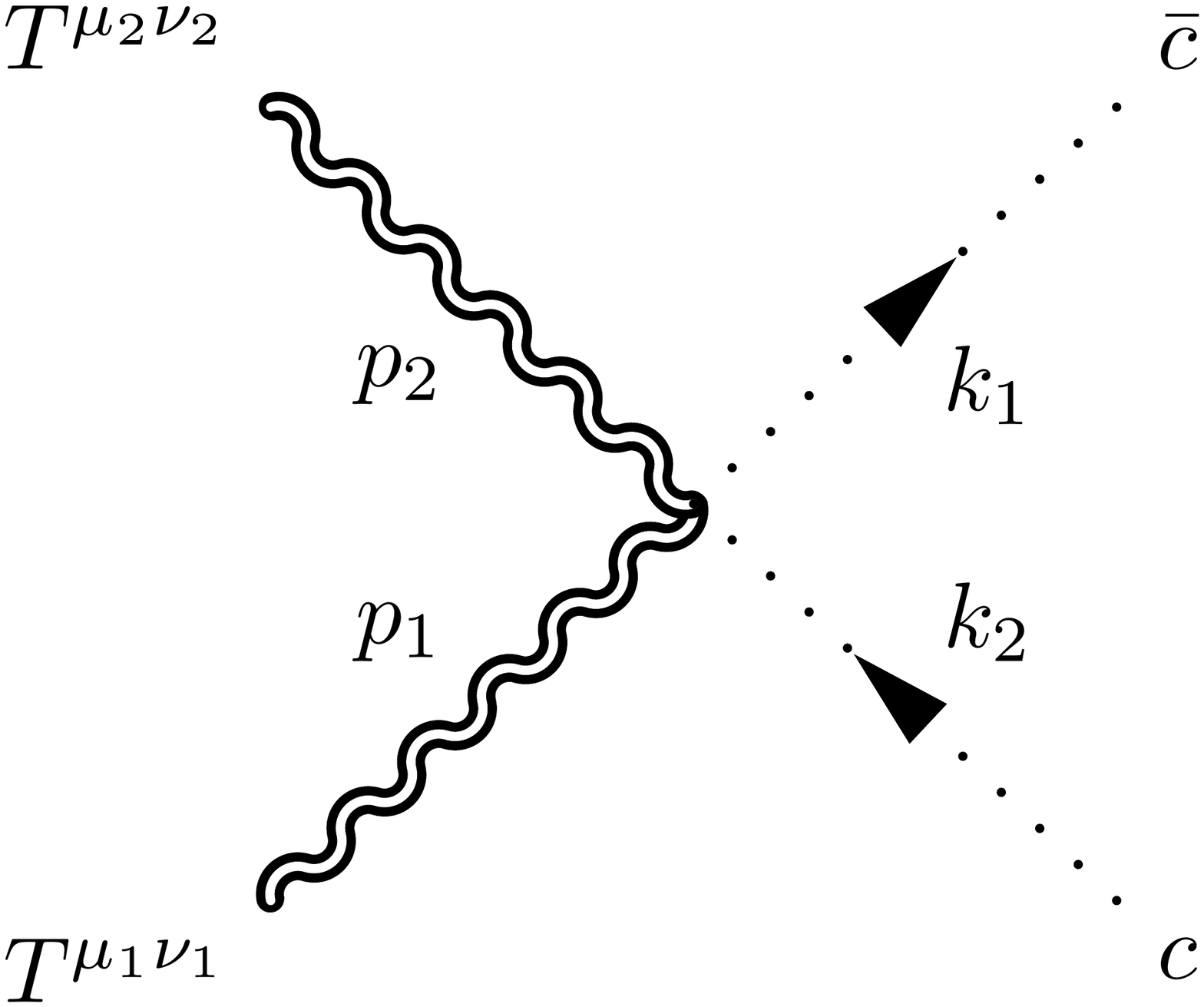}}
	\caption{Vertices used in the Lagrangian realization of the $TTT$ correlator.\label{vertices}}
\end{figure}
\noindent Since we are interested in the most general Lagrangian realization of the $\braket{TTT}$ correlator in the conformal case, this can be obtained only by considering the scalar and fermion sectors in general $d$ dimensions.  

\begin{figure}[t]
	\centering
	\subfigure{\includegraphics[scale=0.2]{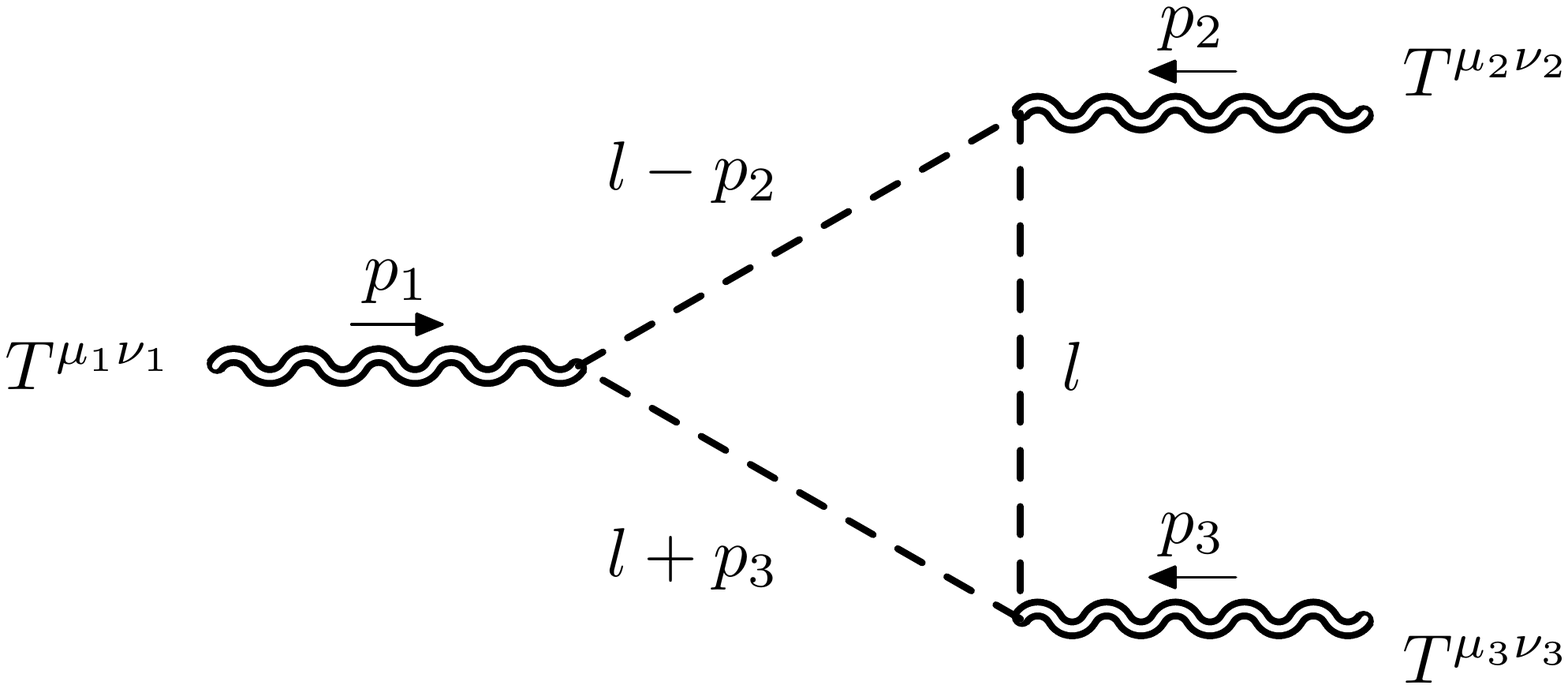}} \hspace{.3cm}
	\subfigure{\includegraphics[scale=0.2]{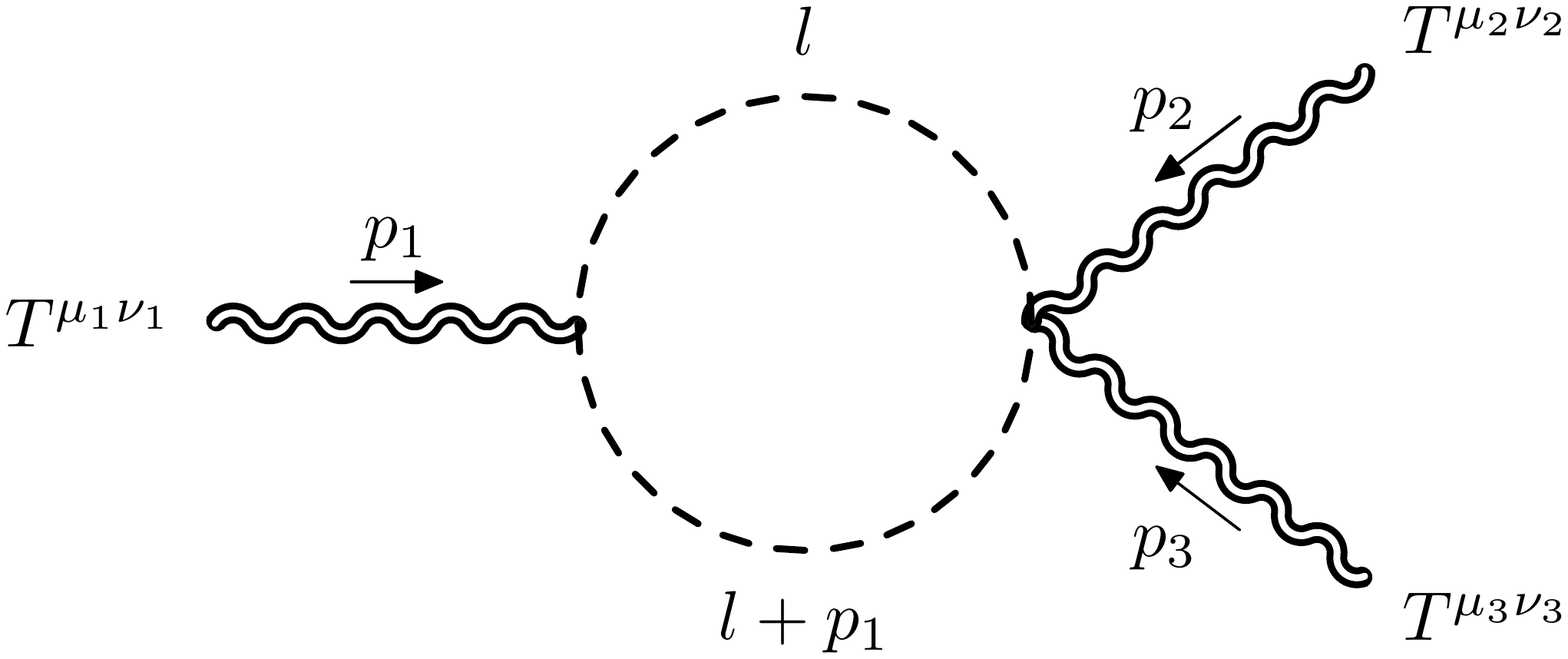}} \hspace{.3cm}
	\raisebox{.12\height}{\subfigure{\includegraphics[scale=0.16]{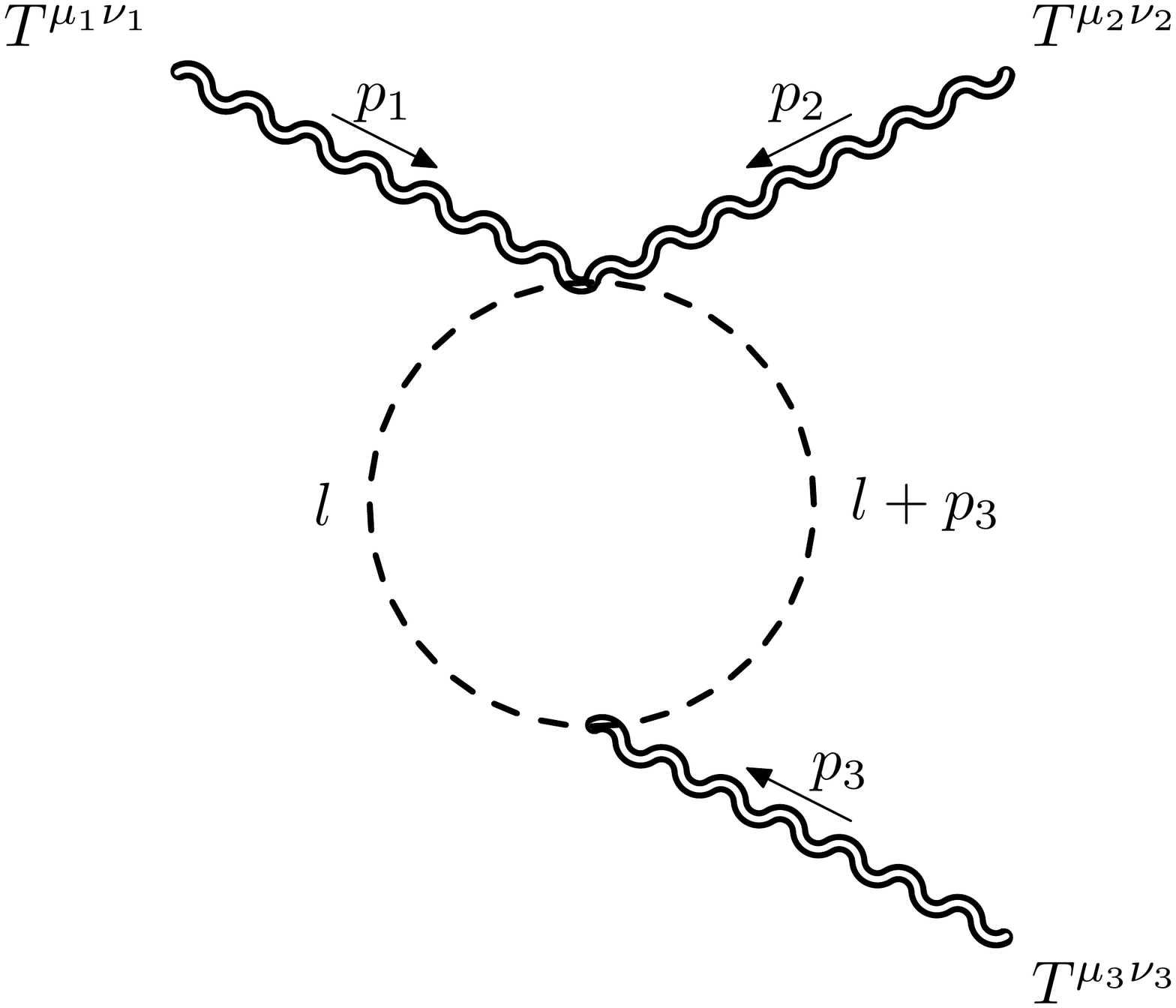}}\hspace{.3cm}}
	\raisebox{.12\height}{\subfigure{\includegraphics[scale=0.16]{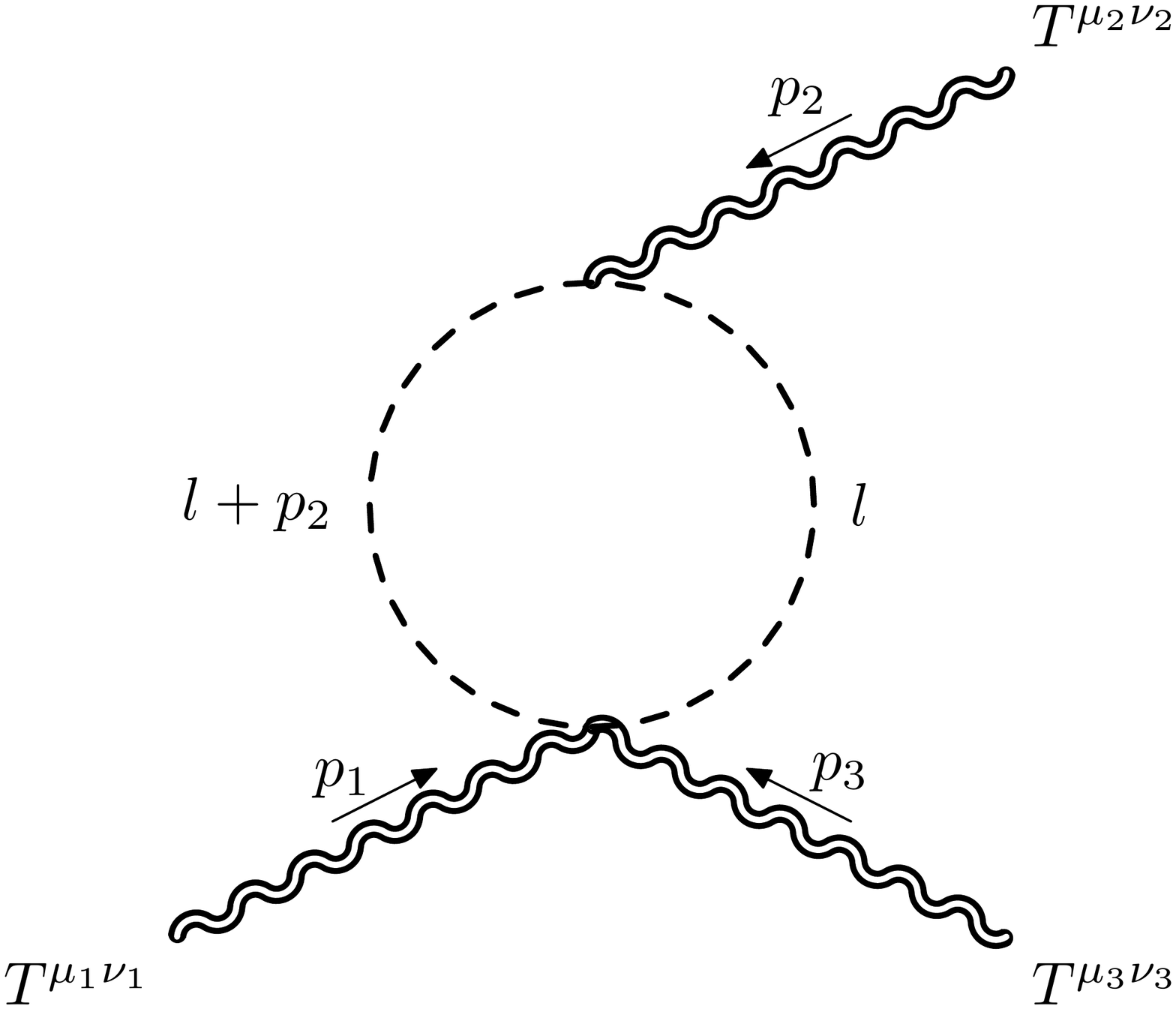}}}
	\vspace{-0.8cm}\caption{One-loop scalar diagrams for the three-graviton vertex.\label{Feynman1}}
\end{figure}
Consider, for instance, the scalar sector.  In the one-loop approximation, the contributions are given by the diagrams in \figref{Feynman1}, with
\begin{align}
\braket{T^{\m_1\n_1}(p_1)T^{\m_2\n_2}(p_2)T^{\m_3\n_3}(p_3)}_S=\, -V_{S}^{\m_1\n_1\m_2\n_2\m_3\n_3}(p_1,p_2,p_3)+\sum_{i=1}^3W_{S,i}^{\m_1\n_1\m_2\n_2\m_3\n_3}(p_1,p_2,p_3)\label{ScalarPart},
\end{align}
where $V_S$ is related to the triangle diagrams in \figref{Feynman1} and $W_{S,i}$ terms are the three bubble contributions labeled by the index $i$, with $i=1,2,3$. 
We act with the projectors $\Pi$ on \eqref{ScalarPart} in order to write the form factors of the transverse and traceless part of the correlator, as in \eqref{tttdec}
\begin{align}
\braket{t^{\m_1\n_1}(p_1)t^{\m_2\n_2}(p_2)t^{\m_3\n_3}(p_3)}_S&=\,\Pi^{\m_1\n_1}_{\a_1\b_1}(p_1)\Pi^{\m_2\n_2}_{\a_2\b_2}(p_2)\Pi^{\m_3\n_3}_{\a_3\b_3}(p_3)\notag\\
&\times\bigg[ -V_{S}^{\a_1\b_1\a_2\b_2\a_3\b_3}(p_1,p_2,p_3)+\sum_{i=1}^3W_{S,i}^{\a_1\b_1\a_2\b_2\a_3\b_3}(p_1,p_2,p_3)\bigg].
\end{align}
In the fermion sector we obtain
\begin{align}
\braket{t^{\m_1\n_1}(p_1)t^{\m_2\n_2}(p_2)t^{\m_3\n_3}(p_3)}_F&=\,\Pi^{\m_1\n_1}_{\a_1\b_1}(p_1)\Pi^{\m_2\n_2}_{\a_2\b_2}(p_2)\Pi^{\m_3\n_3}_{\a_3\b_3}(p_3)\notag\\
&\times\bigg[ -\sum_{j=1}^2V_{F,j}^{\a_1\b_1\a_2\b_2\a_3\b_3}(p_1,p_2,p_3)+\sum_{j=1}^3W_{F,j}^{\a_1\b_1\a_2\b_2\a_3\b_3}(p_1,p_2,p_3)\bigg],
\end{align}
where we have two types of triangle diagrams depending on the direction in which the fermion runs in the the loop. Also in this case the number of fermion families is kept arbitrary and we will multiply the result by a constant $n_F$ to account for it. It will be essential for matching this contribution to the general non-perturbative one. 
As already mentioned, to match the entire anomaly functional, which includes the $F^2$ term after renormalization in $d=4$, we add the third sector
\begin{align}
\braket{t^{\m_1\n_1}(p_1)t^{\m_2\n_2}(p_2)t^{\m_3\n_3}(p_3)}_G&=\,\Pi^{\m_1\n_1}_{\a_1\b_1}(p_1)\Pi^{\m_2\n_2}_{\a_2\b_2}(p_2)\Pi^{\m_3\n_3}_{\a_3\b_3}(p_3)\notag\\
&\times\bigg[ -V_{G}^{\a_1\b_1\a_2\b_2\a_3\b_3}(p_1,p_2,p_3)+\sum_{i=1}^3W_{G,i}^{\a_1\b_1\a_2\b_2\a_3\b_3}(p_1,p_2,p_3)\bigg],
\end{align}
where we have also considered the contributions from the ghost. A direct computation shows that the ghost and the gauge fixing contributions cancel each other. Also in this case the number of gauge fields are kept arbitrary by the inclusion of an overall factor $n_G$.
%%%%%%%%%%%%%%%%%%%%%%%%%%%%%%%%%%%%%%%%%%%%%%%
\subsection{The \texorpdfstring{$d=3$}{d3} and \texorpdfstring{$d=5$}{d5} cases}
%%%%%%%%%%%%%%%%%%%%%%%%%%%%%%%%%%%%%%%%%%%%%%%
In $d=3$ there are significant simplifications and we get
\begin{equation}
{B}_0(p_1^2)=\frac{\p^{3/2}}{\,p_1}\label{B0},
\end{equation}
where $p_1=|p_1|=\sqrt{p_1^2}$, and analogous relations hold for $p_2$ and $p_3$. The explicit expression of ${C}_0$ can be obtained using the star-triangle relation 
\begin{equation}
\int\frac{d^{d}x}{[(x-x_1)^2]^{\a_1}\,[(x-x_2)^2]^{\a_2}\,[(x-x_3)^2]^{\a_3}}\ \  \mathrel{\stackrel{\makebox[0pt]{\mbox{\normalfont\tiny $\sum_i\a_i=d$}}}{=}}\ \  \frac{i\pi^{d/2}\n(\a_1)\n(\a_2)\n(\a_3)}{[(x_2-x_3)^2]^{\frac{d}{2}-\a_1}\,[(x_1-x_2)^2]^{\frac{d}{2}-\a_3}\,[(x_1-x_3)^2]^{\frac{d}{2}-\a_2}}\label{startriangle},
\end{equation}
where
\begin{equation}
\n(x)=\frac{\Gamma\left(\frac{d}{2}-x\right)}{\Gamma(x)},
\end{equation}
that holds only if the condition $\sum_i\a_i=d$ is satisfied. In the case $d=3$ the LHS of \eqref{startriangle} is proportional to the three point scalar integral, and in particular
\begin{align}
{C}_0(p_1^2,p_2^2,p_3^2)&=\int\,\frac{d^d\ell}{\p^\frac{d}{2}}\frac{1}{\ell^2(\ell-p_2)^2(\ell+p_3)^2}=\int\,\frac{d^dk}{\p^\frac{d}{2}}\frac{1}{(k-p_1)^2(k+p_3)^2(k+p_3-p_2)^2}\notag\\[2ex]
&=\frac{\left[\Gamma\left(\frac{d}{2}-1\right)\right]^3}{\,(p_1^2)^{\frac{D}{2}-1}(p_2^2)^{\frac{d}{2}-1}(p_3^2)^{\frac{d}{2}-1}}\ \mathrel{\stackrel{\makebox[0pt]{\mbox{\normalfont\tiny $d=3$}}}{=}}\ \frac{\p^{3/2}}{\,p_1\,p_2\,p_3}.\label{C0}
\end{align}
We obtain
\begin{align}
A_{1}^{d=3}(p_1,p_2,p_3)&=\frac{\p^3(n_S-4n_F)}{60(p_1+p_2+p_3)^6}\Big[p_1^3+6p_1^2(p_3+p_2)+(6p_1+p_2+p_3)\big((p_2+p_3)^2+3 p_2 p_3\big)\Big],\\
A_{2}^{d=3}(p_1,p_2,p_3)&=\frac{\p^3(n_S-4n_F)}{60(p_1+p_2+p_3)^6}\Big[4p_3^2\big(7(p_1+p_2)^2+6p_1 p_2\big)+20p_3^3(p_1+p_2)+4p_3^4\notag\\
&\hspace{4.4cm}+3(5p_3+p_1+p_2)(p_1+p_2)\big((p_1+p_2)^2+p_1 p_2\big)\Big]\notag\\
&+\frac{\p^3\,n_F}{3(p_1+p_2+p_3)^4}\Big[p_1^3+4p_1^2(p_2+p_3)+(4p_1+p_2+p_3)\big((p_2+p_3)^2+p_2 p_3\big)\Big],
\end{align}
\begin{align}
A_{3}^{d=3}(p_1,p_2,p_3)&=\frac{\p^3(n_S-4n_F)\,p_3^2}{240(p_1+p_2+p_3)^4}\Big[28p_3^2(p_1+p_2)+3p_3\big(11(p_1+p_2)^2+6 p_1\,p_2\big)+7p_3^3\notag\\
&+12(p_1+p_2)\big((p_1+p_2)^2+p_1p_2\big)\Big]\notag\\
&+\frac{\p^3n_F\,p_3^2}{6(p_1+p_2+p_3)^3}\Big[3p_2(p_1+p_2)+2\big((p_1+p_2)^2+p_1p_2\big)+p_3^2\Big]\notag\\
&-\frac{\p^3(n_s+4n_F)}{16(p_1+p_2+p_3)^2}\Big[p_1^3+2p_1^2(p_2+p_3)+(2p_1+p_2+p_3)\big((p_2+p_3)^2-p_2 p_3\big)\Big],
\end{align}
\begin{align}
A_{4}^{d=3}(p_1,p_2,p_3)&=\frac{\p^3(n_S-4n_F)}{120(p_1+p_2+p_3)^4}\Big[(4p_3+p_1+p_2)\big(3(p_1+p_2)^4-3(p_1+p_2)^2p_1p_2+4p_1^2p_2^2\big)\notag\\
&+9p_3^2(p_1+p_2)\big((p_1+p_2)^2-3p_1 p_2\big)-3p_3^5-12p_3^4(p_1+p_2)-9p_3^3\big((p_1+p_2)^2+2p_1 p_2\big)\Big]\notag\\
&+\frac{\p^3\,n_F}{6(p_1+p_2+p_3)^3}\Big[(p_1+p_2)\big((p_1+p_2)^2-p_1 p_2\big)(p_1+p_2+3p_3)-p_3^4-3p_3^3(p_1+p_2)\notag\\
&-6p_1p_2p_3^2\Big]-\frac{\p^3(n_s+4n_F)}{8(p_1+p_2+p_3)^2}\Big[p_1^3+2p_1^2(p_2+p_3)+(2p_1+p_2+p_3)\big((p_2+p_3)^2-p_2 p_3\big)\Big],
\end{align}
\begin{align}
A_{5}^{d=3}(p_1,p_2,p_3)&=\frac{\p^3(n_S-4n_F)}{240(p_1+p_2+p_3)^3}\Big[-3(p_1+p_2+p_3)^6+9(p_1+p_2+p_3)^4(p_1 p_2+p_2 p_3+p_1 p_3)\notag\\
&+12(p_1+p_2+p_3)^2(p_1 p_2+p_2p_3+p_3 p_1)^2-33(p_1+p_2+p_3)^2p_1p_2p_3\notag\\
&+12(p_1+p_2+p_3)(p_1p_2+p_2p_3+p_1p_3)p_1p_2p_3+8p_1^2p_2^2p_3^2\Big]\notag\\
&+\frac{\p^3n_F}{12(p_1+p_2+p_3)^2}\Big[-(p_1+p_2+p_3)^5+3(p_1+p_2+p_3)^3(p_1p_2+p_2p_3+p_1p_3)\notag\\
&+4(p_1+p_2+p_3)(p_1p_2+p_2p_3+p_1p_3)^2-11(p_1+p_2+p_3)^2p_1p_2p_3\notag\\
&+4(p_1p_2+p_2p_3+p_1p_3)p_1p_2p_3\Big]-\frac{\p^3(n_S+4n_F)}{16}\Big[p_1^3+p_2^3+p_3^3\Big].
\end{align}
This is in agreement with the expression given by BMS in their work \cite{Bzowski:2013sza} in terms of the constant $\a_1,\a_2$ and $c_T$. The explicit match is given by the relations 
(see \cite{Bzowski:2013sza})
\begin{align}
\a_1=\frac{\p^3(n_S-4n_F)}{480}, \qquad \a_2=\frac{\p^3\,n_F}{6}, \qquad c_T=\frac{3\p^{5/2}}{128}(n_S+4n_F), \qquad c_g=0,
\end{align}
where $c_g$ is a constant introduced in \cite{Bzowski:2013sza} to account for a non-zero functional variation of the stress energy tensor with respect to the metric ($\sim \delta T^{\mu\nu}(x)/\delta g_{\alpha\beta}(y))$ and corresponds to an extra contact term not included in our discussion.
In $d=5$ case instead we have 
%%%%%%%%%%%%%%%%%%%%%%%%%%
\begin{equation}
{C}_0(p_1^2,p_2^2,p_3^2)=\frac{\p^{3/2}}{p_1+p_2+p_3}.
\end{equation}
The ${B}_0$ is calculated in $d=5$ as
\begin{equation}
{B}_0(p_1^2)=-\frac{\p^{3/2}}{4} p_1.
\end{equation}
We give only the expression of the $A_1$ form factor which takes the form 
\begin{align}
A_{1}^{d=5}(p_1,p_2,p_3)&=\frac{\p^4(n_S-4n_F)}{560(p_1+p_2+p_3)^7}\Big[(p_1+p_2+p_3)^2\big((p_1+p_2+p_3)^4+(p_1+p_2+p_3)^2(p_1p_2+p_2p_3+p_1p_3)\notag\\
&\hspace{-1.9cm}+(p_1p_2+p_2p_3+p_1p_3)^2\big)+(p_1+p_2+p_3)\big((p_1+p_2+p_3)^2+5(p_1p_2+p_2p_3+p_1p_3)\big)p_1p_2p_3+10p_1^2p_2^2p_3^2\Big].
\end{align}
All the form factors are in agreement with those given in \cite{Bzowski:2013sza} as far as the general constants (denoted by $\a_1$ and $\a_2$) are matched by the relations
\begin{align}
\a_1=\frac{\p^4(n_S-4n_F)}{560 \times 72}, \qquad \a_2=\frac{\p^4\,n_F}{240}, \qquad c_T=\frac{5\p^{7/2}}{1024}(n_S+8n_F),
\end{align}
as given in \cite{Coriano:2018bsy}.

%%%%%%%%%%%%%%%%%%%

\subsection{Renormalization of the \texorpdfstring{$TTT$}{ttt} in \texorpdfstring{$d=4$}{d4}}
\label{renorm}
In $d=4$ the complete correlation function can be written as
\begin{equation}
\braket{T^{\m_1\n_1}(p_1)T^{\m_2\n_2}(p_2)T^{\m_3\n_3}(p_3)}=\sum_{I=F,G,S}\,n_I\,\braket{T^{\m_1\n_1}(p_1)T^{\m_2\n_2}(p_2)T^{\m_3\n_3}(p_3)}_I,
\end{equation}
also valid for the transverse traceless part of the correlator. In this case we encounter divergences in the forms of single poles in $1/\epsilon$  ($\epsilon=(4-d)/2$). In this section we discuss the structures of such divergences and their elimination in DR using the two usual gravitational counterterms.

Coming to the renormalization of the 3-graviton vertex, this is obtained by adding of 2 counterterms in the defining Lagrangian. In perturbation theory the one loop counterterm Lagrangian is
\begin{equation}
S_{count}=-\sdfrac{1}{\varepsilon}\,\sum_{I=F,S,G}\,n_I\,\int d^dx\,\sqrt{-g}\bigg(\,\b_a(I)\,C^2+\b_b(I)\,E\bigg),
\label{scount}
\end{equation}
corresponding to the Weyl tensor squared and the Euler density, omitting the extra 
$R^2$ operator which is responsible for the $\square R$ term in \eqref{G}, having chosen the local part of anomaly 
$(\sim \beta_c\square R)$ vanishing ($\beta_c=0$). \cite{Coriano:2012wp} addresses in detail this point and the finite renormalization which is needed to get from the general $\beta_c\neq 0$ to the $\beta_c=0$ case.  The counterterm vertices are
\begin{align}
&\braket{T^{\m_1\n_1}(p_1)T^{\m_2\n_2}(p_2)T^{\m_3\n_3}(p_3)}_{count}=\notag\\
&\hspace{2cm}=-\sdfrac{1}{\varepsilon}\sum_{I=F,S,G}n_I\bigg(\b_a(I)\,V_{C^2}^{\m_1\n_1\m_2\n_2\m_3\n_3}(p_1,p_2,p_3)+\b_b(I)\,V_{E}^{\m_1\n_1\m_2\n_2\m_3\n_3}(p_1,p_2,p_3)\bigg),
\end{align} 
where
\begin{align}
V_{C^2}^{\m_1\n_1\m_2\n_2\m_3\n_3}(p_1,p_2,p_3)&=8\int\,d^dx_1\,\,d^dx_2\,\,d^dx_3\,\,d^dx\,\bigg(\sdfrac{\d^3(\sqrt{-g}C^2)(x)}{\d g_{\m_1\n_1}(x_1)\d g_{\m_2\n_2}(x_2)\d g_{\m_3\n_3}(x_3)}\bigg)_{flat}\,e^{-i(p_1\,x_1+p_2\,x_2+p_3\,x_3)}\notag\\
&\equiv 8\big[\sqrt{-g}\,C^2\big]^{\m_1\n_1\m_2\n_2\m_3\n_3}(p_1,p_2,p_3),\\[2ex]
V_{E}^{\m_1\n_1\m_2\n_2\m_3\n_3}(p_1,p_2,p_3)&=8\int\,d^dx_1\,\,d^dx_2\,\,d^dx_3\,\,d^dx\,\bigg(\sdfrac{\d^3(\sqrt{-g}E)(x)}{\d g_{\m_1\n_1}(x_1)\d g_{\m_2\n_2}(x_2)\d g_{\m_3\n_3}(x_3)}\bigg)_{flat}\,e^{-i(p_1\,x_1+p_2\,x_2+p_3\,x_3)}\notag\\
&\equiv 8\big[\sqrt{-g}\,E\big]^{\m_1\n_1\m_2\n_2\m_3\n_3}(p_1,p_2,p_3), \label{count}
\end{align}
which satisfy the relations
\begin{align}
\d_{\mu_1\nu_1}\,V_{C^2}^{\mu_1\nu_1\mu_2\nu_2\mu_3\nu_3}(p_1,p_2,p_3)&=4(d-4)\big[C^2\big]^{\mu_2\nu_2\mu_3\nu_3}(p_2,p_3)\notag\\
&-8\,\bigg([C^2]^{\mu_2\nu_2\mu_3\nu_3}(p_1+p_2,p_3)+[C^2]^{\mu_2\nu_2\mu_3\nu_3}(p_2,p_1+p_3)\bigg),\\[2ex]
\d_{\mu_1\nu_1}\,V_{E}^{\mu_1\nu_1\mu_2\nu_2\mu_3\nu_3}(p_1,p_2,p_3)&=4(d-4)\big[E\big]^{\mu_2\nu_2\mu_3\nu_3}(p_2,p_3),\\[2ex]
p_{1\mu_1}\,V_{C^2}^{\mu_1\nu_1\mu_2\nu_2\mu_3\nu_3}(p_1,p_2,p_3)&=-4\,\bigg(p_2^{\nu_1}[C^2]^{\mu_2\nu_2\mu_3\nu_3}(p_1+p_2,p_3)+p_3^{\nu_1}[C^2]^{\mu_2\nu_2\mu_3\nu_3}(p_2,p_1+p_3)\bigg)\notag\\
&\hspace{-0.7cm}+4\,p_{2\a}\bigg(\d^{\mu_2\nu_1}[C^2]^{\a\nu_2\mu_3\nu_3}(p_1+p_2,p_3)+\d^{\nu_2\nu_1}[C^2]^{\a\mu_2\mu_3\nu_3}(p_1+p_2,p_3)\bigg)\notag\\
&\hspace{-0.7cm}+4\,p_{3\a}\bigg(\d^{\mu_3\nu_1}[C^2]^{\mu_2\nu_2\a\nu_3}(p_2,p_1+p_3)+\d^{\nu_3\nu_1}[C^2]^{\mu_2\mu_2\mu_3\a}(p_2,p_1+p_3)\bigg),\\[2ex]
p_{1\mu_1}\,V_{E}^{\mu_1\nu_1\mu_2\nu_2\mu_3\nu_3}(p_1,p_2,p_3)&=0.
\end{align}

\subsection{Reconstruction of the \texorpdfstring{$TTT$}{ttt} in \texorpdfstring{$d=4$}{d4}}
%%%%%%%%%%%%%%%%%%%%%%%%%%%%%%%%%%%%%%%%%%%%%%%%%%%%%%%
%%%%%%%%%%%%%%%%%%%%%%%%%%%%%%%%%%%%%%%%%%%%%%%%%%%%%%%
We now come to review the renormalization procedure for this vertex and the way 
the massless exchanges emerge from the longitudinal sector. \\
We start from the bare local contributions in the decomposition of the $TTT$ correlator of \eqref{reconstrTTT} in $d$ dimensions which take the form
\begin{align}
\label{loc}
\braket{t_{loc}^{\mu_1\nu_1}T^{\mu_2\nu_2}T^{\mu_3\nu_3}}&=\Big(\mathcal{I}^{\mu_1\nu_1}_{\a_1}(p_1)\,p_{1\b_1}+\frac{\p^{\mu_1\nu_1}(p_1)}{(d-1)}\d_{\a_1\b_1}\Big)\braket{T^{\a_1\b_1}T^{\mu_2\nu_2}T^{\mu_3\nu_3}}\notag\\
&=-\frac{2\,\p^{\mu_1\nu_1}(p_1)}{(d-1)}\Big[\braket{T^{\mu_2\nu_2}(p_1+p_2)T^{\mu_3\nu_3}(p_3)}+\braket{T^{\mu_2\nu_2}(p_2)T^{\mu_3\nu_3}(p_1+p_3)}\Big]
\notag\\
&+\mathcal{I}^{\mu_1\nu_1}_{\a_1}(p_1)\Big\{-p_2^{\a_1}\braket{T^{\mu_2\nu_2}(p_1+p_2)T^{\mu_3\nu_3}(p_3)}-p_3^{\a_1}\braket{T^{\mu_2\nu_2}(p_2)T^{\mu_3\nu_3}(p_1+p_3)}\notag\\
&+p_{2\b}\Big[\d^{\a_1\mu_2}\braket{T^{\b\mu_2}(p_1+p_2)T^{\mu_3\nu_3}(p_3)}+\d^{\a_1\nu_2}\braket{T^{\b\mu_2}(p_1+p_2)T^{\mu_3\nu_3}(p_3)}\Big]\notag\\
&+p_{3\b}\Big[\d^{\a_1\mu_3}\braket{T^{\nu_2\mu_2}(p_2)T^{\b\nu_3}(p_1+p_3)}+\d^{\a_1\nu_3}\braket{T^{\mu_2\nu_2}(p_2)T^{\mu_3\b}(p_1+p_3)}\Big]\Big\},
\end{align}
which develop a singularity for $\epsilon\to 0$, with $\epsilon=(4-d)/2$, just like all the other contributions appearing in \eqref{reconstrTTT}. \\
All the tensor contractions are done in $d$ dimensions and in the final expression we set $d= 4 +\epsilon$. For example, if a projector such as $\Pi^{(d)}$, we will be using the relation
\begin{equation}
\label{pexp}
\P^{(d)\,\mu_1\nu_1\mu_2\nu_2}(p)=\P^{(4)\,\mu_1\nu_1\mu_2\nu_2}(p)-\frac{2}{9}\varepsilon\,\pi^{\mu_1\nu_1}(p)\,\pi^{\mu_2\nu_2}(p)+O(\varepsilon^2), 
\end{equation}
which relates $\Pi^{(d)}$ to $\Pi^{(4)}$, and so on. For instance, a projector such as $\pi^{\mu_1\nu_1}$ with open indices remains unmodified since it has no explicit $d$-dependence, unless it is contracted with a $\delta^{\mu\nu}$. It is apparent from a quick look at the right hand side of \eqref{loc} that the regulated expression of this expression involves a prefactor $1/(d-1)$, which is expanded around $d=4$ and the replacements of all the two-point functions with the regulated expression given by 
\begin{align}
\label{regular}
\braket{T^{\mu_1\nu_1}(p)T^{\mu_2\nu_2}(-p)}_{reg}&=-\frac{\p^2\,p^4}{60\,\varepsilon}\Pi^{(4)\,\mu_1\nu_1\mu_2\nu_2}(p)\left(6 n_F + 12 n_G + n_S\right)\notag\\
&\hspace{-3cm}+\frac{\p^2\,p^4}{270}\p^{\mu_1\nu_1}(p)\p^{\mu_2\nu_2}(p)\left(6 n_F + 12 n_G + n_S\right)-\frac{\p^2\,p^4}{300}\bar{B}_0(p^2)\Pi^{\mu_1\nu_1\mu_2\nu_2}(p)\left(30n_F+60n_G+5n_S\right)\notag\\
&\hspace{-2cm}-\frac{\p^2\,p^4}{900}\Pi^{\mu_1\nu_1\mu_2\nu_2}(p)\left(36n_F-198 n_G+16n_S\right)+O(\varepsilon),
\end{align}
with the insertion of the appropriate momenta.\\
The corresponding counterterm is given by 
\begin{align}
\label{locco}
\braket{t_{loc}^{\mu_1\nu_1}T^{\mu_2\nu_2}T^{\mu_3\nu_3}}_{count}&=\Big(\mathcal{I}^{\mu_1\nu_1}_{\a_1}(p_1)\,p_{1\b_1}+\frac{\p^{\mu_1\nu_1}(p_1)}{(d-1)}\d_{\a_1\b_1}\Big)\braket{T^{\a_1\b_1}T^{\mu_2\nu_2}T^{\mu_3\nu_3}}_{(count)}\notag\\
&\hspace{-3cm}=-\frac{1}{\varepsilon}\frac{(d-4)}{(d-1)}\p^{\mu_1\nu_1}(p_1)\bigg(4[E]^{\mu_2\nu_2\mu_3\nu_3}(p_2,p_3)+4[C^2]^{\mu_2\nu_2\mu_3\nu_3}(p_2,p_3)\bigg)\notag\\
&\hspace{-3cm}+\frac{1}{\varepsilon}\frac{2}{(d-1)}\p^{\mu_1\nu_1}(p_1)\bigg(4[C^2]^{\mu_2\nu_2\mu_3\nu_3}(p_1+p_2,p_3)+4[C^2]^{\mu_2\nu_2\mu_3\nu_3}(p_2,p_1+p_3)\bigg)\notag\\
&\hspace{-3cm}-\frac{1}{\varepsilon}\mathcal{I}^{\mu_1\nu_1}_{\a_1}(p_1)\bigg\{-4p_2^{\a_1}[C^2]^{\mu_2\nu_2\mu_3\nu_3}(p_1+p_2,p_3)-p_3^{\a_1}[C^2]^{\mu_2\nu_2\mu_3\nu_3}(p_2,p_1+p_3)\notag\\
&\hspace{-3cm}+4p_{2\b}\Big[\d^{\a_1\mu_2}[C^2]^{\b\nu_2\mu_3\nu_3}(p_1+p_2,p_3)+\d^{\a_1\nu_2}[C^2]^{\mu_2\b\mu_3\nu_3}(p_1+p_2,p_3)\Big]\notag\\
&\hspace{-3cm}+4p_{3\b}\Big[\d^{\a_1\mu_3}[C^2]^{\mu_2\nu_2\b\nu_3}(p_2,p_1+p_3)+\d^{\a_1\nu_3}[C^2]^{\mu_2\nu_2\mu_3\b}(p_2,p_1+p_3)\Big]\bigg\},
\end{align}
where the dependence on the total contributions has been accounted for by the beta functions $\beta_a$ and $\beta_b$ 
\begin{equation}
\beta_{a,b}\equiv\sum_{I=F,S,G} \beta_{a,b} (I),
\end{equation}
into $[E]$ and $[C^2]$. All the divergent contributions of the local term given in \eqref{loc} are cancelled by the local parts of the counterterm \eqref{locco}.We obtain the renormalized expression 
\begin{align}
\braket{t_{loc}^{\mu_1\nu_1}T^{\mu_2\nu_2}T^{\mu_3\nu_3}}_{Ren}&=\braket{t_{loc}^{\mu_1\nu_1}T^{\mu_2\nu_2}T^{\mu_3\nu_3}}+\braket{t_{loc}^{\mu_1\nu_1}T^{\mu_2\nu_2}T^{\mu_3\nu_3}}_{(count)}\notag\\
&=\mathcal{V}_{loc\,0\,0 } +\braket{t_{loc}^{\mu_1\nu_1}T^{\mu_2\nu_2}T^{\mu_3\nu_3}}^{(4)}_{extra},
\end{align}
where
\begin{align}
&\mathcal{V}_{loc\, 0 \, 0 }=-\frac{2\,\p^{\mu_1\nu_1}(p_1)}{3}\Big[\braket{T^{\mu_2\nu_2}(p_1+p_2)T^{\mu_3\nu_3}(-p_1-p_2)}_{Ren}+\braket{T^{\mu_2\nu_2}(p_2)T^{\mu_3\nu_3}(-p_2)}_{Ren}\Big]
\notag\\
&\quad+\mathcal{I}^{(4)\,\mu_1\nu_1}_{\a_1}(p_1)\Big\{-p_2^{\a_1}\braket{T^{\mu_2\nu_2}(p_1+p_2)T^{\mu_3\nu_3}(-p_1-p_2)}_{Ren}-p_3^{\a_1}\braket{T^{\mu_2\nu_2}(p_2)T^{\mu_3\nu_3}(-p_2)}_{Ren}\notag\\
&\quad+p_{2\b}\Big[\d^{\a_1\mu_2}\braket{T^{\b\mu_2}(p_1+p_2)T^{\mu_3\nu_3}(-p_1-p_2)}_{Ren}+\d^{\a_1\nu_2}\braket{T^{\b\mu_2}(p_1+p_2)T^{\mu_3\nu_3}(-p_1-p_2)}_{Ren}\Big]\notag\\
&\quad+p_{3\b}\Big[\d^{\a_1\mu_3}\braket{T^{\nu_2\mu_2}(p_2)T^{\b\nu_3}(-p_2)}_{Ren}+\d^{\a_1\nu_3}\braket{T^{\mu_2\nu_2}(p_2)T^{\mu_3\b}(-p_2)}_{Ren}\Big]\Big\},
\end{align}
with $\braket{TT}_{Ren}$ given by 
\begin{align}
\braket{T^{\mu_1\nu_1}(p)T^{\mu_2\nu_2}(-p)}_{Ren}&=\braket{T^{\mu_1\nu_1}(p)T^{\mu_2\nu_2}(-p)}+\braket{T^{\mu_1\nu_1}(p)T^{\mu_2\nu_2}(-p)}_{count}\notag\\
&=-\frac{\p^2\,p^4}{60}\bar{B}_0(p^2)\Pi^{\mu_1\nu_1\mu_2\nu_2}(p)\left(6n_F+12n_G+n_S\right)\notag\\
&\quad-\frac{\p^2\,p^4}{900}\P^{\mu_1\nu_1\mu_2\nu_2}(p)\big(126n_F-18n_G+31n_S\big).
\label{tren}
\end{align}
The remainder is an extra contribution coming from the local parts of counterterms given by
\begin{align}
&\braket{t_{loc}^{\mu_1\nu_1}T^{\mu_2\nu_2}T^{\mu_3\nu_3}}^{(4)}_{extra}=\frac{\hat{\p}^{\mu_1\nu_1}(p_1)}{3 \, p_1^2}\bigg(4[E]^{\mu_2\nu_2\mu_3\nu_3}(p_2,p_3)+4[C^2]^{\mu_2\nu_2\mu_3\nu_3}(p_2,p_3)\bigg),
\end{align}
having defined 
\begin{equation}
\hat{\p}^{\mu\nu}(p)=(\delta^{\mu_1\nu_1}p^2 - p^\mu p^\nu),
\end{equation}
which shows the emergence of an anomaly pole, similarly to the $TJJ$ cases  \cite{Giannotti:2008cv,Armillis:2009pq,Coriano:2018zdo}. 

The procedure can be extended to all the other contributions of the correlator. In particular, the contribution with two $t_{loc}$ projections takes the form
\begin{align}
&\braket{t_{loc}^{\mu_1\nu_1}t_{loc}^{\mu_2\nu_2}T^{\mu_3\nu_3}}_{Ren}= \mathcal{V}^{\mu_1\nu_1\mu_2\nu_2\mu_3\nu_3}_{loc\, loc \, 0} +\braket{t_{loc}^{\mu_1\nu_1}t_{loc}^{\mu_2\nu_2}T^{\mu_3\nu_3}}^{(4)}_{extra},
\end{align}
where
\begin{align}
&\mathcal{V}^{\mu_1\nu_1\mu_2\nu_2\mu_3\nu_3}_{loc\, loc \, 0}=\Big(\mathcal{I}^{(4)\,\mu_2\nu_2}_{\a_2}(p_2)\,p_{2\b_2}+\frac{\p^{\mu_2\nu_2}(p_2)}{3}\d_{\a_2\b_2}\Big)\notag\\
&\times\Bigg\{-\frac{2\,\p^{\mu_1\nu_1}(p_1)}{3}\Big[\braket{T^{\a_2\b_2}(p_1+p_2)T^{\mu_3\nu_3}(-p_1-p_2)}_{Ren}+\braket{T^{\a_2\b_2}(p_2)T^{\mu_3\nu_3}(-p_2)}_{Ren}\Big]
\notag\\
&\quad+\mathcal{I}^{(4)\,\mu_1\nu_1}_{\a_1}(p_1)\Big[-p_2^{\a_1}\braket{T^{\a_2\b_2}(p_1+p_2)T^{\mu_3\nu_3}(-p_1-p_2)}_{Ren}-p_3^{\a_1}\braket{T^{\a_2\b_2}(p_2)T^{\mu_3\nu_3}(-p_2)}_{Ren}\notag\\
&\quad+p_{2\b}\Big(\d^{\a_1\a_2}\braket{T^{\b\b_2}(p_1+p_2)T^{\mu_3\nu_3}(-p_1-p_2)}_{Ren}+\d^{\a_1\b_2}\braket{T^{\b\a_2}(p_1+p_2)T^{\mu_3\nu_3}(-p_1-p_2)}_{Ren}\Big)\notag\\
&\quad+p_{3\b}\Big(\d^{\a_1\mu_3}\braket{T^{\b_2\a_2}(p_2)T^{\b\nu_3}(-p_2)}_{Ren}+\d^{\a_1\nu_3}\braket{T^{\a_2\b_2}(p_2)T^{\mu_3\b}(-p_2)}_{Ren}\Big)\Big]\Bigg\},
\end{align}
in which we define 
\begin{align}
\mathcal{I}^{(4)\,\mu\nu}_{\a}(p)\equiv \frac{1}{p^2}\left[2 p^{(\mu}\delta^{\nu)}_\alpha - 
\frac{p_\alpha}{3}\left(\delta^{\mu\nu} +2\,\frac{p^\mu p^\nu}{p^2}\right)\right],
\end{align}
and with an extra term of the form
\begin{align}
\braket{t_{loc}^{\mu_1\nu_1}t_{loc}^{\mu_2\nu_2}T^{\mu_3\nu_3}}^{(4)}_{extra}=\frac{\p^{\mu_1\nu_1}(p_1)}{3}\frac{\p^{\mu_2\nu_2}(p_2)}{3}\d_{\a_2\b_2}\bigg(4[E]^{\a_2\b_2\mu_3\nu_3}(p_2,p_3)+4[C^2]^{\a_2\b_2\mu_3\nu_3}(p_2,p_3)\bigg).
\end{align}
We should also consider the term with three insertions of $t_{loc}$
\begin{align}
&\braket{t_{loc}^{\mu_1\nu_1}t_{loc}^{\mu_2\nu_2}t_{loc}^{\mu_3\nu_3}}_{Ren}= \mathcal{V}^{\mu_1\nu_1\mu_2\nu_2\mu_3\nu_3}_{loc\,loc\,loc} +\braket{t_{loc}^{\mu_1\nu_1}t_{loc}^{\mu_2\nu_2}t_{loc}^{\mu_3\nu_3}}^{(4)}_{extra},
\end{align}
with
\begin{align}
&\mathcal{V}^{\mu_1\nu_1\mu_2\nu_2\mu_3\nu_3}_{loc\,loc\,loc}=\Big(\mathcal{I}^{(4)\,\mu_2\nu_2}_{\a_2}(p_2)\,p_{2\b_2}+\frac{\p^{\mu_2\nu_2}(p_2)}{3}\d_{\a_2\b_2}\Big)\Big(\mathcal{I}^{(4)\,\mu_3\nu_3}_{\a_3}(p_3)\,p_{3\b_3}+\frac{\p^{\mu_3\nu_3}(p_3)}{3}\d_{\a_3\b_3}\Big)\notag\\
&\times\Bigg\{-\frac{2\,\p^{\mu_1\nu_1}(p_1)}{3}\Big[\braket{T^{\a_2\b_2}(p_1+p_2)T^{\a_3\b_3}(-p_1-p_2)}_{Ren}+\braket{T^{\a_2\b_2}(p_2)T^{\a_3\b_3}(-p_2)}_{Ren}\Big]
\notag\\
&\quad+\mathcal{I}^{(4)\,\mu_1\nu_1}_{\a_1}(p_1)\Big[-p_2^{\a_1}\braket{T^{\a_2\b_2}(p_1+p_2)T^{\a_3\b_3}(-p_1-p_2)}_{Ren}-p_3^{\a_1}\braket{T^{\a_2\b_2}(p_2)T^{\a_3\b_3}(-p_2)}_{Ren}\notag\\
&\quad+p_{2\b}\Big(\d^{\a_1\a_2}\braket{T^{\b\b_2}(p_1+p_2)T^{\a_3\b_3}(-p_1-p_2)}_{Ren}+\d^{\a_1\b_2}\braket{T^{\b\a_2}(p_1+p_2)T^{\a_3\b_3}(-p_1-p_2)}_{Ren}\Big)\notag\\
&\quad+p_{3\b}\Big(\d^{\a_1\a_3}\braket{T^{\b_2\a_2}(p_2)T^{\b\b_3}(-p_2)}_{Ren}+\d^{\a_1\b_3}\braket{T^{\a_2\b_2}(p_2)T^{\a_3\b}(-p_2)}_{Ren}\Big)\Big]\Bigg\},
\end{align}
where
\begin{align}
\braket{t_{loc}^{\mu_1\nu_1}t_{loc}^{\mu_2\nu_2}t_{loc}^{\mu_3\nu_3}}^{(4)}_{extra}=\frac{\p^{\mu_1\nu_1}(p_1)\,\p^{\mu_2\nu_2}(p_2)\,\p^{\mu_3\nu_3}(\bar{p}_3)}{27}\d_{\a_2\b_2}\d_{\a_3\b_3}\bigg(4[E]^{\a_2\b_2\a_3\b_3}(p_2,\bar{p}_3)+4[C^2]^{\a_2\b_2\a_3\b_3}(p_2,\bar{p}_3)\bigg).
\end{align}
In conclusion, the renormalization procedure for $\braket{TTT}$ leaves us with an extra term of the form
\begin{align}
\braket{T^{\mu_1\nu_1}T^{\mu_2\nu_2}T^{\mu_3\nu_3}}^{(4)}_{extra}&=\left(\frac{\p^{\mu_1\nu_1}(p_1)}{3}\bigg(4[E]^{\mu_2\nu_2\mu_3\nu_3}(p_2,\bar{p}_3)+4[C^2]^{\mu_2\nu_2\mu_3\nu_3}(p_2,\bar{p}_3)\bigg)+(\text{perm.})\right)\notag\\
&\hspace{-1cm}-\left(\frac{\p^{\mu_1\nu_1}(p_1)}{3}\frac{\p^{\mu_2\nu_2}(p_2)}{3}\d_{\a_2\b_2}\bigg(4[E]^{\a_2\b_2\mu_3\nu_3}(p_2,\bar{p}_3)+4[C^2]^{\a_2\b_2\mu_3\nu_3}(p_2,\bar{p}_3)\bigg) +(\text{perm.})\right)\notag\\
&\hspace{-1cm}+\frac{\p^{\mu_1\nu_1}(p_1)}{3}\frac{\p^{\mu_2\nu_2}(p_2)}{3}\frac{\p^{\mu_3\nu_3}(\bar{p}_3)}{3}\d_{\a_2\b_2}\d_{\a_3\b_3}\Big(4[E]^{\a_2\b_2\a_3\b_3}(p_2,\bar{p}_3)+4[C^2]^{\a_2\b_2\a_3\b_3}(p_2,\bar{p}_3)\Big).
\end{align}
This extra contribution is exactly the anomalous part of the $TTT$. In the flat limit the result takes the form
\begin{align}
\braket{T(p_1)T^{\mu_2\nu_2}(p_2)T^{\mu_3\nu_3}(\bar{p}_3)}_{anomaly}^{(4)}&=\big(4[E]^{\m_2\n_2\m_3\n_3}(p_2,p_3)+4[C^2]^{\m_2\n_2\m_3\n_3}(p_2,\bar{p}_3)\big),\\
\braket{T(p_1)T(p_2)T^{\mu_3\nu_3}(\bar{p}_3)}_{anomaly}^{(4)}&=\d_{\a_2\b_2}\big(4[E]^{\a_2\b_2\m_3\n_3}(p_2,p_3)+4[C^2]^{\a_2\b_2\m_3\n_3}(p_2,\bar{p}_3)\big),\notag\\
\braket{T(p_1)T(p_2)T(\bar{p}_3)}_{anomaly}^{(4)}&=\d_{\a_2\b_2}\d_{\a_3\b_3}\big(4[E]^{\a_2\b_2\a_3\b_3}(p_2,p_3)+4[C^2]^{\a_2\b_2\a_3\b_3}(p_2,\bar{p}_3)\big),
\end{align}
(with $T(p)\equiv \delta_{\m\n}T^{\m\nu}$).
To make it more explicit in momentum space, we need to use the expressions 
\begin{align}
\big[E\big]^{\m_i\nu_i\m_j\nu_j}(p_i,p_j) &=\big[R_{\m\a\n\b}\,R^{\m\a \n\b}\big]^{\m_i\nu_i\m_j\nu_j}
-4\,\big[R_{\m\n}R^{\m\n}\big]^{\m_i\nu_i\m_j\nu_j}
+\big[ R^2\big]^{\m_i\nu_i\m_j\nu_j}\notag\\
&\hspace{-2cm}=\bigg\{\big[R_{\m\a\n\b}\big]^{\m_i\nu_i}(p_i)\big[R^{\m\a \n\b}\big]^{\m_j\nu_j}(p_j)
-4\,\big[R_{\m\n}\big]^{\m_i\nu_i}(p_i)\big[R^{\m\n}\big]^{\m_j\nu_j}(p_j)
+\big[ R\big]^{\m_i\nu_i}(p_i)\big[R\big]^{\m_j\nu_j}(p_j)\bigg\}\notag,\\[1.2ex]
&\hspace{2cm}+\{(\mu_i,\nu_i,p_i)\leftrightarrow (\mu_j,\nu_j,p_j)\}\\[2ex]
\big[C^2\big]^{\m_i\nu_i\m_j\nu_j}(p_i,p_j) &= \big[R_{\m\a\n\b}R^{\m\a \n\b}\big]^{\m_i\nu_i\m_j\nu_j}
-2\,\big[R_{\m\n}R^{\m\n}\big]^{\m_i\nu_i\m_j\nu_j}
+ \sdfrac{1}{3}\,\big[R^2\big]^{\m_i\nu_i\m_j\nu_j}\notag\\
&\hspace{-2cm}=\bigg\{\big[R_{\m\a\n\b}\big]^{\m_i\nu_i}(p_i)\big[R^{\m\a \n\b}\big]^{\m_j\nu_j}(p_j)
-2\,\big[R_{\m\n}\big]^{\m_i\nu_i}(p_i)\big[R^{\m\n}\big]^{\m_j\nu_j}(p_j)
+\frac{1}{3}\big[ R\big]^{\m_i\nu_i}(p_i)\big[R\big]^{\m_j\nu_j}(p_j)\bigg\}\notag\\[1.2ex]
&\hspace{2cm}+\{(\mu_i,\nu_i,p_i)\leftrightarrow (\mu_j,\nu_j,p_j)\},
\end{align}
for which we obtain
\begin{align}
\braket{T^{\mu_1\nu_1}(p_1)T^{\mu_2\nu_2}(p_2)T^{\mu_3\nu_3}(\bar{p}_3)}^{(4)}_{extra}&=\left(\frac{\p^{\mu_1\nu_1}(p_1)}{3}\braket{T(p_1)T^{\mu_2\nu_2}(p_2)T^{\mu_3\nu_3}(\bar{p}_3)}^{(4)}_{anomaly}+(\text{perm.})\right)\notag\\
&\hspace{-1cm}-\left(\frac{\p^{\mu_1\nu_1}(p_1)}{3}\frac{\p^{\mu_2\nu_2}(p_2)}{3}\braket{T(p_1)T(p_2)T^{\mu_3\nu_3}(\bar{p}_3)}^{(4)}_{anomaly}+(\text{perm.})\right)\notag\\
&\hspace{-1cm}+\frac{\p^{\mu_1\nu_1}(p_1)}{3}\frac{\p^{\mu_2\nu_2}(p_2)}{3}\frac{\p^{\mu_3\nu_3}(\bar{p}_3)}{3}\braket{T(p_1)T(p_2)T(\bar{p}_3)}^{(4)}_{anomaly},
\end{align}
noting that this is exactly the anomaly contribution \eqref{fin1} predicted by the anomaly effective action. 
%%%%%%%%%%%%%%%%%%%%%%%%%%%%%%%%%%%%%%%%%%%%%%%%%%%%%%%%%%%%%%%%%%%%%%%%%%%%%%%%
\subsection{The perturbative structure of the \texorpdfstring{$TTT$}{ttt} and the poles separation} 
%%%%%%%%%%%%%%%%%%%%%%%%%%%%%%%%%%%%%%%%%%%%%%%%%%%%%%%%%%%%%%%%%%%%%%%%%%%%%%%%
\begin{figure}[t]
	\centering
	\vspace{2ex}
	\includegraphics[scale=0.57]{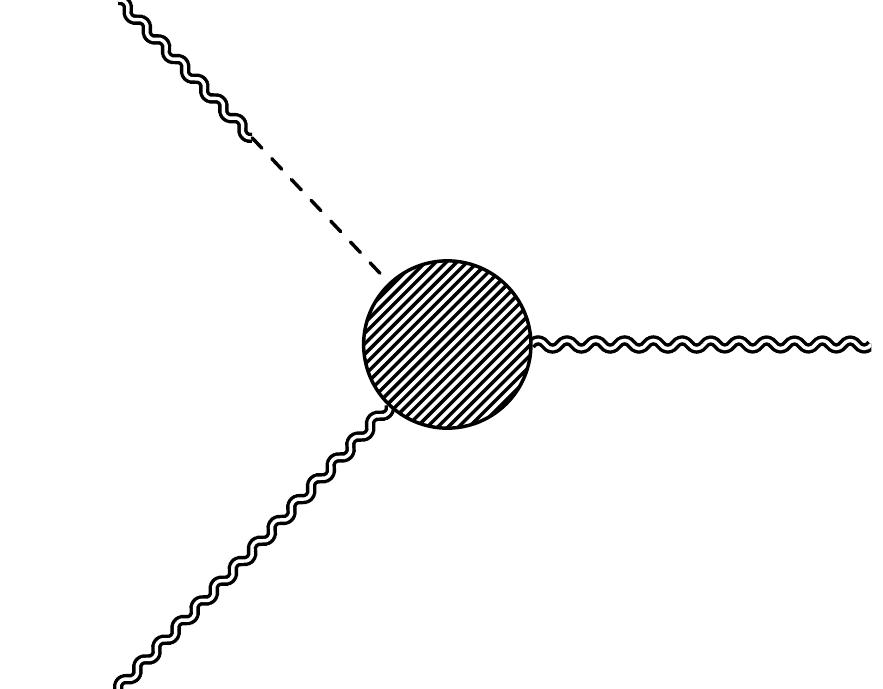}\hspace{1ex}
	\includegraphics[scale=0.57]{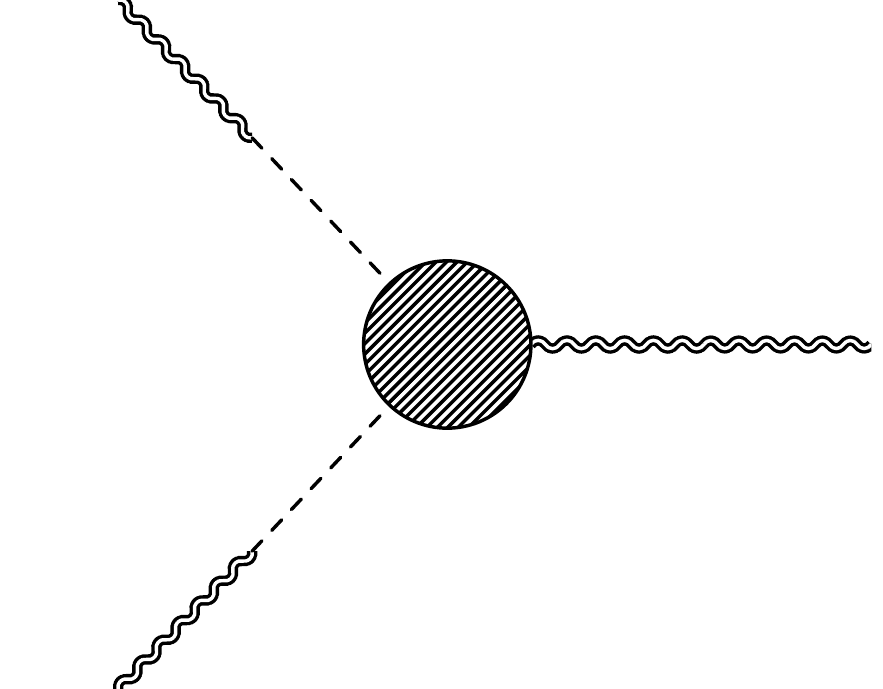}\hspace{1ex}
	\includegraphics[scale=0.57]{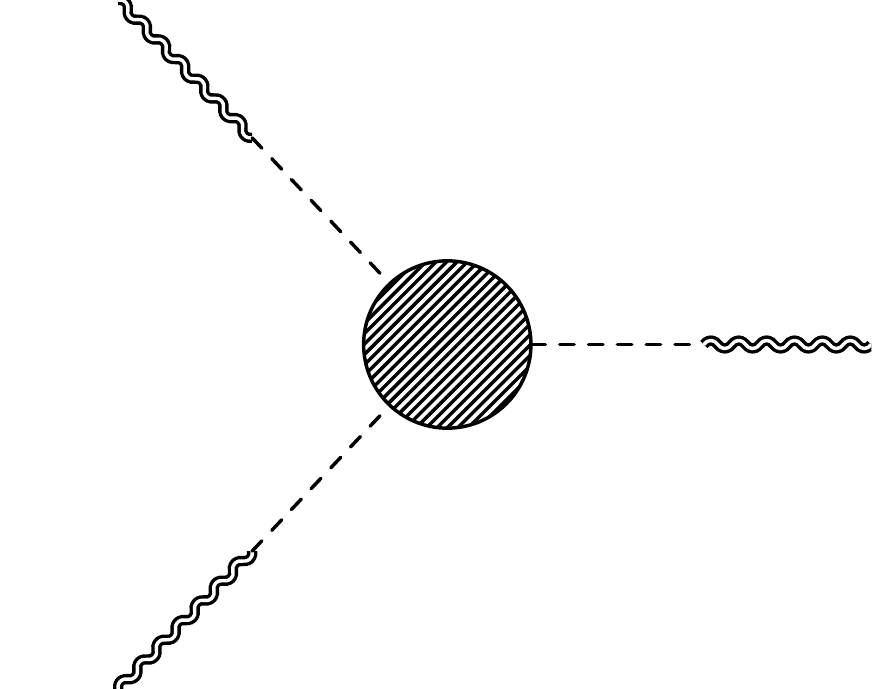}
	\vspace{2ex}
	\caption{Anomaly interactions mediated by the exchange of one, two or three poles. The poles are generated by the renormalization of the longitudinal sector of the $TTT$.}
	\label{dec}
\end{figure}
The structure of the poles in the $TTT$ are summarized in Fig. \ref{dec} where we have denoted with a dashed line the exchange of one or more massless $(\sim 1/p_i^2)$ interactions. In configuration space such extra terms, related to the renormalization of the correlator, are the natural generalization of the typical anomaly pole interaction found, for instance, in the case of the $TJJ$, where the effect of the anomaly is in the generation of a non-local interaction of the form \cite{Giannotti:2008cv,Armillis:2009pq,Armillis:2010qk}
\begin{equation}
\mathcal{S}_{an}\sim \beta(e) \int d^4 x\, d^4 y R^{(1)}(x)  \left(\frac{1}{\square}\right)(x,y) F F(y),
\end{equation}
with $F$ being the QED field strength and $\beta(e)$ the corresponding beta function of the gauge coupling. In the $TTT$ case, as one can immediately figure out from \eqref{expans}, such expressions can be rewritten as contribution to the anomaly action in the form
\begin{equation}
\mathcal{S}_{an}\sim \int d^4 x\, d^4 y R^{(1)}(x)  \left(\frac{1}{\square}\right)(x,y) \left( \beta_b E^{(2)}(y) + \beta_a (C^2)^{(2)}(y)\right),
\end{equation}
and similar for the other terms extracted from \eqref{expans}. Notice that each $\hat\pi$ projector in \eqref{expans} is accompanied by a corresponding anomaly (single) pole of the external invariants, generating contributions of the form $1/p_i^2$,  $1/(p_i^2 p_j^2) (i\neq j)$ and $1/(p_1^2 p_2^2 p_3^2)$, where multiple poles are connected to separate external graviton lines. Each momentum invariant appears as a single pole. One can use the correspondence 
\begin{equation}
\frac{1}{p^2} \hat\pi^{\mu\nu} \leftrightarrow R^{(1)}\frac{1}{\square} ,
\end{equation}
to include such non-zero trace contributions into the anomaly action. This involves a multiplication of the vertex by the external fields together with an integration over all the internal points. As shown in \cite{Coriano:2017mux} such non-zero trace contributions are automatically generated by the non-local conformal anomaly action, which accounts for the entire expression \eqref{expans}. 

The diagrammatic interpretation suggests a possible generalization of this result also to higher point functions, as one can easily guess, in a combination similar to that shown in Fig. \ref{dec}.
Notice that the numerators of such decompositions, which correspond to single, double and triple traces are, obviously, purely polynomial in the external invariants, being derived from the anomaly functional, which is local in momentum space. The extension of this analysis to 4-point functions can be found in \cite{cc}. 
%%%%%%%%%%%%%%%%%%%%%%%%%%%%%%%%%%%%%%%%%%%%%%%
\subsection{An example: the \texorpdfstring{$TJJ$}{tjj} case and the CWI's in QED}
\begin{figure}[t]
	\begin{center}
		\includegraphics[scale=0.8]{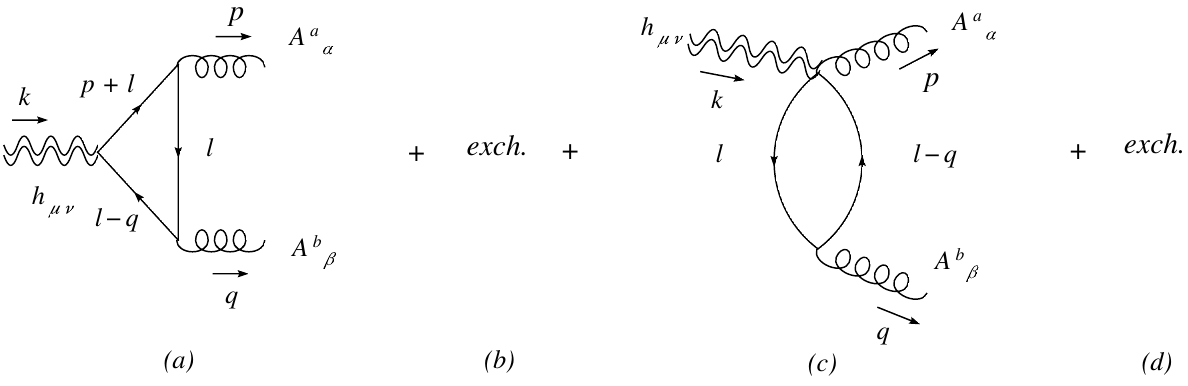}
		\caption{\small The fermionic contributions  with a graviton $h_{\mu\nu}$ in the initial state and two gluons $A^a_\a, A^b_\b$ in the final state in QCD. }
		\label{fermloop}
	\end{center}
\end{figure}

\begin{figure}[t]
	\begin{center}
		\includegraphics[scale=0.7]{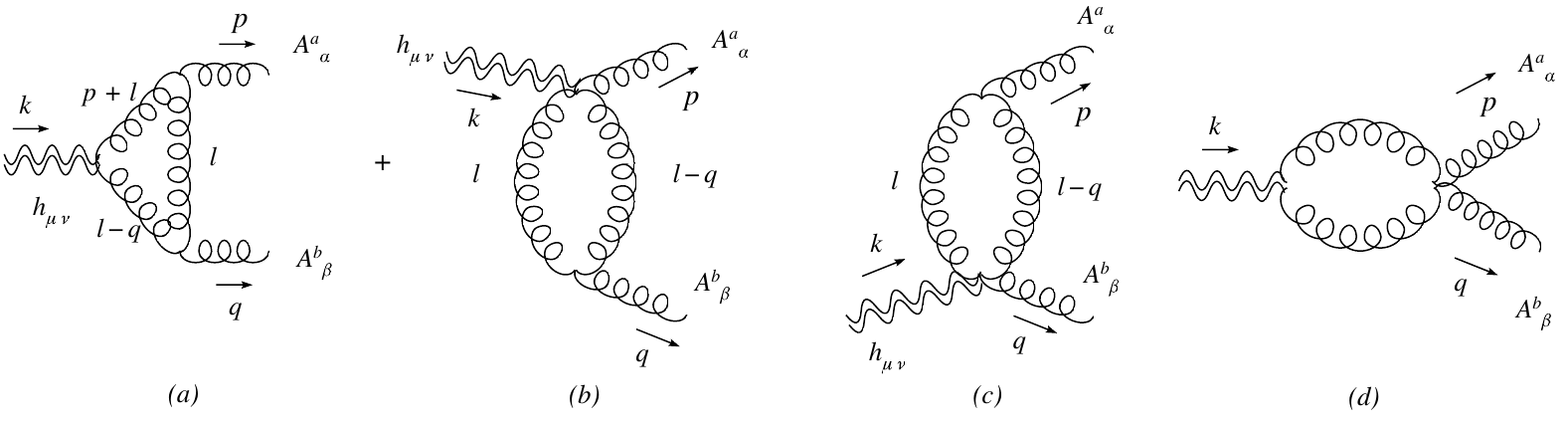}
		\caption{\small The gauge contributions  with a graviton $h_{\mu\nu}$ in the initial state and two gluons $A^a_\a, A^b_\b$ in the final state in QCD. }
		\label{gluonloop}
	\end{center}
\end{figure}

%%%%%%%%%%%%%%%%%%%%%%%%%%%%%%%%%%%%%%%%%%%%%%%
The matching to free field theory provides a significant simplification of the general results for three-point  functions in terms of simple two- and three-point master integrals, from which it is possible to extract significant information on the behavior of the correlators using the Feynman expansion. The approach, therefore, is non-perturbative, but such a matching allows to use perturbation theory, at the last stage, in order to reconstruct the form factors $A_i$ in their simplest forms. It is natural to ask whether this simplification is possible for all the correlators, and the answer is obviously negative. This procedure is possible for correlators involving stress energy tensors, conserved currents and others, whose scaling dimensions can be generated by simple free field theories. The reason is quite straightforward, since the scaling dimensions of a $T$ or a $J$ are $d$ and $d-1$ respectively, which are integers.
In this case we will be needing a number of independent sectors - for instance, scalars, spin-$1$, spin-$1/2$ fermions - running in the loops and interpolating with the gravitational sector by their stress energy tensors (see \figref{fermloop} and \figref{gluonloop}). Therefore, we need to find a number of independent free field theory sectors to match the number of independent constants identified by the solutions of the CWI's.

If we turn to correlators involving $T$, $J$ and, for instance, scalar operators of generic scaling dimensions, this is not possible on general grounds in free field theory. In this respect, the CWI's provide new information which is not available otherwise. In the remaining cases where we deal with correlators containing only $T$ and $J$ and/or scalars of integer dimensions, the free field theory results are equivalent to the most general ones and we don't need to go any further.  As such they provide the simplest match to the non-perturbative solutions. Given the important role played by the stress energy tensor in gravity, the value of such simplification can be hardly underestimated. Especially the anomalous behavior of theories, {\em can be addressed 
by the analysis of simple Feynman integrals}. For instance, one derives  significant and non-perturbative information on the structure of the spectral densities of the anomaly form factors in a straightforward way, as originally done in the $TJJ$ case. Having clarified this point, the perturtbative analysis of correlators in QED or QCD such as the $TJJ$, which is responsible for the most important coupling of gravity to the fields of the Standard Model, remains valid non-perturbatively \cite{Giannotti:2008cv, Coriano:2011zk,Armillis:2010qk,Coriano:2012wp,Coriano:2011ti}.

With these considerations in mind, let's consider the case of the $TJJ$ correlator in $d=4$, expanded in the transverse traceless and longitudinal components in the form
\begin{align}
\braket{T^{\mu_1\nu_1}\,J^{\mu_2}\,J^{\mu_3}}&=\braket{t^{\mu_1\nu_1}\,j^{\mu_2}\,j^{\mu_3}}+\braket{T^{\mu_1\nu_1}\,J^{\mu_2}\,j_{loc}^{\mu_3}}+\braket{T^{\mu_1\nu_1}\,j_{loc}^{\mu_2}\,J^{\mu_3}}+\braket{t_{loc}^{\mu_1\nu_1}\,J^{\mu_2}\,J^{\mu_3}}\notag\\
&\quad-\braket{T^{\mu_1\nu_1}\,j_{loc}^{\mu_2}\,j_{loc}^{\mu_3}}-\braket{t_{loc}^{\mu_1\nu_1}\,j_{loc}^{\mu_2}\,J^{\mu_3}}-\braket{t_{loc}^{\mu_1\nu_1}\,J^{\mu_2}\,j_{loc}^{\mu_3}}+\braket{t_{loc}^{\mu_1\nu_1}\,j_{loc}^{\mu_2}\,j_{loc}^{\mu_3}},
\end{align}
with the transverse traceless sector extracted by acting with the $\Pi$ and $\pi$ projectors on $T$ and $J$ respectively
\begin{align}
\langle t^{\mu_1\nu_1}(p_1)j^{\mu_2}(p_2)j^{\mu_3}(p_3)\rangle& =
{\Pi_1}^{\mu_1\nu_1}_{\alpha_1\beta_1}{\pi_2}^{\mu_2}_{\alpha_2}{\pi_3}^{\mu_3}_{\alpha_3}
\left( A_1\ p_2^{\alpha_1}p_2^{\beta_1}p_3^{\alpha_2}p_1^{\alpha_3} + 
A_2\ \delta^{\alpha_2\alpha_3} p_2^{\alpha_1}p_2^{\beta_1} + 
A_3\ \delta^{\alpha_1\alpha_2}p_2^{\beta_1}p_1^{\alpha_3}\right. \notag\\
& \left. + 
A_3(p_2\leftrightarrow p_3)\delta^{\alpha_1\alpha_3}p_2^{\beta_1}p_3^{\alpha_2}
+ A_4\  \delta^{\alpha_1\alpha_3}\delta^{\alpha_2\beta_1}\right)\label{DecompTJJ},
\end{align}
and parametrised in terms of four form factors $A_i$. A direct analysis shows that the unrenormalized CWI's take the expression \cite{Bzowski:2013sza}
\begin{align}
&0=C_{11}=K_{13}A_1,&&0=C_{21}=K_{23}A_1,\\
&0=C_{12}=K_{13}A_2+2A_1,&&0=C_{22}=K_{23}A_2,\\
&0=C_{13}=K_{13}A_3-4A_1,&&0=C_{23}=K_{23}A_3-4A_1,\\
&0=C_{14}=K_{13}A_3(p_2\leftrightarrow p_3),&&0=C_{24}=K_{23}A_3(p_2\leftrightarrow p_3)+4A_1,\\
&0=C_{15}=K_{13}A_4-2A_3(p_2\leftrightarrow p_3),&&0=C_{25}=K_{23}A_4+2A_3-2A_3(p_2\leftrightarrow p_3).
\label{Primary}
\end{align}
which can be solved in terms of triple-K integrals in the form 
\begin{eqnarray}
A_1 &=&\alpha_1 J_{4[000]}  \nonumber,\\
A_2 &=& \alpha_1 J_{3[100]} +\alpha_3 J_{2[000]}\nonumber ,\\
A_3 &=&2 \alpha_1 J_{3[001]} +\alpha_3 J_{2[000]} \nonumber,\\
A_4 &=&  2 \alpha_1 J_{2[011]} +\alpha_3 J_{1[010]} +\alpha_4 J_{0[000]}.
\label{nonper}
\end{eqnarray}
Variants of these equations are obtained by taking the $T$ operator as a singlet under the action of the spin matrices \cite{Coriano:2018bbe}.\\
In this case, the renormalization procedure involves the operator $F F$, with $F$ the abelian field strength, for removing the singularity of the two-point  function of the two vector currents $JJ$.  If we indicate with $A_R$ the renormalized form factors, the anomalous scaling WIs take the form

\begin{align}
\left(\sum_{i}^{3}p_i\sdfrac{\partial}{\partial p_i}+2\right)\,A_1&=0=-\m\sdfrac{\partial}{\partial \m}A_1,\\
\left(\sum_{i}^{3}p_i\sdfrac{\partial}{\partial p_i}\right)\,A_2^R&=\sdfrac{8\p^2\,e^2}{3}=-\m\sdfrac{\partial}{\partial \m}A_2^R,\\
\left(\sum_{i}^{3}p_i\sdfrac{\partial}{\partial p_i}\right)\,A_3^R(p_2\leftrightarrow p_3)&=\sdfrac{8\p^2\,e^2}{3}=-\m\sdfrac{\partial}{\partial \m}A_3^R,\\
\left(\sum_{i}^{3}p_i\sdfrac{\partial}{\partial p_i}-2\right)\,A_4^R&=-\sdfrac{4}{3}\p^2\,e^2(s-s_1-s_2)=-\m\sdfrac{\partial}{\partial \m}A_4^R,
\end{align}
while for the primary CWI's we obtain
\begin{equation}
\begin{split}
&K_{13}A_1=0,\\
&K_{13}A^R_2=-2A_1,\\
&K_{13}A^R_3=4A_1,\\
&K_{13}A^R_3(p_2\leftrightarrow p_3)=0,\\
&K_{13}A^R_4=2A^R_3(p_2\leftrightarrow p_3)-\sdfrac{16\,\pi^2e^2}{3},
\end{split}
\hspace{1.5cm}
\begin{split}
&K_{23}A_1=0,\\
&K_{23}A^R_2=0,\\
&K_{23}A^R_3=4A_1,\\
&K_{23}A^R_3(p_2\leftrightarrow p_3)=-4A_1,\\
&K_{23}A^R_4=-2A^R_3+2A^R_3(p_2\leftrightarrow p_3).
\end{split}
\end{equation}
In all the equations $e$ is the renormalized charge and can be traded for $\beta(e)$ by the relation $\beta(e)/e=e^2/(12 \pi^2)$. 
For the secondary CWI's the renormalized CWI's one obtains
 \begin{equation}
 \begin{split}
&L_4 A_1+R A_3^R-R A_3^R(p_2\leftrightarrow p_3)=0,\\[1.5ex]
&L_2\,A_2^R-s\,(A_3^R-A_3^R(p_2\leftrightarrow p_3))=\sdfrac{16}{9}\pi^2e^2\left[3s1\,B_0^R(s_1,0,0)-3s_2B_0^R(s_2,0,0)-s_1+s_2\right]+\sdfrac{24}{9}\pi^2e^2s_0,\\[1.5ex]
&L_4\,A^R_3-2R\,A^R_4=\sdfrac{32}{9}\pi^2\,e^2s_2\left[1-3B_0^R(s_2,0,0)\right]+\sdfrac{48}{9}\pi^2\,e^2\,s_0,\\[1.5ex]
&L_4\,A^R_3(p_2\leftrightarrow p_3)+2R\,A^R_4-4\,sA^R_3(p_2\leftrightarrow p_3)=\sdfrac{32}{9}\,\pi^2\,e^2s_1\left[3B_0^R(s_1,0,0)-1\right],\\[1.5ex]
&L'_3\,A_1^R-2R'A_2^R+2R'A_3^R=0,\\[1.5ex]
&L'_1\,A_3^R(p_2\leftrightarrow p_3)+p_2^2(4A^R_2-2A^R_3)+2R'A^R_4=\sdfrac{16}{3}\pi^2\,e^2\,s_1,\\
 \end{split}
 \end{equation}
where we have set $p_1^2=s_0, p_2^2=s_1$ and $p_3^2=s_2$
\begin{align}
L_N&= p_1(p_1^2 + p_2^2 - p_3^2) \frac{\partial}{\partial p_1} + 2 p_1^2\, p_2 \frac{\partial}{\partial p_2} + \big[ (2d - \Delta_1 - 2\Delta_2 +N)p_1^2 + (2\Delta_1-d)(p_3^2-p_2^2)  \big], \\
R &= p_1 \frac{\partial}{\partial p_1} - (2\Delta_1-d) \,. 
\end{align}
and 
\begin{align}
&L'_N=L_N,\quad\text{with}\ p_1\leftrightarrow p_2\ \text{and}\ \D_1\leftrightarrow\D_2,\\
&R'=R,\qquad\text{with}\ p_1\mapsto p_2\ \text{and}\ \D_1\mapsto\D_2.
\end{align}
These expressions depend on the conformal dimensions of the operators involved in the three-point  function under consideration, and additionally on a single parameter $N$ determined by the Ward identity in question. \\
The equations above define the anomalous CWI's for QED at one-loop and take a very simple structure. They remain valid  non-perturbatively in a generic CFT, for being equations which are matched to the non-perturbative solutions, as we have explained. Beyond one-loop classical conformally invariant theories such as QED or QCD  acquire radiative corrections. This is the point where realistic non-conformally invariant theories and CFT's which preserve the structure of such equations at quantum level start diverging.   
\begin{figure}[t]
	\centering
	\vspace{-1cm}
	\subfigure[]{\includegraphics[scale=0.9]{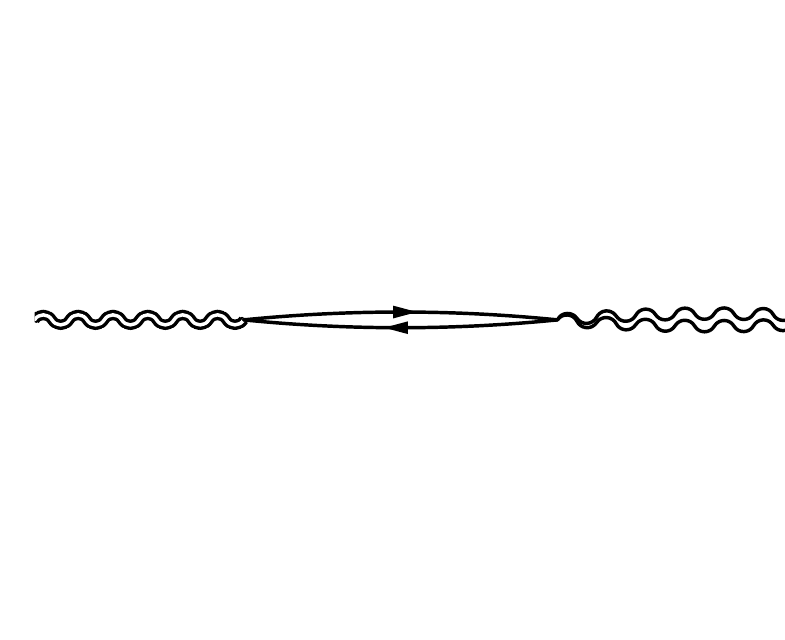}}\hspace{0.5cm}
	\subfigure[]{\includegraphics[scale=0.77]{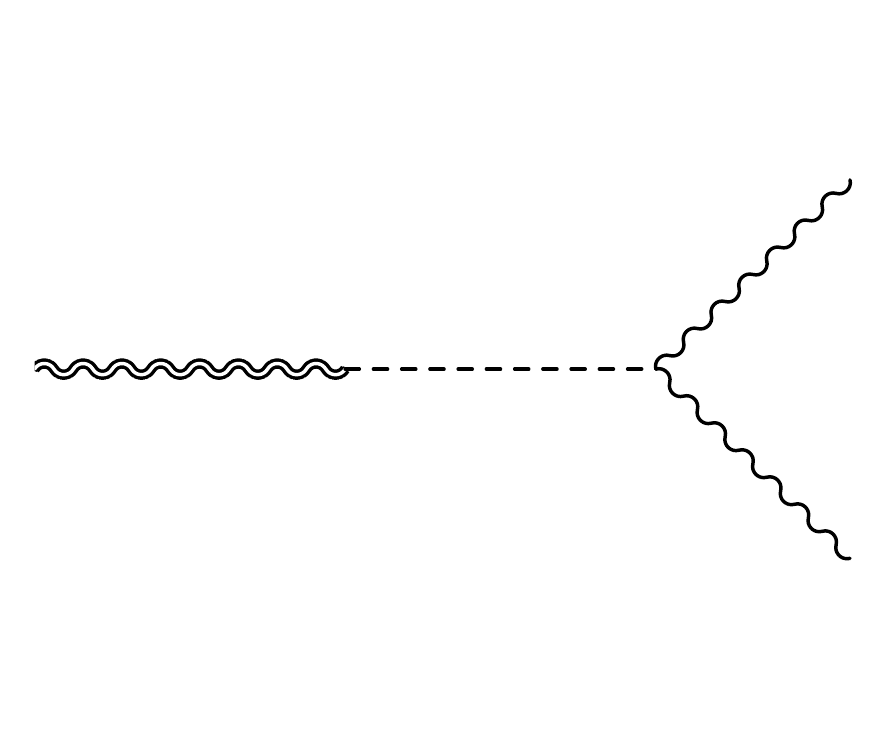}}
	\caption{Fig. (a): Singularity of the spectral density of the $TJJ$ in the anomaly form factor as a spacetime process. Fig. (b): The exchange of a pole as the origin of the conformal anomaly in the $TJJ$, derived in QED and QCD.}
	\label{collinear}
\end{figure}

\section*{Conclusions}
\addcontentsline{toc}{section}{Conclusions}
We have reviewed the momentum space approach to the solution of the CWI's of CFT's in higher dimensions. The goal of our work has been to illustrate the essential steps which are needed in order to build tensor correlators starting from the scalar solutions, for three-point  functions. \\
In the case of four-point functions, our attention has been centered around scalar correlators for which the CWI's are sufficient to isolate the unique solution, if we enhance the symmetry with the addition of a dual conformal symmetry. Dual conformal symmetry in momentum space is obtained once the momentum variables are rewritten in a dual form, as difference of coordinate-like variables and treated as ordinary correlators in such variables, mirroring the action of coordinate space. This enhancement of the symmetry is sufficient to fix the solutions also for such correlators.    
The solution of the conformal constraints are given in terms of triple-$K$ integrals and are expressed in terms of a set of constants, specific for each correlator and spacetime dimension.

We have presented a discussion of the intermediate steps in the description of two non-trivial correlators, the $TTO$ and the $TTT$, in a more pedagogical way, offering  details that could help extend such methods to higher point functions.\\ 
Several parallel studies have widened the goal of this activity, addressing issues such as the use of conformal blocks/CP symmetric blocks (Polyakov blocks) \cite{Isono:2019ihz,Isono:2018rrb, Chen:2019gka}, the operator product expansion in momentum space \cite{Gillioz:2019iye}, as well as light-cone blocks \cite{Gillioz:2019lgs, Gillioz:2018mto,Gillioz:2018kwh}, analytic continuations to Lorentzian spacetimes \cite{Bautista:2019qxj} spinning correlators, and Yangian symmetry \cite{Loebbert:2020hxk,Loebbert:2016cdm} just to mention a few, motivated by CFT in momentum space. Related analyses have explored the link to Witten diagrams within the AdS/CFT correspondence \cite{Anand:2019lkt,Albayrak:2019yve}. Investigations of the structure of higher point functions are contained in 
\cite{sk1,sk2}. 
At the same time, the extension of these investigations to de Sitter space has laid the foundations for new applications in cosmology \cite{Maldacena:2011nz, Arkani-Hamed:2018kmz,Baumann:2019oyu,Arkani-Hamed:2017fdk,Benincasa:2019vqr,Benincasa:2018ssx, Kundu:2014gxa,Almeida:2017lrq,Baumann:2020dch} and in gravitational waves \cite{Almeida:2019hhx}. Finally, investigations of such correlators in Mellin space \cite{Penedones:2010ue,Fitzpatrick:2011ia ,Gopakumar:2016cpb,Gopakumar:2016wkt} offer a new perspective on the bootstrap program both in flat and in curved space \cite{Sleight:2019mgd,Sleight:2019hfp}, providing further insight into the operatorial structure of a given CFT, and connecting in a new way momentum space and Mellin variables.  \\
Undoubtedly, CWI's play a crucial role in this effort, with widespread applications both at zero and at finite temperature \cite{Ohya:2018qkr}.
Among all the possible correlators that one may investigate, those containing stress-energy tensors $(T)$  play a special role, due to the presence of the conformal anomaly \cite{Coriano:2012wp}. \\
Analysis of four-point functions have so far been limited to scalar correlators in flat \cite{Maglio:2019grh, Bzowski:2019kwd} and curved backgrounds \cite{Arkani-Hamed:2018kmz,Baumann:2019oyu}.
Interestingly, this analysis can be performed in  parallel with the ordinary Lagrangian field theory approach, allowing to provide a particle interpretation of the breaking of the conformal symmetry. 

In the second part of our review, we have concentrated on the important issue of the identification of the fundamental structure of the anomaly effective action, which has been widely debated in the former literature. We have shown that the anomaly structure of the three-point  function of stress energy tensors is correctly described by the non-local version of such action. The specific analytic structure and massless poles predicted by the effective action are precisely what is obtained by the reconstruction of the trace parts from the CWI's. This indicates that the anomaly poles in momentum space are a necessary feature of the full $\braket{TTT}$ correlator in CFT. The addition of  local or Weyl invariant terms to the effective action cannot eliminate such contribution. 

By matching general solutions of the CWI's to free field theories, one obtains the general expressions of correlators containing stress energy tensors, conserved currents and scalar operators of integer dimensions. For correlators containing scalar operators of arbitrary scaling dimensions, the non-perturbative solutions of three-point  functions, obviously, cannot be matched by such free field theories.

The extension of the formalism of the CWI's to curved spacetime, especially in the Weyl-flat limit, finds 
significant applications in rather different sectors, from cosmology, for example in the study of the conformal back-reaction to Einstein gravity, to condensed matter physics, where anomaly actions are already playing an increasingly important role in the study of the bulk excitations of topological materials \cite{Chernodub:2021nff}. 

\vspace{1cm}
\centerline{\bf Acknowledgement}
We thank Emil Mottola, Fiorenzo Bastianelli, Paolo Benincasa, Maxim Chernodub, Pietro Colangelo, Olindo Corradini, Luigi Delle Rose, Giovanni Chirilli, Paul  McFadden, Marco Guzzi, Mirko Serino, Kostas Skenderis and Dimosthenis Theofilopoulos for sharing their insight on this and related topics on a long time frame.\\
This work has been partially supported by INFN  Iniziativa Specifica QFT-HEP.
%%%%%%%%%%%%%%%%%%%%%%%%%%%%%%%%%%%%%%%%%%%%%%%%%%%%%%%%%%%%%%%%%%
%%%%%%%%%%%%%%%%%%%%%%%%%%%%%%%%%%%%%%%%%%%%%%%%%%%%%%%%%%%%%%%%%%
%%%%%%%%%%%%%%%%%%%%%%%%%%%%%%%%%%%%%%%%%%%%%%%%%%%%%%%%%%%%%%%%%%\\
%%%%%%%%%%%%%%%%%%%%%%%%%%%%%%%%%%%%%%%%%%%%%%%%%%%%%%%%%%%%%%%%%%
%%%%%%%%%%%%%%%%%%%%%%%%%%%%%%%%%%%%%
\newpage
\appendix

\section{Definitions and conventions} 
We summarize in this appendix our most relevant definitions and conventions used in the main text and some intermediate technical steps in some important derivations. 
For the Riemann tensor we have
\begin{equation}
\label{Tensors}
{R^\lambda}_{\mu\kappa\nu}
=
-\pd_\nu \Gamma^\lambda_{\mu\kappa} + \pd_\kappa \Gamma^\lambda_{\mu\nu}
- \Gamma^\lambda_{\nu\eta}\Gamma^\eta_{\mu\kappa} + \Gamma^\lambda_{\kappa\eta}\Gamma^\eta_{\mu\nu},
\end{equation}
which is opposite to the one used in Birrell and Davies \cite{Birrell:1982ix}.
The Ricci tensor is defined by the contraction $R_{\mu\nu} = {R^{\lambda}}_{\mu\lambda\nu}$ 
and the scalar curvature by $R = g^{\mu\nu}R_{\mu\nu}$.
Weyl tensor
\be
C_{\alpha\beta\gamma\delta} = R_{\alpha\beta\gamma\delta} -
\frac{2}{d-2}( g_{\alpha\gamma} \, R_{\delta\beta} + g_{\alpha\delta} \, R_{\gamma\beta}
- g_{\beta\gamma} \, R_{\delta\alpha} - g_{\beta\delta} \, R_{\gamma\alpha} ) +
\frac{2}{(d-1)(d-2)} \, ( g_{\alpha\gamma} \, g_{\delta\beta} + g_{\alpha\delta} \, g_{\gamma\beta}) R\, ,
\ee
and its square, $F^d$, whose $d=4$ realization, called simply $F$, 
appears in the trace anomaly equation is
\be\label{Geometry1}
C^2\equiv
C^{\alpha\beta\gamma\delta}C_{\alpha\beta\gamma\delta}
=
R^{\alpha\beta\gamma\delta}R_{\alpha\beta\gamma\delta} -\frac{4}{d-2}R^{\alpha\beta}R_{\alpha\beta}+\frac{2}{(d-2)(d-1)}R^2.
\ee
The Euler density is defined as
\be
E =
R^{\alpha\beta\gamma\delta}R_{\alpha\beta\gamma\delta} - 4\,R^{\alpha\beta}R_{\alpha\beta} + R^2\, .
\ee
The functional variations with respect to the metric tensor are computed using the relations
\beqa\label{Tricks}
\delta \sqrt{-g} = -\frac{1}{2} \sqrt{-g}\, g_{\a\b}\,\delta g^{\a \b},\quad &&
\delta \sqrt{-g} = \frac{1}{2} \sqrt{-g}\, g^{\a\b}\,\delta g_{\a \b},  \nonumber \\
\delta g_{\mu\nu} = - g_{\mu\a} g_{\nu\b}\, \delta g^{\a\b} ,\quad&&
\delta g^{\mu\nu} = - g^{\mu\a} g^{\nu\b}\, \delta g_{\a\b}\,.
\eeqa
The following structure has been repeatedly used throughout the calculations
\begin{equation}
\label{Tricks2}
s^{\a\b\g\delta} \, \delta(z,x) \equiv - \frac{\d g^{\a\b}(z)}{\d g_{\g\d}(x)} =
\frac{1}{2}\left[\delta^{\a\g}\delta^{\b\delta} + \delta^{\a\delta}\delta^{\b\g}\right]\delta(z,x)\, .
\end{equation}
\section{Functional derivation of invariant integrals}
\label{FunctionalIntegral}

In this appendix we briefly show how to evaluate the functional variation of the invariant integral
$\mathcal{I}(a,b,c)$, discussed in \cite{Coriano:2012wp}
\be
\mathcal{I}(a,b,c)
\equiv
\int\,d^d x\,\sqrt{-g}\, K\,
\equiv
\int\,d^d x\,\sqrt{-g}\,
\big(a\,R^{\alpha\beta\gamma\delta}R_{abcd} + b\,R^{\alpha\beta}R_{\alpha\beta} + c\, R^2 \big)\, ,
\ee

We have
\beqa
\delta (R^{\alpha\beta\gamma\delta}R_{\alpha\beta\gamma\delta})
&=&
\delta(g_{\alpha\sigma}g^{\beta\eta}g^{\gamma\zeta}g^{\delta\rho}{R^\alpha}_{\beta\gamma\delta}{R^\sigma}_{\eta\zeta\rho}) \nn\\
&=&
\delta (g_{\alpha\sigma}g^{\beta\eta}g^{\gamma\zeta}g^{\delta\rho}){R^\alpha}_{\beta\gamma\delta}{R^\sigma}_{\eta\zeta\rho}
      + g_{\alpha\sigma}g^{\beta\eta}g^{\gamma\zeta}g^{\delta\rho}\delta
       ({R^\alpha}_{\beta\gamma\delta}{R^\sigma}_{\eta\zeta\rho}) \nn\\
&=&
\delta (g_{\alpha\sigma}g^{\beta\eta}g^{\gamma\zeta}g^{\delta\rho}){R^\alpha}_{\beta\gamma\delta}{R^\sigma}_{\eta\zeta\rho}
      + \,2\,\delta ({R^\alpha}_{\beta\gamma\delta}){R_\alpha}^{\beta\gamma\delta}\, ,
\eeqa
The variation can be written at first as
\beqa
\delta \mathcal I(a,b,c)
&=& \int\,d^dx\,\sqrt{-g}\,\bigg\{ \bigg[ \frac{1}{2}g^{\mu\nu}K - 2a\, R^{\mu\alpha\beta\gamma}{R^\nu}_{\alpha\beta\gamma}
   - 2b\,R^{\mu\alpha}{R^\nu}_\alpha - 2c\,R R^{\mu\nu}\bigg]\delta g_{\mu\nu}\nn\\
&&\hspace{20mm}
+ \, 2a\, {R_\alpha}^{\beta\gamma\delta}\delta {R^\alpha}_{\beta\gamma\delta}
+ 2b\, R^{\alpha\beta}\delta R_{\alpha\beta} + 2c\, R g^{\alpha\beta}\delta R_{\alpha\beta}\bigg\}\, .
\eeqa
using the Palatini identities
\be \label{Palatini}
\delta {R^\alpha}_{\beta\gamma\delta}
=
(\delta\Gamma^\alpha_{\beta\gamma})_{;\delta} - (\delta\Gamma^a_{\beta\delta})_{;\gamma} \quad \Rightarrow \quad
\delta R_{\beta\delta}
=
(\delta\Gamma^\lambda_{\beta\lambda})_{;\delta} - (\delta\Gamma^\lambda_{\beta\delta})_{;\lambda}\, ,
\ee
and the Bianchi identities we get
\beqa\label{Bianchi}
R_{\alpha\beta\gamma\delta;\eta} + R_{\alpha\beta\eta\gamma;\delta} + R_{\alpha\beta\delta\eta;\gamma}
&=& 0
\quad \Rightarrow \quad
R_{\beta\delta;\eta} - R_{\beta\eta;\delta} + {R^\gamma}_{\beta\delta\eta;\gamma} =  0 \nn \\
\Rightarrow \quad
R_{;\delta}
&=&
2\,{R^\alpha}_{\delta;\alpha}
\quad \Leftrightarrow \quad
\big( R^{\alpha\beta} - \frac{1}{2}g^{\alpha\beta} R \big)_{;\beta} = 0 \, .
\eeqa
After an integration by parts and a reshuffling of indices we get
\beqa\label{deltaISecond}
\delta \mathcal I(a,b,c)
&=&
\int\,d^dx\,\sqrt{-g}\,\bigg\{\bigg[ \frac{1}{2}g^{\mu\nu}K - 2\big( a\, R^{\mu\alpha\beta\gamma}{R^\nu}_{\alpha\beta\gamma}
                                     + b\,R^{\mu\alpha}{R^\nu}_\alpha + c\,R R^{\mu\nu}\big)\bigg]\delta g_{\mu\nu}\nn\\
&&\hspace{20mm}
+ \, 4a\,g_{\gamma\beta}g^{\delta\eta}(\delta\Gamma^c_{\alpha\delta})_{;\eta}
- (4a+2b)\,(\delta \Gamma^\gamma_{\alpha\beta})_{;\gamma}
+ (4c+2b)\, (\delta \Gamma^\lambda_{\alpha\lambda})_{;\beta}\bigg\}.
\eeqa
The variations of the Christoffel symbols and of their covariant derivatives
in terms of covariant derivatives of the metric tensors variations are
\beqa\label{deltaChristoffel}
\delta \Gamma^\alpha_{\beta\gamma}
&=&
\frac{1}{2}\,g^{\alpha\delta}\big[-(\delta g_{\beta\gamma})_{;\delta}
+ (\delta g_{\beta\delta})_{;\gamma} +(\delta g_{\gamma\delta})_{;\beta} \big]\, ,\nn\\
(\delta\Gamma^\alpha_{\beta\gamma})_{;\delta}
&=&
\frac{1}{2}\,g^{\alpha\eta}\big[-(\delta g_{\beta\gamma})_{;\eta;\delta} + (\delta g_{\beta\eta})_{;\gamma;\delta}
                                + (\delta g_{\gamma\eta})_{;\beta;\delta} \big]\, .
\eeqa
Now we use them to rewrite (\ref{deltaISecond}) as
\beqa\label{Bivio}
\delta \mathcal I(a,b,c)
&=&
\int\,d^dx\,\sqrt{-g}\bigg\{\bigg[ \frac{1}{2}g^{\mu\nu}K - 2\big( a\, R^{\mu\alpha\beta\gamma}{R^\nu}_{\alpha\beta\gamma}
                                  + b\,R^{\mu\alpha}{R^\nu}_\alpha + c\,R R^{\mu\nu}\big)\bigg]\delta g_{\mu\nu}\nn\\
&&\hspace{18,5mm}
+ \, \bigg[2a\,\big[-(\delta g_{\alpha\delta})_ {;\beta;\gamma} +(\delta g_{\alpha\beta})_ {;\gamma;\delta}
+ (\delta g_{\beta\delta})_ {;\alpha;\gamma} \big]\nn\\
&& \hspace{19mm}
- \, (2a + b)\,\big[- (\delta g_{\alpha\beta})_ {;\delta;\gamma} +(\delta g_{\alpha\delta})_ {;\beta;\gamma}
+ (\delta g_{\beta\delta})_ {;\alpha;\gamma} \big] + (2c + b)\, (\delta g_{\gamma\delta})_ {;\alpha;\beta}\nn\\
&& \hspace{19mm}
- \, 2c\, \big[- (\delta g_{\gamma\delta})_ {;\alpha;\beta} +(\delta g_{\alpha\delta})_ {;\gamma;\beta}
+ (\delta g_{\alpha\gamma})_ {;\delta;\beta} \big]\bigg]g^{\gamma\delta}\,R^{\alpha\beta} \bigg\}\, .
\eeqa
The presence of the factor $g^{cd}R^{ab}$ imposes two symmetry constraints on the terms in the last contribution in square brackets. By adding and subtracting  $-(4a+2b)\,(\delta g_{ac})_{;d;b}$ we obtain the expression
\beqa
\delta \mathcal I(a,b,c)
&=& \int\,d^dx\,\sqrt{-g}\,\bigg\{\bigg[\frac{1}{2}g^{\mu\nu}K
- 2\big( a\, R^{\mu\alpha\beta\gamma}{R^\nu}_{\alpha\beta\gamma}
+ b \, R^{\mu\alpha}{R^\nu}_\alpha + c\,R R^{\mu\nu}\big)\bigg]\delta g_{\mu\nu}\nn\\
&&\hspace{18,5mm}
+ \, \bigg[(4a+2b)\,\big[(\delta g_{\alpha\gamma})_ {;\beta;\delta}
- (\delta g_{\alpha\gamma})_ {;\delta;\beta}\big]+
(4a + b)(\delta g_{\alpha\beta})_ {;\gamma;\delta} + (4c + b)\, (\delta g_{\gamma\delta})_ {;\alpha;\beta}\nn\\
&& \hspace{18,5mm}
- \, (4a+2b+4c)\, (\delta g_{\alpha\gamma})_ {;\delta;\beta}\bigg]g^{\gamma\delta}\,R^{\alpha\beta} \bigg\}\, .
\label{HalfFirstFunctional}
\eeqa
The commutation of covariant derivatives allows us to write
\beqa
g^{\gamma\delta} \big[ (\delta g_{\alpha\gamma})_{;\beta;\delta}
- (\delta g_{\alpha\gamma})_{;\delta;\beta} \big]R^{\alpha\beta}
&=&
g^{\gamma\delta} \big[ -\delta g_{\alpha\sigma} {R^\sigma}_{\gamma\delta\beta}
- \delta g_{\gamma\sigma} {R^\sigma}_{\alpha\beta\delta} \big]R^{\alpha\beta}\nn\\
&=&
g^{\gamma\delta} \big[-s^{\mu\nu}_{\alpha\sigma}{R^\sigma}_{\gamma\beta\delta}
- s^{\mu\nu}_{c\sigma}{R^\sigma}_{\alpha\beta\delta}\big]R^{\alpha\beta} \,
\delta g_{\mu\nu}\nn\\
&=&
(- R^{\mu\alpha}{R^\nu}_\alpha + R^{\mu\alpha\nu\beta}R_{\alpha\beta}) \delta g_{\mu\nu}\, .
\eeqa
Inserting this back into (\ref{HalfFirstFunctional}) we get
\beqa
\delta \mathcal I(a,b,c) =
\nonumber\\
&& \hspace{-20mm}
\int\,d^dx\,\sqrt{-g}\,\bigg\{\bigg[ \frac{1}{2}g^{\mu\nu}K
- 2a\, R^{\mu\alpha\beta\gamma}{R^\nu}_{\alpha\beta\gamma} + 4a\,R^{\mu\alpha}{R^\nu}_\alpha
-(4a+2b)\, R^{\mu\alpha\nu\beta}R_{\alpha\beta} - 2c \, R R^{\mu\nu}\bigg]\delta g_{\mu\nu}\nn\\
&&
+ \, \bigg[(4a + b)(\delta g_{\alpha\beta})_ {;\gamma;\delta}
+ (4c + b)\, (\delta g_{\gamma\delta})_ {;\alpha;\beta}
- (4a+2b+4c)\, (\delta g_{\alpha\gamma})_ {;\delta;\beta}\bigg]g^{\gamma\delta}\,R^{\alpha\beta}\bigg\}\, .
\eeqa
If the coefficients are $a = c = 1$ and $b=-4$, i.e. if the integrand is the Euler density,
the last three terms are zero. \\
All that is left to do is a double integration by parts for each one of the last three terms,
to factor out $\delta g_{\mu\nu}$.
This is easily performed and the final result can be written as
\begin{align}
 \label{Magic}
\frac{\delta}{\delta g_{\mu\nu}} \mathcal I(a,b,c)
=&
\frac{\delta}{\delta g_{\mu\nu}} \int\,d^d x\,\sqrt{-g\,}
\big( a\,R^{\alpha\beta\gamma\delta}R_{\alpha\beta\gamma\delta} + b\,R^{\alpha\beta}R_{\alpha\beta}
+ c\,R^2 \big)\notag\\
=&
\sqrt{-g}\, \left[\frac{1}{2}g^{\mu\nu}K
- 2a\, R^{\mu\alpha\beta\gamma}{R^\nu}_{\alpha\beta\gamma}
+ 4a\,R^{\mu\alpha}{R^\nu}_\alpha -(4a+2b)\, R^{\mu\alpha\nu\beta}R_{\alpha\beta} - 2c \, R R^{\mu\nu}
\right.\notag\\
& \hspace{8mm}
+ \, \left. (4a + b)\,\Box{R^{\mu\nu}} + (4c + b)\,g^{\mu\nu}{R^{\alpha\beta}}_{;\alpha;\beta}
- (4a+2b+4c){{R^{\nu\beta}}_{;\beta}}^{;\mu}\right] .
\end{align}
This relation can be used to derive \eqref{nn1}. 
%---------------------------------------------------------------------------------------

\section{List of functional derivatives}
\label{Functionals}

\beqa\label{QuadraticFunctionals}
\big[\Box\,R\big]^{\alpha\beta\rho\sigma}(p,q)
&=& \big[g^{\mu\nu}(\pd_\mu\pd_\nu - \Gamma^\lambda_{\mu\nu}\pd_\lambda)R\big]^{\alpha\beta\rho\sigma}(p,q)\nn\\
&=& i^2 \,(p+q)^2\, \big[R\big]^{\alpha\beta\rho\sigma}(p,q)
- \big\{ i^2\, q^\alpha q^\beta - \delta^{\mu\nu}\big[\Gamma^\lambda_{\mu\nu}\big]^{\alpha\beta}(p)\,i\,
q_\lambda\big\}R^{\rho\sigma}(q)\nn\\
&-& \big\{ i^2\, p^\rho p^\sigma - \delta^{\mu\nu}\big[\Gamma^\lambda_{\mu\nu}\big]^{\rho\sigma}(q)\,i \,
p_\lambda\big\}R^{\alpha\beta}(p)\nn\\
&=&(p+q)^2\bigg\{-\frac{1}{2}\delta^{\alpha\beta}\big(p^\rho q^\sigma + p^\sigma q^\rho + 2\,p^\rho p^\sigma\big)
- \frac{1}{2}\delta^{\rho\sigma}\big(q^\alpha p^\beta + q^\beta q^\alpha + 2\,q^\alpha q^\beta \big)\nn\\
&+&
\frac{1}{2} p \cdot q \, \delta^{\alpha\beta}\delta^{\rho\sigma}
+\frac{1}{4}\big(p^\rho q^\beta \delta^{\alpha\sigma} + p^\rho q^\alpha \delta^{\beta\sigma}
+ p^\sigma q^\beta \delta^{\alpha\rho} + p^\sigma q^\alpha \delta^{\beta\rho}\big)\nn\\
&+&
\frac{1}{2}\bigg[\big(q^\rho p^\beta \delta^{\alpha\sigma} + q^\rho p^\alpha \delta^{\beta\sigma}
+ q^\sigma p^\beta \delta^{\alpha\rho} + q^\sigma p^\alpha \delta^{\beta\rho}\big)\nn\\
&+&
\delta^{\alpha\rho}\big(p^\beta p^\sigma + q^\beta q^\sigma \big)
 + \delta^{\alpha\sigma}\big(p^\beta p^\rho + q^\beta q^\rho \big)
 +\delta^{\beta\rho}\big(p^\alpha p^\sigma + q^\alpha q^\sigma \big)\nn\\
&+&
\delta^{\beta\sigma}\big(p^\alpha p^\rho + q^\alpha q^\rho \big)
-\big(\delta^{\alpha\sigma}\delta^{\beta\rho} + \delta^{\alpha\rho}\delta^{\beta\sigma}\big)
\big(p^2 +q^2 + \frac{3}{2}p \cdot q\big)\bigg]\bigg\}\nn\\
&+&
\frac{1}{2}\big(p^2 \delta^{\alpha\beta} - p^\alpha p^\beta\big)\big(p \cdot q \, \delta^{\rho\sigma}
- (p^\rho q^\sigma + p^\sigma q^\rho) - 2\,p^\rho p^\sigma \big)\nn\\
&+& \frac{1}{2}\big(q^2 \delta^{\rho\sigma} - q^\sigma q^\rho\big)\big(p \cdot q \, \delta^{\alpha\beta}
- (p^\alpha q^\beta + p^\beta q^\alpha) - 2\,q^\alpha q^\beta \big)\, ,
\eeqa
\beqa
\big[R_{\lambda\mu\kappa\nu}R^{\lambda\mu\kappa\nu}\big]^{\alpha\beta\rho\sigma}(p,q)
&=& 2\,\big[R_{\lambda\mu\kappa\nu}\big]^{\alpha\beta}(p)\big[R^{\lambda\mu\kappa\nu}\big]^{\rho\sigma}(q)\nn\\
&=& p \cdot q\, \big[p \cdot q \big(\delta^{\alpha\rho}\delta^{\beta\sigma} + \delta^{\alpha\sigma}\delta^{\beta\rho}\big)
- \big(\delta^{\alpha\rho}p^{\sigma}q^{\beta} + \delta^{\alpha\sigma}p^{\rho}q^{\beta}\nn\\
&+& \delta^{\beta\rho}p^{\sigma}q^{\alpha} + \delta^{\beta\sigma}p^{\rho}q^{\alpha}\big)\big]
+ 2 \, p^{\rho}p^{\sigma}q^{\alpha}q^{\beta}\, , 
\eeqa
\beqa
\big[R_{\mu\nu}R^{\mu\nu}\big]^{\alpha\beta\rho\sigma}(p,q)
&=& 2\,\big[R_{\mu\nu}\big]^{\alpha\beta}(p)\big[R^{\mu\nu}\big]^{\rho\sigma}(q)\nn\\
&=& \frac{1}{4} p \cdot q \big(\delta^{\alpha\rho}p^\beta q^\sigma + \delta^{\alpha\sigma}p^\beta q^\rho
+ \delta^{\beta\rho}p^\alpha q^\sigma + \delta^{\beta\sigma}p^\alpha q^\rho \big)\nn\\
&+&
\frac{1}{2}(p \cdot q)^2 \delta^{\alpha\beta}\delta^{\rho\sigma}
+ \frac{1}{4}p^2 q^2\big(\delta^{\alpha\rho}\delta^{\beta\sigma} + \delta^{\alpha\sigma}\delta^{\beta\rho}\big)\nn\\
&-&
\bigg[\frac{1}{4} p^2\big(q^\alpha q^\rho \delta^{\beta\sigma}+ q^\alpha q^\sigma \delta^{\beta\rho}
+ q^\beta q^\rho \delta^{\alpha\sigma}+ q^\beta q^\sigma \delta^{\alpha\rho}\big)\nn\\
&+&
\frac{1}{2}\delta^{\alpha\beta}\big( p\cdot q\,(p^\rho q^\sigma + p^\sigma q^\rho) - q^2 p^\rho p^\sigma \big)
+ (\alpha,\beta,p)\leftrightarrow(\rho,\sigma,q)\bigg]\, ,
\eeqa
\beqa
\big[R^2\big]^{\alpha\beta\rho\sigma}(p,q)
&=& 2\,\delta^{\mu\nu}\big[R_{\mu\nu}\big]^{\alpha\beta}(p)\delta^{\tau\omega}\big[R_{\tau\omega}\big]^{\rho\sigma}(q)\nn\\
&=& 2\big(p^\alpha p^\beta q^\rho q^\sigma - p^2 q^\rho q^\sigma \delta^{\alpha\beta}
- q^2 p^\alpha p^\beta \delta^{\rho\sigma} + p^2 \, q^2 \delta^{\alpha\beta}\delta^{\rho\sigma} \big)\, .
\eeqa
%%%%%%%%%%%%%%%%%%%%%%%%%%%%%%%%%%%%%%%%%%%%%%%%%%%%%%%%%%%%%%%%%%%%%%%%%%%%%%%%%%%%%%%%%%%%%
\section{Details on the BMS method. Representation of tensor structures}\label{tensstruc}
%%%%%%%%%%%%%%%%%%%%%%%%%%%%%%%%%%%%%%%%%%%%%%%%%%%%%%%%%%%%%%%%%%%%%%%%%%%%%%%%%%%%%%%%%%%%%
In this appendix we present the general method of decomposition of three-point functions involving tensorial operators. This method, presented in \cite{Bzowski:2011ab, Bzowski:2014qja, Bzowski:2017poo, Bzowski:2018fql}, is based on the reconstruction of the full three-point functions involving stress-energy tensors, currents, and scalar operators starting from the expressions of transverse and traceless part only. In order to show all the steps of the method, we will present a fully worked out example, the $\braket{T^{\m_1\n_1}T^{\m_2\n_2}\mO}$ correlation function. 
%%%%%%%%%%%%%%%%%%%%%%%%%%%%%%%%%%%%%%%%
\subsection{The example of the \texorpdfstring{$TTO$}{tto}}
%%%%%%%%%%%%%%%%%%%%%%%%%%%%%%%%%%%%%%%%
As an example consider a three-point  function of two transverse, traceless, symmetric rank-2 operators $T^{\m\n}$ and a scalar operator $\mO$.
By using the transverse and traceless projectors \eqref{Proj}, it is possible to write the most general form of the transverse and traceless part as
\begin{equation}
\braket{t^{\m_1\n_1}(p_1)\,t^{\m_2\n_2}(p_2)\,\mO(p_3)}=\Pi^{\m_1\n_1}_{\a_1\b_1}(p_1)\Pi^{\m_2\n_2}_{\a_2\b_2}(p_2)\,X^{\a_1\b_1\a_2\b_2},\label{decomp}
\end{equation}
where $X^{\a_1\b_1\a_2\b_2}$ is a general tensor of rank-4 built from the metric and the momenta. By using the conservation of the total momentum and the properties of the projectors, one ends up with the general form of our three-point  function 
\begin{equation}
\braket{t^{\m_1\n_1}(p_1)\,t^{\m_2\n_2}(p_2)\,\mO(p_3)}=\Pi^{\m_1\n_1}_{\a_1\b_1}(p_1)\Pi^{\m_2\n_2}_{\a_2\b_2}(p_2)\left[A_1\,p_2^{\a_1}p_2^{\b_1}p_3^{\a_2}p_3^{\b_2}+A_2\,p_2^{\a_1}p_3^{\a_2}\d^{\b_1\b_2}+A_3\,\d^{\a_1\a_2}\d^{\b_1\b_2}\right]\label{TTOdec},
\end{equation}
where we also use the symmetry properties of the projectors in $\m\leftrightarrow\n$, $\a\leftrightarrow\b$, and the coefficient $A_1$, $A_2$ and $A_3$ are the \emph{form factors}, scalar functions of momenta. By Lorentz invariance, these form factors are functions of the momentum magnitudes
\begin{equation}
p_j=\sqrt{p_j^2},\quad j=1,2,3,
\end{equation} 
and in \eqref{TTOdec} we suppressed the dependence of form factors on the momentum magnitudes, writing simply $A_j$ despite of $A_j(p_1,p_2,p_3)$. The symmetry condition under the exchange $(p_1,\mu_1\nu_1)\leftrightarrow(p_2,\mu_2\nu_2)$ reflects in the behavior of the form factors under this permutation and in particular they have the following symmetric properties
\begin{align}
A_i(p_1\leftrightarrow p_2)=A_i, \qquad i=1,2,3. \label{symA}
\end{align}
%%%%%%%%%%%%%%%%%%%%%%%%%%%%%%%%%%%%%%%%%%%%%%%%%%
%%%%%%%%%%%%%%%%%%%%%%%%%%%%%%%%%%%%%%%%%%%%%%%%%%
\subsection{The method of reconstruction}
%%%%%%%%%%%%%%%%%%%%%%%%%%%%%%%%%%%%%%%%%%%%%%%%%%
%%%%%%%%%%%%%%%%%%%%%%%%%%%%%%%%%%%%%%%%%%%%%%%%%%
In this appendix we explain how the BMS reconstruction method works for the entire correlation functions and in particular for the $TTO$ case. We have introduced the transverse and traceless part of $TTO$, that can be defined, in terms of operators as
\begin{equation}
t^{\m\n}(p)\equiv \Pi^{\m\n}_{\a\b}(p)\,T^{\a\b}(p),
\end{equation}
then we define the local part of $T^{\mu\nu}$ as the difference
\begin{equation}
t^{\m\n}_{loc}\equiv\,T^{\m\n}-t^{\m\n}=\,\mathcal I^{\m\n}_{\a\b}\ T^{\a\b},
\end{equation}
where
\begin{equation}
\mathcal{I}^{\m\n}_{\a\b}(p)=\sdfrac{p_\beta}{p^2}\left[2p^{(\m}\d^{\n)}_\a-\sdfrac{p_\a}{d-1}\left(\d^{\m\n}+(d-2)\sdfrac{p^\m p^\n}{p^2}\right)\right]+\sdfrac{\pi^{\m\n}(p)}{d-1}\,\d_{\a\b}.\label{itensor}
\end{equation}
This procedure can be done also for spin-$1$ conserved currents $J^\mu$ as illustrated in \cite{Bzowski:2013sza}. 
We now observe that in a CFT, all terms involving $t_{loc}$ can be computed by means of the transverse and trace Ward identities. One can therefore divide a three-point  function into two parts: the \emph{transverse-traceless} part, and the \emph{semi-local part} (indicated by subscript $loc$) expressible through the transverse Ward identities. This translates, for the case of $TTO$, into considering the decomposition
\begin{equation}
\braket{T^{\m_1\n_1}T^{\m_2\n_2}\mO}=\braket{t^{\m_1\n_1}t^{\m_2\n_2}\mO}+\braket{t_{loc}^{\m_1\n_1}T^{\m_2\n_2}\mO}+\braket{T^{\m_1\n_1}t_{loc}^{\m_2\n_2}\mO}-\braket{t_{loc}^{\m_1\n_1}t_{loc}^{\m_2\n_2}\mO}\label{reconstr},
\end{equation}
where the definition of the transverse and traceless part $\braket{t^{\mu_1\nu_1}t^{\mu_2\nu_2}\mathcal{O}}$ has been given in \eqref{TTOdec}. As previously mentioned, all terms on the right-hand side, apart from the first may be computed by means of Ward identities, but more importantly, all these terms depend on two-point  functions only. Thus, the unknown information about the three-point function is encoded just in the transverse traceless part. 
In the following sections we will obtain more information about the form factors in the tensorial decomposition using the conformal Ward identities. In particular we shall obtain constraints on the form factors in terms of differential equations that will be solved in terms of the triple-K integrals or hypergeometric functions. 
%%%%%%%%%%%%%%%%%%%%%%%%%%%%%%%%%%%%%%%%%%%%%%%%%%%%%%%%%%%%%%%%%%%%%%%%%%%%
\subsection{Transverse and trace Ward identities}\label{TraceTransverse}
%%%%%%%%%%%%%%%%%%%%%%%%%%%%%%%%%%%%%%%%%%%%%%%%%%%%%%%%%%%%%%%%%%%%%%%%%%%%
We show in \appref{transWard} how to obtain the Ward identities starting from the general requirement that the generating functional is invariant under some symmetry transformations. The relevant Ward identities in the presence of external sources are
\begin{equation}
	\begin{split}
	&\nabla_\n\braket{T^{\m\n}}+\partial^\n\phi_0\braket{\mO}=0,\\
	&g_{\m\n}\braket{T^{\m\n}}+(d-\D)\phi_0\braket{\mO}=0,
	\end{split}\label{startpoint}
\end{equation}
that correspond to the transverse and trace Ward identities respectively. Recall that $\nabla_\n$ denotes the covariant derivative with respect to the background metric $g_{\m\n}$. Multiplying these relations by $\sqrt{-g}/2$, we can rewrite them in the form
	\begin{align}
	&\partial_\m\left(\sdfrac{\d Z}{\d g_{\m\n}(x)}\right)+\G^\n_{\m\l}\sdfrac{\d Z}{\d g_{\m\l}(x)}+\sdfrac{1}{2}\partial^\n\phi_0(x)\sdfrac{\d Z}{\d \phi_0(x)}=0,\label{transverse1}\\
	&g_{\m\n}\sdfrac{\d Z}{\d g_{\m\n}(x)}+\sdfrac{1}{2}(d-\D)\phi_0(x)\sdfrac{\d Z}{\d\phi_0}=0,\label{trace}
\end{align}
where $\G$ is the usual Christoffel connection where we have introduced the  expectation value of the energy momentum tensor $(\langle T \rangle)$. In order to derive the transverse and trace Ward identities for the $TT\mO$ correlator, we vary twice \eqref{transverse1} and \eqref{trace} with respect to the metric $g_{\m_2\n_2}(x_2)$ and the scalar field source $\phi_0(x_3)$. After the functional variations, if one switches off the sources, the result will be 
\begin{align}
&\partial_\m\left.\left(\sdfrac{\d^3 Z}{\d g_{\m\n}(x)\d g_{\m_2\n_2}(x_2)\d\phi_0(x_3)}\right)\right|_{g_{\m\n}=\d_{\m\n}}+\sdfrac{1}{2}\big(\partial^\n\d(x-x_3)\big)\left.\left(\sdfrac{\d^2 Z}{\d \phi_0(x)\d g_{\m_2\n_2}(x_2)}\right)\right|_{g_{\m\n}=\d_{\m\n}}\notag\\
&\hspace{2cm}+\left.\left[\big(\,\d^{\n(\n_2}\d^{\m_2)}_\l\partial_\m-\sdfrac{1}{2}\d_\m^{(\m_2}\d^{\n_2)}_\l\d^{\n\e}\partial_\e\big)\d(x-x_2)\right]\left(\sdfrac{\d^2 Z}{\d g_{\m\l}(x)\d\phi_0(x_3)}\right)\right|_{g_{\m\n}=\d_{\m\n}}=0,
\end{align}
and restoring the correlation function through the definitions
	\begin{align}
	\braket{T^{\m\n}(x)T^{\m_2\n_2}(x_2)\,\mO(x_3)}&=4\left.\left(\sdfrac{\d^3 Z}{\d g_{\m\n}(x)\d g_{\m_2\n_2}(x_2)\d\phi_0(x_3)}\right)\right|_{g_{\m\n}=\d_{\m\n}},\label{defTTO}\\
	\braket{T^{\m\n}(x)\,\mO(x_3)}&=2\left.\left(\sdfrac{\d^2 Z}{\d \phi_0(x_3)\d g_{\m\n}(x_2)}\right)\right|_{g_{\m\n}=\d_{\m\n}},
	\end{align}
we obtain the transverse Ward identities in position space
\begin{align}
\partial_\m\braket{T^{\m\n}(x)T^{\m_2\n_2}(x_2)\,\mO(x_3)}&=-2\,\d^{\n(\n_2}\d^{\m_2)}_\l\partial_\m\d(x-x_2)\braket{T^{\m\l}(x)\,\mO(x_3)}\notag\\
&+\partial^\n\d(x-x_2)\braket{T^{\m_2\n_2}(x)\,\mO(x_3)}-\partial^\n\d(x-x_3)\braket{T^{\m_2\n_2}(x_2)\,\mO(x)}\label{transverse2}.
\end{align}

Using the same procedure we obtain the trace Ward identities in position space for the same correlation function
\begin{align}
g_{\m\n}(x)\braket{T^{\m\n}(x)T^{\m_2\n_2}(x_2)\mO(x_3)}=(\D-d)\d(x-x_3)\braket{T^{\m_2\n_2}(x_2)\mO(x)}-2\d(x-x_2)\braket{T^{\m_2\n_2}(x)\mO(x_3)}.\label{trace2}
\end{align}

Multiplying \eqref{transverse2} and \eqref{trace2} with $\exp(ip_1x+ip_2x_2+ip_3x_3)$ and integrating over all $x$, $x_2$ and $x_3$, we obtain the Ward identities in momentum space
	\begin{align}
	p_{1\m_1}\braket{T^{\m_1\n_1}(p_1)T^{\m_2\n_2}(p_2)\mO(p_3)}&=-2p_2^{\n_1}\braket{T^{\m_2\n_2}(p_1+p_2)\mO(p_3)}+2p_3^{\n_1}\braket{T^{\m_2\n_2}(p_2)\mO(p_1+p_3)}\notag\\
	&\hspace{-0.5cm}+2\d^{\n_1\n_2}\,p_{2\m_1}\braket{T^{\m_1\m_2}(p_1+p_2)\mO(p_3)}+2\d^{\n_1\m_2}\,p_{2\m_1}\braket{T^{\m_1\n_2}(p_1+p_2)\mO(p_3)},\\[2ex]
	g_{\m_1\n_1}\braket{T^{\m_1\n_1}(p_1)T^{\m_2\n_2}(p_2)\mO(p_3)}&=2(d-\D)\braket{T^{\m_2\n_2}(p_2)\mO(p_1+p_3)}+4\braket{T^{\m_2\n_2}(p_1+p_2)\mO(p_3)},
	\end{align}
where $\D$ is the conformal dimension of the scalar operator $\mO$. 

Notice that any conformal two-point function involving two operators of different spin or different scale dimensions is zero. For this reason the transverse and trace Ward identities can be expressed in the final form
\begin{equation}
	\begin{split}
	p_{1\m_1}\braket{T^{\m_1\n_1}(p_1)T^{\m_2\n_2}(p_2)\mO(p_3)}&=0,\\
	g_{\m_1\n_1}\braket{T^{\m_1\n_1}(p_1)T^{\m_2\n_2}(p_2)\mO(p_3)}&=0.
	\end{split}\label{finWardTTO}
\end{equation}

In this particular case, the definition of the three-point  function $\braket{TT\mO}$ in \eqref{defTTO} brings to a correlation function that is already transverse-traceless and does not contain longitudinal contributions. 
%%%%%%%%%%%%%%%%%%%%%%%%%%
\subsection{Dilatation Ward identities}
%%%%%%%%%%%%%%%%%%%%%%%%%%
It is simple to rewrite the dilatation Ward identities for the three-point  function $\braket{T^{\m_1\n_1}T^{\m_2\n_2}\mO}$ as
\begin{equation}
\left[\sum_{j=1}^3\D_j-2d-\sum_{j=1}^{2}p_j^\a\sdfrac{\partial}{\partial p_j^\a}\right]\braket{T^{\m_1\n_1}(p_2)T^{\m_2\n_2}(p_2)\mO(p_3)}=0.\label{dilorig}
\end{equation}
We can consider the decomposition of the correlation function in terms of the transverse and traceless part and the semi-local part for which the previous equation can be read as
\begin{align}
&\left[\sum_{j=1}^3\D_j-2d-\sum_{j=1}^{2}p_j^\a\sdfrac{\partial}{\partial p_j^\a}\right]\bigg\{\braket{t^{\m_1\n_1}(p_2)t^{\m_2\n_2}(p_2)\mO(p_3)}+\braket{t_{loc}^{\m_1\n_1}(p_2)t^{\m_2\n_2}(p_2)\mO(p_3)}\notag\\
&\hspace{4cm}+\braket{t^{\m_1\n_1}(p_2)t_{loc}^{\m_2\n_2}(p_2)\mO(p_3)}+\braket{t_{loc}^{\m_1\n_1}(p_2)t_{loc}^{\m_2\n_2}(p_2)\mO(p_3)}\bigg\}=0.\label{dildec}
\end{align}

We are free to apply transverse-traceless projectors \eqref{Proj} to \eqref{dildec}, in order to isolate equations for the form factors appearing in the decomposition of $\braket{t^{\m_1\n_1}t^{\m_2\n_2}\mO}$. Evaluating the action of the differential operator in \eqref{dildec} on the semi-local terms via the formulae in \appref{appendixB}, we find 
\begin{align}
\Pi^{\r_1\s_1}_{\m_1\n_1}(p_1)\,\left(p_1^\l\sdfrac{\partial}{\partial p_1^\l}\,\mathcal{I}^{\m_1\n_1}_{\ \ \ \a_1}p_{1\,\b_1}\right)&=0,\\
\Pi^{\m\n}_{\a\b}(p)\,\mathcal{I}^{\a\b}_{\r\s}(p)&=0,
\end{align}
where the $\mathcal{I}$ tensor is defined in \eqref{itensor} and is used to define the semi-local part of the correlation function. This expression implies that any correlation function with one and more than one insertion of $t_{loc}$ vanish when the dilatation operator and the projectors are applied. For this reason \eqref{dildec} can be easily written as
\begin{equation}
\Pi^{\r_1\s_1}_{\m_1\n_1}(p_1)\,\Pi^{\r_2\s_2}_{\m_2\n_2}(p_2)\,\left[\sum_{j=1}^3\D_j-2d-\sum_{j=1}^{2}p_j^\a\sdfrac{\partial}{\partial p_j^\a}\right]\braket{t^{\m_1\n_1}(p_2)t^{\m_2\n_2}(p_2)\mO(p_3)}=0.\label{dilfin}
\end{equation}
In order to write down the equations of form factors we substitute in \eqref{dilfin} the decomposition of the three-point  function \eqref{TTOdec} 
\begin{align}
&\Pi^{\r_1\s_1}_{\m_1\n_1}(p_1)\,\Pi^{\r_2\s_2}_{\m_2\n_2}(p_2)\,\left[\sum_{j=1}^3\D_j-2d-\sum_{j=1}^{2}p_j^\a\sdfrac{\partial}{\partial p_j^\a}\right]\notag\\
&\hspace{2.5cm}\times\bigg\{\Pi^{\m_1\n_1}_{\a_1\b_1}(p_1)\Pi^{\m_2\n_2}_{\a_2\b_2}(p_2)\left[A_1\,p_2^{\a_1}p_2^{\b_1}p_3^{\a_2}p_3^{\b_2}+A_2\,p_2^{\a_1}p_3^{\a_2}\d^{\b_1\b_2}+A_3\,\d^{\a_1\a_2}\d^{\b_1\b_2}\right]\bigg\}=0,
\end{align}
and using the relations in \appref{appendixB} we find
\begin{equation}
\Pi^{\r_1\s_1}_{\a_1\b_1}(p_1)\Pi^{\r_2\s_2}_{\a_2\b_2}(p_2)\left[\sum_{j=1}^3\D_j-2d-\sum_{j=1}^{2}p_j^\a\sdfrac{\partial}{\partial p_j^\a}\right]\left[A_1\,p_2^{\a_1}p_2^{\b_1}p_3^{\a_2}p_3^{\b_2}+A_2\,p_2^{\a_1}p_3^{\a_2}\d^{\b_1\b_2}+A_3\,\d^{\a_1\a_2}\d^{\b_1\b_2}\right]=0.\label{dilfin2}
\end{equation}
We can act with the differential operator in \eqref{dilfin2} observing that there is no change in the independent tensor structures.  The result of the action of the differential operator will be that the decomposition structure will not be altered except in the presence of coefficients involving derivatives of the form factors. Thus, it is possible to obtain a set of equations for all the form factors from the vanishing of the coefficients of the independent tensor structures in \eqref{dilfin2}. In these equations there will be terms like
\begin{equation}
\sum_{j=1}^2\,p_j^\a \sdfrac{\partial}{\partial p_j^\a}\,A_n(p_1,p_2,p_3),\qquad n=1,2,3\label{terms},
\end{equation}
but the form factors are purely functions of the momenta magnitude by the Lorentz invariance. The action of momentum derivatives on form factors may be obtained using the chain rules,
\begin{equation}
	\begin{split}
	\sdfrac{\partial}{\partial p_{1\m}}&=\sdfrac{\partial p_1}{\partial p_{1\m}}\sdfrac{\partial}{\partial p_1}+\sdfrac{\partial p_2}{\partial p_{1\m}}\sdfrac{\partial}{\partial p_2}+\sdfrac{\partial p_3}{\partial p_{1\m}}\sdfrac{\partial}{\partial p_3}=\sdfrac{p_1^\m}{p_1}\sdfrac{\partial}{\partial p_1}+\sdfrac{p_1^\m+p_2^\m}{p_3}\sdfrac{\partial}{\partial p_3},\\
	\sdfrac{\partial}{\partial p_{2\m}}&=\sdfrac{p_2^\m}{p_2}\sdfrac{\partial}{\partial p_2}+\sdfrac{p_1^\m+p_2^\m}{p_3}\sdfrac{\partial}{\partial p_3},
	\end{split}\label{chainrule}
\end{equation}
noting that $p_3$ is fixed by the conservation of the total momentum $p^\m_3=-p^\m_1-p^\m_2$. Using these relations we may re-expressed \eqref{terms} purely in terms of the momentum magnitudes 
\begin{equation}
\sum_{j=1}^2\,p_j^\a \sdfrac{\partial}{\partial p_j^\a}\,A_n(p_1,p_2,p_3)=\sum_{j=1}^3\,p_j \sdfrac{\partial}{\partial p_j}\,A_n(p_1,p_2,p_3),\qquad n=1,2,3.
\end{equation}

Therefore it is possible to rewrite the dilatation Ward identity \eqref{dilorig} for a three-point function of three conformal primary operator of any tensor structure in terms of its form factors as
\begin{equation}
\left[2d+N_n-\sum_{i=1}^3\D_i+\sum_{i=1}^{3}\,p_i\sdfrac{\partial}{\partial p_i}\right]\,A_n(p_1,p_2,p_3)=0,\label{dilfin3}
\end{equation}
{where $N_n$ is the number of momenta that $A_n$ multiply in the decomposition \eqref{TTOdec}, and as previously $\D_j$ denote the conformal dimensions of the operator in the three-point  function: in this case $\D_{1,2}=d$ for a stress-energy tensor and $\D_3$ depending on the particular scalar operator chosen. From \eqref{dilfin3} it is clear that the form factor $A_n$ has scaling degree
\begin{equation}
\deg(A_n)=\D_t-2d-N_n,\label{deg}
\end{equation}
where $\D_t=\D_1+\D_2+\D_3$, as also pointed out in \cite{Bzowski:2013sza}. 
%%%%%%%%%%%%%%%%%%%%%%%%%%%%%%%%%%%%%%%%%%%%%%%%%%%%%%%%%%%%%%%%%%%%%
\subsection{Special conformal Ward identities}\label{specialconfward}
%%%%%%%%%%%%%%%%%%%%%%%%%%%%%%%%%%%%%%%%%%%%%%%%%%%%%%%%%%%%%%%%%%%%%
We now extract scalar equations for the form factors in the same way of the previous section but use the special conformal Ward identities (SCWI's). Considering the SCWI's for the three-point  function $\braket{TT\mO}$  we obtain
\begin{align}
0&=\sum_{j=1}^{2}\left[2(\D_j-d)\sdfrac{\partial}{\partial p_j^\k}-2p_j^\a\sdfrac{\partial}{\partial p_j^\a}\sdfrac{\partial}{\partial p_j^\k}+(p_j)_\k\sdfrac{\partial}{\partial p_j^\a}\sdfrac{\partial}{\partial p_{j\a}}\right]\braket{\,T^{\m_1\n_1}(p_1)T^{\m_2\n_2}(p_2)\mO(p_3)}\notag\\
&\hspace{0.5cm}+4\left(\d^{\k(\m_1}\sdfrac{\partial}{\partial p_1^{\a_1}}-\d^\k_{\a_1}\d^{\l(\m_1}\sdfrac{\partial}{\partial p_1^\l}\right)\braket{\,T^{\n_1)\a_1}(p_1)T^{\m_2\n_2}(p_2)\mO(p_3)}\notag\\
&\hspace{0.5cm}+4\left(\d^{\k(\m_2}\sdfrac{\partial}{\partial p_2^{\a_2}}-\d^\k_{\a_2}\d^{\l(\m_2}\sdfrac{\partial}{\partial p_2^\l}\right)\braket{\,T^{\n_2)\a_2}(p_2)T^{\m_1\n_1}(p_1)\mO(p_3)}\equiv \mathcal{K}^\k\braket{\,T^{\m_1\n_1}(p_1)T^{\m_2\n_2}(p_2)\mO(p_3)},
\end{align}
where we have defined the $\mathcal K^\k$ operator for simplicity. As previously we can consider the decomposition of the three-point  function to obtain
\begin{align}
0&=\mathcal{K}^\k\braket{\,T^{\m_1\n_1}(p_1)T^{\m_2\n_2}(p_2)\mO(p_3)}\notag\\
&=\mathcal{K}^\k\braket{\,t^{\m_1\n_1}t^{\m_2\n_2}\mO}+\mathcal{K}^\k\braket{\,t_{loc}^{\m_1\n_1}t^{\m_2\n_2}\mO}+\mathcal{K}^\k\braket{\,t^{\m_1\n_1}t_{loc}^{\m_2\n_2}\mO}+\mathcal{K}^\k\braket{\,t_{loc}^{\m_1\n_1}t_{loc}^{\m_2\n_2}\mO}.\label{specConf}
\end{align}
In order to isolate equations for the form factors appearing in the decomposition, we are free to apply transverse-traceless projectors. Using the properties in \appref{appendixB} and through a direct calculation we find that
	\begin{align}
	\Pi^{\r_1\s_1}_{\m_1\n_1}(p_1)\,\Pi^{\r_2\s_2}_{\m_2\n_2}(p_2)\,\mathcal{K}^\k\braket{\,t_{loc}^{\m_1\n_1}t^{\m_2\n_2}\mO}&=\sdfrac{4d}{p_1^2}\,\Pi^{\r_1\s_1\k}_{\hspace{0.7cm}\m_1}(p_1)\,\bigg(p_{1\n_1}\braket{T^{\m_1\n_1}(p_1)t^{\r_2\s_2}(p_2)\mO(p_3)}\bigg),\\
	\Pi^{\r_1\s_1}_{\m_1\n_1}(p_1)\,\Pi^{\r_2\s_2}_{\m_2\n_2}(p_2)\,\mathcal{K}^\k\braket{\,t^{\m_1\n_1}t_{loc}^{\m_2\n_2}\mO}&=\sdfrac{4d}{p_2^2}\,\Pi^{\r_2\s_2\k}_{\hspace{0.7cm}\m_2}(p_2)\,\bigg(p_{2\n_2}\braket{t^{\r_1\s_1}(p_1)T^{\m_2\n_2}(p_2)\mO(p_3)}\bigg),\\
	\Pi^{\r_1\s_1}_{\m_1\n_1}(p_1)\,\Pi^{\r_2\s_2}_{\m_2\n_2}(p_2)\,\mathcal{K}^\k\braket{\,t_{loc}^{\m_1\n_1}t_{loc}^{\m_2\n_2}\mO}&=0,
	\end{align}
and \eqref{specConf} takes the form
\begin{align}
0&=\Pi^{\r_1\s_1}_{\m_1\n_1}(p_1)\,\Pi^{\r_2\s_2}_{\m_2\n_2}(p_2)\,\bigg\{\mathcal K^\k\braket{\,t^{\m_1\n_1}(p_1)t^{\m_2\n_2}(p_2)\mO(p_3)}\notag\\
&\hspace{1cm}+\sdfrac{4d}{p_1^2}\,\d^{\k\m_1}\,p_{1\r_1}\braket{T^{\n_1\r_1}(p_1)T^{\m_2\n_2}(p_2)\mO(p_3)}+\sdfrac{4d}{p_2^2}\,\d^{\k\m_2}\,p_{2\r_2}\braket{T^{\m_1\n_1}(p_1)T^{\m_2\r_2}(p_2)\mO(p_3)}\ \bigg\}.\label{specC}
\end{align}
The last two terms may be re-expressed in terms of two-point  functions via the transverse Ward identities, but by using \eqref{finWardTTO} these terms vanish. The remaining task is to rewrite the transverse and traceless component in terms of form factors and extract the conformal Ward identities for this particular scalar function of the magnitudes of the momenta.\\
By a direct calculation we find that the first term of \eqref{specC}, $\mathcal K^\k\braket{\,t\,t\,\mO}$, is transverse and traceless in the covariant indices with respect to the corresponding momenta, i.e.
\begin{equation}
\begin{split}
\d_{\m_1\n_1}[\mathcal K^\k\braket{\,t^{\m_1\n_1}(p_1)t^{\m_2\n_2}(p_2)\mO(p_3)}]=0,&\qquad\d_{\m_2\n_2}[\mathcal K^\k\braket{\,t^{\m_1\n_1}(p_1)t^{\m_2\n_2}(p_2)\mO(p_3)}]=0,\\
p_{1\m_1}[\mathcal K^\k\braket{\,t^{\m_1\n_1}(p_1)t^{\m_2\n_2}(p_2)\mO(p_3)}]=0,&\qquad p_{\m_2}[\mathcal K^\k\braket{\,t^{\m_1\n_1}(p_1)t^{\m_2\n_2}(p_2)\mO(p_3)}]=0.
\end{split}
\end{equation}
Using this result, and following the discussion in \appref{tensstruc}, we can write the most general form of $\mathcal K^\k\braket{\,t\,t\,\mO}$ as
\begin{align}
&\Pi^{\r_1\s_1}_{\m_1\n_1}(p_1)\,\Pi^{\r_2\s_2}_{\m_2\n_2}(p_2)\,\mathcal K^\k\braket{\,t^{\m_1\n_1}(p_1)t^{\m_2\n_2}(p_2)\mO(p_3)}\notag\\
&=\Pi^{\r_1\s_1}_{\m_1\n_1}(p_1)\,\Pi^{\r_2\s_2}_{\m_2\n_2}(p_2)\,\bigg[\,p_1^\k\left(C_{11}\,p_2^{\m_1}p_2^{\n_1}p_3^{\m_2}p_3^{\n_2}+C_{12}\,p_2^{\m_1}p_3^{\m_2}\d^{\n_1\n_2}+C_{13}\,\d^{\m_1\m_2}\d^{\n_1\n_2}\right)\notag\\
&\hspace{1.3cm}+p_2^\k\left(C_{21}\,p_2^{\m_1}p_2^{\n_1}p_3^{\m_2}p_3^{\n_2}+C_{22}\,p_2^{\m_1}p_3^{\m_2}\d^{\n_1\n_2}+C_{23}\,\d^{\m_1\m_2}\d^{\n_1\n_2}\right)\notag\\
&\hspace{1.3cm}+\d^{\k\m_1}\left(C_{31}p_2^{\n_1}p_3^{\m_2}p_3^{\n_2}+C_{32}\d^{\m_2\n_1}p_3^{\n_2}\right)+\d^{\k\m_2}\left(C_{41}p_2^{\n_1}p_2^{\m_1}p_3^{\n_2}+C_{42}\d^{\m_1\n_2}p_2^{\n_1}\right)\bigg]\label{CWI},
\end{align}
where now $C_{ij}$ are scalar differential equations involving the form factors $A_i$, $i=1,2,3$, written in terms of the momentum magnitudes $p_j$.
The coefficients $C_{jk}$ in \eqref{CWI} are not all independent, indeed, $C_{1j}$ and $C_{2j}$, as well as $C_{3j}$ and $C_{4j}$, are pairwise equivalent, due to the symmetry under the exchange of the stress energy tensors in the correlator. The difference in the coefficients is manifest in the order of the corresponding differential equations. In all the cases of three-point function, the coefficients multiplying the momentum $p_i^\kappa$ ($\k$ is the special index related to the conformal operator $\mathcal{K}^\k$) are second order partial differential equations, and they are called \emph{primary} conformal Ward identities (CWI's). The coefficients of the structure $\delta^{\kappa\,\mu_i}$ are always first order partial differential equations, called \emph{secondary} CWI's in \cite{Bzowski:2013sza}. 
We observe that the other terms in \eqref{specC}, differently from \eqref{CWI}, involve the $\delta^{\kappa\,\mu_i}$ structure, then they will contribute to the secondary CWI's. Furthermore, the primary CWI's are equivalent to the vanishing of the coefficients $C_{1j}$ and $C_{2j}$. 
%%%%%%%%%%%%%%%%%%%%%%%%%%%%%%%
\subsection{Primary conformal Ward identities}
%%%%%%%%%%%%%%%%%%%%%%%%%%%%%%%
In order to write the primary CWI's in a simpler form, we have to rearrange the expression of $C_{jk}$ using the dilatation Ward identities and some differential identities. We will show the explicit procedure for the first coefficient $C_{11}$ and the other coefficients follow the same prescription. In an explicit form the coefficient $C_{11}$ is expressed as
\begin{equation}
C_{11}= - \frac{2}{p_3} \left[ p_1 \frac{\partial^2}{\partial p_1 \partial p_3 } + p_2 \frac{\partial^2}{\partial p_2 \partial p_3 } \right] A_1 + \frac{d-1}{p_1} \frac{\partial}{\partial p_1} A_1 - \frac{\partial^2}{\partial p_1^2} A_1 + \frac{d-9}{p_3} \frac{\partial}{\partial p_3} A_1 - \frac{\partial^2}{\partial p_3^2} A_1\,.\label{C11}
\end{equation}
Taking the dilatation Ward identities \eqref{dilfin3} for the $A_1$, and deriving it with respect to the magnitude of the momentum $p_3$ we obtain
\begin{equation}
\frac{\partial}{\partial p_3}\left(\sum_{i=1}^3\,p_i\,\frac{\partial}{\partial\,p_i}\, A_n\right) =\,\deg(A_1)\,\frac{\partial\,A_1}{\partial p_3},
\end{equation}
where the degree of the form factor is defined in \eqref{deg} and for $A_1$ is $\deg(A_1)=\Delta_t-2d-4=\Delta_3-4$. Then, after some simplification, we get
\begin{equation}
\left[  p_1 \frac{\partial^2}{\partial p_1 \partial p_3} +  p_2 \frac{\partial^2}{\partial p_3 \partial p_2}  +  p_3 \frac{\partial^2}{\partial p_3 \partial p_3} \right] A_n =\,\bigg(\deg(A_1)-1\bigg)\,\frac{\partial\,A_1}{\partial p_3}\,.
\label{eq:dilWI}
\end{equation}
By using \eqref{eq:dilWI}, we can re-expressed the first term in \eqref{C11} as
\begin{equation}
-\frac{2}{p_3}  \left[ p_1 \frac{\partial^2}{\partial p_1 \partial p_3} +  p_2 \frac{\partial^2}{\partial p_3 \partial p_2}  \right]  A_1 = \frac{\left( 2 - 2 \deg(A_1) \right)}{p_3} \frac{\partial}{\partial p_3} A_1 +2  \frac{\partial^2}{\partial p_3^2} A_1\,,
\end{equation}
and inserting this result into \eqref{C11} we simplify the form of the differential equation $C_{11}$ as
\begin{equation}
C_{11}= \left[ -\frac{\partial^2}{\partial p_1^2} + \frac{d-1}{p_1} \frac{\partial}{\partial p_1} \right] A_1 + \left[ \frac{\partial^2 }{\partial p_3^2}  + \frac{d+1-2\Delta_3}{p_3} \frac{\partial}{\partial p_3} \right] A_1 \,.  
\label{eq:C11quasi}  
\end{equation}
In order to write the primary CWI's in a simple way, we define the following fundamental differential operators
\begin{equation}
	\begin{split}
	\textup{K}_i &= \frac{\partial^2}{\partial p_i^2} + \frac{d+1-2\Delta_i}{p_i} \frac{\partial}{\partial p_i},  \qquad i=1,2,3\, ,   \\ 
	\textup{K}_{ij} &= \textup{K}_i - \textup{K}_j \,,
	\end{split}\label{Koper}
\end{equation}
where $\D_j$ is the conformal dimension of the j-th operator in the three-point  function under consideration. Through this definition the $C_{11}$ is re-expressed as
\begin{equation}
C_{11} = (\textup{K}_3 - \textup{K}_1 ) A_1 = \textup{K}_{31}A_1 \,.
\end{equation}
The procedure presented above allows to obtain a simple second-order differential equations and it can be applied in the same way to all the $C_{1j}$'s, $j=1,2,3$. 
Usually, while performing the lengthy computations, one may encounter the term
\begin{equation}
-\frac{2}{p_3}  \left[ p_1 \frac{\partial^2}{\partial p_1 \partial p_3} +  p_2 \frac{\partial^2}{\partial p_3 \partial p_2}  \right]  A_n = \frac{\left( 2 - 2 \deg(A_n) \right)}{p_3} \frac{\partial}{\partial p_3} A_n +2  \frac{\partial^2}{\partial p_3^2} A_n \,,
\end{equation}
where also in this case $\textup{deg}(A_n)= \Delta_1 + \Delta_2+ \Delta_3 -2 d - N_n \,$.
Here, $N_n$ represents the tensorial dimension of $A_n$, i.e. the number of momenta multiplying the form factors $A_n$ and the projectors $\Pi$. 

In the $\braket{TT\mO}$ case, for instance, $\Delta_1=\Delta_2=d$, whereas $\Delta_3$ remains implicit because of the unknown nature of the generic scalar operator $\mathcal{O}(p_3)$ (e.g. if $\mathcal{O}=\phi^2$ one has $\Delta_3=d-2$). 
In this case we have
\begin{equation}
-\frac{2}{p_3}  \left[ p_1 \frac{\partial^2}{\partial p_1 \partial p_3} +  p_2 \frac{\partial^2}{\partial p_3 \partial p_2}  \right]  A_n = \frac{\left( 2 - 2 (\Delta_3 - N_n)\right)}{p_3} \frac{\partial}{\partial p_3} A_n +2  \frac{\partial^2}{\partial p_3^2} A_n \,,
\end{equation}
where
\begin{align} 
N_1&=4    \qquad\textup{for} \qquad   A_1(p_1,p_2,p_3)\,\, p_2^{\alpha_1} p_2^{\beta_1} p_3^{\alpha_2} p_3^{\beta_2}, \\
N_2&=2 \qquad\textup{for} \qquad A_2(p_1,p_2,p_3)\,\, p_2^{\alpha_1} p_3^{\alpha_2} \delta^{\beta_1\beta_2},  \\
N_3&=0    \qquad \textup{for}  \qquad  A_3(p_1,p_2,p_3)\,\, \delta^{\alpha_1\alpha_2} \delta^{\beta_1\beta_2}. 
\end{align}
In this way we can simplify all the primary coefficients $C_{i,j}$, $i=1,2$ and $j=1,2,3$.
The primary CWI's are obtained, as previously discussed, when the coefficients $C_{1j}$ and $C_{2j}$ are equal to zero. One obtains
\begin{equation}
\begin{matrix}
K_{31}\,A_1=0,&\qquad K_{13}A_2=8A_1,&\qquad K_{13}A_3=2A_2,\\[1.3ex]
K_{23}A_1=0,&\qquad K_{23}A_2=8A_1,&\qquad K_{23}A_3=2A_2.
\end{matrix}
\end{equation} 
Note that, from the definition \eqref{Koper}, we have
\begin{equation}
K_{ii}=0,\qquad K_{ji}=-K_{ij},\qquad K_{ij}+K_{jk}=K_{ik},
\end{equation}
for any $i,j,k\,\in\{1,2,3\}$. One can therefore subtract corresponding pairs of equations and obtain the following system of independent partial differential equations
\begin{equation}
\begin{matrix}
K_{13}\,A_1=0,&\qquad K_{13}A_2=8A_1,&\qquad K_{13}A_3=2A_2,\\[1.3ex]
K_{12}A_1=0,&\qquad K_{12}A_2=0,&\qquad K_{12}A_3=0.
\end{matrix}\label{primaryCWI}
\end{equation} 
%%%%%%%%%%%%%%%%%%%%%%%%%%%%%%%%%%%
\subsection{Secondary conformal Ward identities}
%%%%%%%%%%%%%%%%%%%%%%%%%%%%%%%%%%%
As previously mentioned, the secondary conformal Ward identities are first-order partial differential equations. In order to write them compactly, we define the two differential operators
\begin{align}
\textup{L}_N&= p_1(p_1^2 + p_2^2 - p_3^2) \frac{\partial}{\partial p_1} + 2 p_1^2 p_2 \frac{\partial}{\partial p_2} + \left[ (2d - \Delta_1 - 2\Delta_2 +N)p_1^2 + (2\Delta_1-d)(p_3^2-p_2^2)  \right]\label{Ldef} ,\\
\textup{R} &= p_1 \frac{\partial}{\partial p_1} - (2\Delta_1-d) \label{Rdef},\, 
\end{align}
as well as their symmetric versions
\begin{align}
&L'_N=L_N,\quad\text{with}\ p_1\leftrightarrow p_2\ \text{and}\ \D_1\leftrightarrow\D_2,\\
&R'=R,\qquad\text{with}\ p_1\mapsto p_2\ \text{and}\ \D_1\mapsto\D_2.
\end{align}
In the $\braket{TT\mO}$ case one finds for the coefficients $C_{31}$ and $C_{32}$ the results
\begin{align}
C_{31}&= \left[ - \frac{1}{p_1} \left( p_1^2 + p_2^2 - p_3^2 \right) \frac{\partial}{\partial p_1} - 2 p_2 \frac{\partial}{\partial p_2}  + \frac{d}{p_1^2} (p_2^2-p_3^2) + (d-2)  \right]A_1 + \left[ \frac{d}{p_1^2} - \frac{1}{p_1} \frac{\partial}{\partial p_1}   \right] A_2   \nonumber \\&= -\frac{1}{p_1^2}\left( \textup{L}_2\, A_1 + \textup{R}\, A_2 \right), \\[1ex] 
C_{32}&= \left[  \frac{1}{4 p_1} \left( p_1^2 + p_2^2 - p_3^2 \right) \frac{\partial}{\partial p_1} + \frac{1}{2} p_2 \frac{\partial}{\partial p_2}  + \frac{d}{4 p_1^2} (p_3^2-p_2^2) + \frac{(2-d)}{4}  \right]A_2 + \left[\frac{1}{p_1} \frac{\partial}{\partial p_1}   - \frac{d}{p_1^2}  \right] A_3   \nonumber \\&= \frac{1}{4 p_1^2}\left( \textup{L}_2\, A_2 + 4 \textup{R}\, A_3 \right),
\end{align}
and for the last two coefficient $C_{41}$ and $C_{42}$
\begin{align}
C_{41}&= \left[ - \frac{1}{p_2} \left( p_1^2 + p_2^2 - p_3^2 \right) \frac{\partial}{\partial p_2} - 2 p_1 \frac{\partial}{\partial p_1}  + \frac{d}{p_1^2} (p_1^2-p_3^2) + (d-2)  \right]A_1 + \left[ \frac{d}{p_2^2} - \frac{1}{p_2} \frac{\partial}{\partial p_2}   \right] A_2   \nonumber \\&= -\frac{1}{p_2^2}\left( \textup{L'}_2\, A_1 + \textup{R'}\, A_2 \right), \\[1ex] 
C_{42}&= \left[  \frac{1}{4 p_2} \left( p_1^2 + p_2^2 - p_3^2 \right) \frac{\partial}{\partial p_2} + \frac{1}{2} p_1 \frac{\partial}{\partial p_1}  + \frac{d}{4 p_2^2} (p_3^2-p_1^2) + \frac{(2-d)}{4}  \right]A_2 + \left[\frac{1}{p_2} \frac{\partial}{\partial p_2}   - \frac{d}{p_2^2}  \right] A_3   \nonumber \\&= \frac{1}{4 p_2^2}\left( \textup{L'}_2\, A_2 + 4 \textup{R'}\, A_3 \right) \,.
\end{align}
	The secondary CWI's are equivalent to the vanishing of these coefficient because in \eqref{specC} the only term that is not zero is the transverse traceless part. This can be seen  by using the transverse and trace Ward identities \eqref{finWardTTO}, and one finds two independent secondary CWI's, namely
\begin{equation}
	\begin{split}
	\textup{L}_2\, A_1 + \textup{R}\, A_2&=0,\\
	\textup{L}_2\, A_2 + 4 \textup{R}\, A_3&=0.
	\end{split}\label{secondary}
\end{equation}
In fact, due to the symmetry properties of the form factor in \eqref{symA}, one realizes that the coefficients $C_{4i}$ are equivalent to the $C_{3i}$, with $i=1,2$.

%%%%%%%%%%%%%%%%%%%%%%%%%%%%%%%%%%%%%%%%%%%%%%%%%%%%%%%%%%%%%%%%
\section{Solutions of the CWI's}\label{solution}
%%%%%%%%%%%%%%%%%%%%%%%%%%%%%%%%%%%%%%%%%%%%%%%%%%%%%%%%%%%%%%%%
We have shown in \secref{Sol3Point} how to solve the conformal constraints for the scalar three-point functions. In particular, the solutions can be equivalently given either in terms of hypergeometric functions or of triple-K integrals, as discussed in \cite{Coriano:2013jba, Bzowski:2013sza} respectively. For tensorial three-point functions, the solutions can be found, equivalently, by both methods  \cite{Bzowski:2013sza,Coriano:2018bsy,Coriano:2018bbe,Bzowski:2015pba}. 
%%%%%%%%%%%%%%%%%%%%%%%%%%%%%%%%%%%%
\section{Ward identities from the formalism of the effective action}\label{transWard}
%%%%%%%%%%%%%%%%%%%%%%%%%
In this appendix we illustrate the procedure for obtaining the canonical Ward identities related to the three local symmetries that one can have in the calculation of correlation functions of scalar, conserved current and conserved and traceless stress energy tensor operators. The first step is to couple the system to some background fields, and then require that the resulting generating functional is invariant  under diffeomorphisms, gauge and Weyl transformations. \\
In particular, we know that under a diffeomorphism $\x^\m$ the sources in the generating functional transform as
\begin{align}
\d g^{\m\n}&=-(\nabla^\m\x^\n+\nabla^\n\x^\m),\\
\d A^a_\m&=\x^\n\nabla_\n A_\m^a+\nabla_\m\x^\n\,A_\n^a,\\
\d\phi_0^I&=\x^\n\partial_\n\phi_0^I,
\end{align}
where $\nabla$ is a Levi-Civita connection. Under a gauge symmetry transformation with parameter $\a^a$, the sources transform as
\begin{align}
\d g^{\m\n}&=0,\\
\d A^a_\m&=-D_\m^{ac}\a^c=-\partial_\m\a^a-f^{abc}A_\m^b\a^c,\\
\d\phi_0^I&=-i\a^a(T^a_R)^{IJ}\phi^J_0,
\end{align}
where $T^a_R$ are matrices of a representation $R$ and $f^{abc}$ are structure constants of the group $G$. The gauge field transforms in the adjoint representation, while $\phi^I$ may transform in any representation $R$. The covariant derivative is $D_\m^{IJ}=\d^{IJ}\partial_\m-iA_\m^a(T_R^a)^{IJ}$.\\
The Ward identities follow from the requirement that the generating functional $Z$ is invariant under these transformations and in particular under the variation
\begin{align}
\d_\x&=\int\,d^dx\left[-(\nabla^\m\x^\n+\nabla^\n\x^\m)\sdfrac{\d}{\d g^{\m\n}}+(\x^\n\nabla_\n A_\m^a+\nabla_\m\x^\n\,A_\n^a)\sdfrac{\d}{\d A_\m^a}+\x^\n\partial_\n\phi_0^I\sdfrac{\d}{\d\phi_0^I}\right],\\
\d_\a&=-\int\,d^dx\left[(\partial_\m\a^a-f^{abc}A_\m^b\a^c)\sdfrac{\d}{\d A_\m^a}+i\a^a(T^a_R)^{IJ}\phi^J_0\sdfrac{\d}{\d\phi_0^I}\right],
\end{align}
so that the canonical Ward identities for the diffeomorphism and gauge transformations are respectively given by 
\begin{equation}
\d_\x Z=0,\qquad \d_\a Z=0.
\end{equation}
Using the definitions of the $1$-point functions
\begin{align}
\braket{T^{\mu\nu}(x)}=-\frac{2}{\sqrt{-g(x)}}\frac{\delta Z}{g_{\mu\nu}(x)}\Bigg|_{g=\d},\qquad\braket{J^{a\mu}(x)}=-\frac{1}{\sqrt{-g(x)}}\frac{\delta Z}{A^a_{\mu}(x)}\Bigg|_{A=0},\qquad \braket{\mO^I(x)}=-\frac{1}{\sqrt{-g(x)}}\frac{\delta Z}{\phi_0^I(x)}\Bigg|_{\phi=0},
\end{align}
the arbitrariness of the parameter $\a^a$, integrating by parts and also using the property $\sdfrac{1}{\sqrt{-g}}\partial_\m\sqrt{-g}=\G^\l_{\l\m}$ we find
\begin{align}
\d_\a Z=0&=-\int\,d^dx\left[(\partial_\m\a^a-f^{abc}A_\m^b\a^c)\sdfrac{\d}{\d A_\m^a}+i\a^a(T^a_R)^{IJ}\phi^J_0\sdfrac{\d}{\d\phi_0^I}\right]Z\notag\\
&=\int\,d^dx\sqrt{-g}\,\left[(\partial_\m\a^a-f^{abc}A_\m^b\a^c)\sdfrac{-1}{\sqrt{-g}}\sdfrac{\d}{\d A_\m^a}+i\a^a(T^a_R)^{IJ}\phi^J_0\,\,\sdfrac{-1}{\sqrt{-g}}\sdfrac{\d}{\d\phi_0^I}\right]Z\notag\\
&=\int\,d^dx\sqrt{-g}\,\a^a\,\left[-\G^\l_{\m\l}\langle J^{\m a}(x)\rangle-(\partial_\m\d^{ab}+f^{acb}A_\m^c)\langle J^{\m b}(x)\rangle+i(T^a_R)^{IJ}\phi^J_0\,\,\langle \mathcal{O}_I(x)\rangle\right],
\end{align}
from which we get the first Ward identity related to the gauge symmetry, expressed as
\begin{align}
0&=D^{ab}_\m\langle J^{\m a}\rangle+\G^\l_{\m\l}\langle J^{\m a}(x)\rangle-i(T^a_R)^{IJ}\phi_0^J\langle\mathcal{O}_I\rangle\notag\\
&=\nabla_\m\langle J^{\m a}\rangle+f^{abc}A_\m^b\langle J^{\m c}\rangle-i(T^a_R)^{IJ}\phi_0^J\langle\mathcal{O}_I\rangle\label{Ward1}.
\end{align}
The other Ward identities related to the diffeomorphism invariance will be
\begin{align}
0=\d_\x Z&=\int\,d^dx\left[-(\nabla^\m\x^\n+\nabla^\n\x^\m)\sdfrac{\d}{\d g^{\m\n}}+(\x^\n\nabla_\n A_\m^a+\nabla_\m\x^\n\,A_\n^a)\sdfrac{\d}{\d A_\m^a}+\x^\n\partial_\n\phi_0^I\sdfrac{\d}{\d\phi_0^I}\right]Z\notag\\
&=\int\,d^dx\,\sqrt{-g}\left[\sdfrac{1}{2}(\nabla^\m\x^\n+\nabla^\n\x^\m)\,\sdfrac{-2}{\sqrt{-g}}\sdfrac{\d}{\d g^{\m\n}}-(\x^\n\nabla_\n A_\m^a+\nabla_\m\x^\n\,A_\n^a)\sdfrac{-1}{\sqrt{-g}}\sdfrac{\d}{\d A_\m^a}-\x^\n\partial_\n\phi_0^I\sdfrac{-1}{\sqrt{-g}}\sdfrac{\d}{\d\phi_0^I}\right]Z\notag\\
&=\int\,d^dx\,\sqrt{-g}\left[\sdfrac{1}{2}(\nabla^\m\x^\n+\nabla^\n\x^\m)\,\langle T_{\m\n}(x)\rangle-(\x^\n\nabla_\n A_\m^a+\nabla_\m\x^\n\,A_\n^a)\langle J^{\m a}(x)\rangle-\x^\n\partial_\n\phi_0^I\langle \mathcal{O}_I(x)\rangle\right]\notag\\
&=\int\,d^dx\,\sqrt{-g}\x^\n\,\left[-\nabla^\m\,\langle T_{\m\n}(x)\rangle-\nabla_\n A_\m^a\langle J^{\m a}(x)\rangle+\nabla_\m\,(A_\n^a\langle J^{\m a}(x)\rangle)-\partial_\n\phi_0^I\,\langle \mathcal{O}_I(x)\rangle\right],
\end{align}
that leads to 
\begin{align}
0&=\nabla^\m\,\langle T_{\m\n}(x)\rangle+\nabla_\n A_\m^a\,\langle J^{\m a}(x)\rangle-\nabla_\m\,A_\n^a\,\langle J^{\m a}(x)\rangle-A_\n^a\,\nabla_\m\,\langle J^{\m a}(x)\rangle+\partial_\n\phi_0^I\,\langle \mathcal{O}_I(x)\rangle\notag\\
&=\nabla^\m\,\langle T_{\m\n}(x)\rangle-F_{\m\n}^a\,\langle J^{\m a}(x)\rangle-f^{abc}A_\n^a A_\m^b\langle J^{\m c}\rangle-A_\n^a\,\nabla_\m\,\langle J^{\m a}(x)\rangle+\partial_\n\phi_0^I\,\langle \mathcal{O}_I(x)\rangle.\label{Ward2}
\end{align}

Using the Ward identity related to the gauge symmetry we obtain the final result 
\begin{align}
0&=\nabla^\m\,\langle T_{\m\n}(x)\rangle-F_{\m\n}^a\,\langle J^{\m a}(x)\rangle-f^{abc}A_\n^a A_\m^b\langle J^{\m c}\rangle-A_\n^a\,\nabla_\m\,\langle J^{\m a}(x)\rangle+\partial_\n\phi_0^I\,\langle \mathcal{O}_I(x)\rangle\notag\\
&=\nabla^\m\,\langle T_{\m\n}(x)\rangle-F_{\m\n}^a\,\langle J^{\m a}(x)\rangle-f^{abc}A_\n^a A_\m^b\langle J^{\m c}\rangle+A_\n^a\,\left[f^{abc}A_\m^b\langle J^{\m c}\rangle-i(T^a_R)^{IJ}\phi_0^J\langle\mathcal{O}_I\rangle\right]+\partial_\n\phi_0^I\,\langle \mathcal{O}_I(x)\rangle\notag\\
&=\nabla^\m\,\langle T_{\m\n}(x)\rangle-F_{\m\n}^a\,\langle J^{\m a}(x)\rangle+D_\n^{IJ}\,\phi_0^I\,\langle \mathcal{O}_J(x)\rangle\label{transWardF}.
\end{align}
The equations \eqref{Ward1} and \eqref{Ward2} are the Ward identities for 1-point functions with sources turned on. These equations may then be differentiated with respect to the sources, with the aim of obtaining the corresponding Ward identities for higher point functions. 
%%%%%%%%%%%%%%%%%%%%%
\subsection{Trace Ward identities}
%%%%%%%%%%%%%%%%%%%%%

We have shown that the Lagrangian of a conformally invariant theory cannot contain dimensionful coupling constants. This constraint can be circumvented if we allow position dependent couplings, i.e. background fields and if we prescribe the correct transformation properties under Weyl transformation. Assume the operator $\mathcal{O}$ has a conformal dimension $\D$. For the term
\begin{equation}
\int\,d^dx\,\,\mathcal{O}^{\m_1\dots\m_m}_{\n_1\dots\n_n}\ \phi^{\n_1\dots\n_n}_{\m_1\dots\m_m},
\end{equation}
to be invariant under rescaling, we must have
\begin{equation}
\phi^{\n_1\dots\n_n}_{\m_1\dots\m_m}=(d-\D+m-n)\phi^{\n_1\dots\n_n}_{\m_1\dots\m_m}\,\s.
\end{equation}
For the metric, the gauge field and the scalar source we have the ordinary transformation rules,
\begin{align}
\d_\s g_{\m\n}&=2g_{\m\n}\s,\\
\d_\s A_\m^a&=0,\\
\d_\s\phi_0&=(d-\D)\phi_0\,\s.
\end{align}
Let's then consider the case when the generating functional is free of the Weyl anomaly, for which
\begin{equation}
\d_\s Z=0.\label{Weyl}
\end{equation}
The variation of the generating functional is realised by the following operator,
\begin{equation}
\d_\s=\int\,d^dx\ \,\s\left[2g^{\m\n}\sdfrac{\d}{\d g^{\m\n}}+(d-\D)\phi^I\,\sdfrac{\d}{\d\phi^I_0}\right],
\end{equation}
and to be more specific, we can expand \eqref{Weyl} as
\begin{align}
0=\d_\s Z&=\int\,d^dx\ \,\s\left[2g^{\m\n}\sdfrac{\d}{\d g^{\m\n}}+(d-\D)\phi^I\,\sdfrac{\d}{\d\phi^I_0}\right]Z\notag\\
&=\int\,d^dx\ \sqrt{-g}\,\s\left[-g_{\m\n}\langle T^{\m\n}(x)\rangle-(d-\D)\phi^I\,\langle\mathcal{O}^I(x)\rangle\right].
\end{align}
In this case we find the following trace, or Weyl, Ward identity in the presence of sources 
\begin{equation}
\langle T^\m_{\ \m}(x)\rangle=(\D-d)\phi_0^I\langle\mathcal{O}^I(x)\rangle.
\end{equation}
Also in this case, we can differentiate with respect to the sources in order to obtain the trace Ward identities for $n$-point functions. 

\subsection{The anomalous CWI's using conformal Killing vectors}
\label{heres}
The expressions of the anomalous conformal WIs can be derived in an alternative way following the formulation of  \cite{Coriano:2017mux}, that here we are going to extend to the four-point function case.\\
The derivation of such identities relies uniquely on the effective action and can be obtained as follows. We illustrate it first in the $TT$ case, and then move to the 4T.

We start from the conservation of the conformal current as derived in \eqref{iso}
\begin{equation}
\int d^d x \sqrt{g}\, \nabla^\alpha \left( \epsilon_\alpha \frac{2}{\sqrt{g}}\frac{\delta \sm}{\delta g_{\mu\alpha}}\right)=\int d^4 x \sqrt{g}\, \nabla_\mu\braket{\epsilon_\alpha T^{\alpha \nu}} =0.  
\end{equation}
In the $TT$ case the derivation of the special CWIs is simplified, since there is no trace anomaly if the counterterm action is defined as in \eqref{counter}, a point that we will address in \secref{renren}. We rely on the fact that the conservation of the conformal current $J^\mu_{(K)}$ implies the conservation equation 
\begin{align}
0=\int\,d^dx\,\sqrt{-g}\, \,\nabla_\mu\,\braket{J^\mu_{(K)}(x)\,T^{\mu_1\nu_1}(x_1)}.
\end{align}
By making explicit the expression $J^\mu(x)=K_\nu(x)\,T^{\mu\nu}(x)$, with $\epsilon \to K$ in the flat limit,  the previous relation takes the form
 \begin{align}
0=\int\,d^dx\,\bigg(\partial_\mu K_\nu\,\braket{T^{\mu\nu}(x)\,T^{\mu_1\nu_1}(x_1)}+ K_\nu\,\partial_\mu\,\braket{T^{\mu\nu}(x)\,T^{\mu_1\nu_1}(x_1)}\bigg).\label{cons}
 \end{align}
We recall that $K_\nu$ satisfies the conformal Killing equation in flat space
\begin{align}
\label{flatc}
\partial_\mu K_\nu+\partial_\nu K_\mu=\frac{2}{d}\delta_{\mu\nu}\,\left(\partial\cdot K\right),
\end{align}
and by using this equation \eqref{cons} can be re-written in the form
\begin{align}
	0=\int\,d^dx\,\bigg(K_\nu\partial_\mu\,\braket{T^{\mu\nu}(x)\,T^{\mu_1\nu_1}(x_1)}+\frac{1}{d}\big(\partial\cdot K\big)\,\braket{T(x)\,T^{\mu_1\nu_1}(x_1)}\bigg).\label{newcons}
\end{align}
We can use in  this previous expression the conservation and trace Ward identities for the two-point function $\braket{TT}$, that in the flat spacetime limit are explicitly given by
\begin{align}
\partial_\mu\braket{T^{\mu\nu}(x)T^{\mu_1\nu_1}(x_1)}&=\bigg(\delta^{(\mu_1}_\nu\delta^{\nu_1)}_\lambda\partial^\mu\delta(x-x_1)-2\delta^{\mu(\mu_1}\delta^{\nu_1)}_\nu\partial_\lambda\delta(x-x_1)\bigg)\braket{T^{\lambda\nu}(x)},\label{consTT}\\
\delta_{\mu\nu}\braket{T^{\mu\nu}(x)T^{\mu_1\nu_1}(x_1)}&\equiv\braket{T(x)T^{\mu_1\nu_1}(x_1)}=-2\delta(x-x_1)\braket{T^{\mu_1\nu_1}(x)}\label{traceTT},
\end{align}
and the explicit expression of the Killing vector $K^{(C)}_\nu$ for the special conformal transformations 
\begin{equation}
\begin{split}
K^{(C)\,\kappa}_\mu&=2x^\kappa\,x_\mu-x^2\delta^\kappa_\mu,\\
\partial\cdot  K^{(C)\,\kappa}&=2d\,x^\kappa,
\end{split}\label{spc}
\end{equation}
where $\kappa=1,\dots,d$. By using \eqref{spc} in the integral \eqref{newcons}, we can rewrite that expression as
\begin{align}
	0=\int\,d^dx\,\bigg[\big(2x^\kappa\,x_\nu-x^2\delta^\kappa_\nu\big)\partial_\mu\,\braket{T^{\mu\nu}(x)\,T^{\mu_1\nu_1}(x_1)}+2\,x^\kappa\,\braket{T(x)\,T^{\mu_1\nu_1}(x_1)}\bigg],
\end{align}
A final integrating by parts finally gives the relations 
\begin{align}
&\left(2d\,x_1^\kappa+2x_1^\kappa\,x^{\mu}_1\frac{\partial}{\partial x_1^\mu}+x_1^2\frac{\partial}{\partial x_{1\kappa}}\right)\braket{T^{\mu_1\nu_1}(x_1)}\notag\\
&\quad+2\bigg(x_{1\lambda}\,\delta^{\mu_1\kappa}-x_1^{\mu_1}\delta^\kappa_\lambda\bigg)\braket{T^{\lambda\nu_1}(x_1)}+2\bigg(x_{1\lambda}\,\delta^{\nu_1\kappa}-x_1^{\nu_1}\delta^\kappa_\lambda\bigg)\braket{T^{\mu_1\lambda}(x_1)}=0,
\end{align}
that are  the special CWIs for the 1-point function $\braket{T^{\mu_1\nu_1}(x_1)}$. The derivation above can be extended to $n$-point functions, starting from the identity 
\begin{equation} 
\int d^d x \sqrt{g} \nabla_\alpha(x) \langle J^\alpha_c(x)T^{\mu_1\nu_1}(x_1)\ldots T^{\mu_n\nu_n}(x_n)
\label{div}
\rangle=0.
\end{equation}
We have used the conservation of the conformal current in d dimensions under variations of the metric, induced by the conformal Killing vectors.

In absence of an anomaly, the conservation of the current $J^\mu_c$ follows from the conservation of the stress energy tensor plus the zero trace condition. 
As in the example illustrated above, we consider \eqref{div} in the flat limit 
\begin{align}
\int dx^d \,\partial_\nu\bigg[K_\mu(x)\braket{T^{\mu\nu}(x)T^{\mu_1\nu_1}(x_1)\dots T^{\mu_4\nu_4}(x_4)}\bigg]=0,\label{Killing1}
\end{align}
where we are assuming that the surface terms vanish, due to the fast fall-off behavior of the correlation function at infinity. Expanding \eqref{Killing1} we obtain an expression similar to \eqref{newcons}

\begin{align}
0=\int d^dx\left\{K_{\mu}(x)\partial_\nu\braket{T^{\mu\nu}(x)T^{\mu_1\nu_1}(x_1)\dots T^{\mu_4\nu_4}(x_4)}+\frac{1}{d}\big(\partial\cdot K\big)\delta_{\mu\nu}\braket{T^{\mu\nu}(x)T^{\mu_1\nu_1}(x_1)\dots T^{\mu_4\nu_4}(x_4)}
\right\}.\label{killing01}
\end{align}
Starting  from this expression, the dilatation CWI is obtained by the choice of the CKV characterizing the dilatations  
\begin{equation}
K^{(D)}_\mu(x)=x_\mu,\qquad\partial\cdot K^{(D)}=d,
\end{equation}
and \eqref{killing01} becomes
\begin{align}
0=\int d^d x\bigg\{x_\mu\,\partial_\nu\braket{T^{\mu\nu}(x)T^{\mu_1\nu_1}(x_1)\dots T^{\mu_4\nu_4}(x_4)}+\delta_{\mu\nu}\braket{T^{\mu\nu}(x)T^{\mu_1\nu_1}(x_1)\dots T^{\mu_4\nu_4}(x_4)}\label{anomDil}
\bigg\}.
\end{align}
At this stage, we use the conservation and trace Ward identities in $d=4$ for the four-point function written as
\begin{align}
&\partial_\nu\braket{T^{\mu\nu}(x)T^{\mu_1\nu_1}(x_1)\dots T^{\mu_4\nu_4}(x_4)}=\notag\\
=&-8\bigg\{\left[\Gamma^{\mu}_{\nu\lambda}(x)\right]^{\mu_1\nu_1\mu_2\nu_2\mu_3\nu_3}(x_1,x_2,x_3)\braket{T^{\lambda\nu}(x)T^{\mu_4\nu_4}(x_4)}+(14)+(24)+(34)\bigg\}\notag\\
&-4\bigg\{\left[\Gamma^{\mu}_{\nu\lambda}(x)\right]^{\mu_1\nu_1\mu_2\nu_2}(x_1,x_2)\braket{T^{\lambda\nu}(x)T^{\mu_3\nu_3}(x_3)T^{\mu_4\nu_4}(x_4)}+(13)+(23)+(14)+(24)+(34)\bigg\}\notag\\
&-2\bigg\{\left[\Gamma^{\mu}_{\nu\lambda}(x)\right]^{\mu_1\nu_1}(x_1)\braket{T^{\lambda\nu}(x)T^{\mu_2\nu_2}(x_2)T^{\mu_3\nu_3}(x_3)T^{\mu_4\nu_4}(x_4)}+(12)+(13)+(14)\bigg\}\label{5ptcons},
\end{align}
and
\begin{align}
\delta_{\mu\nu}\braket{T^{\mu\nu}(x)T^{\mu_1\nu_1}(x_1)\dots T^{\mu_4\nu_4}(x_4)}&=-2\bigg\{\delta_{xx_1}\braket{T^{\mu_1\nu_1}(x)T^{\mu_2\nu_2}(x_2)\dots T^{\mu_4\nu_4}(x_4)}+(12)+(13)+(14)\bigg\}\notag\\
&\hspace{2cm}+2^4\big[\mathcal{A}(x)\big]^{\mu_1\nu_1\dots\mu_4\nu_4}(x_1,\dots,x_4).\label{5pttrace}
\end{align}
to finally derive the dilatation WI from \eqref{anomDil} in the form
\begin{align}
\left(4d+\sum_{j=1}^4\,x_j^\alpha\frac{\partial}{\partial x_j^\alpha}\right)\braket{T^{\mu_1\nu_1}(x_1)\dots T^{\mu_4\nu_4}(x_4)}=2^4\int dx \big[\mathcal{A}(x)\big]^{\mu_1\nu_1\dots\mu_4\nu_4}(x_1,\dots,x_4)\label{DilAnom},
\end{align}
where $d=4$. It is worth mentioning that \eqref{DilAnom} is valid in any even spacetime dimension if we take into account the particular structure of the trace anomaly in that particular dimension.

The special CWIs correspond to the $d$ special conformal Killing vectors in flat space given in \eqref{spc}, as in the $TT$ case. Also in this case we derive the identity 
\begin{align}
0=&\int d^d x\bigg\{\big(2x^\kappa\,x_\mu-x^2\delta^\kappa_\mu\big)\,\partial_\nu\braket{T^{\mu\nu}(x)T^{\mu_1\nu_1}(x_1)\dots T^{\mu_4\nu_4}(x_4)}\notag\\
&\hspace{4cm}+2x^\kappa\delta_{\mu\nu}\braket{T^{\mu\nu}(x)T^{\mu_1\nu_1}(x_1)\dots T^{\mu_4\nu_4}(x_4)}\label{anomSpecWI}
\bigg\}.
\end{align}
By using the relations \eqref{5ptcons} and \eqref{5pttrace} and performing the integration over $x$ explicitly in the equation above,  the anomalous special CWIs for the four-point function take the form
\begin{align}
&\sum_{j=1}^4\left[2x_j^\kappa\left(d+x_j^\alpha\frac{\partial}{\partial x_j^\alpha}\right)-x_j^2\,\delta^{\kappa\alpha}\frac{\partial}{\partial x_j^\alpha}\right]\braket{T^{\mu_1\nu_1}(x_1)\dots T^{\mu_4\nu_4}(x_4)}\notag\\
&+2\sum_{j=1}^4\left(\delta^{\kappa\mu_j}x_{j\,\alpha}-\delta^\kappa_\alpha x_j^{\mu_j}\right)\braket{T^{\mu_1\nu_1}(x_1)\dots T^{\nu_j\alpha}(x_j)\dots T^{\mu_4\nu_4}(x_4)}\notag\\
&+2\sum_{j=1}^4\left(\delta^{\kappa\nu_j}x_{j\,\alpha}-\delta^\kappa_\alpha x_j^{\nu_j}\right)\braket{T^{\mu_1\nu_1}(x_1)\dots T^{\mu_j\alpha}(x_j)\dots T^{\mu_4\nu_4}(x_4)}=2^5\,\int dx\,x^\kappa\big[\mathcal{A}(x)\big]^{\mu_1\nu_1\dots\mu_4\nu_4}(x_1,\dots,x_4),
\end{align}
where the presence of the anomaly term comes from the inclusion of the trace  WI, exactly as in the $TT$ case. Obviously, in the $TT$, the anomaly is prescription dependent, and can be set to zero, but the approach in that case and 
in the case of the higher point functions is identical. \\
At this stage, these equations can be transformed to momentum space, giving the final expressions of the CWIs in the form 
\begin{align}
\left(d-\sum_{j=1}^3\,p_j^\alpha\frac{\partial}{\partial p_j^\alpha}\right)\braket{T^{\mu_1\nu_1}(p_1)T^{\mu_2\nu_2}(p_2)T^{\mu_3\nu_3}(p_3)T^{\mu_4\nu_4}(\bar{p}_4)}=2^4\,\mathcal{A}^{\mu_1\nu_1\mu_2\nu_2\mu_3\nu_3\mu_4\nu_4}(p_1,p_2,p_3,\bar{p}_4),
\end{align} 
for the dilatation, and 
\begin{align}
&\sum_{j=1}^3\left(p_j^\kappa\frac{\partial^2}{\partial p_j^\alpha \partial p_{j\alpha}}-2p_j^\alpha\frac{\partial}{\partial p_j^\alpha\partial p_{j\kappa}}\right)\braket{T^{\mu_1\nu_1}(p_1)T^{\mu_2\nu_2}(p_2)T^{\mu_3\nu_3}(p_3) T^{\mu_4\nu_4}(\bar{p}_4)}\notag\\
&+2\sum_{j=1}^3\left(\delta^{\kappa\mu_j}\frac{\partial}{\partial p_{j\,\alpha}}-\delta^\kappa_\alpha \frac{\partial}{\partial p_j^{\mu_j}}\right)\braket{T^{\mu_1\nu_1}(p_1)\dots T^{\nu_j\alpha}(p_j)\dots T^{\mu_4\nu_4}(\bar{p}_4)}\notag\\
&+2\sum_{j=1}^3\left(\delta^{\kappa\nu_j}\frac{\partial}{\partial p_{j\,\alpha}}-\delta^\kappa_\alpha \frac{\partial}{\partial p_j^{\nu_j}}\right)\braket{T^{\mu_1\nu_1}(p_1)\dots T^{\mu_j\alpha}(p_j)\dots T^{\mu_4\nu_4}(\bar{p}_4)}\notag\\
&=-2^5\,\left[\frac{\partial}{\partial p_{4\kappa}}\,\mathcal{A}^{\mu_1\nu_1\mu_2\nu_2\mu_3\nu_3\mu_4\nu_4}(p_1,p_2,p_3,p_4)\right]_{\bar{p_4}=-p_1-p_2-p_3},
\end{align}
for the special CWI's, having used the definition \eqref{Avardef}. \\
When this procedure is applied to the $n$-point function, one finds that the anomalous CWIs are written as
\begin{align}
\left(d-\sum_{j=1}^{n-1}\,p_j^\alpha\frac{\partial}{\partial p_j^\alpha}\right)\braket{T^{\mu_1\nu_1}(p_1)\dots T^{\mu_n\nu_n}(\bar{p}_n)}=2^n\,\mathcal{A}^{\mu_1\nu_1\dots\mu_n\nu_n}(p_1,\dots,\bar{p}_n),
\end{align}
for the dilatation and 
	\begin{align}
		&\sum_{j=1}^{n-1}\left(p_j^\kappa\frac{\partial^2}{\partial p_j^\alpha \partial p_{j\alpha}}-2p_j^\alpha\frac{\partial}{\partial p_j^\alpha\partial p_{j\kappa}}\right)\braket{T^{\mu_1\nu_1}(p_1)\dots T^{\mu_n\nu_n}(\bar{p}_n)}\notag\\
		&+2\sum_{j=1}^{n-1}\left(\delta^{\kappa\mu_j}\frac{\partial}{\partial p_{j\,\alpha}}-\delta^\kappa_\alpha \frac{\partial}{\partial p_j^{\mu_j}}\right)\braket{T^{\mu_1\nu_1}(p_1)\dots T^{\nu_j\alpha}(p_j)\dots T^{\mu_n\nu_n}(\bar{p}_n)}\notag\\
		&+2\sum_{j=1}^{n-1}\left(\delta^{\kappa\nu_j}\frac{\partial}{\partial p_{j\,\alpha}}-\delta^\kappa_\alpha \frac{\partial}{\partial p_j^{\nu_j}}\right)\braket{T^{\mu_1\nu_1}(p_1)\dots T^{\mu_j\alpha}(p_j)\dots T^{\mu_n\nu_n}(\bar{p}_n)}\notag\\
		&=-2^{n+1}\left[\frac{\partial}{\partial p_{n\kappa}}\,\mathcal{A}^{\mu_1\nu_1\dots\mu_n\nu_n}(p_1,\dots,p_n)\right]_{p_n=\bar{p}_n},
	\end{align}
for the special conformal Ward identities, where $\bar{p}_n=-\sum_{i=1}^{n-1}p_i$ and we have used the definition \eqref{Avardef}.

\subsection{Triple-K integrals}
%%%%%%%%%%%%%%%%%%%%%%%%%%%%%%%%%%%%
In this appendix we use the formulation of the triple-K integrals, because they reflect naturally the symmetry properties of the correlation functions and also because of their analytical properties. 
We recall the definition of the general triple-K integral
\begin{equation}
I_{\a\{\b_1\b_2\b_3\}}(p_1,p_2,p_3)=\int dx \,x^\a\prod_{j=1}^{3}\,p_j^{\b_j}\,K_{\b_j}(p_jx),\label{3Kint}
\end{equation}
where $K_\n$ is a modified Bessel function of the second kind defined as
\begin{equation}
K_\n(x)=\sdfrac{\pi}{2}\sdfrac{I_{-\n}(x)-I_\n(x)}{\sin(\n\pi)},\ \n\notin\mathbb Z,\qquad I_\n(x)=\left(\sdfrac{x}{2}\right)^{\n}\sum_{k=0}^\infty\sdfrac{1}{\G(k+1)\,\G(\n+1+k)}\,\left(\sdfrac{x}{2}\right)^{2k}.\label{Kdef}
\end{equation} 
with the property 
\begin{equation}
K_n(x)=\lim_{\e\to0}K_{n+\e}(x),\quad n\in\mathbb Z.
\end{equation}
The triple-K integral in \eqref{3Kint} depends on four parameters: the power $\a$ of the integration variable $x$, and the three Bessel function indices $\b_j$. The argument of this integral are magnitudes of momenta $p_j$, $j=1,2,3$. One can notice the integral is invariant under the exchange $(p_j,\beta_j)\leftrightarrow(p_i,\beta_i)$, and we will see that this properties will reflect the symmetry properties of the correlation functions. We will use also the reduced version $J_{N\{k_1,k_2,k_3\}}$ of the triple-K integral defined as
\begin{equation}
J_{N\{k_j\}}=I_{\frac{d}{2}-1+N\{\D_j-\frac{d}{2}+k_j\}},\label{defJintegr}
\end{equation}
that are mapped to the expression \eqref{3Kint} via the substitutions
\begin{equation}
\a=\sdfrac{d}{2}-1+N,\qquad \b_j=\D_j-\sdfrac{d}{2}+k_j,\ j=1,2,3,
\end{equation}
where we have used the condensed notation $\{k_j\}=\{k_1\,k_2\,k_3\}$. These triple-K integrals may also be re-expressed using the Feynman parametrization as
\begin{equation}
I_{\a\{\b_1\b_2\b_3\}}(p_1,p_2,p_3)=2^{\a-3}\G\left(\sdfrac{\a-\b_t+1}{2}\right)\G\left(\sdfrac{\a+\b_t+1}{2}\right)\int_{[0,1]^3}dX\,D^{\frac{1}{2}(\b_t-\a-1)}\prod_{j=1}^3\,x_j^{\frac{1}{2}(\a-1-\b_t)+\b_j}\label{Feyn},
\end{equation}
where $\b_t=\b_1+\b_2+\b_3$ and the integration extends over the unit interval $[0,1]$ for each of the $x_j$, $j=1,2,3$ with the standard measure $dX=dx_1\,dx_2\,dx_3\,\d(1-x_1-x_2-x_3)$ and with
\begin{equation}
D=p_1^2\,x_2\,x_3+p_2^2\,x_1\,x_3+p_3^2\,x_1\,x_2,
\end{equation}
being the usual denominator appearing in the Feynman representation. Furthermore, the expression of the triple-K integral is linked to the hypergeometric functions through \eqref{3K}. 

In order to study the convergences of the triple-K integral we assume that all the parameters in \eqref{3Kint} are real. At large $x$, the Bessel functions have the asymptotic expansions
\begin{equation}
I_\n(x)=\sdfrac{1}{\sqrt{2\pi}}\sdfrac{e^x}{\sqrt{x}}+\dots,\quad K_\n(x)=\sqrt{\sdfrac{\pi}{2}}\,\sdfrac{e^{-x}}{\sqrt{x}}+\dots,\ \n\in\mathbb R, 
\end{equation}
and one readily observes that  inserting the expansions above in \eqref{3Kint}, the integral converges at large $x$ for physical configurations of the momenta, with $p_1+p_2+p_3>0$. However, there may still be a divergence at $x=0$. Considering the asymptotic expansion of the Bessel functions at small $x$ as
\begin{equation}
I_{\nu}(x)=\frac{x^\nu}{2^\nu\ \Gamma(\nu+1)}+\dots,\qquad K_\nu(x)= \frac{2^{\nu-1}\Gamma(\nu)}{x^\nu}+\dots,
\end{equation}
in \eqref{3Kint}, one observes that the triple-K integral converges at small $x$ only if
\begin{equation}
\a>\sum_{j=1}^{3}|\b_j|+1,\qquad p_1,p_2,p_3>0.\label{region}
\end{equation}
If $\a$ does not satisfy this inequality, the integrals must be defined by an analytic continuation. The quantity
\begin{equation}
\d\equiv \sum_{j=1}^3|\b_j| -1-\a
\end{equation}
is the expected degree of divergence. When
\begin{equation}
\a+1\pm\b_1\pm\b_2\pm\b_3=-2k,\label{condition}
\end{equation}
for some non-negative integer $k$ and any choice of the $\pm$ sign, the analytic continuation of the triple-K integral generally has poles in the regularization parameter. From the representation \eqref{Feyn} these poles may either be present in the gamma function or in the Feynman parametrization of the integral multiplying them, or both. As one can see from \eqref{Feyn} the Feynman integrals are finite if
\begin{equation}
\sdfrac{1}{2}(\a-1-\b_t)+\b_j>1,
\end{equation}
as $d\to4$ for all $j=1,2,3$, but generally diverge otherwise. 
More details about the convergence of these kind of integrals and on their regularization procedure, can be found in \cite{Bzowski:2015yxv,Bzowski:2014qja, Bzowski:2018fql, Bzowski:2015pba}.
%%%%%%%%%%%%%%%%%%%%%%%%%%%%%%%%%%%%%%%
\subsection{Dilatation Ward identities}
%%%%%%%%%%%%%%%%%%%%%%%%%%%%%%%%%%%%%%%<

We will provide the solution to the dilatation Ward identities in terms of triple-K integrals. First notice that using the relations \eqref{identityBess} we obtain
\begin{align}
\sum_{j=1}^3p_j\sdfrac{\partial }{\partial p_j}\,I_{\a\{\b_k\}}&=-\sum_{j=1}^3\,p_j^2\,I_{\a+1\{\b_k-\d_{jk}\}}=-\sum_{j=1}^3\,p_j^2\,I_{\a+1\{\b_k+\d_{jk}\}}+2\sum_{j=1}^3\,\b_j\,I_{\a\{\b_k\}}\notag\\
&=-\sum_{j=1}^3\,p_j^2\,I_{\a+1\{\b_k+\d_{jk}\}}+2\,\b_t\,I_{\a\{\b_k\}},\label{DCWIsol}
\end{align}
and by using
\begin{equation}
\int_0^\infty dx\,x^{\a+1}\,\sdfrac{\partial}{\partial x}\left(\prod_{j=1}^{3}p_j^{\b_j}K_{\b_j}(p_j\,x)\right)=-\sum_{j=1}^3\,p_j^2\,I_{\a+1\{\b_k+\d_{jk}\}}+\,\b_t\,I_{\a\{\b_k\}},
\end{equation}
we can re-expressed \eqref{DCWIsol}, after an integration by parts as
\begin{equation}
\sum_{j=1}^3p_j\sdfrac{\partial }{\partial p_j}\,I_{\a\{\b_k\}}=\int_0^\infty dx\,\sdfrac{\partial}{\partial x}\left(x^{\a+1}\,\prod_{j=1}^{3}p_j^{\b_j}K_{\b_j}(p_j\,x)\right)-(\a+1-\b_t)\,I_{\a\{\b_k\}}.\label{DCWIsol2}
\end{equation}
The first term on the right-hand side leads to a boundary term at $x=0$. In the region of convergence \eqref{region}, all integrals in this expression are well-defined and the boundary term vanishes. Then the triple-K integral satisfies an equation analogous to the dilatation equation with scaling degree
\begin{equation}
\deg\Big(I_{\a\{\b_k\}}\Big)=\b_t-\a-1,
\end{equation}
and for $J_{N\{k_j\}}$ we obtain
\begin{equation}
\deg\Big({J_{N\{k_j\}}}\Big)=\D_t+k_t-2d-N.
\end{equation}
It is worth mentioning that these results are valid in the case $\a+1\pm\b_1\pm\b_2\pm\b_3\ne-2k$ for some non-negative $k$ and independent choice of signs.

From this analysis it is simple to relate the form factors to the triple-K integrals. In the $\braket{TT\mO}$ case, we know from \eqref{dilfin3} that the form factors have degrees $\deg(A_n)=\D_t-2d-N_n$. In general, if $A_n=\a_\n\,J_{N\{k_j\}}$, where $\a_N\in \mathbb R$ is a constant value, then we can determine the value of $N=N(A_n)$ just through a comparison of the degree, i.e.  
\begin{equation}
\begin{matrix}
\hspace{2.5ex}\deg\big(A_n\big)&\hspace{-4.5ex}=\D_t-2d-N_n,\\[1.2ex]
\deg\Big(\,J_{N\{k_j\}}\Big)&\hspace{-1.5ex}=\D_t+k_t-2d-N,
\end{matrix}\quad\implies\quad
N(A_n)=N_n+k_t,
\end{equation}
where $k_t=k_1+k_2+k_3$. The dilatation Ward identities will fix the values of $N$ in the triple-K integral for which a first general expression for the form factors can be given as
\begin{equation}
	\begin{split}
	A_1&=\a_1\,J_{4+k_t\{k_j\}},\\
	A_2&=\a_2\,J_{2+k_t\{k_j\}},\\
	A_3&=\a_3\,J_{k_t\{k_j\}}.
	\end{split}\label{result1}
\end{equation}
These relations will be more constrained by imposing the primary and secondary CWI's. 
%%%%%%%%%%%%%%%%%%%%%%%%%
\subsection{Primary CWI's}\label{primarysol}
%%%%%%%%%%%%%%%%%%%%%%%%%
We have analysed the basic properties of the triple-K integral. We now want to use the results in the previous section in order to write a solution of the primary CWI's. Using the relation \eqref{Fund} in \appref{AppendixJ} we show that for any $n,m=1,2,3$ 
\begin{equation}
K_{nm}\,J_{N\{k_j\}}=-2k_n\ J_{N+1\{k_j-\d_{jn}\}}+2k_m\ J_{N+1\{k_j-\d_{jm}\}},\label{sol}
\end{equation}
with $k_1,\,k_2,\,k_3,\,N\in\mathbb R$, and where $K_{nm}$ is the conformal scalar operator defined in \eqref{Koper} and $\delta_{jn}$ is the Kronecker delta. To make the above equation clearer, we consider the case with $K_{13}$ 
\begin{align}
K_{13}\,J_{N\{k_j\}}&=-2k_1\ J_{N+1\{k_j-\d_{j1}\}}+2k_3\ J_{N+1\{k_j-\d_{j3}\}}\notag\\
&=-2k_1\ J_{N+1\{k_1-1,k_2,k_3\}}+2k_3\ J_{N+1\{k_1,k_2,k_3-1\}},
\end{align}
from which one can write the two following equations
\begin{align}
&K_{13}\,J_{N\{0,k_2,0\}}=0,
\end{align}
for any $k_2$, then also for $k_2=0$. We can generalize this result as
\begin{equation}
K_{nm}\,J_{N\{0,0,0\}}=0,
\end{equation}  
for any $n,m=1,2,3$. Let us consider the primary CWIs for the $\braket{TT\mO}$ given in \eqref{primaryCWI} and we are going to solve these equations in terms of triple-K integrals. The approach that allows us to find the general solutions to these constraints, is to construct a set of linear homogeneous differential equations starting from \eqref{primaryCWI}, and applying on them recursively operators $K_{nm}$ as follows
\begin{equation}
\begin{matrix}
K_{12}A_1=0,&\qquad K_{13}A_1=0,&\qquad\\[1.2ex]
K_{12}A_2=0,&\qquad K_{13}^2A_2=0,&\qquad K_{12}K_{13}A_2=0,\\[1.2ex]
K_{12}A_3=0,&\qquad K_{13}^3A_3=0,&\qquad K_{12}K^2_{13}A_3=0,&\qquad K_{12}K_{13}A_3=0.\\
\end{matrix}\label{homog}
\end{equation}
Considering the expression of the form factors, partially obtained in \eqref{result1}, we can add more constraints on the form factors by the homogeneous differential equations \eqref{homog}, and, in the case of $A_1$, we have
\begin{equation}
\begin{matrix}
&K_{12}A_1=\a_1K_{12}\ J_{4+k_t\{k_j\}}=0,\\[1.2ex]
&K_{13}A_1=\a_1K_{13}\ J_{4+k_t\{k_j\}}=0,
\end{matrix}
\,\quad \implies\quad k_1=k_2=k_3=0,
\end{equation}
giving the solution $A_1=\a_1\,J_{4\{000\}}$. Observe that if we impose only one homogeneous equation, say $K_{13}A_1=0$, then the most general solution in terms of the triple-K integrals is $\a\,J_{N\{0,k_2,0\}}$ for any $\a,\,N,\,k_3\in\mathbb{R}$. For the second form factors $A_2$ we can resolve the first equation
\begin{equation}
0=K_{12}A_2=\a_2K_{12}\ J_{2+k_t\{k_j\}}=-2\a k_1\,J_{3+k_t\{k_1-1,k_2,k_3\}}+ 2\a k_2\,J_{3+k_t\{k_1,k_2-1,k_3\}},\\[1.2ex]
\end{equation}
that give the constraint $k_1=k_2=0$. Using this information we can resolve the second homogeneous equation for $A_2$ as
\begin{equation}
0=K_{13}^2A_2=\a_2K^2_{13}\ J_{2+k_3\{00 k_3}=4\a\,k_3(k_3-1)\,J_{4+k_3\{00,k_3-2\}},
\end{equation}
given two solutions $k_3=0$ and $k_3=1$. The last equation 
\begin{equation}
K_{12}K_{13}A_2=\a_2K_{12}K_{13}\ J_{2+k_3\{0,0,k_3\}}=0
\end{equation}
is identically satisfied and does not give any further informations. Finally, we can write down the most general solution for the second form factor $A_2$ in terms of the triple-K integrals as
\begin{equation}
A_2=\a_2\,J_{2\{000\}}+\a_{21}\,J_{3\{001\}},
\end{equation}
where we have introduced another constant $\a_{12}\in\mathbb R$, related to the solution $k_3=1,\,k_1=k_2=0$. In the same way, we can solve the equations for the $A_3$ form factor in the form
\begin{align}
&K_{12}A_3=0\,\quad\implies\quad k_1=k_2=0,\\[1.4ex]
&K_{13}^3A_3=8\a_3\,k_3(k_3-1)(k_3-2)\,J_{k_3+3\{00,k_3-3\}}=0,\quad\implies\quad k_3=0,\ k_3=1,\ k_3=2,
\end{align}
with the other equations identically satisfied. To summarize, we can write down the most general solutions obtained in our approach in the form
	\begin{align}
	A_1&=\a_1\,J_{4\{000\}},\\[1.1ex]
	A_2&=\a_2\,J_{2\{000\}}+\a_{21}\,J_{3\{001\}},\\[1.1ex]
	A_3&=\a_3\,J_{0\{000\}}+\a_{31}\,J_{1\{001\}}+\a_{32}\,J_{2\{002\}},
	\end{align}
where all the $\a$ are numerical constants. Finally, the inhomogeneous parts of \eqref{primaryCWI} fix some of these constants. When the solution above is substituted into the primary CWI's \eqref{primaryCWI} 
	\begin{align}
	K_{13}A_2&=8A_1\ \implies\ 2\a_{21}J_{4\{000\}}=8\a_1\,J_{4\{000\}},\\
	K_{13}A_3&=2A_2\ \implies \ 2\a_{31}\,J_{2\{000\}}+4\a_{32}\,J_{3\{001\}}=2\a_2\,J_{2\{000\}}+2\a_{21}\,J_{3\{001\}},
	\end{align}
they imply that
\begin{equation}
\a_{21}=4\a_1,\quad \a_{31}=\a_2,\quad\a_{32}=2\a_1.
\end{equation}
In conclusions, we have analyzed the primary CWI's for the $\braket{TT\mO}$ correlation function and finally we have found that the solutions can be expressed in the form 
\begin{equation}
	\begin{split}
	A_1&=\a_1\,J_{4\{000\}},\\[1.1ex]
	A_2&=\a_2\,J_{2\{000\}}+4\a_{1}\,J_{3\{001\}},\\[1.1ex]
	A_3&=\a_3\,J_{0\{000\}}+\a_{2}\,J_{1\{001\}}+2\a_{1}\,J_{2\{002\}},
	\end{split}\label{primarysolution}
\end{equation}
depending on three undetermined constants $\a_1,\,\a_2,\,\a_3\in\mathbb R$. The method to solve the primary CWI's is applicable to generic three-point  functions and we will see that the secondary CWI's - in this case for the $TTO$, but equivalently in all the other cases - will reduce the number of undetermined constants to just one.
%%%%%%%%%%%%%%%%%%%%%%%%%%%%%%%
\subsection{The analysis of the secondary CWI's}
%%%%%%%%%%%%%%%%%%%%%%%%%%%%%%%
We have shown that the primary CWI's fix the functional structure of the form factors and allow to write the solutions as a linear combinations of triple-K integrals with some undetermined constants. The role of the secondary CWI's, that are first order partial differential equations, is to fix the algebraic relation between the undetermined constants in order to give a final solution given as a linear combination of minimal and independent number of constants.

We can now impose the secondary CWI's for the solutions obtained in \eqref{primarysolution}. If we substitute the full solutions to the primary CWI's into the secondary in order to extract more information about the constants, we may encounter some troubles in the computation that is purely algebraic. However, we can examine the relations in the zero-momentum limit and this procedure can simplify our algebraic equations.

In the zero-momentum limit $p_{3\m}\to0$, $p_{1\m}=-p_{2\m}=p_{\m}$, there are different expansions for the Bessel function. In particular, depending on the parameter $\nu$ one can expand as
\begin{align}
\,K_{0}(p_3 x)&= \log\left(\frac{2}{p_3\,x}\right)-\gamma_E,\\
p_3^{\nu}\,K_{\nu}(p_3 x),&
=\left[\sdfrac{2^{{\nu}-1}\G(\nu)}{x^{\nu}}+O(p_3^2)\right]+p_3^{2{\nu}}\left[2^{-{\nu}-1}\G(-{\nu})\,x^\nu+O(p_3^2)\right],\quad \nu\notin\mathbb{Z},\\
p_3^{n}\,K_{n}(p_3 x),&
=\left[\sdfrac{2^{{n}-1}\G(n)}{x^{n}}+O(p_3^2)\right]+p_3^{2{n}}\left[\frac{(-1)^{n+1}}{2^n\,\Gamma(n+1)}\,x^n\,\log\,p_3+f(\log(x))+O(p_3^2)\right],\quad n\in\mathbb{N},
\end{align}
obtained from the expression of the modified Bessel functions K given in \cite{Prudnikov}, indeed for the case $\nu=n\in\mathbb{N}$ the Bessel function assumes the following form
\begin{align}
K_{n}(x)=\lim_{\nu\to n}\,K_{\nu}(x)&=(-1)^{n+1}\,I_{n}(x)\,\log\frac{x}{2}+\frac{1}{2}\sum_{k=0}^{n-1}(-1)^k\frac{(n-k-1)!}{k!}\,\left(\frac{x}{2}\right)^{2k-n}\notag\\
&\qquad +\frac{(-1)^n}{2}\sum_{k=0}^\infty\,\frac{\psi(k+n+1)+\psi(k+1)}{(n+k)!k!}\,\left(\frac{x}{2}\right)^{n+2k},
\end{align}
where $I_n(x)$ is the modified Bessel function of the first kind. For the further discussion, we will assume that $\beta_3>0$ that translates, in terms of scaling dimensions, in the condition $\D_3>\sdfrac{d}{2}$. The latter is always satisfied when the correlation function is constructed with conserved currents and stress-energy tensor. It is worth noting that the relation $K_{-\nu}(x)=K_{\nu}(x)$ is valid in any case. 
With the assumptions discussed above, we can compute the zero-momentum limit of the triple-K integrals as
\begin{equation}
\lim_{p_3\to0}I_{\a\{\b_j\}}(p,p,p_3)=2^{\b_3-1}\G(\b_3)p^{\b_1+\b_2}\int_0^\infty\,dx\,x^{\,\a-\b_3}\,K_{\b_1}(px)\,K_{\b_2}(px),
\end{equation}
and using the relation in \cite{Prudnikov} 
\begin{align}
\int_0^\infty\,dx\,x^{\,\a-1}\,K_{\n}(cx)\,K_{\m}(cx)&=\sdfrac{2^{\a-3}}{\G(\a)\,c^\a}\,\Gamma\left(\frac{\alpha+\mu+\nu}{2}\right)\Gamma\left(\frac{\alpha+\mu-\nu}{2}\right)\Gamma\left(\frac{\alpha-\mu+\nu}{2}\right)\Gamma\left(\frac{\alpha-\mu-\nu}{2}\right),\notag\\
&\text{with}\quad \Re(c)>0,\,\Re(\alpha)>|\Re(\mu)|+|\Re(\nu)|,
\end{align}
we obtain the final result
\begin{equation}
\lim_{p_3\to0}I_{\a\{\b_j\}}(p,p,p_3)=p^{\b_t-\a-1}\ \ell_{\a\{\b_j\}},\label{zeromom}
\end{equation}
where
\begin{equation}
\ell_{\a\{\b_j\}}=\sdfrac{2^{\alpha-3}\Gamma(\beta_3)}{\G(\a-\beta_3+1)}\,\Gamma\left(\frac{\alpha+\beta_t+1}{2}-\beta_3\right)\Gamma\left(\frac{\alpha-\beta_t+1}{2}+\beta_1\right)\Gamma\left(\frac{\alpha-\beta_t+1}{2}+\beta_2\right)\Gamma\left(\frac{\alpha-\beta_t+1}{2}\right)\label{ldef},
\end{equation}
which is valid away from poles of the gamma function with the conditions $\a>\b_t-1$, $p>0$. 
%%%%%%%%%%%%%%%%%%%%%%%%%%%%%%%%%%%%%%%%%%%%%%%%%%%%%%%%%%%%
\subsection{Solving the secondary equations}
%%%%%%%%%%%%%%%%%%%%%%%%%%%%%%%%%%%%%%%%%%%%%%%%%%%%%%%%%%%%
As previously mentioned, the secondary CWI's allow to establish a relation among all the constants in the solutions of the primary CWI's. For this reason, we perform the zero momentum limit in order to extract these algebraic equations involving all the constants. In this section we illustrate this procedure for the $\braket{TTO}$. \\
First of all we have to explicit the secondary CWI's \eqref{secondary} using the expression of the form factors given by the primary \eqref{primarysolution}, and using the relations \eqref{1dJ}-\eqref{Fund} we obtain
\begin{align}
0&=L_2A_2+4RA_3=\a_2\bigg[-p_1^2(p_1^2+p_2^2-p_3^2)\,I_{\frac{d}{2}+2\,\left\{\frac{d}{2}-1,\,\frac{d}{2},\,\D_3-\frac{d}{2}\right\}}-2p_1^2p_2^2\,I_{\frac{d}{2}+2\,\left\{\frac{d}{2},\,\frac{d}{2}-1,\,\D_3-\frac{d}{2}\right\}}\notag\\
&+d(p_3^2-p_2^2-p_1^2)\,I_{\frac{d}{2}+1\,\left\{\frac{d}{2},\,\frac{d}{2},\,\D_3-\frac{d}{2}\right\}}+2p_1^2\,I_{\frac{d}{2}+1\,\left\{\frac{d}{2},\,\frac{d}{2},\,\D_3-\frac{d}{2}\right\}}\bigg]
+4\a_1\bigg[-p_1^2(p_1^2+p_2^2-p_3^2)\,I_{\frac{d}{2}+3\,\left\{\frac{d}{2}-1,\,\frac{d}{2},\,\D_3-\frac{d}{2}+1\right\}}\notag\\
&-2p_1^2p_2^2\,I_{\frac{d}{2}+3\,\left\{\frac{d}{2},\,\frac{d}{2}-1,\,\D_3-\frac{d}{2}+1\right\}}+d(p_3^2-p_2^2-p_1^2)\,I_{\frac{d}{2}+2\,\left\{\frac{d}{2},\,\frac{d}{2},\,\D_3-\frac{d}{2}+1\right\}}+2p_1^2\,I_{\frac{d}{2}+2\,\left\{\frac{d}{2},\,\frac{d}{2},\,\D_3-\frac{d}{2}+1\right\}}\bigg]\notag\\
&+4\a_3\,\bigg(-p_1^2\,I_{\frac{d}{2}\,\left\{\frac{d}{2}-1,\,\frac{d}{2},\,\D_3-\frac{d}{2}\right\}}-d\,I_{\frac{d}{2}-1\,\left\{\frac{d}{2},\,\frac{d}{2},\,\D_3-\frac{d}{2}\right\}}\bigg)+4\a_2\,\left(-p_1^2\,I_{\frac{d}{2}+1\,\left\{\frac{d}{2}-1,\,\frac{d}{2},\,\D_3-\frac{d}{2}+1\right\}}-d\,I_{\frac{d}{2}\,\left\{\frac{d}{2},\,\frac{d}{2},\,\D_3-\frac{d}{2}+1\right\}}\right)\notag\\
&+8\a_1\,\left(-p_1^2\,I_{\frac{d}{2}+2\,\left\{\frac{d}{2}-1,\,\frac{d}{2},\,\D_3-\frac{d}{2}+2\right\}}-d\,I_{\frac{d}{2}+1\,\left\{\frac{d}{2},\,\frac{d}{2},\,\D_3-\frac{d}{2}+2\right\}}\right),\label{firstsecondary}
\end{align}
where we have used the definition of the reduced version of the triple-K integral 
\begin{equation}
J_{N\{k_j\}}=I_{\frac{d}{2}-1+N\{\D_j-\frac{d}{2}+k_j\}}.
\end{equation}
Equally, the other secondary CWI's will have the form
\begin{align}
0&=L_2A_1+RA_2\notag\\
&=\a_1\bigg[-p_1^2(p_1^2+p_2^2-p_3^2)\,I_{\frac{d}{2}+4\,\left\{\frac{d}{2}-1,\,\frac{d}{2},\,\D_3-\frac{d}{2}\right\}}-2p_1^2p_2^2\,I_{\frac{d}{2}+4\,\left\{\frac{d}{2},\,\frac{d}{2}-1,\,\D_3-\frac{d}{2}\right\}}+d(p_3^2-p_2^2-p_1^2)\,I_{\frac{d}{2}+3\,\left\{\frac{d}{2},\,\frac{d}{2},\,\D_3-\frac{d}{2}\right\}}\notag\\
&\hspace{1.6cm}+2p_1^2\,I_{\frac{d}{2}+3\,\left\{\frac{d}{2},\,\frac{d}{2},\,\D_3-\frac{d}{2}\right\}}\bigg] +\a_2\left(-p_1^2\,I_{\frac{d}{2}+2\,\left\{\frac{d}{2}-1,\,\frac{d}{2},\,\D_3-\frac{d}{2}\right\}}-d\,I_{\frac{d}{2}+1\,\left\{\frac{d}{2},\,\frac{d}{2},\,\D_3-\frac{d}{2}\right\}}\right)\notag\\
&\hspace{0.7cm}+4\a_1\,\left(-p_1^2\,I_{\frac{d}{2}+3\,\left\{\frac{d}{2}-1,\,\frac{d}{2},\,\D_3-\frac{d}{2}+1\right\}}-d\,I_{\frac{d}{2}+2\,\left\{\frac{d}{2},\,\frac{d}{2},\,\D_3-\frac{d}{2}+1\right\}}\right)\ .\label{secondsecondary}
\end{align}
We now consider the zero-momentum limit \eqref{zeromom} of these equations, obtaining the relation
	\begin{align}
	0&=\lim_{p_3\to0}L_2A_2+4RA_3\notag\\[1.5ex]
	&=-2p^{\D_3}\bigg\{ 4\a_1	 \bigg[d\,\ell_{\frac{d}{2}+1\,\left\{\frac{d}{2},\,\frac{d}{2},\,\D_3-\frac{d}{2}+2\right\}}+\,\ell_{\frac{d}{2}+2\,\left\{\frac{d}{2}-1,\,\frac{d}{2},\,\D_3-\frac{d}{2}+2\right\}}+\,\ell_{\frac{d}{2}+3\,\left\{\frac{d}{2}-1,\,\frac{d}{2},\,\D_3-\frac{d}{2}+1\right\}}+\,\ell_{\frac{d}{2}+3\,\left\{\frac{d}{2},\,\frac{d}{2}-1,\,\D_3-\frac{d}{2}+1\right\}}\notag\\
	&+(d-1)\,\ell_{\frac{d}{2}+2\,\left\{\frac{d}{2},\,\frac{d}{2},\,\D_3-\frac{d}{2}+1\right\}}\bigg]\,+\,\a_2\bigg[\,2\,\ell_{\frac{d}{2}+1,\left\{\frac{d}{2}-1,\,\frac{d}{2}-1,\,\D_3-\frac{d}{2}+1\right\}}+\,\ell_{\frac{d}{2}+2\,\left\{\frac{d}{2}-1,\,\frac{d}{2},\,\D_3-\frac{d}{2}\right\}}+\,\ell_{\frac{d}{2}+2\,\left\{\frac{d}{2},\,\frac{d}{2}-1,\,\D_3-\frac{d}{2}\right\}}\notag\\
	&+(d-1)\,\ell_{\frac{d}{2}+1\,\left\{\frac{d}{2},\,\frac{d}{2},\,\D_3-\frac{d}{2}\right\}}+2d\,\ell_{\frac{d}{2}\,\left\{\frac{d}{2},\,\frac{d}{2},\,\D_3-\frac{d}{2}+1\right\}}\bigg]\,+\,2\a_3\,\bigg[d\,\ell_{\frac{d}{2}-1\,\left\{\frac{d}{2},\,\frac{d}{2},\,\D_3-\frac{d}{2}\right\}}+\,\ell_{\frac{d}{2}\,\left\{\frac{d}{2}-1,\,\frac{d}{2},\,\D_3-\frac{d}{2}\right\}}\bigg]\ \bigg\}\label{firstE},\\[1.7ex]
	0&=\lim_{p_3\to0}L_2A_1+RA_2\notag\\[1.5ex]
	&=p^{\D_3-2}\bigg\{\, -2\a_1\,\bigg[\,2d\,\ell_{\frac{d}{2}+2\,\left\{\frac{d}{2},\,\frac{d}{2},\,\D_3-\frac{d}{2}+1\right\}}+2\,\ell_{\frac{d}{2}+3\,\left\{\frac{d}{2}-1,\,\frac{d}{2},\,\D_3-\frac{d}{2}+1\right\}}+(d-1)\,\ell_{\frac{d}{2}+3\,\left\{\frac{d}{2},\,\frac{d}{2},\,\D_3-\frac{d}{2}\right\}}\notag\\
	&\qquad+\,\ell_{\frac{d}{2}+4\,\left\{\frac{d}{2}-1,\,\frac{d}{2},\,\D_3-\frac{d}{2}\right\}}+\,\ell_{\frac{d}{2}+4\,\left\{\frac{d}{2},\,\frac{d}{2}-1,\,\D_3-\frac{d}{2}\right\}}\bigg]\,-\,\a_2\,\bigg[\,d\,\ell_{\frac{d}{2}+1\,\left\{\frac{d}{2},\,\frac{d}{2},\,\D_3-\frac{d}{2}\right\}}+\,\ell_{\frac{d}{2}+2\,\left\{\frac{d}{2}-1,\,\frac{d}{2},\,\D_3-\frac{d}{2}\right\}}\bigg]\ \bigg\}.\label{secondE}
	\end{align}
The two secondary CWI's in \eqref{firstE} and \eqref{secondE} scale homogeneously as $p^{\D_3}$ and $p^{\D_3-2}$ respectively. Thus, expanding the result, by using the definition \eqref{ldef}, we derive the relations
\begin{align}
0&=\lim_{p_3\to0}L_2A_2+4RA_3=p^{\D_3}\,2^{\frac{d}{2}-4}\,\G\left(-\sdfrac{\D_3}{2}\right)\,\left[\G\left(\sdfrac{d-\D_3}{2}\right)\right]^2\,\G\left(d-\sdfrac{\D_3}{2}\right)\,\G\left(\D_3-\sdfrac{d}{2}\right)\sdfrac{1}{\G\left(d-\D_3\right)}\notag\\
&\hspace{-0.4cm}\times\bigg\{-2\a_3(2d-\D_3)-2\a_1(2d-\D_3)(\D_3+2)(d-\D_3-2)(d-2\D_3)+\a_2\left(d-\sdfrac{\D_3}{2}\right)\big(d(\D_3+4)-\D_3(\D_3+8)\big)\bigg\},\\
0&=\lim_{p_3\to0}L_2A_1+RA_2=p^{\D_3-2}\,2^{\frac{d}{2}-4}\,\G\left(1-\sdfrac{\D_3}{2}\right)\left[\G\left(\sdfrac{d-\D_3}{2}\right)\right]^2\G\left(d-\sdfrac{\D_3}{2}+2\right)\G\left(\D_3-\sdfrac{d}{2}\right)\sdfrac{(d-\D_3)^2}{\G(d-\D_3+2)}\notag\\
&\times\bigg\{\ \a_1(\D_3+2)(d-\D_3-2)-\a_2\bigg\}.
\end{align}
In order to satisfy the above equations, the only way is to set the coefficient in curly brackets to vanish, obtaining
\begin{align}
\a_2&=\a_1(\D_3+2)(d-\D_3-2),\\
\a_3&=\sdfrac{\a_1}{4}\D_3(\D_3+2)(d-\D_3-2)(d-\D_3).
\end{align}
From this analysis we observe that the final solution depends only on a single constant factor $\a_1$. 

%%%%%%%%%%%%%%%%%%%%%%%%%%%%%%%%%%%%%%%%%%%%%%%%%%%%%%%%%%%%
\subsection{Regularisation and renormalization}\label{reg}
%%%%%%%%%%%%%%%%%%%%%%%%%%%%%%%%%%%%%%%%%%%%%%%%%%%%%%%%%%%%
{ In this section we briefly discuss the procedure of regularization of the triple-K integrals when some divergences occur. For more details about this procedure see \cite{Bzowski:2015yxv,Bzowski:2014qja, Bzowski:2017poo}.
	
We will now focus on the special cases where the triple-K integral is singular, i.e., when the dimensions of operators satisfy one or more of the conditions
\begin{equation}
\a+1\pm\b_1\pm\b_2\pm\b_3=-2k,\label{cond}
\end{equation}
for some non-negative integer $k$. In this region of the parameters, the general triple-K integral defined in \eqref{3Kint} diverges and we have to take in account a regularization procedure. In order to do that, we introduce the regulated parameters}
\begin{equation}
\a\mapsto\bar{\a}=\a+u\e,\qquad \b_j\mapsto\bar{\b_j}=\b_j+v_j\e,\label{shift}
\end{equation}
where $u$ and $v_j$, $j=1,2,3$ are fixed but arbitrary numbers that specify the direction of the shift. In this way we regard the regulated triple-K integral 
\begin{equation}
I_{\a\{\b_k\}}(p_1,p_2,p_3)\mapsto I_{\bar{\a}\{\bar{\b_k}\}}(p_1,p_2,p_3),
\end{equation}
as a function of the regulator $\e$ with all the momenta fixed. The divergences of the integrals manifest as a pole at $\e=0$. { It is worth mentioning that not all the choices of the shifts $u$, $v_j$ actually regulate the integral, but some useful choices exist. For instance, if we consider the scheme $u=v_1=v_2=v_3$, this is equivalent to perform a shift of the physical dimensions of the same quantity $2u\e$. In fact, the general mapping \eqref{shift} can be a viewed as shift of the spacetime and conformal dimensions as
\begin{equation}
d\mapsto d+2u\e,\qquad \D_j\mapsto\D_j+(u+v_j)\e,\ \ j=1,2,3,
\end{equation}
where with the choice mentioned above we have an equal scaling of $d$ and $\Delta_j$. Another particular and useful choice is to consider $v_j=0$ for all $j=1,2,3$. This scheme preserves the $\b_j$ and hence the indices of the Bessel functions in the triple-K integral, and then the expansion to extract the pole structures in $\e$ will be easier to perform.\\ 
We now consider the case of the $\braket{TT\mO}$ correlation function. The solutions for the form factors from the conformal Ward identities are written in terms of triple-K integrals as in  \eqref{primarysolution}. Looking at the first form factor $A_1$, given just in terms of $J_{4\{000\}}$, we observe that it can diverge, by using \eqref{cond}, if one or more of the conditions
\begin{equation}
\frac{d}{2}+4\pm\frac{d}{2}\pm\frac{d}{2}\pm\left(\D_3-\frac{d}{2}\right)=-2k,\qquad k\in\mathbb{Z}^+\label{diverge}
\end{equation}
are satisfied. One can use a special notation in order to refer to all the cases in which a singularity shows up. If we denote by $(\pm\pm\pm)$ the possible choices of the signs in \eqref{diverge}, there are six cases in total. However, in order to have a negative even number on the LHS of \eqref{diverge}, there are just two possible choices of signs, which are $(++-)$ and $(+++)$, with
\begin{equation}
\d_{+}=4-d+\D_3,\qquad\d_{-}=4-\D_3,\label{delta+}
\end{equation}
where $\d$ is the left-hand side of \eqref{diverge} and the subscript $\pm$ indicate the sign choice  of the $\D_3-d/2$ term. We observe that in the case $\D_3=4+2k$ or $\D_3=d-4-2k$ the particular triple-K integral $J_{4\{000\}}$ has divergences and needs to be regulated. One needs to repeat this analysis for all the integrals appearing in the final form of all the form factors.\\
To present a practical example, we consider the scalar operator in the three-point function $\braket{TT\mO}$ with conformal dimension $\D_3=1$ and in a $d=3$ dimensional CFT. Under these conditions, one can immediately see that the form factor $A_1$ does not contain divergences and indeed by using \eqref{delta+} we find $\delta_{\pm}\ne-2k$, with $k\in\mathbb{Z}^+$. The same does not happen for the other two form factors $A_2$ and $A_3$ in \eqref{primarysolution}.   
Using the results in \cite{Bzowski:2015yxv, Bzowski:2013sza} we see that the only convergent integrals are $J_{4\{000\}}=I_{\frac{9}{2}\left\{\frac{3}{2},\frac{3}{2},-\frac{1}{2}\right\}}$ and $J_{3\{001\}}=I_{\frac{7}{2}\left\{\frac{3}{2},\frac{3}{2},\frac{1}{2}\right\}}$. The remaining integrals require a regularization procedure. To do this, we choose the particular scheme $u=1$ and $v_j=0$, $j=1,2,3$ for which we can calculate all the shifted integrals of the form $J_{N+\e\{k_j\}}$ using \eqref{halfinteg}, and we can expand the result around $\e=0$. For instance, if we take the integral
	\begin{equation}
	J_{2+\e\{000\}}=I_{\frac{5}{2}+\e\left\{\frac{3}{2},\frac{3}{2},-\frac{1}{2}\right\}}=\left(\frac{\pi}{2}\right)^{\frac{3}{2}}\frac{1}{\e\,p_3}+O(\e^0),
	\end{equation}
since the regulator $\e$ in the triple-K integrals cannot be removed, we cannot easily take the zero-momentum limit to resolve the secondary CWI's. In order to obtain a finite expression of the form factors, one has to assume that the constants $\a_j$ in \eqref{primarysolution} depend on the regulator $\e$ as well. We make explicit the $\e$ dependence of the coefficient $\a_j$ by the definition
\begin{equation}
\a_j=\sum_{n=-\infty}^{\infty}\,\a_j^{(n)}\e^n,\qquad j=1,2,3\label{expans}.
\end{equation}
However, the constant $\a_1$ does not depend on the regulator because it multiplies an integral which is already finite, and for this reason we will choose $\a_1=\a^{(0)}_1$. In the next section we will present all the details of the analysis of the secondary CWI's for the case $\braket{TT\mO}$ when there are divergences. 
}

\subsection{Secondary CWI's with divergent triple-K integral} 
We analyse in details the method to extract information from the secondary CWI's when the relation \eqref{cond} holds and the triple-K integrals manifest a singular behavior. 
Consider the solution to the primary CWI's for the $\braket{TT\mO}$ point function 
	\begin{align}
	A_1&=\a_1\,J_{4\{000\}},\\
	A_2&=\a_2\,J_{2\{000\}}+4\a_{1}\,J_{3\{001\}},\\
	A_3&=\a_3\,J_{0\{000\}}+\a_{2}\,J_{1\{001\}}+2\a_{1}\,J_{2\{002\}},\label{TTOsol}
	\end{align}
and the secondary CWI's
\begin{align}
L_2A_1+RA_2=0,\qquad L_2A_2+4RA_3=0.
\end{align}
We consider the case previously mentioned with $\D_3=1$, $d=3$, recalling that the secondary CWI's are}
\begin{align}
0&=L_2A_1+RA_2,\\
0&=L_2A_2+4RA_3.
\end{align}
{
These form factors contain the integrals $J_{2\{000\}}$, $J_{1\{001\}}$, $J_{0\{000\}}$, $J_{1\{-100\}}$ that are divergent and can be regulated by using the prescription in \appref{reg}. Then, expanding in power of $\epsilon$ we find
\begin{equation}
	\begin{split}
	J_{2+\e\{002\}}&=I_{\frac{5}{2}+\e\left\{\frac{3}{2},\frac{3}{2},\frac{3}{2}\right\}}=-\left(\sdfrac{\pi}{2}\right)^\frac{3}{2}\sdfrac{p_1^3+2p_1^2(p_2+p_3)+2p_1(p_2^2+p_2\,p_3+p_3^2)+(p_2+p_3)(p_2^2+p_2p_3+p_3^2)}{(p_1+p_2+p_3)^2}+O(\e),\\
	J_{2+\e\{000\}}&=I_{\frac{5}{2}+\e\left\{\frac{3}{2},\frac{3}{2},-\frac{1}{2}\right\}}=\left(\sdfrac{\pi}{2}\right)^\frac{3}{2}\sdfrac{1}{\e p_3}+O(\e^0),\\
	J_{1+\e\{001\}}&=I_{\frac{3}{2}+\e\left\{\frac{3}{2},\frac{3}{2},\frac{1}{2}\right\}}=-\left(\sdfrac{\pi}{2}\right)^\frac{3}{2}\sdfrac{p_3}{\e}+O(\e^0),\\
	J_{0+\e\{000\}}&=I_{\frac{1}{2}+\e\left\{\frac{3}{2},\frac{3}{2},-\frac{1}{2}\right\}}=-\left(\sdfrac{\pi}{2}\right)^\frac{3}{2}\sdfrac{p_1^2+p_2^2-p_3^2}{2\e p_3}+O(\e^0).
	\end{split}\label{integrals}.
\end{equation}
All the other integrals in the form factors are convergent and are explicitly given by
\begin{align}
J_{4\{000\}}&=I_{\frac{9}{2}\left\{\frac{3}{2},\frac{3}{2},-\frac{1}{2}\right\}}=\left(\sdfrac{\pi}{2}\right)^\frac{3}{2}\sdfrac{3p_1^3+4p_1(3p_2+p_3)+(p_2+p_3)(3p_2+p_3)}{(p_1+p_2+p_3)^4}\label{J4000},\\
J_{3\{001\}}&=I_{\frac{7}{2}\left\{\frac{3}{2},\frac{3}{2},\frac{1}{2}\right\}}=\left(\sdfrac{\pi}{2}\right)^\frac{3}{2}\sdfrac{(p_1+p_2+p_3)(p_1+2p_2+p_3)+p_1(p_1+3p_2+p_3)}{(p_1+p_2+p_3)^3}\label{J3001}.
\end{align}
Using the explicit expressions written above, the secondary CWI's turn into 
\begin{align}
0&=L_2A_1+RA_2\notag\\
&=-\a_1\left(\sdfrac{\pi}{2}\right)^\frac{3}{2}\sdfrac{9}{p_3}-\a_2\left(\sdfrac{\pi}{2}\right)^\frac{3}{2}\sdfrac{p_1^2(p_1+3p_2+p_3)}{p_3\,(p_1+p_2+p_3)^3}\notag\\
&\quad-3\left(\sdfrac{\pi}{2}\right)^\frac{3}{2}\frac{\a_2}{p_3}\left(\sdfrac{(p_1+p_2)(p_1+p_2+p_3)+p_1p_2)}{(p_1+p_2+p_3)^2}-\log\left(p_1+p_2+p_3\right)+\frac{1}{\e}-\gamma_E\right)+O(\e)\label{F7},\\
0&=L_2A_2+4RA_3\notag\\[2ex]
&=36\a_1\left(\sdfrac{\p}{2}\right)^\frac{3}{2}\,p_3+\frac{\alpha_2}{p_3}\left(\sdfrac{\p}{2}\right)^\frac{3}{2}\bigg[\frac{(15p_3^2-3p_2^2-p_1^2)}{\e}+f_1\Big(p_1,p_2,p_3,\g_E\Big)\bigg]\notag\\
&\qquad+\alpha_3\left(\sdfrac{\p}{2}\right)^\frac{3}{2}\Bigg[\frac{2(p_1^2+3p_2^2-3p_3^2)}{p_3\,\e}+f_2\Big(p_1,p_2,p_3,\g_E\Big)\Bigg]+O(\e)\label{F8},
\end{align}
where $f_1$ and $f_2$ are two analytic functions of the momenta in the physical region $p_1+p_2+p_3>0$, with $\gamma_E$ denoting the Euler constant. Notice that  
the $f'$s will not be relevant in the discussion that will follow, since they will be multiplied by positive powers of the regulator $\epsilon$.
In order to take care of the divergences we make explicit the $\e$ dependence of the coefficients $\alpha_j$ through the definition \eqref{expans}. Referring to the solution \eqref{TTOsol} of the form factors, we observe that they must not depend on the regulator $\e$. Therefore we impose the conditions \cite{Bzowski:2013sza}
\begin{equation}
\begin{split}
&\a_1=\a_1^{(0)},\\
&\a_2^{(n)}=0,\\
&\a_3^{(n)}=0,
\end{split}
\qquad n=-1,-2,-3,\dots
\end{equation}
Then, collecting all the terms in \eqref{F7} with respect to the powers of $\e$ and imposing that they are zero order by order we find the conditions
\begin{align}
\a_2^{(0)}&=0,&&\text{$\e^{-1}$ order}\label{Cond1},\\
\a_2^{(1)}&=-3\a_1^{(0)},&&\text{$\e^0$ order}\label{Cond2},\\
&\dots\,,\label{Condn}
\end{align}
where we have defined $p_{123}=p_1+p_2+p_3$. 
The same procedure may be applied to the remaining secondary CWI's \eqref{F8} giving the conditions
\begin{align}
\a_3^{(0)}&=0,&&\text{$\e^{-1}$ order}\label{Cond10},\\
\a_3^{(1)}&=-\frac{3}{2}\a_1^{(0)},&&\text{$\e^{0}$ order}\label{Cond11},\\
&\dots,
\end{align}
where we have used the results in \eqref{Cond1}-\eqref{Condn}. It is worth mentioning that the coefficients $\alpha_2^{(k)},\,\alpha_3^{(k)}$, $k=2,3,\dots$, do not contribute to the final regularized expression of the form factors when the regulator $\e$ is set to zero. Finally, we can give the regulated expression for the form factors in $d=3$ and $\Delta=1$ as
\begin{align}
A_1&=\a_1^{(0)}\,J_{4\{000\}},\\
A_2&=-3\a_2^{(0)}\,J^{Reg}_{2\{000\}}+4\a_{1}^{(0)}\,J_{3\{001\}},\\
A_3&=-\sdfrac{3}{2}\a_1^{(0)}\,J^{Reg}_{0\{000\}}-3\a_{1}^{(0)}\,J^{Reg}_{1\{001\}}+2\a_{1}^{(0)}\,J^{Reg}_{2\{002\}},
\end{align}
where $J_{4\{000\}}$ and $J_{3\{000\}}$ are given by \eqref{J4000} and \eqref{J3001} respectively, and the other regularized integrals are written explicitly as
\begin{align}
J^{Reg}_{2\{002\}}&=-\left(\sdfrac{\pi}{2}\right)^\frac{3}{2}\sdfrac{p_1^3+2p_1^2(p_2+p_3)+2p_1(p_2^2+p_2\,p_3+p_3^2)+(p_2+p_3)(p_2^2+p_2p_3+p_3^2)}{(p_1+p_2+p_3)^2},\\
J^{Reg}_{2\{000\}}&=\left(\sdfrac{\pi}{2}\right)^\frac{3}{2}\sdfrac{1}{p_3},\\
J^{Reg}_{0\{000\}}&=-\left(\sdfrac{\pi}{2}\right)^\frac{3}{2}\,p_3,\\
J^{Reg}_{1\{001\}}&=-\left(\sdfrac{\pi}{2}\right)^\frac{3}{2}\sdfrac{p_1^2+p_2^2-p_3^2}{2p_3}.
\end{align}
Also in this case all the form factors are completely fixed modulo an overall constant $\alpha_1^{(0)}$. We notice that these results are in agreement with the relations between the constants obtained in the general case 
\begin{align}
\a_2&=\a_1(\D_3+2)(d-\D_3-2),\\
\a_3&=\sdfrac{\a_1}{4}\D_3(\D_3+2)(d-\D_3-2)(d-\D_3).
\end{align}
With the conditions $\Delta_3=1$ and $d=3+\e$, they are written as
\begin{equation}
\a_2=-3\e\,\a_1^{(0)},\qquad\a_3=-\sdfrac{3}{2}\,\e\,\a_1^{(0)},
\end{equation}
which, order by order in $\e$, give the same constraints \eqref{Cond1}-\eqref{Cond2} and \eqref{Cond10}-\eqref{Cond11} discussed in this section. 
We have illustrated a method \cite{Bzowski:2013sza} for extracting the algebraic dependencies among the integration constants, using the information of the primary and secondary CWI's. The analysis has been performed in the general case, where the regulator can be removed from all the triple-K integrals involved since we are simply avoiding unphysical singularities. In the following section we will discuss the case in which the regulator cannot be removed, which takes to the ordinary regularization of the integrals and to the generation of the conformal anomaly.

As we have just discussed, the primary CWI's can be solved using triple-K integrals and the solutions are substituted into the secondary CWI's. In general, the secondary CWI's lead to linear algebraic equations between the various constants appearing in the solutions of the primary conformal Ward identities. The precise form of the secondary CWI's depends on the information provided by the transverse Ward identities. In the case of the $\braket{TT\mO}$ we have shown that the transverse Ward identity does not contribute to the secondary CWI's. In general this is not the case in the $\braket{TTT}$. 

Furthermore, we have also illustrated how to extract a set of algebraic equations for the constants, obtained from the secondary CWI's. This set of equations may be extracted through an analysis of the zero-momentum limit of the same equations when the regulator can be removed from all triple-K integrals. Finally, we have also shown how to regulate the integrals when the zero-momentum limit generates divergences. This procedure brought to another set of equations that, once solved, give relations among the constants. At the end, the final result of the three-point  function is unique up to one overall constant, which can be matched to a free field theory with a particular particle content. This approach is fully addressed in the case of the $TTT$ correlator in \appref{secTTTgen}.

%%%%%%%%%%%%%%%%%%%%%%%%%%%%%%%%%%%%%%%%%%%%%%%%% 
\section{The  \texorpdfstring{$TTT$}{ttt} }\label{secTTTgen}
%%%%%%%%%%%%%%%%%%%%%%%%%%%%%%%%%%%%%%%%%%%%%%%%%
\subsection{Reconstruction method and primary CWI's}
We start with the analysis of the transverse and trace Ward identities. These relations follow from the analysis in \appref{TraceTransverse}. By taking two functional derivatives of \eqref{transverse1},\eqref{trace} and then in the limit $g_{\m\n}=\d_{\m\n}$ we obtain the canonical Ward identities for the $\braket{TTT}$ in position space
	\begin{align}
	\partial_{\n_1}\braket{T^{\m_1\n_1}(p_1)T^{\m_2\n_2}(p_2)T^{\m_3\n_3}(p_3)}&=\big[\partial_{\,\n_1}\d(x_1-x_2)\big]\mathcal{B}^{\m_1\n_1\m_2\n_2}_{\hspace{1.1cm}\a\b}\braket{T^{\a\b}(x_1)T^{\m_3\n_3}(x_3)}\notag\\
	&+\big[\partial_{\n_1}\d(x_1-x_3)\big]\mathcal{B}^{\m_1\n_1\m_3\n_3}_{\hspace{1.1cm}\a\b}\braket{T^{\a\b}(x_1)T^{\m_2\n_2}(x_2)},\notag\\
	\d_{\m_1\n_1}\braket{T^{\m_1\n_1}(p_1)T^{\m_2\n_2}(p_2)T^{\m_3\n_3}(p_3)}&=-2\braket{T^{\m_2\n_2}(p_2+p_1)T^{\m_3\n_3}(x_3)}-2\braket{T^{\m_2\n_2}(p_2)T^{\m_3\n_3}(p_3+p_1)},\notag
	\end{align}
where $\mathcal B$ is a constant tensor defined as
\begin{equation}
\mathcal{B}^{\m_1\n_1\m_j\n_j}_{\hspace{1.2cm}\a\b}\equiv -2\, \d^{\m_1(\m_j}  \d{}^{\n_j)}{}_{(\a}\d^{\n_1}{}_{\b)} 
+ \d^{\m_1\n_1}\,\d^{(\m_j}{}_{\a}  \d^{\n_j)}{}_{\b}.
\end{equation}

By Fourier transforming we obtain these canonical Ward identities in momentum space as
	\begin{align}
	p_{1\n_1}\braket{T^{\m_1\n_1}(p_1)T^{\m_2\n_2}(p_2)T^{\m_3\n_3}(p_3)}&=p_{2\,\n_1}\mathcal{B}^{\m_1\n_1\m_2\n_2}_{\hspace{1.1cm}\a\b}\braket{T^{\a\b}(p_1+p_2)T^{\m_3\n_3}(p_3)}\notag\\
	&+p_{3\n_1}\mathcal{B}^{\m_1\n_1\m_3\n_3}_{\hspace{1.1cm}\a\b}\braket{T^{\a\b}(p_1+p_3)T^{\m_2\n_2}(p_2)}\label{transvttt},\\
	\d_{\m_1\n_1}\braket{T^{\m_1\n_1}(p_1)T^{\m_2\n_2}(p_2)T^{\m_3\n_3}(p_3)}&=-2\braket{T^{\m_2\n_2}(p_2+p_1)T^{\m_3\n_3}(x_3)}-2\braket{T^{\m_2\n_2}(p_2)T^{\m_3\n_3}(p_3+p_1)}\label{tracettt},
	\end{align}
{
We can now apply the general method presented above in order to reconstruct the $\braket{T^{\m_1\n_1}T^{\m_2\n_2}T^{\m_3\n_3}}$. For the decomposition of the transverse and traceless part of this correlation function we follow the analysis of \appref{tensstruc}. The decomposition of the $TTT$ can be written as
\begin{align}
\braket{T^{\mu_1\nu_1}\,T^{\mu_2\n_2}\,T^{\mu_3\n_3}}&=\braket{t^{\mu_1\nu_1}\,t^{\mu_2\n_2}\,t^{\mu_3\n_3}}+\braket{T^{\mu_1\nu_1}\,T^{\mu_2\n_2}\,t_{loc}^{\mu_3\n_3}}+\braket{T^{\mu_1\nu_1}\,t_{loc}^{\mu_2\n_2}\,T^{\mu_3\n_3}}\notag\\
&+\braket{t_{loc}^{\mu_1\nu_1}\,T^{\mu_2\n_2}\,T^{\mu_3\n_3}}-\braket{T^{\mu_1\nu_1}\,t_{loc}^{\mu_2\n_2}\,t_{loc}^{\mu_3\n_3}}-\braket{t_{loc}^{\mu_1\nu_1}\,t_{loc}^{\mu_2\n_2}\,T^{\mu_3\n_3}}\notag\\
&-\braket{t_{loc}^{\mu_1\nu_1}\,T^{\mu_2\n_2}\,t_{loc}^{\mu_3\n_3}}+\braket{t_{loc}^{\mu_1\nu_1}\,t_{loc}^{\mu_2\n_2}\,t_{loc}^{\mu_3\n_3}},\label{reconstrTTT}
\end{align}
where the transverse traceless part consists of five form factors,
\begin{align}
&\braket{t^{\m_1\n_1}(p_1)t^{\m_2\n_2}(p_2)t^{\m_3\n_3}(p_3)}
= \Pi^{\m_1\n_1}_{\a_1\b_1}(p_1)
\Pi^{\m_2\n_2}_{\a_2\b_2}(p_2)\Pi^{\m_3\n_3}_{\a_3\b_3}(p_3)\notag\\
&\hspace{0.7cm}\times\Big[ A_1\,p_2^{\a_1} p_2^{\b_1} p_3^{\a_2} p_3^{\b_2} p_1^{\a_3} p_1^{\b_3}+ A_2\,\d^{\b_1\b_2} p_2^{\a_1} p_3^{\a_2} p_1^{\a_3} p_1^{\b_3} 
+ A_2\,(p_1 \leftrightarrow p_3)\, \d^{\b_2\b_3}  p_3^{\a_2} p_1^{\a_3} p_2^{\a_1} p_2^{\b_1} \notag\\
&\hspace{1.3cm}+ A_2\,(p_2\leftrightarrow p_3)\, \d^{\b_3\b_1} p_1^{\a_3} p_2^{\a_1}  p_3^{\a_2} p_3^{\b_2}+ A_3\,\d^{\a_1\a_2} \d^{\b_1\b_2}  p_1^{\a_3} p_1^{\b_3} + A_3(p_1\leftrightarrow p_3)\,\d^{\a_2\a_3} \d^{\b_2\b_3}  p_2^{\a_1} p_2^{\b_1} \notag\\
&\hspace{2cm}
+ A_3(p_2\leftrightarrow p_3)\,\d^{\a_3\a_1} \d^{\b_3\b_1}  p_3^{\a_2} p_3^{\b_2} + A_4\,\d^{\a_1\a_3} \d^{\a_2\b_3}  p_2^{\b_1} p_3^{\b_2} + A_4(p_1\leftrightarrow p_3)\, \d^{\a_2\a_1} \d^{\a_3\b_1}  p_3^{\b_2} p_1^{\b_3} \notag\\
&\hspace{3.5cm}+ A_4(p_2\leftrightarrow p_3)\, \d^{\a_3\a_2} \d^{\a_1\b_2}  p_1^{\b_3} p_2^{\b_1} + A_5  \d^{\a_1\b_2}  \d^{\a_2\b_3}  \d^{\a_3\b_1}\Big],\label{tttdec}
\end{align}
and all the local parts depend on the two-point functions $TT$ via the canonical Ward identities. One notices that the form of the transverse traceless part in \eqref{tttdec} is manifestly invariant under the permutation group of the set $\{1,2,3\}$. 
}
In this case the special conformal Ward identities reads
\begin{align}
0=&\mathcal{K}^\kappa\braket{T^{\mu_1\nu_1}(p_1)T^{\mu_2\nu_2}(p_2)T^{\mu_3\nu_3}(p_3)}\notag\\
=&\sum_{j=1}^{2}\left(2(\Delta_j-d)\frac{\partial}{\partial p_{j\,\kappa}}-2p_j^\alpha\frac{\partial}{\partial p_j^\alpha}\frac{\partial}{\partial p_{j\,\kappa}}+(p_j)^\kappa\frac{\partial}{\partial p_j^\alpha}\frac{\partial}{\partial p_{j\,\alpha}}\right)\braket{T^{\mu_1\nu_1}(p_1)T^{\mu_2\nu_2}(p_2)T^{\mu_3\nu_3}(p_3)}\notag\\
&\hspace{1cm}+4\left(\delta^{\kappa(\mu_1}\frac{\partial}{\partial p_1^{\alpha_1}}-\delta^\kappa_{\alpha_1}\d_\l^{(\m_1}\frac{\partial}{\partial p_{1\,\l}}\right)\braket{ T^{\nu_1)\alpha_1}(p_1)T^{\mu_2\nu_2}(p_2)T^{\mu_3\nu_3}(p_3)}\notag\\
&\hspace{1cm}+4\left(\delta^{\kappa(\m_2}\frac{\partial}{\partial p_2^{\a_2}}-\delta^\kappa_{\a_2}\d^{(\m_2}_\l\frac{\partial}{\partial p_{2\,\l}}\right)\braket{ T^{\mu_2)\a_2}(p_2)T^{\mu_1\nu_1}(p_1)T^{\mu_3\nu_3}(p_3)}.
\end{align}

Considering the decomposition $T^{\m\n}=t^{\m\n}+t^{\m\n}{\hspace{-2ex}\vspace{-0.5ex}}_{\substack{\scalebox{0.5}{loc}}}$, we find that the action of the special conformal operator $\mathcal K^\k$ on the transverse and traceless part of the correlator is still transverse and traceless. For this reason we are free to apply the transverse-traceless projectors in order to isolate equations for the form factors appearing in the decomposition of $\braket{t^{\m_1\n_1}t^{\m_2\n_2}t^{\m_3\n_3}}$. Using the relations in \appref{appendixB} one finds
\begin{align}
0=&\ \Pi^{\mu_1\nu_1}_{\alpha_1\beta_1}(p_1)\Pi^{\mu_2\nu_2}_{\alpha_2\beta_2}(p_2)\Pi^{\mu_3\nu_3}_{\alpha_3\beta_3}(p_3)\mathcal{K}^\kappa\braket{ t^{\alpha_1\beta_1}(p_1)t^{\alpha_2\beta_2}(p_2)t^{\alpha_3\beta_3}(p_3)}\notag\\
&+\frac{4d}{p_1^2}\Pi^{\mu_1\nu_1}_{\alpha_1\beta_1}(p_1)\left[\delta^{\alpha_1\kappa}p_{1\gamma_1}\braket{ T^{\beta_1\gamma_1}(p_1)t^{\mu_2\nu_2}(p_2)t^{\mu_3\nu_3}(p_3)}\right]\notag\\
&+\frac{4d}{p_2^2}\Pi^{\mu_2\nu_2}_{\alpha_2\beta_2}(p_2)\left[\delta^{\alpha_2\kappa}p_{2\gamma_2}\braket{t^{\mu_1\nu_1}(p_1)T^{\beta_2\gamma_2}(p_2)t^{\mu_3\nu_3}(p_3)}\right]\notag\\
&+\frac{4d}{p_3^2}\Pi^{\mu_3\nu_3}_{\alpha_3\beta_3}(p_3)\left[\delta^{\alpha_3\kappa}p_{3\gamma_3}\braket{t^{\mu_1\nu_1}(p_1)t^{\mu_2\nu_2}(p_2)T^{\beta_3\gamma_3}(p_3)}\right].
\end{align}
The last three terms are semi-local and may be re-expressed in terms of two-point  functions via \eqref{transvttt}. From the previous relation one can extract the CWI's as in \appref{specialconfward} and it is possible to write the most general form of the result as 
\begin{align}
0&=\Pi^{\mu 1\nu 1}_{\alpha_1\beta_1}(p1) \Pi^{\mu 2\nu 2}_{\alpha_2\beta_2}(p2)\Pi^{\mu_3\nu_3}_{\alpha_3\beta_3}(p_3) \mathcal{K}^{\kappa} \braket{ \braket{t^{\alpha_1 \beta_1} (p_1) t^{\alpha_2 \beta_2} (p_2) t^{\alpha_3 \beta_3} (p_3)  }} \notag\\
&= \Pi^{\mu 1\nu 1 }_{\alpha_1 \beta_1} (p_1) \Pi^{\mu 2\nu 2 }_{\alpha_2 \beta_2} (p_2)\Pi^{\mu_3\nu_3}_{\alpha_3\beta_3}(p_3) \times \Bigl\{ p_1^{\kappa} \Bigl( C_{1,1} \, p_2^{\alpha_1} p_2^{\beta_1} p_3^{\alpha_2} p_3^{\beta_2}p_1^{\alpha_3} p_1^{\beta_3} + C_{1,2}\, p_2^{\a_1} p_1^{\alpha_3} p_1^{\beta_3} p_3^{\alpha_2}\delta^{\b_1 \beta_2} \notag\\
&\quad+ C_{1,3}\, p_2^{\beta_1} p_2^{\alpha_1} p_1^{\alpha_3} p_3^{\a_2}\delta^{\b_2 \beta_3}+ C_{1,4}\, p_2^{\a_1} p_3^{\alpha_2} p_1^{\alpha_3} p_3^{\beta_2}\delta^{\b_1 \beta_3}+ C_{1,5}\, p_1^{\beta_3} p_2^{\beta_1} \delta^{\alpha_1 \alpha_2}\delta^{\alpha_3 \beta_2}\notag\\
&\quad+ C_{1,6}\, p_1^{\beta_3} p_1^{\alpha_3}\delta^{\alpha_1 \a_2}\delta^{\b_1 \beta_2}+ C_{1,7}\, p_1^{\beta_3} p_3^{\alpha_2}\delta^{\alpha_3 \a_1}\delta^{\b_1 \beta_2}+ C_{1,8}\, p_2^{\beta_1} p_2^{\alpha_1}\delta^{\alpha_2 \a_3}\delta^{\b_2 \beta_3}\notag\\
&\quad+ C_{1,9}\, p_3^{\beta_2} p_3^{\alpha_2}\delta^{\alpha_1 \a_3}\delta^{\b_1 \beta_3}+ C_{1,10}\, \delta^{\alpha_1 \beta_2}\delta^{\alpha_2 \beta_3}\delta^{\alpha_3 \beta_1}\Bigr)+ \big(p_1^\kappa\leftrightarrow p_2^\kappa; \ C_{1,j}\leftrightarrow C_{2,j}\big) + \big( p_1^\kappa\leftrightarrow p_3^\kappa; \ C_{1,j}\leftrightarrow C_{3,j}\big)\notag\\
& \quad\  + \delta^{\kappa \alpha_1} \Bigl( C_{3,1}\, p_{3}^{\alpha_2}p_{3}^{\beta_2} p_{1}^{\alpha_3}p_{1}^{\beta_3}p_{2}^{\beta_1} + C_{3,2}\, \delta^{\alpha_2 \beta_3} p_{3}^{\beta_2}  p_{1}^{\alpha_3} p_{2}^{\beta_1} + C_{3,3}\, \delta^{\alpha_2 \beta_1} p_{1}^{\alpha_3}  p_{3}^{\beta_2} p_{1}^{\beta_3}+ C_{3,4}\, \delta^{\alpha_3 \beta_1} p_{3}^{\alpha_2}  p_{3}^{\beta_2} p_{1}^{\beta_3}\notag\\
&\qquad\qquad+ C_{3,5}\, \delta^{\alpha_2 \beta_1} \delta^{\alpha_3 \beta_2} p_{1}^{\beta_3}+C_{3,6}\, \delta^{\alpha_2 \beta_3} \delta^{\alpha_3 \beta_1} p_{3}^{\beta_2}+C_{3,7}\, \delta^{\alpha_2 \beta_3} \delta^{\alpha_3 \beta_2} p_{2}^{\beta_1}\Bigr)\notag\\
&\qquad\qquad+[(\a_1,\b_1,p_2)\leftrightarrow (\a_2,\b_2,p_3)]+[(\a_1,\b_1,p_2)\leftrightarrow (\a_3,\b_3,p_1)]\Bigr\},\label{specialTTT}
\end{align}
where the coefficients $C_{j,k}$ are differential equations involving the form factors $A_1$, $A_2$, $A_3$, $A_4$, $A_5$. Each CWI's can the be presented in terms of the momentum magnitudes $p_j$. The primary CWI's appear as the coefficients of transverse ore transverse-traceless tensors containing $p_1^\k$, $p_2^\k$. The remaining equations, following from all other transverse or transverse-traceless terms, are then the secondary CWI's. Using the definitions of the operators $L_N$ and $R$ from \eqref{Ldef} and \eqref{Rdef} we can write the primary Ward identities
\begin{equation}\label{PrimaryTTT}
\begin{aligned}[c]
&\textup{K}_{13}A_1=0 ,\\[1.5ex]
&\textup{K}_{13}A_2=8 A_1, \\[1.5ex]
& \textup{K}_{13}A_3=2A_2, \\[1.5ex]
& \textup{K}_{13}A_4=-4A_2(p_{2\leftrightarrow 3}), \\[1.5ex]
& \textup{K}_{13}A_5=2\left[A_4-A_4(p_{3\leftrightarrow 1}) \right],\\[1.5ex]
\end{aligned}
\qquad
\begin{aligned}[c]
&\textup{K}_{32}  A_1=0 ,\\[1.5ex]
&\textup{K}_{32}A_2=-8 A_1, \\[1.5ex]
&\textup{K}_{32}A_3=-2A_2, \\[1.5ex]
&\textup{K}_{32}A_4= 4A_2(p_{3\leftrightarrow 1}),\\[1.5ex]
&\textup{K}_{32}A_5=2A_4(p_{3\leftrightarrow 2})-2A_4 ,\\[1.5ex]
\end{aligned}\qquad
\begin{aligned}[c]
&\textup{K}_{12}A_1=0 ,\\[1.5ex]
&\textup{K}_{12}A_2=0,\\[1.5ex]
& \textup{K}_{12}A_3=0,\\[1.5ex]
& \textup{K}_{12}A_4=4\left[A_2(p_{1\leftrightarrow 3})-A_2(p_{2\leftrightarrow 3}) \right],\\[1.5ex]
& \textup{K}_{12}A_5=2\left[A_4(p_{2\leftrightarrow 3})-A_4(p_{3\leftrightarrow 1}) \right].\\[1.5ex]
\end{aligned}
\end{equation}
The solution follow from the analysis in \appref{primarysol},
	\begin{subequations}
	\begin{align}
	A_1&=\a_1\,J_{6\{000\}},\\
	A_2&=4\a_1\,J_{5\{001\}}+\a_2\,J_{4\{000\}},\\
	A_3&=2\a_1\,J_{4\{002\}}+\a_2\,J_{3\{001\}}+\a_3\,J_{2\{000\}},\\
	A_4&=8\a_1\,J_{4\{110\}}-\a_2\,J_{3\{001\}}+\a_4\,J_{2\{000\}},\\
	A_5&=8\a_1\,J_{3\{111\}}+2\a_2\,\big(\,J_{2\{110\}}+J_{2\{101\}}+J_{2\{011\}}\big)+\a_5\,J_{0\{000\}},
	\end{align}\label{PrimSolTTT}
	\end{subequations}
where $a_j$, $j=1,\dots,5$ are constants. If the integrals diverge, we should we need to regulate them. 
The independent secondary Ward identities are written in the form
\begin{align}
C_{31}&= -\frac{2}{p_1^2}\Bigl( \textup{L}_6\, A_1 + \textup{R}\bigl[\, A_2-A_2(p_3\leftrightarrow p_2)\bigr]\Bigr)\notag\\
&=-\frac{4d}{p_1^2}\text{coeff. of } p_3^{\mu_2}p_3^{\nu_2}p_1^{\mu_3}p_1^{\nu_3}p_2^{\mu_1} \text{ in } p_{1\nu_1}\braket{ T^{\mu_1\nu_1}(p_1)T^{\mu_2\nu_2}(p_2)T^{\mu_3\nu_3}(p_3)},\\
C_{32}&= \frac{2}{p_1^2}\Bigl( \textup{L}_4\, A_2(p_3\leftrightarrow p_1) + \textup{R}\bigl[\, A_4(p_3\leftrightarrow p_2)-A_4\bigr]\Bigr)-4\left(A_2-A_2(p_3\leftrightarrow p_2)\right)\nonumber\\
&=\frac{4d}{p_1^2}\text{coeff. of } \delta^{\mu_2\mu_3}p_3^{\nu_2}p_1^{\nu_3}p_2^{\mu_1} \text{ in } p_{1\nu_1}\braket{ T^{\mu_1\nu_1}(p_1)T^{\mu_2\nu_2}(p_2)T^{\mu_3\nu_3}(p_3)},
\end{align}
\begin{align}
C_{33}&= \frac{1}{p_1^2}\Bigl( \textup{L}_6\, A_2 + 2\textup{R}\bigl[\,2A_3-A_4(p_3\leftrightarrow p_1)\bigr]\Bigr)\notag\\
&=\frac{4d}{p_1^2}\text{coeff. of } \delta^{\mu_2\mu_1}p_1^{\mu_3}p_1^{\nu_3}p_3^{\nu_2} \text{ in } p_{1\nu_1}\braket{ T^{\mu_1\nu_1}(p_1)T^{\mu_2\nu_2}(p_2)T^{\mu_3\nu_3}(p_3)},\\
C_{35}&= -\frac{1}{p_1^2}\Bigl( \textup{L}_4\, A_4(p_3\leftrightarrow p_2)  - 2\textup{R}A_5\Bigr)+8A_3-2A_4(p_3\leftrightarrow p_1)\notag\\
&=-\frac{4d}{p_1^2}\text{coeff. of } \delta^{\mu_2\mu_1}\delta^{\mu_3\nu_2}p_1^{\nu_3} \text{ in } p_{1\nu_1}\braket{ T^{\mu_1\nu_1}(p_1)T^{\mu_2\nu_2}(p_2)T^{\mu_3\nu_3}(p_3)},\\
C_{36}&= \frac{1}{p_1^2}\Bigl( \textup{L}_4\, A_4- 2\textup{R}A_5(p_3\leftrightarrow p_2)\Bigr)+8A_3(p_3\leftrightarrow p_2)-2A_4(p_3\leftrightarrow p_1)\notag\\
&=\frac{4d}{p_1^2}\text{coeff. of } \delta^{\mu_3\mu_1}\delta^{\mu_2\nu_3}p_3^{\nu_2} \text{ in } p_{1\nu_1}\braket{ T^{\mu_1\nu_1}(p_1)T^{\mu_2\nu_2}(p_2)T^{\mu_3\nu_3}(p_3)},\\
C_{37}&= -\frac{2}{p_1^2} \textup{L}_2\, A_3(p_3\leftrightarrow p_1)  -2A_4+2A_4(p_3\leftrightarrow p_2)\notag\\
&=-\frac{4d}{p_1^2}\text{coeff. of } \delta^{\mu_2\mu_3}\delta^{\nu_3\nu_2}p_2^{\mu_1} \text{ in } p_{1\nu_1}\braket{ T^{\mu_1\nu_1}(p_1)T^{\mu_2\nu_2}(p_2)T^{\mu_3\nu_3}(p_3)}.
\end{align}
The equations $C_{31}$ and $C_{32}$ are trivially satisfied in all the cases and do not impose any additional conditions on primary constants. In order to show this statement we can now render explicit the contributions on the right-hand sides of the secondary CWI's. We need the expression of the transverse WI's to do so. Starting from \eqref{transvttt} and writing explicitly the tensor coefficient $\mathcal{B}$, one obtains 
\begin{align}
&p_{1\n_1}\braket{T^{\m_1\n_1}(p_1)T^{\m_2\n_2}(p_2)T^{\m_3\n_3}(p_3)}=p_2^{\,\m_1}\braket{T^{\m_2\n_2}(p_1+p_2)T^{\m_3\n_3}(p_3)}+p_3^{\,\m_1}\braket{T^{\m_2\n_2}(p_2)T^{\m_3\n_3}(p_3+p_1)}\notag\\
&\quad-2\d^{\m_1(\m_2}p_{2\a}\braket{T^{\n_2)\a}(p_1+p_2)T^{\m_3\n_3}(p_3)}-2\d^{\m_1(\m_3}p_{3\a}\braket{T^{\n_3)\a}(p_1+p_3)T^{\m_2\n_2}(p_2)},
\end{align}
and using the explicit form for the $TT$ two-point  function we find
\begin{align}
&p_{1\n_1}\braket{T^{\m_1\n_1}(p_1)T^{\m_2\n_2}(p_2)T^{\m_3\n_3}(p_3)}=p_2^{\,\m_1}\big[\Pi^{\m_2\n_2\m_3\n_3}(p_3)\,A_{TT}(p_3)-\Pi^{\m_2\n_2\m_3\n_3}(p_2)\,A_{TT}(p_2)\big]\notag\\
&\hspace{2cm}-2\big[\d^{\m_1(\m_2}p_{2\a}\Pi^{\n_2)\a\m_3\n_3}(p_3)\,A_{TT}(p_3)-\d^{\m_1(\m_3}p_{3\a}\Pi^{\n_3)\a\m_2\n_2}(p_2)\,A_{TT}(p_2)\big],\label{TTTtran}
\end{align}
where $A_{TT}(p_i)$ depends on the normalization of the coefficient of the two-point  function $TT$, which is defined as
\begin{equation}
A_{TT}(p)=c_T\,\begin{cases}
p^d&\text{if}\ d=3,5,7,\dots\\
p^d(-\log(p^2)+\text{local})&\text{if}\ d=2,4,6,\dots\,.
\end{cases}\label{ATT}
\end{equation}
In \eqref{specialTTT} the transverse Ward identities multiply three projection operators. Keeping in mind this information, we can drop from \eqref{TTTtran} those terms that vanish, if contracted with a projector. In this way we obtain
\begin{align}
p_{1\n_1}\braket{T^{\m_1\n_1}(p_1)T^{\m_2\n_2}(p_2)T^{\m_3\n_3}(p_3)}&=2\,p_2^{\m_1}\d^{\n_2\m_3}\d^{\m_2\n_3}\big[A_{TT}(p_3)-A_{TT}(p_2)\big]\notag\\
&+4\big[p_1^{\n_3}\d^{\m_1\m_2}\d^{\n_2\m_3}A_{TT}(p_3)+p_3^{\n_2}\d^{\m_1\m_3}\d^{\n_3\m_2}A_{TT}(p_2)\big].
\end{align}
From the previous expression we derive an explicit form of the secondary WI's as
	\begin{align}
	&\textup{L}_6\, A_1 + \textup{R}\bigl[\, A_2-A_2(p_3\leftrightarrow p_2)\bigr]=0,\label{Tsec1}\\[1.5ex] 
	&\textup{L}_4\, A_2(p_3\leftrightarrow p_1) + \textup{R}\bigl[\, A_4(p_3\leftrightarrow p_2)-A_4\bigr]+2p_1^2\left[A_2(p_3\leftrightarrow p_2)-A_2\right]=0,\\[1.5 ex] 
	&\textup{L}_6\, A_2 + 2\textup{R}\bigl[\,2A_3-A_4(p_3\leftrightarrow p_1)\bigr]=0,\\[1.5 ex] 
	&\textup{L}_4\, A_4(p_3\leftrightarrow p_2)  - 2\textup{R}A_5+2p_1^2\left[A_4(p_3\leftrightarrow p_1)-4A_3\right]=8d\,A_{TT}(p_3),\\[2 ex] 
	&\textup{L}_4\, A_4- 2\textup{R}A_5(p_3\leftrightarrow p_2)  +2p_1^2\left[A_4(p_3\leftrightarrow p_1)-4A_3(p_3\leftrightarrow p_2)  \right]=8d\,A_{TT}(p_2),\\[2 ex]  
	&\textup{L}_2\, A_3(p_3\leftrightarrow p_1) +p_1^2\left[A_4-A_4(p_3\leftrightarrow p_2)\right]=4d\,\big[A_{TT}(p_3)-A_{TT}(p_2)\big].\label{F14}
	\end{align}
From the definition of $A_{TT}$ in \eqref{ATT} one observes that the secondary CWI's simplify in zero-momentum limit $p_3\to0$, $p_1=-p_2=p$. 

\subsection{Divergences}
We discuss the divergences of the solutions \eqref{PrimSolTTT} for the form factors that satisfy the primary CWI's \eqref{PrimaryTTT}. We have discussed the cases in which the triple-K integrals are divergent and in the $TTT$ we obtain 
\begin{equation}
\left(\frac{d}{2}+N\right)\pm\left(\frac{d}{2}+k_1\right)\pm\left(\frac{d}{2}+k_2\right)\pm\left(\frac{d}{2}+k_3\right)=-2n,\qquad n\in\mathbb{N},\label{TTTdiv}
\end{equation}
where the parameters $N,k_1,k_2,k_3$ are those appearing in the solutions of the form factors written as reduced triple-K integrals $J_{N,\{k_1,k_2,k_3\}}$. 
As pointed out in \cite{Bzowski:2017poo}, the triple-K integrals appearing in \eqref{PrimSolTTT} satisfy the divergent condition \eqref{TTTdiv} in two different ways, represented as $(---)$ and $(--+)$ (and permutation thereof), where the signs $\pm$ are those in \eqref{TTTdiv}. The difference between these two types of singularities lies in the proper way of removing them, and in particular for the $(--+)$ there are no counterterms which allow to obtain a finite result. The only form factor in \eqref{PrimSolTTT} affected by this kind of divergences is $A_5$ and in particular through the integrals
\begin{equation}
J_{2\{110\}}: (--+),\quad
J_{2\{101\}}:(-+-),\quad
J_{2\{011\}}:(+--),
\end{equation}
\begin{equation}
J_{0\{000\}}:(+--),\ (+--),\ (+--),
\end{equation}
which manifest these types of divergences independently of the specific dimensions $d$. These divergences emerge as poles in $1/\e(u-v)$ or $1/(u-v)$, where $u$ and $v$ are the parameters that regularize the integrals. 
Furthermore, in the case of odd dimensions, from \eqref{divFormFactor}, the form factors do not have any physical \emph{ultralocal} singularities $(---)$ but only \emph{semilocal} ones of the type $(--+)$ that can be removed by a choice of the parameters $\a_i$. In particular, we observe that after an expansion in the zero momentum limit of the divergent part of $A_5$, we obtain
\begin{align}
&2\a_2\,\big(\,J_{2\{110\}}+J_{2\{101\}}+J_{2\{011\}}\big)+\a_5\,J_{0\{000\}}\notag\\
&=2^{\frac{d}{2}+u\e-3}\,p^{d-u\epsilon+3v\e}\Gamma\left(\frac{d}{2}+v\e\right)\Gamma\left(\frac{d}{2}+\frac{(u-v)\e}{2}\right)\Gamma\left(1+\frac{(u-v)\e}{2}\right)\notag\\
&\Gamma\left(-\frac{d}{2}+\frac{(u-3v)\e}{2}\right)\Gamma\left(\frac{(u-v)\e}{2}\right)\Bigg\{\a_5\bigg[\e(u-v)+1\bigg]+\a_2\bigg[2d^2+\e\left(\frac{5d^2}{2}(u-v)+3du+5dv\right)
\notag\\
&+\e^2\big((3d+1)u^2+4(d+1)uv+(3-7d)v^2\big)+O(\e^3)\bigg]\Bigg\}.
\end{align}
Notice that the pole $1/((u-v)\e)$, coming from the gamma function $\Gamma[(u-v)\e/2]$, is present independently of the choice of the dimensions. In the case of odd dimensions $d=2n+1$, $1/((u-v)\e)$ is the only spurious pole that, once the expansion of the coefficients \eqref{expans} is taken in account, can be removed by the choice
\begin{equation}
\begin{split}
\a_5^{(0)}&=-2d^2\,\a_2^{(0)},\\
\a_5^{(1)}&=-2d\left(4v\,\a_2^{(0)}+d\a_2^{(1)}\right)+(u-v)\a_5^{\prime\,(1)},
\end{split}\label{redef1}
\end{equation}
where $\a_5^{\prime(1)}$ redefines the undetermined constant,  that will be determined by the secondary CWI's. With this choice, the limit $u\to v$ is well defined and does not produce any divergence, leaving $A_5$ with its physical divergences expressed only in terms of $1/\e$.
  
Conversely, in the case of even spacetime dimensions $d=2n$, other divergences arise from the gamma function $\Gamma(-d/2+(u-3v)\e/2)$ for which the redefinition of the constants will be
\begin{equation}
\begin{split}
\a_5^{(0)}&=-2d^2\,\a_2^{(0)},\\
\a_5^{(1)}&=-2d\left(4v\,\a_2^{(0)}+d\a_2^{(1)}\right)+(u-v)\a_5^{\prime\,(1)},\\
\a_5^{(2)}&=-8v^2\,\a_2^{(0)}-8v\,d\,\a_2^{(1)}-2d^2\a_2^{(2)}+(u-v)\a_5^{\prime(2)},
\end{split}\label{redef2}
\end{equation}
allowing the cancellation of the semilocal divergences. 
In the next appendix we study the secondary CWI's in the two cases of even and odd spacetime dimensions, in order to deal with the two different types of divergences that arise.  
The other type of divergence $(---)$, as previously mentioned, can be removed by adding counterterms constructed out of the metric in a covariant way. In particular, by inspection of \eqref{TTTdiv}, we list the possible cases in which the form factors manifest $(---)$ divergences as
\begin{equation}
\begin{split}
&A_1\hspace{1.3cm}\text{diverges for} \,d=6+2n,\\
&A_2\hspace{1.3cm}\text{diverges for} \,d=4+2n,\\
&A_3,\,A_4 \hspace{0.6cm}\text{diverge for} \,d=2+2n,\\
&A_5 \hspace{1.3cm}\text{diverges for} \,d=2n,
\end{split}\qquad n\in\mathbb{N}, \label{divFormFactor}
\end{equation}
which clearly depend on the dimensions $d$. 
%%%%%%%%%%%%%%%%%%%%%%%%%%%%%%%%%%%%%%%%%%%%%%%%%%%%%
\subsection{The secondary CWI's for the \texorpdfstring{$TTT$}{ttt} correlator}\label{appendixH}
%%%%%%%%%%%%%%%%%%%%%%%%%%%%%%%%%%%%%%%%%%%%%%%%%%%%%
We have previously discussed the redefinitions of the constants in order to remove the semilocal divergences. In this appendix, we discuss the solution of the secondary CWI's in two different cases, for odd and for even spacetime dimensions respectively. We will be using the zero momentum limit in the correlators, in which the triple-K integral behaves as
\begin{equation}
\lim_{p_3\to 0}I_{\alpha\{\beta_1,\beta_2,\beta_3\}}(p_1,p_2,p_3)=p^{\beta_t-\alpha-1}\ell_{\alpha\{\beta_1,\beta_2,\beta_3\}}\label{limit},
\end{equation}
where $\beta_t=\beta_1+\beta_2+\beta_3$ and with \eqref{ldef}. It is worth noting that the three independent differential equations 
\begin{align}
&\textup{L}_6\, A_2 + 2\textup{R}\bigl[\,2A_3-A_4(p_3\leftrightarrow p_1)\bigr]=0,\\[1.5 ex] 
&\textup{L}_4\, A_4(p_3\leftrightarrow p_2)  - 2\textup{R}A_5+2p_1^2\left[A_4(p_3\leftrightarrow p_1)-4A_3\right]=8d\,A_{TT}(p_3),\\[2 ex]  
&\textup{L}_2\, A_3(p_3\leftrightarrow p_1) +p_1^2\left[A_4-A_4(p_3\leftrightarrow p_2)\right]=4d\,\big[A_{TT}(p_3)-A_{TT}(p_2)\big],
\end{align}
reduce the number of undetermined constants from five to two. By applying the approach previously illustrated for the $\braket{TT\mO}$ to this case, and with the redefinitions \eqref{redef1} or \eqref{redef2}, the secondary CWI's give the condition
\begin{align}
\a_3^{(0)}&=-d\left[2(d+2)\a_1^{(0)}+\a_2^{(0)}\right]-\left(\frac{\pi}{2}\right)^{-\frac{3}{2}}\frac{2d(-1)^n\,c_T}{(d-2)!!},\\
\a_4^{(0)}&=2\a_3^{(0)}+(3d+2)\a_2^{(0)},\\
\a_5^{\prime(1)}&=-d\left[2d(d+2)\a_1^{(0)}+\frac{1}{2}(d+6)\a_2^{(0)}\right]-\left(\frac{\pi}{2}\right)^{-\frac{3}{2}}\frac{2d(-1)^{n}c_T}{(d-2)!!},
\end{align}
for odd spacetime dimensions $d=2n+1$, $(n=1,2,3,\dots)$, and 
\begin{align}
\a_3^{(0)}&=-d\left[2(d+2)\a_1^{(0)}+\a_2^{(0)}\right]-\frac{2^{3-n}(-1)^n\,c_T}{(n-1)!},\\
\a_4^{(0)}&=2\a_3^{(0)}+(3d+2)\a_2^{(0)},\\
\a_3^{(1)}&=-2u\big(4(d+1)\a_1^{(0)}+\a_2^{(0)}\big)-d\big(2(d+2)\a_1^{(1)}+\a_2^{(1)}\big)+\frac{(-1)^{n}2^{3-n}u}{(n-1)!}\bigg[c_t(H_{n-1}+\ln 2-\g_e)-c_T^{(0)}\bigg],\\
\a_4^{(1)}&=2a_3^{(1)}+6u\a_2^{(0)}+(3d+2)\a_2^{(1)},\\
\a_5^{\prime\,(1)}&=-d\left[2d(d+2)\a_1^{(0)}+\frac{1}{2}(d+6)\a_2^{(0)}\right]-\left(\frac{\pi}{2}\right)^{-\frac{3}{2}}\frac{2^{4-n}(-1)^{n}c_T}{(n-1)!!},\\
\a_5^{\prime\,(2)}&=-2u\bigg(2d(3d+4)\a_1^{(0)}+(d+3)\a_2^{(0)}\bigg)-d\bigg(2d(d+2)\a_1^{(1)}+\frac{d+6}{2}\a_2^{(1)}\bigg)\notag\\
&\hspace{1cm}+\frac{(-1)^n2^{4-n}nu}{(n-1)!}\bigg[c_T\big(H_{n-1}-\frac{1}{n}+\ln 2-\g_e\big)-c_T^{(0)}
\bigg],
\end{align}
for even spacetime dimensions $d=2n$, $(n=2,3,\dots)$, as dicussed in \cite{Bzowski:2017poo}. In this case we have also expanded the contributions proportional to the constant $c_T$ of the two-point function, because of the appearance of divergences in the two-point function. $H_n=\sum_{k=1}^n1 /k$ is the n-th harmonic number.

As a final consideration, by looking at the solutions of the secondary CWI's, we notice that the three-point function $\braket{TTT}$  both for even and odd dimensions ($d>2$) depends only on two undetermined constants. This property is the 
cornerstone to prove the correspondence between the general approach and the perturbative realizations in \secref{Perturbative}.
%%%%%%%%%%%%%%%%%%%%%%%%%%%%%%

 \section{Simplifications in the \texorpdfstring{$n$}{n}-\texorpdfstring{$p$}{p} basis} 
 \label{nnp}
Simplifications in the structure of the renormalized four-point function are possible once we re-express all the contributions of the previous section in terms of the tensorial basis formed by the $n^\mu$ and $p_1,p_2$ and $p_3$ four-vectors
 \begin{equation}
n^{\mu}=\epsilon^{\m \a \b\g}p_{i,\a}p_{j,\b}p_{k,\g}, \quad i\neq j \neq k =1,2,3,4,
\end{equation}
with $(n^\mu, p_i,p_j, p_k)$ forming a tetrad that can be used as a basis of expansion in Minkowski space. We just recall that in the computation of the residue of the $\sim V_E$ counterterm, we 
need to parametrize the Kronecker $\delta_{\mu\nu}$ in this basis in the form 
\begin{equation}
\d^{(4)}_{\m \n}=\sum_{i,j}^4 p_i^{\m} p_j^{\n} (Z^{-1})_{j i},
\end{equation}
where $(Z^{-1})_{j i}$ is the inverse of the Gramm matrix, defined as $Z=[p_i\cdot p_j]_{i,j=1}^d$.\\
Using the expression of $\delta^{\mu\nu}$ in the $n$-$p$ basis, denoted as $\delta^{(4)}_{\mu\nu}$ 
\begin{align}\label{deltaepsilon}
\d^{(4)}_{\m \n}=&\frac{1}{n^2}\Bigg(2p_1^{(\mu } p_2^{\nu) } \left(p_3^2
   p_1\cdot p_2-p_1\cdot p_3 p_2\cdot p_3\right)+2p_1^{(\nu } p_3^{\mu )}
   \left(p_2^2 p_1\cdot p_3-p_1\cdot p_2 p_2\cdot p_3\right) \notag \\ 
 & \qquad \qquad  +2p_2^{(\nu } p_3^{\mu) }
   \left(p_1^2 p_2\cdot p_3-p_1\cdot p_2 p_1\cdot p_3\right)\notag +p_1^{\mu }
   p_1^{\nu } \left((p_2\cdot p_3)^2-p_2^2 p_3^2\right)\notag \\
   & \qquad \qquad +p_2^{\mu } p_2^{\nu }\left((p_1\cdot p_3)^2-p_1^2 p_3^2\right) +p_3^{\mu } p_3^{\nu }\left((p_1\cdot p_2)^2-p_1^2 p_2^2\right)+n^{\mu } n^{\nu } \Bigg),
   \end{align}
one derives several relations in $d=4$. We easily derive the constraint
\begin{align}\label{projtwon}
\Pi^{\m_i \n_i}_{\a_i \b_i}(p_i)n^{\a_i } n^{\b_i }=& -\Pi^{\m_i \n_i}_{\a_i \b_i}(p_i)\Bigg(2p_j^{(\b_i  } p_k^{\a_i) }
   \left(p_i^2 p_j\cdot p_k-p_i\cdot p_k p_i\cdot p_j\right)\notag \\
   & +p_j^{\a_i } p_j^{\b_i  }\left((p_i\cdot p_k)^2-p_i^2 p_k^2\right) +p_k^{\a_i } p_k^{\b_i  }\left((p_i\cdot p_j)^2-p_i^2 p_j^2\right) \Bigg),
\end{align}
while the relation $\d^{(4)}_{\a_i \b_i}\Pi(p_i)^{(4)\m_i \n_i}_{\a_i \b_i}=0$ is obviously satisfied in the new basis. Using these relations, it is possible to show the vanishing of 
$\left[ V^{\mu_1\nu_1\ldots \mu_n\nu_n}\right] $, which is obviously defined at $d=4$. 
An explicit check has been discussed in \cite{Serino:2020pyu} for $n=4$, using the $n$-$p$ decomposition. 
%%%%%%%%%%%%%%%%%%%%%%%%%%%%%%
\section{Identities with projectors \label{appendixB}}
%%%%%%%%%%%%%%%%%%%%%%%%%%%%%%
The projectors are defined as
\begin{align}
\pi^\m_\a(p)&=\d^\m_\a-\sdfrac{p^\m p_\a}{p^2},\\
\Pi^{\m\n}_{\a\b}(p)&=\pi^{(\m}_\a\pi^{\n)}_\b-\sdfrac{1}{d-1}\pi^{\m\n}\pi_{\a\b},\label{Proj}\\
\Pi^{\m\n\r\s}(p)&=\d^{\r\a}\d^{\s\b}\,\Pi^{\m\n}_{\a\b}(p),
\end{align}
where we define $\pi^{(\m}_\a\pi^{\n)}_\b=1/2(\pi^{\m}_\a\pi^{\n}_\b+\pi^{\n}_\a\pi^{\m}_\b)$. One derives the following identities 
\begin{align}
p_\m\pi^{\m\n}(p)=p_\m\Pi^{\m\n}_{\a\b}(p)&=0,\\
\d_{\m\n}\pi^{\m\n}(p)=\pi^\m_\m(p)&=d-1,\\
\Pi^{\m\n\r}_{\ \ \ \ \ \r}(p)=\d_{\r\s}\Pi^{\m\n\r\s}(p)&=\pi_{\r\s}(p)\Pi^{\m\n\r\s}(p)=0,\\
\Pi^{\m\r\n}_{\ \ \ \ \ \r}(p)=\d_{\r\s}\Pi^{\m\r\n\s}(p)=\pi_{\r\s}(p)\Pi^{\m\r\n\s}(p)&=\sdfrac{(d+1)(d-2)}{2(d-1)}\pi^{\m\n}(p),\\
\Pi^{\m\n}_{\r\s}(p)\Pi^{\r\s}_{\m\n}(p)&=\sdfrac{1}{2}(d+1)(d-2),\\
\pi^\m_\a(p)\pi^\a_\m(p)&\pi^\m_\n(p),\\
\Pi^{\m\n}_{\a\b}(p)\Pi^{\a\b}_{\r\s}(p)&=\Pi^{\m\n}_{\r\s}(p),\\
\Pi^{\m\n\r}_{\ \ \ \ \ \a}(p)\pi^{\a\s}(p)&=\Pi^{\m\n\r\s}(p),\\
\Pi^{\m\n}_{\a\b}(p)\Pi^{\a\r\b\s}(p)&=\sdfrac{d-3}{2(d-1)}\Pi^{\m\n\r\s}(p).
\end{align}
Denoting with $\partial_\m$ the derivative with respect to the $p$ momentum we find
\begin{align}
\partial_\k\,\pi_{\m\n}(p)&=-\sdfrac{p_\m}{p^2}\pi_{\n\k}(p)-\sdfrac{p_\n}{p^2}\pi_{\m\k}(p),\\
\partial_\k\Pi_{\m\n\r\s}&=-\sdfrac{p_\m}{p^2}\Pi_{\k\n\r\s}-\sdfrac{p_n}{p^2}\Pi_{\m\k\r\s}-\sdfrac{p_\r}{p^2}\Pi_{\m\n\k\s}-\sdfrac{p_\s}{p^2}\Pi_{\m\n\r\k},\\
\pi^\m_\a\partial_\k\pi^\a_\n&=-\sdfrac{p_\n}{p^2}\pi^\m_\k,\\
\pi^{\m\k}\partial_\a\pi^\a_\n-\pi^{\m\a}\partial_\a\pi^\k_\n&=-(d-2)\sdfrac{p_\n}{p^2}\pi^{\m\k}+\sdfrac{p^\k}{p^2}\pi^\m_\n,\\
\Pi^{\m\n}_{\a\b}\partial_\k\Pi^{\a\b}_{\r\s}&=-\sdfrac{p_\r}{p^2}\Pi^{\m\n}_{\k\s}-\sdfrac{p_\s}{p^2}\Pi^{\m\n}_{\r\k},\\
\Pi^{\m\n}_{\k\b}\partial_\a\Pi^{\a\b}_{\r\s}-\Pi^{\m\n\a}_{\quad\  \b}\partial_\a\Pi^{\b}_{\ \ \k\r\s}&=-\sdfrac{1}{2}\sdfrac{d-1}{p^2}[p_\r\Pi^{\m\n}_{\k\s}+p_\s\Pi^{\m\n}_{\r\k}]+\sdfrac{p_\k}{p^2}\Pi^{\m\n}_{\r\s},\\
\Pi^{\m\n}_{\a\b}\,p^\l\partial_\l\Pi^{\a\b}_{\r\s}&=0.
\end{align}
Analogous expressions with two derivatives are
\begin{align}
\pi^\m_\a\partial^2\pi^\a_\n&=-\sdfrac{2}{p^2}\pi^\m_\n,\\
p^\a\pi^\m_\b\partial_\a\partial_\k\pi^\b_\n&=\sdfrac{p_\n}{p^2}\pi^\m_\k,\\
\Pi^{\m\n}_{\a\b}\partial^2\Pi^{\a\b}_{\r\s}&=-\sdfrac{4}{p^2}\Pi^{\m\n}_{\r\s},\\
p^\g\Pi^{\m\n}_{\a\b}\partial_\g\partial_\k\Pi^{\a\b}_{\r\s}&=\sdfrac{p_\r}{p^2}\Pi^{\m\n}_{\k\s}+\sdfrac{p_\s}{p^2}\Pi^{\m\n}_{\r\k}.
\end{align}
For the semi-local operators we find
\begin{align}
\Pi^{\m\n}_{\a\b}t^{\a\b}_{loc}&=0,\\
\Pi^{\m\n}_{\a\b}\partial_\k t^{\a\b}_{loc}&=\sdfrac{2}{p^2}\Pi^{\m\n}_{\a\k}\,p_\b T^{\a\b},\\
\Pi^{\m\n\k}_{\quad\ \k}\partial_\a t^{\a\b}_{loc}-\Pi^{\m\n\a}_{\quad\ \b}\partial_\a\,t^{\k\b}_{loc}&=\sdfrac{d-2}{p^2}\Pi^{\m\n\k}_{\quad\ \a} p_\b\,T^{\a\b}+\sdfrac{p^\b}{p^2}\Pi^{\m\n\k}_{\quad\ \a}\partial_\b (p_\r T^{\a\r})-\sdfrac{p^\k}{p^2}\Pi^{\m\n\a}_{\quad\ \b}\partial_\a(p_\r T^{\b\r}),\\
\Pi^{\m\n}_{\a\b}\partial^2 t^{\a\b}_{loc}&=\sdfrac{4}{p^2}\Pi^{\m\n\a}_{\quad\ \b}\partial_\a(p_\r\,T^{\r\b}),\\
p^\g\Pi^{\m\n}_{\a\b}\partial_\g\partial_\k\,t^{\a\b}_{loc}&=-\sdfrac{4}{p^2}\Pi^{\m\n}_{\a\k}p_\b\,T^{\a\b}+\sdfrac{2}{p^2}\Pi^{\m\n}_{\a\k}\,p^\b\partial_\b(p_\r T^{\a\r}).
\end{align}
%%%%%%%%%%%%%%%%%%%%%%%%%%%%%%%%%%
\section{Generalized hypergeometrics, 3K and 4K integrals}
%%%%%%%%%%%%%%%%%%%%%%%%%%%%%%%%%%
Appell's hypergeometric functions $F_1(x,y)$, $F_2(x,y)$, $F_3(x,y)$, $F_4(x,y)$ are defined by the  hypergeometric series
\begin{eqnarray} \label{appf1}
\app 1{a;\;b_1,b_2}{c}{x,\,y} \equal \sum_{n=0}^{\infty} \sum_{m=0}^{\infty}
\frac{(a)_{n+m}\,(b_1)_n\,(b_2)_m}{(c)_{n+m}\;n!\,m!}\,x^n\,y^m,\\ \label{appf2}
\app 2{a;\;b_1,b_2}{c_1,c_2}{x,\,y} \equal \sum_{n=0}^{\infty} \sum_{m=0}^{\infty}
\frac{(a)_{n+m}\,(b_1)_n\,(b_2)_m}{(c_1)_n\,(c_2)_m\;n!\,m!}\,x^n\,y^m,\\ \label{appf3}
\app 3{\!a_1,a_2;\,b_1,b_2}{c}{x,\,y} \equal \sum_{n=0}^{\infty} \sum_{m=0}^{\infty}
\frac{(a_1)_n(a_2)_m(b_1)_n(b_2)_m}{(c)_{n+m}\;n!\,m!}\,x^n\,y^m,\\ \label{appf4}
F_4(a,b,c_1,c_2; x,y)\equiv\app 4{a;\;b}{c_1,c_2\,}{x,\,y} \equal \sum_{n=0}^{\infty} \sum_{m=0}^{\infty}
\frac{(a)_{n+m}\,(b)_{n+m}}{(c_1)_n\,(c_2)_m\;n!\,m!}\,x^n\,y^m,
\end{eqnarray}
and are bivariate generalizations of the Gauss hypergeometric series
\begin{equation} \label{gausshpg}
\hpg21{A,\,B}{C}{z} = \sum_{n=0}^{\infty} 
\frac{(A)_{n}\,(B)_n}{(C)_n\,n!}\,z^n,
\end{equation}
with the (Pochhammer) symbol $(\alpha)_{k}$ given by
\begin{equation}
(\alpha)_{k}\equiv (\alpha,k)\equiv\frac{\Gamma(\alpha+k)}{\Gamma(\alpha)}=\alpha(\alpha+1)\dots(\alpha+k-1).\label{Pochh}
\end{equation}
A summary of many of the properties of such functions and a discussion of the univariate cases, obtained when the two variables coincide, can be found in \cite{Vidunas1}. They are solutions of equations generalizing Euler's hypergeometric equation 
\begin{equation} \label{eq:euler}
z(1-z)\,\frac{d^2y(z)}{dz^2}+
\big(C-(A+B+1)z\big)\frac{dy(z)}{dz}-A\,B\,y(z)=0,
\end{equation}
whose solution is denoted as $\hpgo21$, written in (\ref{gausshpg}).
This is a Fuchsian equation with singularities at $z=0$, $z=1$ and $z=\infty$. 
When the two arguments $x,y$ of the Appell functions are related, they 
are referred to as univariate functions, satisfying a Fuchsian ordinary
differential equations.\\
In \cite{Coriano:2013jba} and \cite{Bzowski:2013sza} it has been shown, independently, that the CWIs of three-point  functions are equivalent to hypergeometric systems of equations. We recall that, in the case of Appell functions of type $F_4$ given in \eqref{appf4}, such functions are solutions of the system of differential equations
\begin{equation}
\label{F4diff.eq}
\left\{
\begin{aligned}
&\\[-1ex]
&\bigg[ x(1-x) \frac{\partial^2}{\partial x^2} - y^2 \frac{\partial^2}{\partial y^2} - 2 \, x \, y \frac{\partial^2}{\partial x \partial y} +  \left[ \gamma - (\alpha + \beta + 1) x \right] \frac{\partial}{\partial x}- (\alpha + \beta + 1) y \frac{\partial}{\partial y}  - \alpha \, \beta \bigg] F(x,y) = 0 \,, \nn \\[2ex]
&\bigg[ y(1-y) \frac{\partial^2}{\partial y^2} - x^2 \frac{\partial^2}{\partial x^2} - 2 \, x \, y \frac{\partial^2}{\partial x \partial y} +  \left[ \gamma' - (\alpha + \beta + 1) y \right] \frac{\partial}{\partial y} - (\alpha + \beta + 1) x \frac{\partial}{\partial x}  - \alpha \, \beta \bigg] F(x,y) = 0 \,, 
\\[-1ex]
&
\end{aligned} 
\right.
\end{equation}
where $F(x,y)$ can be in the most general case a linear combinations of 4 independent functions $F_4$, hypergeometric of two variables $x$ and $y$. The univariate limits of the solutions are relevant for studying of the behavior of the corresponding correlation functions in special kinematics. See \cite{Coriano:2019nkw}, for instance, for four-point functions.

%%%%%%%%%%%%%%%%%%%%%%%%%%%%%%%%%%%%%%%%%%%%%%%%%%%%%%%%%%%%%%%%%%%%%%%%%%%%%%%%%%%%%%%%%%%%%
%%

\subsection{Lauricella's as 4-K integrals}
	%%%%%%%%%%%%%%%%%%%%%%%%%%%%%%%%%
	%%%%%%%%%%%%%%%%%%%%%%%%%%%%%%%%%
Parametric representations of the solutions of the CWI's in terms of 3K integrals \eqref{trekappa} 
in \cite{Bzowski:2013sza} carries some advantages, since it is possible to implement the symmetry constraints of a three-point  function more simply. Alternatively, one has to proceed from the four fundamental solutions introduced in \eqref{sol} (see the discussion in \cite{Coriano:2013jba}). 4K integral, in the case of four-point functions, appear to be the natural generalization of such previous approach developed for the 3K. This also allows to discuss the kinematical limits in which  a four-point function reduces to a three-point  one in some special kinematical limits. \\  
Hypergeometric functions of 3-variables, which belong to the class of Lauricella functions, with the addition of a third variable, can be related to 4K integrals. \\
 The key identity necessary to obtain the relation between the Lauricella functions and the 4K integral is given by	
 \begin{align}
\int_0^\infty dx\,x^{\a-1}\prod_{j=1}^3\,J_{\m_j}(a_j\,x)\,K_{\nu}(c\,x)&=2^{\a-2}\,c^{-\a-\l}\,\Gamma\left(\frac{\a+\l-\n}{2}\right)\Gamma\left(\frac{\a+\l+\n}{2}\right)\notag\\
&\hspace{-3cm}\times\prod_{j=1}^3\,\frac{a_j^{\m_j}}{\Gamma(\m_j+1)}F_C\left(\frac{\a+\l-\n}{2},\frac{\a+\l+\n}{2},\m_1+1,\m_2+1,\m_3+1;-\frac{a_1^2}{c^2},-\frac{a_2^2}{c^2},-\frac{a_3^2}{c^2}\right)\notag\\
&\centering\,\bigg[\l=\sum_{j=1}^3\,\m_j\,;\, \Re(\a+\l)>|\Re(\n)|,\,\Re(c)>\sum_{j=1}^{3}|\Im\,a_j|\bigg]\label{Prudnikov}.
\end{align}
If we rewrite the solutions of the systems of scalar four-point functions in the form
	\begin{align}
	I_{\a-1\{\n_1,\n_2,\n_3,\n_4\}}(a_1,a_2,a_3,a_4)&=\int_0^\infty\,dx\,x^{\a-1}\,\prod_{i=1}^4(a_i)^{\n_i}\,K_{\n_i}(a_i\,x),
	\label{4Kintegral}
	\end{align} 
	with the Bessel functions $I_\nu,J_\nu, K_\nu$ related by the identities
	\begin{align}
	I_\nu(x)&=i^{-\n}\,J_{\n}(i\,x),\\
	K_\nu(x)&=\frac{\pi}{2\sin(\pi\,\n)}\bigg[I_{-\n}(x)-I_\n(x)\bigg]=\frac{1}{2}\bigg[i^\nu\, \Gamma(\n)\Gamma(1-\n)\,J_{-\n}(i\,x)+i^{-\n}\,\Gamma(-\n)\Gamma(1+\n)\,J_\n(i\,x)\bigg],\label{Kscomp}
	\end{align}
	having used the properties of the Gamma functions
	\begin{equation}
	\frac{\pi}{\sin(\pi\n)}=\Gamma(\n)\,\Gamma(1-\n),\qquad-\frac{\pi}{\sin(\pi\n)}=\Gamma(-\n)\,\Gamma(1+\n),	\end{equation}
	the dilatation Ward identities  can be written as
	\begin{equation}
	\bigg[(\D_t-3d)-\sum_{i=1}^4p_i\frac{\partial}{\partial p_i}\bigg]I_{\a\{\b_1,\b_2,\b_3,\b_4\}}(p_1,p_2,p_3,p_4)=0.
	\end{equation}
Using some properties of 4K integrals \cite{Maglio:2019grh} it is possible to derive the relation
	\begin{align}
	(\a-\b_t+1+\D_t-3d)I_{\a\{\b_1,\b_2,\b_3,\b_4\}}(p_1,p_2,p_3,p_4)=0,
	\end{align}
	where $\Delta_t=\sum_{i=1}^4 \Delta_i$. This is identically satisfied if the $\a$ exponent is equal to $\tilde{\a}$
	\begin{equation}
	\tilde{\a}=\b_t+3d-\D_t-1.
	\end{equation}
	The conformal Ward identities are then expressed in the form
	\begin{equation}
	\left\{
	\begin{aligned}
	\textup{K}_{14}I_{\tilde{\a}\{\b_1,\b_2,\b_3,\b_4\}}&=0,\\
	\textup{K}_{24}I_{\tilde{\a}\{\b_1,\b_2,\b_3,\b_4\}}&=0,\\
	\textup{K}_{34}I_{\tilde{\a}\{\b_1,\b_2,\b_3,\b_4\}}&=0,
	\end{aligned}
	\right.
	\end{equation}
	obtaining the relations
	\begin{equation}
	\left\{
	\begin{aligned}
	(d-2\D_4+2\b_4)I_{\tilde{\a}+1\{\b_1,\b_2,\b_3,\b_4-1\}}-(d-2\D_1+2\b_1)I_{\tilde{\a}+1\{\b_1-1,\b_2,\b_3,\b_4\}}&=0,\\
	(d-2\D_4+2\b_4)I_{\tilde{\a}+1\{\b_1,\b_2,\b_3,\b_4-1\}}-(d-2\D_2+2\b_2)I_{\tilde{\a}+1\{\b_1,\b_2-1,\b_3,\b_4\}}&=0,\\
	(d-2\D_4+2\b_4)I_{\tilde{\a}+1\{\b_1,\b_2,\b_3,\b_4-1\}}-(d-2\D_3+2\b_3)I_{\tilde{\a}+1\{\b_1,\b_2,\b_3-1,\b_4\}}&=0.\\
	\end{aligned}
	\right.
	\end{equation}
	These are satisfied if
	\begin{align}
	\b_i=\D_i-\frac{d}{2},\qquad i=1,\dots,4,
	\end{align}
	giving
	\begin{equation}
	\tilde{\a}=d-1.
	\end{equation}
The final solution takes the form
	\begin{align}
	\phi(p_1,p_2,p_3,p_4)&=\bar{\bar{\a}}\, I_{d-1\left\{\D_1-\frac{d}{2},\D_2-\frac{d}{2},\D_3-\frac{d}{2},\D_4-\frac{d}{2}\right\}}(p_1,p_2,p_3,p_4)=\int_0^\infty\,dx\,x^{d-1}\,\prod_{i=1}^4(p_i)^{\D_i-\frac{d}{2}}\,K_{\D_i-\frac{d}{2}}(p_i\,x)
	,\label{4Kfin}
	\end{align}
	with $\bar{\bar{\a}}$ an undetermined constant. \\
	A useful relation is 
	\begin{equation}
	I_{\a\{\b_1,\b_2,\b_3,\b_4\}}(p_1,p_2,p_3,p_4)=\int_0^\infty\,dx\,x^\a\,\prod_{i=1}^4(p_i)^{\b_i}\,K_{\b_i}(p_i\,x),
\end{equation}
from which one derives the identity
\begin{equation}
p_i\frac{\partial}{\partial p_i}I_{\a\{\b_j\}}=-p_i^2\,I_{\a+1\{\b_j-\d_{ij}\}},\qquad i,j=1,\dots,4.
\end{equation}
Other relations satisfied by these types of integrals can be found in \cite{Maglio:2019grh}. For example, using
\begin{equation}
\int_0^\infty\,x^{\a+1}\frac{\partial}{\partial x}\left[\prod_{i=1}^4\,p_i^{\b_i}\,K_{\b_i}(p_i\,x)\right]=-\int_0^\infty\,\left[\frac{\partial x^{\a+1}}{\partial x}\right]\prod_{i=1}^4\,p_i^{\b_i}\,K_{\b_i}(p_i\,x),
\end{equation}
one obtains the identity
\begin{equation}
\sum_{i=1}^{4}p_i^2I_{\a+1\{\b_j-\d_{ij}\}}=(\a-\b_t+1)\,I_{\a\{\b_j\}},\qquad j=1,\dots,4,
\end{equation}
where $\b_t=\b_1+\b_2+\b_3+\b_4$. 

\section{Properties of triple-K integrals \label{AppendixJ}}

The modified Bessel function of the second kind is defined by
\begin{equation}
K_\n(x)=\sdfrac{\pi}{2\sin(\n x)}[I_{-\n}(x)-I_{\n}(x)],\ \ \n\in\mathbb Z.
\end{equation}
If $\n=\frac{1}{2}+n$, for $n$ integer, the Bessel function reduced to elementary functions
\begin{equation}
K_\n(x)=\sqrt{\sdfrac{\pi}{2}}\,\sdfrac{e^{-x}}{\sqrt{x}}\,\sum_{j=0}^{\lfloor\,|\n|-1/2\rfloor}\ \frac{(|\n|-1/2+j)!}{j!(|\n|-1/2-j)!}\sdfrac{1}{(2x)^j},\ \ \n+1/2\in\mathbb{Z},
\end{equation}
where we have used the floor function. In particular
\begin{eqnarray}
&\hspace{-2cm}K_{\frac{1}{2}}(x)=\sqrt{\sdfrac{\pi}{2}}\sdfrac{e^{-x}}{\sqrt{x}},\quad &K_{\frac{3}{2}}(x)=\sqrt{\sdfrac{\pi}{2}}\sdfrac{e^{-x}}{\sqrt{x^3}}(1+x),\notag\\
&K_{\frac{5}{2}}(x)=\sqrt{\sdfrac{\pi}{2}}\sdfrac{e^{-x}}{\sqrt{x^5}},(x^2+3x+3),\quad
&K_{\frac{7}{2}}(x)=\sqrt{\sdfrac{\pi}{2}}\sdfrac{e^{-x}}{\sqrt{x^5}}(x^3+6x^3+15x+5),
\end{eqnarray}
Using this expressions the triple-K integrals can be calculate in a very simple way. We can derive a useful expression in order regularise the triple-K integral in the region of non-convergence. Considering $\b_i$ half-integers the triple-K integral read
\begin{align}
I_{\a\{\b_1\,\b_2,\b_3\}}&=\int_0^\infty\,dx\,x^\a\,p_1^{\b_1}\,p_2^{\b_2}\,p_3^{\b_3}\,K_{\b_1}(p_1x)\,K_{\b_2}(p_2x)\,K_{\b_3}(p_3x)\notag\\[1.5ex]
&\hspace{-1cm}=\sum_{k_1=0}^{|\b_1|-\frac{1}{2}}\ \,\sum_{k_2=0}^{|\b_2|-\frac{1}{2}}\ \,\sum_{k_3=0}^{|\b_3|-\frac{1}{2}}\ p_1^{\b_1-\frac{1}{2}-k_1}\,p_2^{\b_2-\frac{1}{2}-k_2}\,p_3^{\b_3-\frac{1}{2}-k_3}\,C_{k_1}(\b_1)\,C_{k_2}(\b_2)\,C_{k_3}(\b_3)\,\int_0^\infty\,dx\,x^{\a-k_t-\frac{3}{2}}\,e^{-p_t\,x}\notag\\
&\hspace{-1cm}=\sum_{k_1=0}^{|\b_1|-\frac{1}{2}}\ \,\sum_{k_2=0}^{|\b_2|-\frac{1}{2}}\ \,\sum_{k_3=0}^{|\b_3|-\frac{1}{2}}\ p_1^{\b_1-\frac{1}{2}-k_1}\,p_2^{\b_2-\frac{1}{2}-k_2}\,p_3^{\b_3-\frac{1}{2}-k_3}\,p_t^{k_t-\a-\frac{1}{2}}\,C_{k_1}(\b_1)\,C_{k_2}(\b_2)\,C_{k_3}(\b_3)\,\G\left(\a-k_t-\sdfrac{1}{2}\right)\label{halfinteg},
\end{align}
where $k_t=k_1+k_2+k_3$ and $p_t=p_1+p_2+p_3$ and we have define $C_{k_i}(\b_i)$ as
\begin{equation}
C_{k_i}(\b_i)\equiv \sqrt{\frac{\pi}{2^{2k_i+1}}}\,\frac{\left(|\b_i|-1/2+k_i\right)\,!}{k_i\,!\,\left(|\b_i|-1/2-k_i\right)\,!},
\end{equation}
and we have used the definition of the gamma function in order to write the integral
\begin{equation}
\int_0^\infty\,dx\,x^{\a-k_t-\frac{3}{2}}\,e^{-p_t\,x}=p_t^{k_t-\a+\frac{1}{2}}\int_0^\infty\,dy\,y^{\a-k_t-\frac{3}{2}}\,e^{-\,y}=p_t^{k_t-\a+\frac{1}{2}}\,\G\left(\a-k_t-\frac{1}{2}\right),
\end{equation}
Using \eqref{halfinteg} we can calculate for instance the integrals
\begin{align}
I_{\frac{9}{2}\left\{\frac{3}{2},\frac{3}{2},-\frac{1}{2}\right\}}&=\left(\sdfrac{\pi}{2}\right)^{3/2}\sdfrac{3(p_1^2+p_2^2)+p_3^2+12p_1\,p_2+4p_3(p_1+p_2)}{p_3(p_1+p_2+p_3)^4},\\[1.2ex]
I_{\frac{7}{2}\left\{\frac{3}{2},\frac{3}{2},\frac{1}{2}\right\}}&=\left(\sdfrac{\pi}{2}\right)^{3/2}\sdfrac{2(p_1^2+p_2^2)+p_3^2+6 p_1\,p_2+3p_3(p_1+p_2)}{p_3(p_1+p_2+p_3)^3},
\end{align}
and any integrals with half-integer $\b_j$, $j=1,2,3$.

We now analyse the basic properties of the triple-K integrals. We have
\begin{equation}
	\begin{split}
	\sdfrac{\partial}{\partial p_n}I_{\a\{\b_j\}}&=-p_n\,I_{\a+1\{\b_j-\d_{jn}\}},\\
	I_{\a\{\b_j+\d_{jn}\}}&=p_n^2\,I_{\a\{\b_j-\d_{jn}\}}+2\b_n\,I_{\a-1\{\b_j\}},\\
	I_{\a\{\b_1\b_2,-\b_3\}}&=p_3^{-2\b_3}I_{\a\{\b_1\b_2,\b_3\}},
	\end{split}\label{identityBess}
\end{equation}
for any $n=1,2,3$, as follows from the basic Bessel function relations
	\begin{align}
	\sdfrac{\partial}{\partial a}[a^\n K_\n(ax)]&=-x\,a^\n K_{\n-1}(ax),\\
	K_{\n-1}(x)+\sdfrac{2\n}{x}K_\n(x)&=K_{\n+1}(x),\\
	K_{\n-1}(p\,x)+\sdfrac{2\n}{p\,x}K_\n(p\,x)&=K_{\n+1}(p\,x),\\
	K_{-\n}(x)&=K_{\n}(x),\\
	\sdfrac{\partial}{\partial x}[a^\n K_\n(ax)]&=-\sdfrac{1}{2}a^{\n+1} (K_{\n-1}(ax)+K_{\n+1}(ax))\notag\\[0.8ex]
	&=-a^{\n+1}K_{\n+1}(ax)+\sdfrac{\n\,a^\n}{x}\,K_{\n}(ax).
	\end{align}

The properties of the reduced triple-K integral is directly obtained from \eqref{identityBess}, in fact noticing that its definition given in \appref{solution} 
\begin{equation}
J_{N\{k_j\}}=I_{\frac{d}{2}-1+N\{\D_j-\frac{d}{2}+k_j\}},
\end{equation}
is just a redefinition of the indices, we can obtain similar equations to \eqref{identityBess}. It is not difficult to show that
\begin{align}
\sdfrac{\partial}{\partial p_n}\,J_{N\{k_j\}}&=-p_n\,J_{N+1\{k_j-\d_{jn}\}},\label{1dJ}\\
\,J_{N\{k_j+\d_{jn}\}}&=p^2_n\ J_{N\{k_j-\d_{jn}\}}+2\left(\D_n-\sdfrac{d}{2}+k_n\right)\ J_{N-1\{k_j\}},\\
\,J_{N+2\{k_j\}}&=p^2_n\ J_{N+2\{k_j-2\d_{jn}\}}+2\left(\D_n-\sdfrac{d}{2}+k_n-1\right)\ J_{N+1\{k_j-\d_{jn}\}},\\
\sdfrac{\partial^2}{\partial p_n^2}\,J_{N\{k_j\}}&=J_{N+2\{k_j\}}-2\left(\D_n-\sdfrac{d}{2}+k_n-\sdfrac{1}{2}\right)\ J_{N+1\{k_j-\d_{jn}\}},\\
K_n\,J_{N\{k_j\}}&\equiv\left(\sdfrac{\partial^2}{\partial p_n^2}+\sdfrac{(d+1-2\D_n)}{p_n}\sdfrac{\partial}{\partial p_n}\right)J_{N\{k_j\}}=J_{N+2\{k_j\}}-2\,k_n\,J_{N+1\{k_j-\d_{jn}\}}\label{Fund}.
\end{align}
%%%%%%%%%%%%%%%%%%%%%%%%%%%%%%%%%%%%
\subsection{Vertices}\label{Appendix1}
%%%%%%%%%%%%%%%%%%%%%%%%%%%%%%%%%%%%
We have shown in \figref{vertices} a list of all the vertices which are needed for the  perturbative expansion of the $TTT$ correlator. Here we list their explicit expressions. We use the letter $V$ to indicate the vertex, the subscript denotes the fields involved and the Greek indices are linked to the Lorentz structure of the space-time. Referring to \figref{vertices}, we take all the graviton momenta incoming as well as the momentum indicated as $k_1$, while $k_2$ is outgoing. In order to simplify the notation, we introduce the tensor components
\begin{align}
A^{\m_1\n_1\m\n}&\equiv\d^{\mu_1\nu_1}\d^{\m\n}-2\d^{\m(\m_1}\d^{\n_1)\n},\\
B^{\m_1\n_1\m\n}&\equiv\d^{\mu_1\nu_1}\d^{\m\n}-\d^{\m(\m_1}\d^{\n_1)\n},\\
C^{\m_1\n_1\m_2\n_2\m\n}&\equiv\d^{\m(\m_1}\d^{\n_1)(\m_2}\d^{\n_2)\n}+\d^{\m(\m_2}\d^{\n_2)(\m_1}\d^{\n_1)\n},\\
\tilde{C}^{\m_1\n_1\m_2\n_2\m\n}&\equiv\d^{\m(\m_1}\d^{\n_1)(\m_2}\d^{\n_2)\n},\\
D^{\m_1\n_1\m_2\n_2\m\n}&\equiv\d^{\m_1\n_1}\d^{\m(\m_2}\d^{\n_2)\n}+\d^{\m_2\n_2}\d^{\m(\m_1}\d^{\n_1)\n},\\
E^{\m_1\n_1\m_2\n_2\m\n}&\equiv\d^{\m_1\n_1}B^{\m_2\n_2\m\n}+C^{\m_1\n_1\m_2\n_2\m\n},\\
F^{\a_1\a_2\m\n}&\equiv\d^{\a_1[\m}\d^{\n]\a_2},\\
\tilde{F}^{\a_1\a_2\m\n}&\equiv\d^{\a_1(\n}\d^{\m)\a_2},\\
\tilde{F}^{\a_1\a_2}_{\m\n}&\equiv\d^{(\a_1}_{\n}\d_{\m}^{\a_2)},\\
G^{\m_1\n_1\a_1\a_2\m\n}&\equiv\d^{\m[\n}\d^{\a_2](\m_1}\d^{\n_1)\a_1}+\d^{\a_1[\a_2}\d^{\n](\m_1}\d^{\n_1)\m},\\
H^{\m_1\n_1\m_2\n_1\a_1\a_2\m\n}&\equiv A^{\m_1\n_1\m\a_1}\tilde{F}^{\m_2\n_2\n\a_2}-A^{\m_2\n_2\m\a_1}\tilde{F}^{\m_1\n_1\n\a_2},\\
I^{\m_1\n_1\m_2\n_2\a_1\a_1\m\n}&\equiv\d^{\m_1\n_1}D^{\m\a_1\n\a_2\m_2\n_2}-\sdfrac{1}{2}\d^{\a_1\m}\d^{\a_2\n}A^{\m_1\n_1\m_2\n_2},
\end{align}
where the round brackets denote symmetrization and the square brackets anti-symmetrization of the corresponding indices
\begin{align}
\d^{\m(\m_1}\d^{\n_1)\n}&\equiv\sdfrac{1}{2}\bigg(\d^{\m\m_1}\d^{\n_1\n}+\d^{\m\n_1}\d^{\m_1\n}\bigg),\\
\d^{\m[\m_1}\d^{\n_1]\n}&\equiv\sdfrac{1}{2}\bigg(\d^{\m\m_1}\d^{\n_1\n}-\d^{\m\n_1}\d^{\m_1\n}\bigg).
\end{align}
In the scalar sectors we obtain
\begin{align}
V^{\mu_1\nu_1}_{T\phi\phi}(k_1,k_2)&=\sdfrac{1}{2}A^{\m_1\n_1\m\n}\,k_{1\nu}\,k_{2\m}+\c\,B^{\m_1\n_1\m\n}\,(k_1-k_2)_\mu\,(k_1-k_2)_\nu,\\[2ex]
V^{\mu_1\nu_1\mu_2\nu_2}_{TT\phi\phi}(p_2,k_1,k_2)&=\left(\sdfrac{1}{4}A^{\m_1\n_1\m_2\n_2}\d^{\m\n}+C^{\m_1\n_1\m_2\m_2\m\n}-\sdfrac{1}{2}D^{\m_1\n_1\m_2\n_2\m\n}\right)\,k_{1\nu}\,k_{2\m}\notag\\
&\hspace{-2cm}+\sdfrac{\c}{2}\bigg[\sdfrac{1}{2}\,\Big(E^{\m_1\n_1\m_2\n_2\m\n}-D^{\m_2\n_2\m\n\m_1\n_1}\Big)\,p_{2\m}p_{2\n}+\sdfrac{1}{2}\,\Big(E^{\m_1\n_1\m_2\n_2\m\n}-D^{\m_1\n_1\m\n\m_2\n_2}\Big)\,p_{2\m}(k_2-k_1)_\n\notag\\
&\hspace{1cm}+\Big(C^{\m_1\n_1\m_2\n_2\m\n}-D^{\m_1\n_1\m\n\m_2\n_2}\Big)(k_2-k_1)_\m(k_2-k_1)_\n\bigg].
\end{align}
In the fermion sector the relevant vertices are
\begin{align}
V^{\mu_1\nu_1}_{T\bar\psi\psi}(k_1,k_2)&=\sdfrac{1}{4}\,B^{\m_1\n_1\m\n}\,\gamma_\n\,(k_1+k_2)_\m,\\[2ex]
V^{\mu_1\nu_1\m_2\n_2}_{TT\bar\psi\psi}(p_2,k_1,k_2)&=\sdfrac{1}{8}\bigg[\d^{\m\n}A^{\m_1\n_1\m_2\n_2}-D^{\m_1\n_1\m_2\n_2\m\n}+C^{\m_1\n_1\m_2\n_2\m\n}+\tilde{C}^{\m_2\n_2\m_1\n_1\m\n}\bigg]\,\gamma_\n\,(k_1+k_2)_\m\notag\\
&\hspace{1cm}+\sdfrac{1}{32}\tilde{C}^{\m_1\n_1\m_2\n_2\n\m}\,p_2^\sigma\,\left\{\gamma_\n,\left[\gamma_\m,\gamma_\sigma\right]\,\right\},
\end{align}
where we the last term in the expression above is related to the spin connection. However, one can prove that this term does not contribute to the pinched two-graviton diagrams. 

In the gauge sector we separate the gauge fixing contributions (GF) from the others, denoted with a subscript $M$, obtaining 
\begin{align}
V^{\m_1\n_1\a_1\a_2}_{TAA,\,M}(k_1,k_2)&=\bigg(\d^{\m_1\n_1}F^{\a_1\m\n\a_2}+2\,G^{\m_1\n_1\a_1\a_2\m\n}\bigg)\,k_{1\m}\,k_{2\n},\\[2ex]
V^{\m_1\n_1\m_2\n_2\a_1\a_2}_{TTAA,\,M}(k_1,k_2)&=\bigg[-\sdfrac{1}{2}A^{\m_1\n_1\m_2\n_2}F^{\m\a_1\n\a_2}+\d^{\m_2\n_2}\,G^{\m_1\n_1\a_1\a_2\m\n}+\d^{\m_1\n_1}\,G^{\m_2\n_2\a_1\a_2\m\n}\notag\\
&\hspace{-2cm}-\big(\d^{\a_1\a_2}C^{\m_1\n_1\m_2\n_2\m\n}+\d^{\m\n}C^{\m_1\n_1\m_2\n_2\a_1\a_2}-\d^{\a_1\n}C^{\m_1\n_1\m_2\n_2\a_2\m}-\d^{\a_2\m}C^{\m_1\n_1\m_2\n_2\a_1\n}\big)\notag\\
&\hspace{-2cm}-\big(\tilde{F}^{\m\n\m_1\n_1}\tilde{F}^{\m_2\n_2\a_1\a_2}+\tilde{F}^{\m\n\m_2\n_2}\tilde{F}^{\m_1\n_1\a_1\a_2}-\tilde{F}^{\m\a_2\m_1\n_1}\tilde{F}^{\m_2\n_2\a_1\n}-\tilde{F}^{\m\a_2\m_2\n_2}\tilde{F}^{\m_1\n_1\n\a_1}\big)\,\bigg]\,k_{1\m}\,k_{2\n},
\end{align}
\begin{align}
V^{\m_1\n_1\a_1\a_2}_{TAA,\,GF}(k_1,k_2)&=-\sdfrac{1}{2\xi}\bigg[-\d^{\m_1\n_1}\d^{\a_1\m}\d^{\a_2\n}\,k_{1\m}\,k_{2\n}+\big(\d^{\m_1\n_1}\tilde{F}^{\a_1\a_2\m\n}-2\tilde{C}^{\m_1\n_1\m\n\a_1\a_2}\big)\,k_{2\m}\,k_{2\n}\notag\\
&\hspace{2cm}+\big(\d^{\m_1\n_1}\tilde{F}^{\a_1\a_2\m\n}-2\tilde{C}^{\m_1\n_1\m\n\a_2\a_1}\big)\,k_{1\m}\,k_{1\n}\,\bigg],
\end{align}
\begin{align}
V^{\m_1\n_1\m_2\n_2\a_1\a_2}_{TTAA,\,GF}(p_2,k_1,k_2)&=-\sdfrac{1}{2\xi}\bigg[I^{\m_1\n_1\m_2\n_2\a_1\a_2\m\n}\,k_{1\m}\,k_{2\n}+H^{\m_2\n_2\m_1\n_1\a_2\a_1\m\n}\,p_{2\m}k_{1\n}\notag\\
&\hspace{-3cm}+H^{\m_1\n_1\m_2\n_2\a_1\a_2\m\n}\,p_{2\m}k_{2\n}-\big(I^{\m_1\n_1\m_2\n_2\a_1\a_2\m\n}+A^{\m_2\n_2\a_2\n}\tilde{F}^{\m_1\n_1\m\a_1}-2\d^{\a_2\n}C^{\m_1\n_1\m_2\n_2\m\a_1}\big)\,k_{2\m}\,k_{2\n}\notag\\
&\hspace{-3cm}-\big(I^{\m_1\n_1\m_2\n_2\a_2\a_1\m\n}+A^{\m_2\n_2\a_1\n}\tilde{F}^{\m_1\n_1\m\a_2}-2\d^{\a_1\n}C^{\m_1\n_1\m_2\n_2\m\a_2}\big)\,k_{1\m}\,k_{1\n}+\big(4\,\tilde{F}^{\m_1\n_1\n(\a_1}\tilde{F}^{\a_2)\m\m_2\n_2}\notag\\
&\hspace{-2cm}-2\d^{\m_1\n_1}\tilde{C}^{\a_1\a_2\m_2\n_2\n\m}-2\d^{\m_2\n_2}\tilde{C}^{\a_1\a_2\m_1\n_1\m\n}+\d^{\m_1\n_1}\d^{\m_2\n_2}\tilde{F}^{\m\n\a_1\a_2}\big)\,p_{2\m}\,(p_2-k_2+k_1)_\n
\bigg].
\end{align}
In the ghost sector we obtain
\begin{align}
V^{\m_1\n_1}_{T\bar cc}(k_1,k_2)&=\sdfrac{1}{2}\,A^{\m_1\n_1\m\n}\,k_{1\m}\,k_{2\n},\\
V^{\m_1\n_1\m_2\n_2}_{TT\bar cc}(k_1,k_2)&=\bigg(\sdfrac{1}{4}\d^{\m_2\n_2}\,A^{\m_1\n_1\m\n}-\sdfrac{1}{2}\,D^{\m_1\n_1\m\n\m_2\n_2}+C^{\m_1\n_1\m_2\n_2\m\n}\bigg)\,k_{1\m}\,k_{2\n}.
\end{align}

In these expressions one should include a complex factor of $i$ coming from the generating functional. In the Fourier transforms, for instance, we conventionally introduce in these expressions a factor  $\exp[-i(px-qy)]$ if $p$ is an incoming and $q$ an outgoing factor. 
%%%%%%%%%%%%%%%%%%%%
%%%%%%%%%%%%%%%%%%%%

\end{document}